# CAN WE TEACH GENERATIVE ARTIFICIAL INTELLIGENCE THE DESIGN LANGUAGE OF ENGINEERED LIVING MATERIALS?

Andrés Díaz Lantada [1,2*], José A. Yáñez [1], William Solórzano-Requejo [1], Monsur Islam [1]

[1] Mechanical Engineering Department, Universidad Politécnica de Madrid, c/ José Gutiérrez Abascal 2, 28006 Madrid, Spain.

[2] IMDEA Materials Institute, c/ Eric Kandel 2, 28906 Getafe, Spain.

*Contact: andres.diaz@upm.es / andres.diaz@imdeamaterials.org

**Abstract**

This study presents a versatile ontology and a useful codification scheme for describing all kinds of engineered living materials (ELMs). The different components of the ontology, namely: families according to the taxonomy for ELMs, industrial applications and synthesis or processing methods, are systematically organized, enumerated, classified, codified and explained. The methodic application of the ontology to a set of 100 relevant examples of ELMs helps to demonstrate its utility and adaptability to many different types of ELMs with a wide range of industrial applications and obtained through numerous synthesis and processing methods. This proves that the developed ontology and codification schemes, with the glossary provided to support its implementation and application, can serve as a comprehensive classification tool for the emergent field of ELMs. Furthermore, the usability of the ELMs ontology and codification by a generative artificial intelligence (AI) is explored and validated by different means, checking that both natural language and the codification are understandable for describing ELMs, verifying that the generative AI adequately codifies examples of ELMs according to the ontology, and validating the synergic applicability of the ontology and codification with generative AI tools for illustrating novel ELMs and supporting their conceptual design. This study is expected to provide a universal language to facilitate communication in the ELMs field and to foster the discovery of new ELMs and related innovations, hoping it may accelerate scientific and technological discoveries.

## 1. Introduction

Engineered living materials have emerged as a new family of materials characterized by their biohybrid nature. Foundational references describe them as materials, in which living cells, including eukaryotic, prokaryotic and synthetic cell types, and biomaterials as scaffolding structures coexist. Usually ELMs are divided into biological, made of cells and their self-generated extracellular matrices (ECMs), and biohybrid, based on cells colonizing artificial EMCs.[1–6] ELMs have emerged thanks to decades of research in converging fields including: tissue engineering and biofabrication [7,8], micro-physiological systems and organs-on-chips [9,10], multi-cellular engineered living systems (MCELS),[11,12] biohybrid robotics,[13,14] and even "*Baubotanik*" or living plant construction,[15,16] to cite a few. These ELMs constitute a contemporary revolution in materials science and engineering, and pioneer studies have pointed out that we are currently living their defining moment.[17] Indeed, relevant contributions have focused on organizing the research field of ELMs,[18] providing a taxonomy for ELMs,[19] and reviewing the more recent advances in an organized manner oriented to their engineering design and promote their societal impacts.[20,21]

Being a nascent field, specific engineering design approaches are required for overcoming challenges linked to their reliability, strength, autonomy, and sustainable development. Relevant advances have been carried out recently in connection with achieving higher performances thanks to ceramic and carbon chassis,[22,23] expanding their autonomy and lifespan by combining cell types,[24,25] or mapping their ethical and societal challenges for enabling ethics-guided ELMs research.[26] Still, engineering design methodologies purposefully developed for ELMs are required, as well as tools for systematizing ELMs research and empowering innovators to create new ELMs and derived technologies navigating the turbulent waters of such a multidisciplinary and rapidly evolving field.

In parallel, advances in generative artificial intelligence (AI), especially since the advent of large language models and supporting algorithms for generative knowledge extraction, graph-based representation, and multimodal intelligent graph reasoning, are importantly accelerating scientific discovery with intriguing implications for designing materials and structures, as an inspiring reference has pointed out.[27] Indeed, generative artificial intelligence is redefining the innovation cycle and reformulating traditional approaches to materials engineering in connection with product design. To mention some examples, the application of innovative biomaterials for designing new structures, buildings, vehicles, medical devices and everyday products has been extensively explored and systematized employing "text-to-image" resources.[28–30] In the current state, the interactions with generative AIs lead to conceptual designs and promote the screening of ideas, in connection with complex-shaped geometries usually requiring additive manufacturing technologies for their materialization, from the beginning of the innovation cycle. In addition, the interaction with these AIs may prove useful along the whole design cycle, helping to refine the concepts,[31] to approach manufacturing, for instance by means of 3D and 4D printing,[32,33] to plan their functional integration, and even to validate nature-inspired designs *in silico*.[34] In this context, the combination of ontologies, codification schemes and generative AIs has been also explored by our team for designing stimuli-responsive or smart (but not living) materials and structures and derived actuators.[35]

In our hypothesis, generative AI is bound to have a highly transformative role for ELMs research, paving the way towards straightforward, effective and safe solutions in an enormous variety of industrial fields, once specific supporting resources and design methods are developed. For developing these resources, a new design language for ELMs may be required, which may facilitate interactions with generative AIs in connection with ELMs. With this objective in mind, our study presents the development of a codified ontology for ELMs, to the authors' best knowledge the first available, and demonstrates its synergic employment with generative AI tools.

To this end, expanding the taxonomy for ELMs,[19] a complete ontology with an accompanying codification scheme is developed in this study for describing all kinds of ELMs. The different components of the ontology: families according to the mentioned taxonomy, industrial applications and synthesis or processing methods, are systematically organized, classified, coded and explained. Then, the ontology and codification schemes are methodically applied to a set of 100 relevant examples of ELMs for demonstrating their utility and adaptability. This proves that the developed ontology and codification schemes, with the glossary provided to support its implementation and application, serve as a reliable classification tool for the emergent field of ELMs. Finally, the usability of the ELMs ontology and codification by a generative artificial intelligence (AI) is explored and validated by different means: checking that both natural language and the codification are understandable for describing ELMs, verifying that the generative AI adequately codifies examples of ELMs according to the ontology, and validating the synergic applicability of the ontology and codification with generative AI tools for illustrating novel ELMs and supporting their conceptual design.

The materials and methods employed are presented in the following section, before describing and discussing the main results obtained and analyzing the detected current limitations and proposed research directions, hoping to contribute to this nascent field.

## 2. Ontology and knowledge graph representation of engineered living materials

### 2.1. Search criteria for the references employed for evaluating the ELMs ontology

Employing Google Scholar, searching for "engineered living materials" within the temporal framework of years 2018 and 2025, a total of 150 relevant references on ELMs are found and chronologically ordered. For different reasons (redundance, lack of detail, non-living entity) 30 references are discarded. From the remaining 120, the first 100 are employed for evaluating the developed ELMs ontology and the applicability of the proposed codification scheme, and for training the generative AIs employed (see Sections 2.3 and 3). The 20 more recent ones are used for validation purposes, so that the trained generative AIs are presented with examples never seen before during the training, following the procedure explained in subsection 2.3 and discussed in more detail in section 3.

### 2.2. Ontology and codification scheme for ELMs

Starting from the Linnean-style taxonomy for ELMs developed by our team,[19] in order to reach a truly applicable ontology for supporting the engineering design of new living materials, additional fundamental components should be considered, including industrial application fields and synthesis processes or production methods. Furthermore, facilitating the interaction with generative AIs and reaching a higher level of abstraction for conceptual design requires the development of a coding scheme and codification, which can be considered a new language for engineering living materials through synergic interactions with generative AIs. In agreement with the basic taxonomy for ELMs, families are organized and hierarchically classified employing higher rank taxa (domains and kingdoms) that take inspiration from life's taxonomy, considering the cell types within ELMs as the most fundamental features of living materials, which determine their essence and key functions. The lower rank taxa (classes and below) depend on the material types used for the living materials' "skeletons" or scaffolds, providing structural support to the living cell colonies. In this way, the classical life's taxonomy is merged with common materials classifications.

Importantly, expanding from the available taxonomy for ELMs, Table 1 presents the hierarchical organization of ELMs families, including taxonomic ranks or taxa, categories, descriptions, and codifications for engineered living materials (ELMs).

In addition, for the first time the taxonomy is revised, expanded with detailed descriptions for the different categories and provided with a biunivocal coding (three-letter codes) for all the ranked categories. It can be advanced that the descriptions prove fundamental in the process of training the generative AI, towards an understanding of taxa and categories, and their essence and boundaries, as analyzed in section 3. At the same time, the codes contribute to achieving a sort of genomic representation for ELMs, which is expected to leverage the capabilities of AIs as creativity promotion tools in this field (for instance, employing already classical genetic algorithms and more recent large language models). Besides, Table 2 lists the more common production methods and industrial applications for ELMs, as well as their descriptions and codifications, as a complement to Table 1, towards a more comprehensive codified ontology for ELMs.

Once developed, the ontology and codification scheme are applied to describing, classifying and coding the gathered (see subsection 2.1) set of 100 pioneering examples of ELMs, achieving one of the most comprehensive overviews of the current panorama in the field of ELMs. Table 3 presents a subset of 10 examples following the rules of the ontology and codification and providing detailed textual descriptions for the different examples, which is necessary for the subsequent interactions with the AI. For a full overview of the 100 examples, please consult Table S1 in the Supplementary Material S1. References [See the references in supplementary information] detailing the engineering of the different examples of living materials are also provided as part of Supplementary Material S1, while the subset of 10 examples from Table 3 also includes the relevant references.[36–45]

For visual exploration and communication of the dataset structure, RAWGraphs 2.0[46] is employed to represent the 100 codified examples. Based on the comprehensive review, Figure 2A**¡Error! No se encuentra el origen de la referencia.** shows a dendritic dendrogram organizing the dataset of 100 examples in families and subfamilies according to the proposed ontology and codification scheme. Colors correspond to the five main domains of ELMs: archaeal (purple), bacterial (orange), eukaryotic (green), cross-domain (yellow) and synthetic (blue). To complement this, a linear dendrogram as well as graphic representations of relationships between species, application areas and production methods, are also provided in Figure S1.1 and Figure S1.2 in Supplementary Material S1.

To introduce the potential of generative AI for ELMs innovation, Figure 1 presents AI-generated illustrations for a variety ELMs, obtained by employing the textual descriptions of different examples from Table 3 as “prompts”. Microsoft Copilot was employed as accessible and easy to use generative artificial intelligence capable of generating images from textual instructions.

Table 1. Taxonomic ranks or taxa and their descriptions and codifications for engineered living materials (ELMs). Adapted, updated and expanded from "Taxonomy for engineered living materials", Cell Reports Physical Science, 3(4), 100807, 2022.

| ***Taxonomic ranks or taxa for engineered living materials (ELMs)*** | ***Categories*** | ***Description*** | ***Codification*** |
|---|---|---|---|
| **Domain** | Archaeal ELMs | Engineered living materials (ELMs) populated by archaea, that is prokaryotic cells from the biological domain archaea. | ARC |
| | Bacterial ELMs | Engineered living materials (ELMs) populated by bacteria, that is prokaryotic cells from the biological domain bacteria. | BAC |
| | Eukaryotic ELMs | Engineered living materials (ELMs) populated by eukaryotic cells, that is cells with nucleus from the biological domain eukarya. | EUK |
| | Synthetic ELMs | Engineered living materials (ELMs) populated by synthetic cells. | SDM |
| | Cross-domain ELMs | Engineered living materials (ELMs) populated by different cellular domains. | CRD |
| **Super kingdom** | Biological engineered living materials (bio-ELMs) | ELMs made only of cells and their extracellular matrices, without requiring or employing supporting synthetic biomaterials or synthetic structures. | BIO |
| | Hybrid living materials (HLMs) | ELMs combining cells and an abiotic synthetic biomaterial or synthetic structure, often referred to as scaffold. | HYB |
| **Kingdom** | Archaebacterial ELMs | ELMs of the archaeal domain populated by cells from the biological kingdom archaebacteria. | ARA |
| | Eubacterial ELMs | ELMS of the bacterial domain populated by cells from the biological kingdom eubacteria. | EUB |
| | Protist ELMs | ELMs of the eukaryotic domain populated by eukaryotic cells, which are not animal cells, plant cells or fungal cells. | PRO |
| | Fungal ELMs | ELMs of the eukaryotic domain populated by cells from the biological kingdom fungi, heterotrophic with chitin cell walls. | FUN |
| | Plant ELMs | ELMs of the eukaryotic domain populated by eukaryotic cells from the biological kingdom plantae, predominantly photosynthetic and with cellulose cell walls. | PLN |
| | Animal ELMs | ELMs of the eukaryotic domain populated by eukaryotic cells from the biological kingdom animalia. | ANI |
| | Synthetic / artificial cells ELMs | Engineered living materials (ELMs) populated by synthetic or artificial cells. | SKM |
| | Cross-kingdom ELMs | Engineered living materials (ELMs) combining cells from different kingdoms. | CRK |
| **Phylum** | 0D ELMs | ELMs with a quasi-punctual structure, usually cells coupled to micro- / nano-particles. | 0D |
| | 1D ELMs | ELMs with a linear or one-dimensional structure, usually based on cells populating or generating fibers as extracellular matrices. | 1D |
| | 2D ELMs | ELMs with a planar or bi-dimensional structure, usually biofilms or based on cells populating a planar biomaterial or generating a planar extracellular matrix. | 2D |
| | 3D ELMs | ELMs with a three-dimensional structure, usually cells within a three-dimensional biomaterial or three-dimensional scaffold. | 3D |
| | 4D ELMs | ELMs with a four-dimensional structure, mainly for cases in which the temporal dimension, involving metamorphoses, plays a relevant functional role. | 4D |
| | Fractal ELMs | ELMs with a irregular, fractured or self-similar structure that cannot be defined by integer dimensional values. | FR |
| | Cross-phylum ELMs | ELMs with structures combining different of the 0D, 1D, 2D, 3D, 4D and fractal phyla, besides, multi-scale ELMs whose structures are not fractal. | CP |
| **Class** | Living alloys | ELMs with extracellular matrices made predominantly of alloys. Examples of families within the class of living alloys include: Fe-based, Ti-based, Au-based, Cu-based, Ag-based, to cite a few. | LAL |
| | Living carbons | ELMs with extracellular matrices made predominantly of carbons. Examples of families within the class of living carbons include: allotropic forms of C, such as diamond, graphite, graphene, carbon nanotubes, glassy carbon, carbon dots and fullerenes to cite a few. | LCA |
| | Living ceramics | ELMs with extracellular matrices made predominantly of ceramics. Examples of families within the class of living ceramics include: oxides, silicates, carbides, nitrides, to cite a few. | LCE |
| | Living composites | ELMs with extracellular matrices made predominantly of composites. Examples of families within the class of living composites include metal-matrix composites, polymer-matrix composites, ceramic-matrix composites, to cite a few. | LCO |
| | Living polymers | ELMs with extracellular matrices made predominantly of polymers. Examples of families within the class of living polymers include: thermosets, thermoplastics, elastomers, hydrogels, liquid crystals, to cite a few. | LPO |
| | Cross-class ELMs | ELMs with extracellular matrices made of various materials classes, which cannot be described by the traditional term composite. | CRF |

Table 2. Production methods and industrial applications and their descriptions and codifications for engineered living materials (ELMs), as a complement to Table 1, for approaching a more comprehensive codified ontology for ELMs. Categories may be combined by using "+" in the codification of an ELM.

| *Additional attributes for ELMs* | *Categories* | *Description* | *Codification* |
|---|---|---|---|
| **Industrial applications for engineered living materials (ELMs)** | Biohybrid art and product design | ELMs employed for artistic and design purposes. | ART |
| | Biohybrid computers and electronics | ELMs employed as constituents of biohybrid computers and biohybrid components for electronics. | BCE |
| | Biohybrid machines and robots | ELMs employed as constituents of biohybrid machines and biohybrid robots and their subsystems. | BMR |
| | Biohybrid medical devices | ELMs employed for healthcare within advanced medical devices with capabilities such as self-healing, improved biointeraction, drug delivery, self-cleaning, among others. | BMD |
| | Bioremediation | ELMs employed for processing, recycling or revalorizing waste and debris from products, processes and systems. | BRE |
| | Biosensing | ELMs employed as transducers for detection purposes. | BSE |
| | Biotechnology and bioprocessing | ELMs employed within biotechnological processes. | BTP |
| | Catalysis and other chemical processes | ELMs employed as catalysts in chemical processes. | CAT |
| | Education and research | ELMs employed for innovative educational activities and related pedagogical research. | EDU |
| | Energy production and harvesting | ELMs employed for energy production and harvesting. | ENP |
| | Environmental monitoring | ELMs employed as advanced sensors for environmental monitoring, usually linked to ecosystems health. | ENV |
| | Fashion and biohybrid wearables | ELMs employed as innovative materials for fashion, usually linked to stimuli-responsive or autonomous wearables. | FAS |
| | Materials production and biofunctionalization | ELMs employed as factories of materials or as support to materials functionalization. | MAP |
| | Military and defense | ELMs employed as solutions for the military industry. | WAR |
| | Smart and living buildings | ELMs employed as subsystems of smart and living buildings, usually for improved sensing or actuation linked to improved functionalities or healthier environments. | SLB |
| | Smart and living materials and structures (multipurpose) | ELMs employed as transducers for enhanced actuation and sensing in a wide set of applications apart from those already covered by the categories above. | SLM |
| | Smart and living surfaces and biointerfaces | ELMs employed as smart, active, stimuli-responsive, living surfaces and biointerfaces for improved interaction with the surrounding environment or for an enhanced integration of engineering subsystems. | SLS |
| | Space exploration and colonization | ELMs employed for space exploration and colonization, as a complement to space technologies in the broadest sense. | SPE |
| | Synthetic biology | ELMs employed as resources for the synthetic biology toolkit aimed at the creation of new biological entities or the redesign of existing ones. | SYB |
| | Tissue engineering and regenerative medicine | ELMs employed as advanced healthcare solutions for reconstructing biological tissues and structures, deeply connected to tissue engineering and regenerative medicine solutions based on scaffolds with cells. | TER |
| | Transport and related infrastructures | ELMs employed as components or subsystems in transport applications (vehicles, ships, drones, airplanes) and related infrastructures (f.e. biocement for self-healing roads). | TRI |
| **Manufacturing and production methods for engineered living materials (ELMs)** | Engineering cells to produce ELMs | Artificially manipulated cells lead to special extracellular matrix production and to materials synthesizing materials. | ECP |
| | Extracellular matrix (ECM) grown in culture | Cultured cells grow an extracellular matrix in some cases aided by a biomaterial acting as substrate or support, the process applies both to hybrid living materials (HLMs) and to biological engineered living materials (bio-ELMs). | EMG |
| | Extracellular matrix (ECM) bioaccumulation | Cultured cells foster chemical reactions, typically redox, capable of reducing ions leading usually to metallic deposits or aggregates as part of their extracellular matrices, related to processes like biomineralization but with metals. | EMB |
| | Intracellular synthesis, intracellular transport or invagination | Materials production inside cells in culture, normally involving the intracellular transport or the invagination of a material to modify cellular states and functionalities to trigger the synthesis, trapping or deposit, usually aimed at polymeric production or the generation of metallic nanoparticles within the cells (bio-ELMs). | ICS |
| | Biofilm formation and related extracellular matrix (ECM) growth | Cultured cells form a biofilm, and an extracellular matrix is grown, usually for biological engineered living materials (bio-ELMs) grown in Petri dishes or bioreactors. | BFF |
| | Biofilm based trapping or anchoring of inorganic particles or materials and biofilms interfacing with inorganic materials | Cultured cells form a biofilm, and an extracellular matrix is grown, trapping or anchoring inorganic micro / nano-particles or a biofilm interfaces with inorganic materials, usually leading to living composites, mainly HLMs. | BTI |
| | Surface bio-patterning for selective cell culture / extracellular matrix (ECM) growth | Selective surface patterning creates special anchoring regions for cells capable of growing an extracellular matrix that generates a pattern of ELM upon the original substrate. | SBP |
| | Biomineralization | Process by which living cells generate minerals for creating stiff mineralized tissues, for magnetic sensing, for storing elements and for protection, a process that is present in all biological kingdoms of life and proves interesting both for HLMs and bio-ELMs. | BMN |
| | Molecular self-assembly | Process by which molecules spontaneously aggregate and form ordered structures without human intervention. | MSA |
| | 3D bioprinting of cells and biomaterials | Additive manufacturing processes involving the processing of biomaterials and living cells for creating living three-dimensional constructs, usually applied to HLMs. | 3DB |
| | 4D bioprinting of cells and biomaterials | Similar to 3D bioprinting but with the incorporation of a relevant temporal dimension linked to the fact that 4D bioprinted ELMs (usually HLMs) exhibit one or various metamorphoses during their life. | 4DB |
| | Scaffold (synthetic) structure 3D printing and cell colonization | Additive manufacturing processes aimed at the creation of three-dimensional structures or scaffolds (as in tissue engineering) adequately mimicking extracellular matrices for subsequent cell colonization leading to HLMs. | 3DP |
| | Scaffold (synthetic) structure 4D printing and cell colonization | Similar to 3D printed scaffolds for cell colonization but with the incorporation of a relevant temporal dimension linked to the fact that 4D printed scaffolds exhibit one or various metamorphoses during their life, hence leading to shape-morphing HLMs. | 4DP |
| | Solvent casting or melt molding, eventual particulate leaching and cell colonization or solvent casting with cells | Process involving the casting of a solvent with soluble particles within a mold, the subsequent leaching of the soluble particles to reach a porous biomaterial or synthetic scaffold and the final cell colonization to reach HLMs. | SCP |
| | Gas foaming of porous structures and cell colonization | As in the previous process but using gas for foaming and thus generating the porous biomaterial or synthetic scaffold for HLMs. | GFP |
| | Fiber based technologies and cell colonization or processing of fiber-encapsulated microorganisms | Different spinning processes for generating fibers, for weaving textiles and for obtaining non-wovens, as biomaterials or synthetic scaffolds for cell colonization leading to HLMs. | FBT |
| | Phase separation processes and cell colonization | Processes aimed at the creation of two or more phases from an originally single homogeneous mixture, usually leading to a porous biomaterials or structure for HLMs. | PSP |
| | Pyrolysis of polymeric precursors and cell colonization | Process focused on the generation of glassy carbon or pyrolytic carbon structures as scaffolds for cell colonization achieved by pyrolysis of polymeric precursors, usually oriented to the generation of living carbons (HLMs with synthetic carbon matrix). | PPP |
| | Composite manufacturing methods and cell colonization | Processes from the composites industry leading to woven, porous or functionally graded structures as synthetic scaffolds for cells, again for HLMs. | CMM |
| | Other production methods or combinations | Other processes or combinations of the above described and encapsulation of cells within engineered constructs for the generation of bio-ELMs and HLMs. | OPM |

Table 3. Subset of pioneering examples of ELMs described, classified according to the proposed ELMs ontology and codified following the presented codification scheme. Supplementary material S1 provides Table S1.1 as complete set of pioneering examples of ELMs employed for this study.

| *Pioneering examples of engineered living materials (ELMs)* | *Classification according to proposed expanded ontology* | | | | | | | *Codification* | *Refs.* |
|---|---|---|---|---|---|---|---|---|---|
| *Textual descriptions of ELMs* | *Domain* | *Superkingdom* | *Kingdom* | *Phylum* | *Class* | *Application field* | *Production method* | *Codified ELMs examples* | |
| Archaeal cultures for intracellular polymer synthesis aimed at the production of thermoplastics like PHA and PHB | Archaeal | Biological ELMs | Archaebacterial ELMs | 3D ELMs | Living polymers | Materials production and biofunctionalization | Intracellular synthesis | ARC-BIO-ARA-3D-LPO-MAP-ICS | 36 |
| Bacterial cultures for discrete extracellular gold nanocrystals or gold nanoparticles production with antibacterial properties | Bacterial | Biological ELMs | Eubacterial ELMs | 0D ELMs | Living alloys | Materials production and biofunctionalization | Extracellular matrix bioaccumulation | BAC-BIO-EUB-0D-LAL-MAP-EMB | 37 |
| 3D bioprinted photosynthetic cyanobacteria encapsulated within hydrogels forming a biofilm for a bioelectricity generation platform through photosynthesis | Bacterial | Hybrid living materials | Eubacterial ELMs | 2D ELMs | Living polymers | Energy production and harvesting | 3D bioprinting of cells and biomaterials | BAC-HYB-EUB-2D-LPO-ENP-3DB | 38 |
| Bacteria grown on marine agar synthesize silk proteins and lead to a microbial cell factory as a green and sustainable production platform for proteins and biopolymers, including spider silks | Bacterial | Hybrid living materials | Eubacterial ELMs | 3D ELMs | Living polymers | Materials production and biofunctionalization | Extracellular matrix (ECM) bioaccumulation | BAC-HYB-EUB-3D-LPO-MAT-EMB | 39 |
| Bacterial concrete based on calcite-precipitating bacteria as a crack-healing agent in concrete for construction and civil engineering | Bacterial | Hybrid living materials | Eubacterial ELMs | 3D ELMs | Living ceramics | Smart and living buildings | Biomineralization | BAC-HYB-EUB-3D-LCE-SLB-BMN | 40 |
| Musculoskeletal cells attached to hair for biohybrid micro-mechanical systems with application as micro-tweezers and muscle-powered micro-cantilevers | Eukaryotic | Hybrid living materials | Animal ELMs | 1D ELMs | Living polymers | Biohybrid machines and robots | Fiber based technologies and cell colonization or processing of fiber-encapsulated microorganisms | EUK-HYB-ANI-1D-LPO-BMR-FBT | 41 |
| PDMS medusoids with living tissues for biohybrid robots mimicking structural design, stroke kinematics and animal-fluid interactions in jellyfish propulsion | Eukaryotic | Hybrid living materials | Animal ELMs | 2D ELMs | Living polymers | Biohybrid machines and robots | Surface bio-patterning for selective cell culture / extracellular matrix (ECM) growth | EUK-HYB-ANI-2D-LPO-BMR-SBP | 42 |
| Plant cells encapsulated and grown within silica (glass-like) matrices employable for for long-term plant cell entrapment applications | Eukaryotic | Hybrid living materials | Plant ELMs | 3D ELMs | Living ceramics | Smart and living materials and structures (multipurpose) | Other production methods or combinations | EUK-HYB-PLA-3D-LPO-SLM-OPM | 43 |
| Polymeric layers with *Penicilium roqueforti* creating biofilms upon synthetic substrates, which may be sandwiched or stacked for achieving smart and living surfaces for architecture, design, packaging and biotechnology | Eukaryotic | Hybrid living materials | Fungal ELMs | 2D ELMs | Living polymers | Smart and living surfaces and biointerfaces | Biofilm formation and related extracellular matrix (ECM) growth | EUK-HYB-FUN-2D-LPO-SLS-BFF | 44 |
| 3D printed corals hosting microalgae as hybrid photosynthetic biomaterials and photobioreactors for scalable bioenergy and bioproduct generation | Eukaryotic | Hybrid living materials | Protist ELMs | 3D ELMs | Living composites | Energy production and harvesting | Scaffold (synthetic) structure 3D printing and cell colonization | EUK-HYB-PRO-3D-LCO-ENP-3DP | 45 |

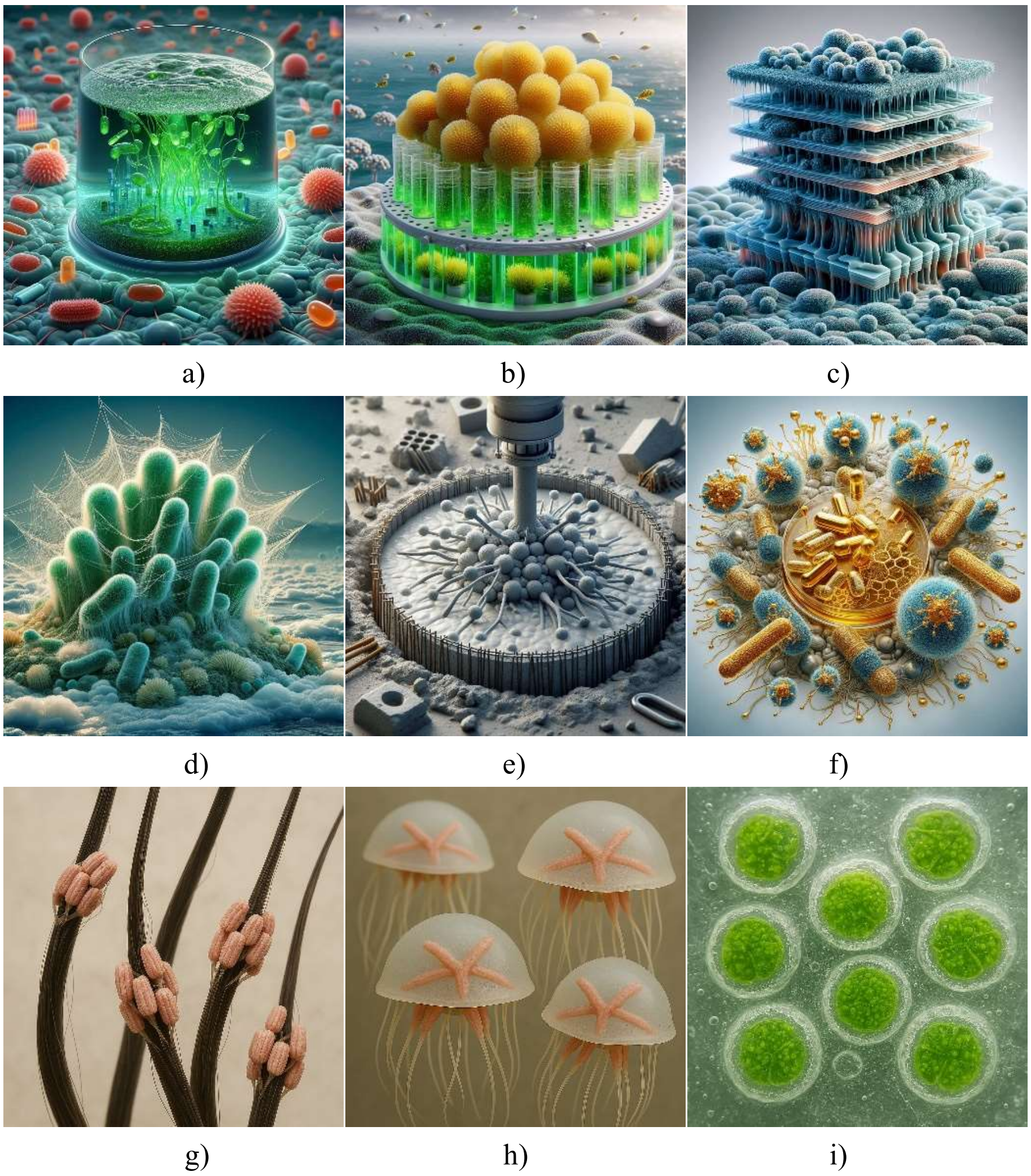

Figure 1. Selected examples of ELMs AI-generated illustrations employing the following "prompts" or textual descriptions of different examples from Table 3: a) 3D bioprinted photosynthetic cyanobacteria encapsulated within hydrogels forming a biofilm for a bioelectricity generation platform through photosynthesis. b) 3D printed corals hosting microalgae as hybrid photosynthetic biomaterials and photobioreactors for scalable bioenergy and bioproduct generation. c) An artistic representation of polymeric layers with Penicillium roqueforti forming biofilms on synthetic surfaces. The layers are stacked or sandwiched to create smart, living surfaces for applications in architect. d) Bacteria grown on marine agar synthesize silk proteins and lead to a microbial cell factory as a green and sustainable production platform for proteins and biopolymers, including spider silks. e) Bacterial concrete based on calcite-precipitating bacteria as a crack-healing agent in concrete for construction and civil engineering. f) Bacterial cultures for discrete extracellular gold nanocrystals or gold nanoparticles production with antibacterial properties. g) Musculoskeletal cells attached to hair for biohybrid micro-mechanical systems with application as micro-tweezers and muscle-powered micro-cantilevers realistic style. h) PDMS medusoids with living tissues for biohybrid robots mimicking structural design, stroke kinematics and animal-fluid interactions in jellyfish propulsion realistic style. i) Plant cells encapsulated and grown within silica (glass-like) matrices employable for long-term plant cell entrapment applications realistic style.

## 2.3. Knowledge graph representation of ELM ontology database

While the proposed ontology establishes a novel formal semantics for ELMs, ontology-based classification alone is insufficient to capture how design choices co-occur, dominate, or remain underexplored across available knowledge. The proposed ontology defines the admissible semantic space of ELM design; however, it does not represent the empirical structure of the design space as observed in the literature. To support evidence-grounded design and innovation workflows, the ontology-coded ELM dataset (subset in Table 3, complete version as Supplementary Material S1) was therefore instantiated as a knowledge graph (KG) derived entirely from data extracted from peer-reviewed journal articles. Graph-based reasoning methods, especially the pioneering works by Prof. Dr. Markus J. Buehler,[27] inspired us in this direction, in our case focusing on engineered living matter.

Each ELM case, encoded through a seven-slot code aligned with the seven ontology ranks or taxons, is represented as a case node in the graph. The ontology codes used to annotate the ELM cases are also represented as nodes, yielding a bipartite graph generated in Python 3.13.7 employing the NetworkX library.[47] The resulting graph therefore consists of two partitions: (i) ELM cases reported in the literature, and (ii) ontology-derived category nodes spanning Domain, Super kingdom, Kingdom, Phylum, Class, Applications and Production Methods. An edge is created whenever an ontology code is assigned to a given ELM case.

This modeling choice preserves the semantic structure of the ontology while making the ELM case base computationally queryable. Importantly, edges are not defined between ontology category nodes in this representation. As a result, the graph codes only the explicit use of ontology categories in published ELM designs and use cases. It should therefore be interpreted as a partial observational projection of the ontological design space, reflecting reported evidence rather than the full set of ontologically admissible ELM configurations. Consequently, the absence of a configuration in the graph indicates a lack of reported evidence, not ontological or biological impossibility. This distinction is critical for the subsequent design and innovation analyses.

The KG provides a structured, computable representation of bibliographic evidence that complements the provided ontology tables and code lists (Tables 1-3 and S1). It enables the quantitative characterization of empirical design trends, sparsity patterns, and potential biases in the reported ELM literature, while preserving ontological semantics. In this study, the KG serves as an operational semantic base for: (i) retrieving relevant ELM cases for design tasks, (ii) ontology-constrained data augmentation, and (iii) application-conditioned gap detection to support innovative design workflows. In this way, the ontology, codification scheme, and KG jointly define a reusable design language for advancing ELM research and enabling its downstream integration with generative AI approaches.

Complementary graph-level exploration is performed using Gephi 0.10.1,[48] which is used for KG visualization and exploratory inspection of the dataset. This environment enables interactive analysis of network structure, facilitating the manual identification of recurrent node usage, emerging literature trends and potential biases. Figure 2a presents the global graph (dendritic dendrogram or "***tree of life of ELMs***"), with node color distinguishing partitions and ontology slots (ELM case nodes are shown in light purple), and node size proportional to degree (number of edges connected to a node). The visualization highlights a non-uniform occupation of the ontology design space: a small set of recurrent ontology categories form high-connectivity hubs (e.g., HYB, LPO and 3D), whereas many categories remain weakly connected. In addition, Figure 2b ranks ontology category nodes by degree, making the uneven distribution of bibliographic evidence across ontology-coded categories more explicit.

Quantitative network statistics further characterize the structural properties of the KG. The graph contains 160 nodes, including 100 ELM case nodes and 60 ontology categories nodes, connected by 738 edges. The graph consists of a single connected component, indicating that all ELM cases are indirectly linked through shared ontology categories. In the undirected network representation, density is 0.058 and the average degree is 9.225.

ELM case nodes exhibit a mean degree of 7.38 (range: 7-9), reflecting their assignment across the seven ontology ranks, with a limited number of cases showing multiple assignments within individual slots. Ontology category nodes instead show a long-tailed degree distribution (mean 12.3, median 5.5, maximum 82), indicating that bibliographic evidence is strongly concentrated around a limited number of recurrent categories. The most connected category nodes include Superkingdom HYB (82), Class LPO (70), Phylum 3D (62), and Domain BAC / Kingdom EUB (49 each). In contrast, several ontology category nodes appear only once or a few times, indicating sparsely evidenced regions of the empirically observed ELM design space; in particular, 4DB, 4DP, CP, PPP, and PSP are among the least represented categories.

To translate annotation co-occurrence evidence into actionable "design-by-evidence" guidance, association rules are additionally mined. Rules are reported using support (the proportion of cases in which the pattern appears), confidence (the probability that the consequent appears when the antecedent is present), and lift (the strength of the association relative to what would be expected by chance), and should be interpreted as co-occurrence patterns rather than causal relations, hence providing additional insights about the current state-of-the-art in ELMs.

Representative rules with high confidence restricted to single-item and cross-slot associations, that can be used as quick empirical priors when selecting structural classes, are further detailed in Supplementary Material S1. Rule subsets with production method as consequents and application field as antecedents are also reported providing direct evidence-based guidance for: (i) manufacturing selection based on defined ontological code and (ii) ontological code selection based on target application. In this sense, association rules complement graph-level analysis by converting recurrent patterns in literature into interpretable design heuristics.

Considering the current panorama of ELMs, the quantitative analysis of the graph highlight the actual relevance of hybrid living materials, a superkingdom in which synthetic materials are colonized by the living entities. In hybrid living materials, the possibility of designing the extracellular matrix considering the morphological requirements of the final application brings in some advantages, as compared to biological living materials, in which the cells in culture generate their own extracellular matrix forming more amorphous and difficult to tune geometries. Hybrid living materials in a way evolve from the concept of scaffold-based solutions for tissue engineering and are already supported by a plethora of advanced manufacturing technologies enabling multi-scale and multi-material approaches. Furthermore, the living polymers class stands out because most biological living materials generate polymeric extracellular matrices and as a consequence of the versatile employment of polymers for biohybrid systems in general.

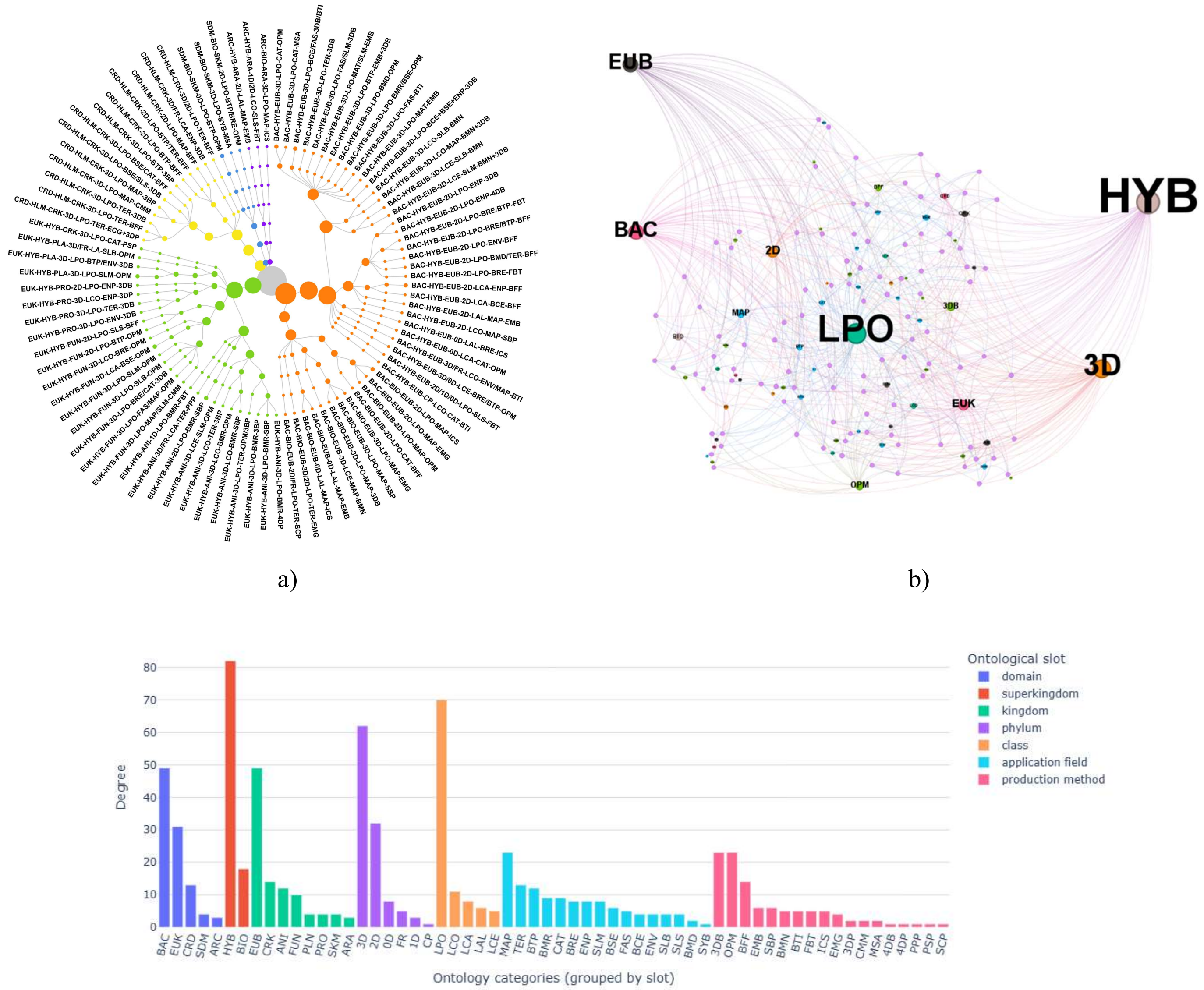

a) b) c)

Figure 2. a) Dendritic dendrogram organizing the dataset ELM examples in families and subfamilies according to the proposed ontology and codification scheme. Colors correspond to the five main domains of ELMs: archaeal (purple), bacterial (orange), eukaryotic (green), cross-domain (yellow) and synthetic (blue). b) Global bipartite ELM knowledge graph; c) degree-by-slot summary.

## 3. Ontology- and knowledge graph-grounded generative AI framework for designing and innovating engineered living materials

The framework is built on the developed coding scheme and its bipartite knowledge graph, which contains literature-derived ELMs and incorporates ChatGPT 5.2 Thinking as a reasoning engine for this challenging task. This framework is created to aid and foster creativity in the selection, generation, and design of ELMs, either directly from ontological codes or in response to specific application requirements. Consisting of five pipelines and six prompts, the framework may be used as either a set of independent modules or as part of multi-stage processes for: (i) database augmentation, (ii) application-conditioned candidate selection, (iii) gap-driven innovation, (iv) downstream design generation, and (v) visual representation of ELM candidates. Each pipeline operates through a prompt-response-analysis loop to ensure that generative outputs are critically assessed by the expert interacting with the AI in terms of feasibility, novelty and translational relevance, rather than being accepted unquestioningly.

In our approach, derived from previous studies by our team,[19,28,35] the generative AI is considered a supporting colleague along the creativity promotion steps, and quasi-human dialogues employing natural language are lived through during the grounding, evaluation and pipeline-based application of the model.

Figure 3 illustrates the framework and its pipeline instances in schematic form. Meanwhile, Table 4 summarizes the proposed prompts, indicating their function, as well as their expected inputs and outputs. Full prompt texts are provided in a directly reusable form for interaction with ChatGPT in the Supplementary Material S2, supporting reproducibility. Among these, Prompt F is shared across most pipelines and is aimed at generating a visual representation of the designed ELM. It takes either a 7-slot ontology code or a complete ontology-derived instantiation as input, optionally supplemented with user-defined visualization preferences including style (e.g., schematic, multiscale, cutaway, mechanism, workflow, ontology iconography) and formatting controls (e.g. aspect ratio, view, abstraction, annotations). The output is a single, living-matter-centered image that explicitly depicts the ELM and, where relevant, its application context. The individual pipelines are described in detail in the following sections.

Table 4. Summary of the set of prompts for the framework pipelines.

| ***Prompt name*** | ***Pipeline*** | ***Task*** | ***Input*** | ***Output*** |
|---|---|---|---|---|
| Prompt A | Pipeline A | Validate ontology grounding and consistency. | (i) Codified ontology;<br>(ii) GraphML knowledge graph. | Decoded ELM examples; structured validation report. |
| Prompt B | Pipeline B | Perform bibliography-based database augmentation. | (i) Article abstract or<br>(ii) full text. | ELM identification result and 7-slot ontology code; concept description. |
| Prompt C | Pipeline C | Identify ontology-valid but absent or underrepresented 5-slot structural combinations in the ELM knowledge graph; formulate novel ELM hypotheses grounded in ontology space, either generically or application-conditioned. | (i) Gap detection mode (generic or application conditioned);<br>(ii) optional target application domain;<br>(iii) ELM knowledge graph. | Structured list of ELM hypotheses, including a 7-slot ontology code, concept description, and explicit plausibility justification. |
| Prompt D | Pipeline D | Select and refine up to five ontology-valid ELM candidates for a defined application context by filtering dataset-derived or gap-derived candidates. | (i) Target application domain and its boundary conditions;<br>(ii) candidate ELMs originated from data augmentation and/or gap detection. | Structured list of up to five refined ELM candidates, each including selection rationale, complete 7-slot ontology code, application-oriented concept description and justification. |
| Prompt E | Pipeline E | Instantiate each provided complete 7-slot ELM code into a fully specified, scientifically coherent, and technically plausible ELM concept. | Complete 7-slot ELM candidates, provided as:<br>(i) application-conditioned candidate blocks,<br>(ii) gap-detection hypotheses or<br>(iii) raw ontology codes. | Full ELM instantiation. |
| Prompt F | - | Generate a scientifically interpretable image of an ontology-defined ELM from either a 7-slot code or a fully instantiated ELM description, using the selected visualization style. | (i) Complete 7-slot ELM code or ontology-derived instantiation, plus optional format controls and style mode. | Generated image of the provided ELM. |

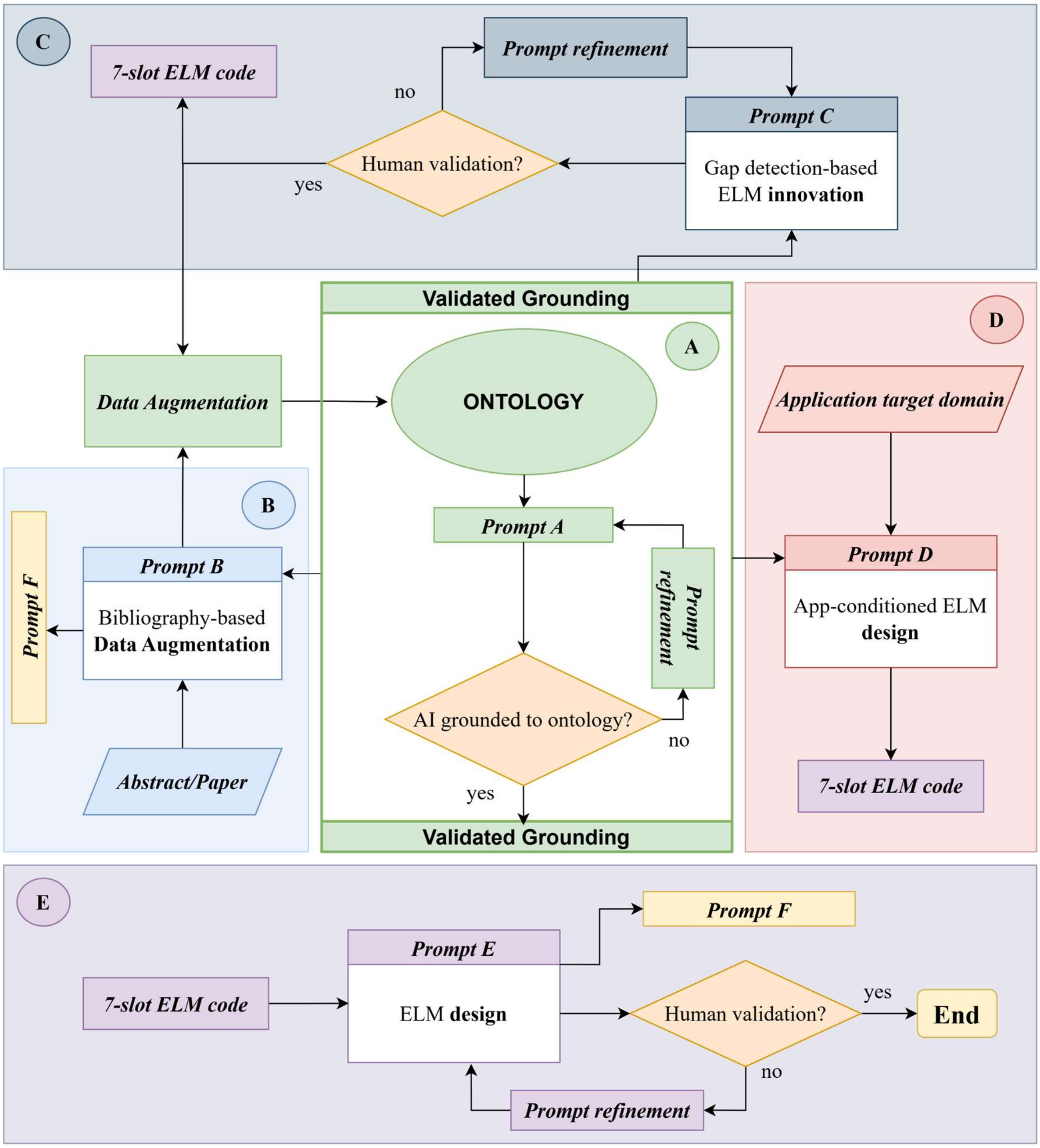

Figure 3. Scheme of the proposed ontology-grounded AI framework based on a multi-stage pipeline for designing and innovating ELMs.

### 3.1. Pipeline A: AI ontology semantic grounding validation

Pipeline A is a prerequisite for all subsequent generative and inferential tasks. At this stage, the generative AI employed, ChatGPT in our case, is constrained by the ELM ontology, whose coding scheme defines valid classes, codes and combinations within the design space, explicitly forbidding the extension, reinterpretation or invention of codes. This shared semantic basis is established to ensure that downstream outputs remain interpretable, reproducible, and ontologically consistent. As shown in Figure 3, if the model does not demonstrate an adequate understanding of the ontology, this step may be iterative until satisfactory grounding is achieved by refining and reapplying Prompt A. In this sense, ontology functions as both a semantic constraint and a communication interface between the user and the generative AI. Only after ontological grounding has been achieved can the remaining pipelines be implemented.

The knowledge graph is treated as bibliographic evidence rather than as a constraint on admissible ELM ontological configurations. The model is then required to demonstrate its ontology comprehension by extracting ontology codes and definitions; reconstructing the codification syntax; providing one or two ELM examples using only ontology definitions; and identifying inconsistencies, ambiguities, or deviations in code usage and codification.

### 3.2. Pipeline B: Bibliography-based data augmentation

As is widely recognized, ELMs as field of knowledge is continuously and very rapidly expanding and evolving. Therefore, the knowledge graph containing examples of existing ELMs should grow accordingly. ChatGPT (or other generative AIs) can be employed for bibliography-based data augmentation through Pipeline B using Prompt B, under two possible input strategies: (i) the scientific paper abstract or (ii) the full text. In both cases, the source data is processed through the same ontology-constrained decoding workflow. First, the relevance of the evidence to ELMs is confirmed, and then the evidence is translated into seven-slot ELM codes. This enables the graph to be expanded while preserving the integrity of the coding scheme.

Beyond its role in data augmentation, Pipeline B also functions as a secondary validation procedure for determining the ontology grounding and coding performance of the AI model on unseen bibliographic evidence through comparison with human codification. The details of this evaluation and its results are presented in Use case 1.

### 3.3. Pipeline C: Graph-based gap detection for ELM innovation

Pipeline C performs graph-based gap detection to enable unbiased hypothesis generation for ELMs. The pipeline operates in two modes: (i) generic, in which the application domain is not used as a constraint, and (ii) application-conditioned, in which gap prioritization is guided by a specified target application domain while preserving application-independent novelty criteria. Based on the identified gaps, ChatGPT -or the generative AI of choice- proposes ELM candidates guided by biological, structural, and technological plausibility criteria.

This pipeline consists of two main steps. First, a structural analysis of the knowledge graph is performed to characterize sparsity and bias in the observed literature design space. This analysis serves exclusively to inform the interpretation and prioritization of candidate gaps and is not treated as a constraint on biological admissibility. Second, structure-informed gap detection is carried out within the ontology-defined space by enumerating biologically admissible combinations across the structural part of the code (comprising: Domain, Superkingdom, Kingdom, Phylum, and Class), and identifying combinations that are absent or strongly underrepresented in the knowledge graph.

Importantly, innovation is defined strictly at this structural level, such that gaps are detected exclusively within these five slots. For each selected gap, the model generates a complete 7-slot

ELM hypothesis, in which Application and Production Method, as secondary taxons, are introduced solely to complete a coherent and testable hypothesis. In the application-conditioned mode, the target application domain is used only as a downstream prioritization factor, whereas gap definition and novelty remain strictly determined by these five structural slots.

All generated candidates must remain living-matter-centric and satisfy explicit plausibility constraints along three complementary axes: (i) biological plausibility, referring to the compatibility between the living system and the organizational scale or matrix environment; (ii) structural/material plausibility, referring to the coherence between Phylum/Class and the expected physical or material organization; and (iii) technological feasibility, referring to whether the concept is realizable in principle using known or foreseeable fabrication approaches. Furthermore, the reasoning process must explicitly consider practical implications, such as manufacturability, scalability, experimental feasibility and integration potential.

Pipeline C produces a structured list of 7-slot ELM candidates. Each one is accompanied by a concise concept description and bullet-point justifications. These candidate designs are then subjected to expert validation. If an output is not approved by the expert, the prompt is refined iteratively, and a new output is generated until a coherent response is obtained. Moreover, this pipeline could support dataset augmentation by incorporating ELMs derived from gaps and validated by experts into the current database.

### 3.4. Pipeline D: Application-conditioned ELM retrieval

Pipeline D enables the selection of ELMs for a specific application by using ChatGPT to filter, categorize, and contextualize up to five concepts from the KG. In this strictly goal-directed, non-generative procedure, prompt D identifies the most suitable candidate from ELMs derived from: (i) the initial bibliographic references, (ii) the data augmentation procedure in Pipeline B, and (iii) the gap-derived generation process in Pipeline C.

Selection is carried out under user-defined application boundary conditions that serve as acceptance and rejection criteria. The five-slot structural core is preserved while the Application and Production Method slots are completed during the reasoning process. The final output is the AI model response, which includes the full seven-slot code, a description of the selected ELM concept, and justification of its biological, structural, and technological plausibility.

### 3.5. Pipeline E: Ontology-based AI-driven ELM full instantiation

Pipeline E aids the creation of ELMs from a 7-slot ontological code that has been previously defined, by translating it into an output that is both scientifically coherent and technically plausible. Regardless of its origin, the input incorporates always a complete seven-slot code. This code might come from: (i) hypotheses generated in Pipeline C, (ii) candidates selected in Pipeline D based on the application, or (iii) raw ontological codes provided without additional context.

Prompt E first verifies the integrity of each input code and then converts the ontological specification into a design proposal by translating each slot into constraints relating to biology, materials, geometry, applications and processes. Based on this, it provides concrete information on the biological entities involved while preserving consistency with the predefined code: compatible matrix materials, geometries consistent with the structural slot, deployment scenarios associated with the application slot and feasible workflows aligned with the production method slot. Reasoning is restricted to approaches centered on (engineered) living matter and considers manufacturability, scalability, experimental feasibility and system integration. Each major design decision must be justified and at least two candidate options must be proposed for the biological and material components. This prepares the concept for experimental planning, simulation or optimization.

## 4. Ontology-grounded AI-driven framework assessment via use cases for innovating new ELMs: on the understanding of the design language by a trained generative AI

This section presents the performance of the proposed ELM design framework. Rather than being organized as a single linear workflow, the framework is evaluated through a set of selected workflows representative of its modular capabilities. This choice reflects the modular nature of the proposed methodology: each pipeline constitutes a self-contained, reproducible component with well-defined inputs, outputs, and constraints, thereby enabling multiple workflow configurations to address distinct research objectives. Thus, the framework does not prescribe a unique execution path; instead, different combinations of pipelines can be assembled depending on the target application and the specific design objective.

Two target application domains, one more realistic or pragmatic, one more futuristic, are considered in this study: (i) self-sustaining buildings in desert environments and (ii) ELM-enabled habitats and infrastructure for the long-term colonization of Titan. These applications were selected as use case to evaluate the framework under markedly different boundary conditions, which are detailed in Table 5.

From the range of possible pipeline combinations, a subset of workflows was selected to illustrate and assess framework performance. Accordingly, the results are structured around three representative workflow-based use cases as illustrated in Figure 4:

i. Use case 1: data augmentation from bibliographic sources, together with coding-performance validation on 20 unseen articles, in order to assess the ontology-grounded extraction capabilities of the model.
ii. Use case 2: application-driven ELM innovation for target application domain 1 (self-sustaining buildings in desert environments), implemented through workflow C (ELM Innovation) combining application-conditioned gap detection and ELM detailed design generation. Workflow C is executed multiple times to evaluate the stability and variability of the generated outputs across repeated runs.
iii. Use case 3: ELM design for target application domain 2 (long-term colonization of Titan), for which workflow D (ELM retrieval) is executed multiple times to evaluate stability and variability of the generated outputs across repeated runs.

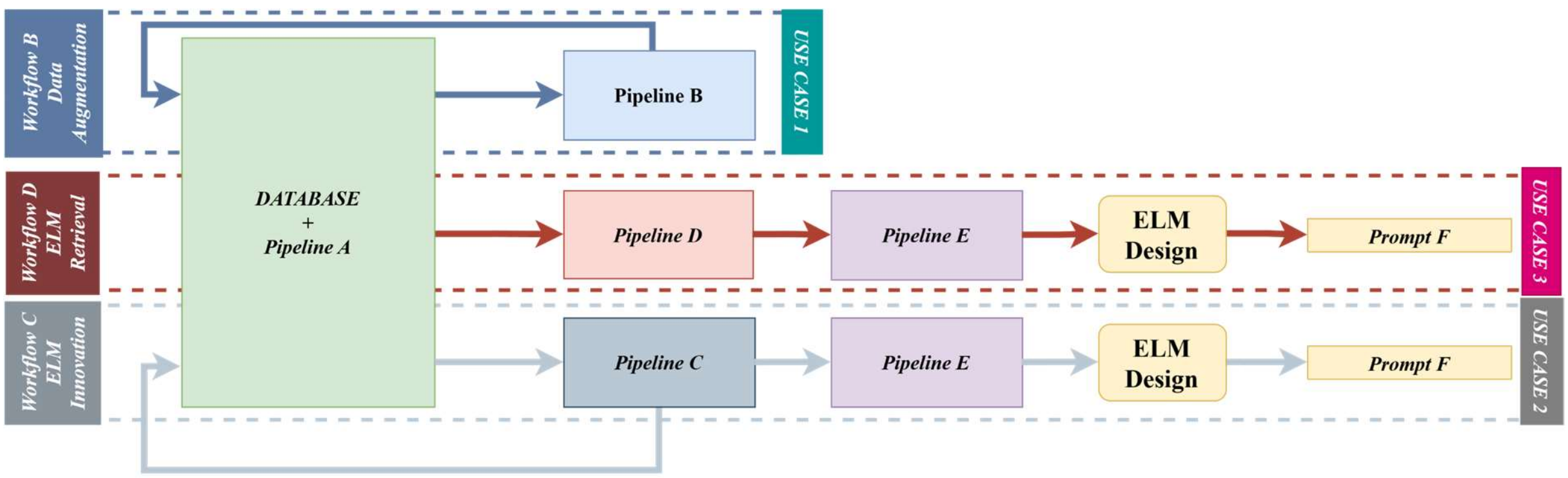

Figure 4. Proposed workflow-based use cases.

Table 5. Target application domains and their boundary conditions.

| ***Boundary conditions*** | ***Application domains*** | |
|---|---|---|
| | ***Self-sustaining building*** | ***Colonizing Titan*** |
| Environmental conditions | - Extreme heat and strong day–night temperature swings.<br>- Intense UV radiation.<br>- Very low humidity.<br>- Sand and dust abrasion with wind exposure.<br>- Scarce freshwater and limited resupply. | - Cryogenic temperatures.<br>- Dense nitrogen-rich atmosphere with methane and hydrocarbon aerosols.<br>- Extremely low solar energy availability.<br>- No stable liquid water at the surface.<br>- Hydrocarbon precipitation and exposure to chemically unusual surface media.<br>- Remote operation with severe resupply and maintenance constraints. |
| Hard constraints | ELMs must be:<br>- Safe and biologically contained.<br>- Functional under desert stress or realistically shielded.<br>- Scalable and construction-compatible.<br>- Field-repairable and low maintenance. | ELMs must be:<br>- Safe and biologically contained.<br>- Realistically functional under Titan-relevant conditions or operated within shielded, controlled subsystems.<br>- Compatible with modular habitat, ISRU (in situ resource utilization), and infrastructure systems.<br>- Resource-efficient in water, energy, nutrients, and imported inputs.<br>- Robust over long mission durations with low-maintenance or autonomous operation. |
| Functional contribution | Each candidate must support at least one:<br>- Water harvesting, storage, purification, or recycling.<br>- Passive cooling, insulation, or thermal buffering.<br>- Energy-supporting or energy-enabling surfaces or systems.<br>- Structural self-healing or soil/sand stabilization.<br>- Dust filtration or localized air/$CO_2$ control.<br>- Resource or nutrient cycling.<br>- Any additional function clearly supporting building self-sufficiency. | Each candidate must support at least one of the following:<br>- Atmospheric processing or gas generation for habitat or industrial support.<br>- Resource extraction, conversion, storage, or recycling from locally available feedstocks.<br>- Waste treatment and closed-loop material or nutrient cycling.<br>- Biomanufacturing of useful materials, coatings, membranes, or composites.<br>- Insulating, protective, self-healing, or anti-fouling functions in infrastructure.<br>- Any other function that clearly supports sustained human or robotic presence on Titan.<br><br>ELMs act as **components of habitats, ISRU systems, bioprocessing units, or material platforms**, not as standalone solutions for planetary-scale transformation. |

Prior to any AI-driven ELM design, generative AI is grounded to the developed ontology by following the methods described in subsection 2.3, employing the ontology and codification scheme, as presented in Table 1 and Table 2, and the knowledge on ELMs gathered in Table 3 and S1, as an ontology-grounded graph-based reasoning context. The model understanding and adherence to the ontology and codification scheme are evaluated using pipeline A. The ontology grounding performance of the generative AI model was evaluated using Prompt A and the model response is shown in Supplementary Material S3. The grounding validation step confirms that the AI model correctly interprets ontology dimensions, codification rules, and formal semantics. This validation enforces alignment with the ontology.

### 4.1. Use case 1: bibliographic data augmentation and coding-performance validation

Use case 1 consists of a bibliography-based data augmentation procedure implemented through Pipeline B. By expanding the database, this step increases the scope of candidate ELMs available for downstream selection in target-specific applications. This same data augmentation function is also used as a validation procedure to assess the AI model's ontology-grounded coding performance. Performance was compared using full-text articles versus abstracts to assess whether abstract-only input is sufficient for ELM encoding.

For this validation, the model is asked to codify 20 additional abstracts[49–68] and full papers describing ELM examples that are not included in the current database (selection of Table 3 from the complete 100 dataset of S1) and are incorporated specifically for this evaluation. The resulting model-generated codifications are then compared against human codifications of the same examples. This comparison provides an external assessment of whether the model can reliably encode unseen bibliographic evidence into ontology-consistent ELM representations, rather than assuming the correctness of its outputs a priori. Table 6 presents the ELM codes obtained from generative AI-based abstract and paper encoding. Detailed information on the responses is provided in Supplementary Material S4, which illustrate the reliability and explanatory ability of the generative AI, even providing rationales for ambiguous codifications.

Table 6. GenAI abstract and paper ELM coding of unseen examples and subsequent comparison with human codification.

| *Article* | *Abstract-only input* | *Full-text article input* |
|---|---|---|
| 49 | BAC-HYB-EUB-3D-LCA-ENP-OPM | BAC-HYB-EUB-3D-LCA-ENP-OPM |
| 50 | BAC-BIO-EUB-UNK-UNK-SYB-ECP+MSA | BAC-BIO-EUB-3D-UNK-SYB-ECP+MSA |
| 51 | EUK-HYB-FUN-4D-LCO-SLM-OPM | EUK-HYB-FUN-4D-LCO-SLM-OPM |
| 52 | BAC-BIO-EUB-UNK-LPO-MAP-ECP | BAC-BIO-EUB-2D-LPO-MAP-ECP+BFF+MSA+SCP |
| 53 | BAC-HYB-EUB-3D-UNK-BRE-3DB | BAC-HYB-EUB-3D-LCO-BRE-3DB |
| 54 | BAC-HYB-EUB-UNK-UNK-BMD-OPM | BAC-HYB-EUB-0D-LPO-BMD-OPM |
| | | BAC-HYB-EUB-0D-LPO-BMD-ECP+OPM |
| 55 | UNK-HYB-UNK-UNK-UNK-BSE+ENV-OPM | EUK-HYB-FUN-3D-LPO-BSE+ENV-OPM |
| | | BAC-HYB-EUB-3D-LPO-BSE+ENV-OPM |
| 56 | BAC-HYB-EUB-2D-LCE-SLS+MAP-ECP+BTI+BMN | BAC-HYB-EUB-2D-LPO+LCE-SLS-ECP+BFF+BTI+BMN |
| 57 | BAC-BIO-EUB-2D-LPO-BSE+BRE-ECP+BFF | BAC-BIO-EUB-2D-LPO-BSE+BRE+ENV-ECP+BFF |
| | BAC-HYB-EUB-UNK-LCA-BRE-ECP+BTI | BAC-HYB-EUB-3D-LCA-BRE-ECP+BFF+BTI |
| 58 | UNK-BIO-UNK-4D-LPO-MAP-OPM | BAC-BIO-EUB-3D-LPO-BMD-EMG |
| 59 | BAC-HYB-EUB-3D-LPO-CAT+BTP-3DB | BAC-HYB-EUB-3D-LPO-CAT-3DB |
| | EUK-HYB-PRO-3D-LPO-CAT+BTP-3DB | EUK-HYB-PRO-3D-LPO-CAT-3DB |
| 60 | EUK-HYB-FUN-3D-LCO-MAP-3DP | EUK-HYB-FUN-3D-LPO-SLM-OPM |
| 61 | EUK-HYB-PRO-0D-UNK-BMD-OPM | EUK-BIO-PRO-0D-UNK-BMD+BMR-ECP+OPM |
| 62 | BAC-HYB-EUB-3D-LPO-TER-OPM | BAC-HYB-EUB-3D-LPO-TER-ECP |
| 63 | BAC-HYB-EUB-3D-LPO-BTP-OPM | BAC-HYB-EUB-3D-LPO-BTP-OPM |
| | EUK-HYB-FUN-3D-LPO-BTP-OPM | EUK-HYB-FUN-3D-LPO-BTP-3DB |
| 64 | BAC-HYB-EUB-3D-LPO-BMD+BTP-OPM | BAC-HYB-EUB-3D-LPO-BMD+BTP-OPM |
| 65 | EUK-BIO-FUN-2D-LPO-FAS-EMG | EUK-BIO-FUN-2D-LPO-FAS+TRI-EMG+BFF |
| 66 | UNK-HYB-UNK-3D-LPO-BMR+MAP-3DB | BAC-HYB-EUB-3D-LPO-MAP+BMR-3DB |
| | | BAC-HYB-EUB-3D-LPO-BTP-3DB |
| | | EUK-HYB-FUN-3D-LPO-MAP+BMR-3DB |
| 67 | BAC-HYB-EUB-3D-LPO-BRE-3DB | BAC-HYB-EUB-3D-LPO-BRE-3DB |
| 68 | EUK-BIO-FUN-2D-LPO-BSE-ECP+EMG | EUK-BIO-FUN-2D-LPO-BSE+ENV-ECP+EMG |

## 4.2. Use case 2: Application-conditioned gap detection and downstream ELM design

Use case 2 corresponds to Workflow C in Figure 4 and combines Pipelines C and E for target application domain 1 (self-sustaining buildings in desert environments). The workflow is implemented on the initial dataset obtained from bibliography annotations. Application-conditioned gap detection is then performed using Pipeline C to identify biologically admissible but absent or underrepresented structural combinations in the knowledge-graph database, while explicitly orienting this process toward target application domain 1. For each selected structural gap, Pipeline C produces a completed 7-slot ELM hypothesis together with its associated description, which are subsequently passed to Pipeline E for full design generation.

To assess robustness and exploratory diversity, the workflow was implemented following a cascaded multi-iteration strategy. First, Pipeline C was executed in three independent iterations. Based on the outcomes of these runs, three ELM candidates were selected, one from each iteration, in order to preserve inter-run variability and capture potentially complementary opportunity spaces.

Subsequently, Pipeline E was applied to each of the three selected ELMs. For each input ELM, Pipeline E was again executed in three independent iterations, thereby extending the exploration while maintaining consistency with the iterative evaluation framework adopted in the previous stage. From the resulting set of outputs, three ELMs generated through Pipeline E were selected for downstream analysis and as input for Prompt F, which was also conducted in three independent iterations per selected ELM to support final visualization and comparative interpretation of the generated designs. For every Pipeline output, the selected ELM candidates are ELM candidate 1 of iteration 1; ELM candidate 2 of iteration 2; ELM candidate 3 of iteration 3.

This cascaded execution scheme allows the analysis to determine whether the workflow converges toward recurrent solution patterns or, alternatively, produces distinct yet application-relevant opportunities across iterations. Table 7 and Table 8 summarizes the results obtained from Pipeline C and Pipeline E for use case 2, whereas the complete generative AI outputs are provided in Supplementary Material S5 (S5_1 providing the generative AI output following prompt C and S5_2 giving the generative AI response to prompt E). Table 9 shows the generated images from prompt F. As output of workflow C, three ELMs fully instantiated are proposed for target application domain 1, after hierarchical analysis performed by the generative AI and our team about the different possibilities generated and discussed along the iterations:

- ELM candidate 1: BAC–BIO–EUB–FR–LCE–SLB–BFF+BMN
- ELM candidate 2: ARC–HYB–ARA–3D–LCE–SLB+BRE–CMM
- ELM candidate3: CRD–HYB–CRK–CP–LCO–SLB–CMM+BMN

Intriguingly, the generative AI is capable of proposing novel engineered living materials (a total of 9 in the first stage if we consider the 3 iterations for the 3 candidates of Table 7), with a justification of their biological and structural plausibility and providing indications towards their technological materialization. In the second stage (results shown in Table 8), for the selected candidates, progressively more detailed and reliable descriptions are provided toward the practical implementation of these ELMs. Both qualitative and quantitative indications are provided for their engineering design, practical implementation and operation. The level of detail achieved and the explained design decisions demonstrate that the generative AI, after being trained with our described ontology and knowledge graph and guided with our collection of prompts, constitutes a remarkable companion for the discovery and co-creation of highly innovative ELMs with outstanding industrial potential. The detailed information of these discoveries is provided in Tables 7 and 8, in supplementary materials S5_1 and S5_2, and the resulting ELMs schematically illustrated in Table 9.

Table 7. Use case 2 Pipeline C output: app-conditioned gap-derived ELM candidates for different runs.

| ELM candidate | Component | Iteration 1 | Iteration 2 | Iteration 3 |
|---|---|---|---|---|
| **ELM candidate 1** | Name | Mineralizing repair / sand-stabilization bioceramic. | Bacterial adaptive living-ceramic façade insert. | Bacterial living ceramic block for graywater polishing and crack sealing. |
| | Gap-defining 5-slot combination | BAC–BIO–EUB–FR–LCE | BAC–HYB–EUB–4D–LCE | BAC–HYB–EUB–3D–LCE |
| | Full 7-slot hypothesis | BAC–BIO–EUB–FR–LCE–SLB–BFF+BMN | BAC–HYB–EUB–4D–LCE–SLB+SLS–4DP+BMN | BAC–HYB–EUB–3D–LCE–SLB+BRE–BMN+3DP |
| | Concept description | A branched bacterial mineralizing bioceramic layer that grows inside cracks or in near-surface sand networks to seal defects and stabilize wind-erodible ground around desert buildings. | A sealed mineralized façade insert containing nonpathogenic spore-forming bacteria in protected pores, designed to undergo useful day–night shape change while self-sealing microcracks after rare wetting events. | A modular living ceramic service block containing biomineralizing bacteria for local graywater polishing, mineral retention, and autonomous sealing of microcracks in building service zones. |
| | Justification | • Biological plausibility<br>Spore-forming or mineralizing eubacteria are compatible with intermittent activation, dormancy, and reactivation after rare moisture inputs, which fits desert operation better than continuously metabolically active systems.<br>• Structural / material plausibility<br>The FR phylum matches irregular crack paths and branched pore/sand networks, while LCE is coherent with mineral deposition forming a stiff ceramic-like repair phase rather than a soft biofilm only.<br>• Technological feasibility and practical implications<br>This can be integrated as a low-maintenance repair grout, dry powder patch, or perimeter stabilization treatment activated by recycled water or brief wetting events; it is construction-compatible and field-repairable without complex hardware. | • Biological plausibility<br>Spore-forming eubacteria are compatible with long dry periods, intermittent activation, and protected pore habitats; the living fraction can remain dormant under desert stress and revive after dew, maintenance wetting, or recycled-water exposure.<br>• Structural / material plausibility<br>4D is appropriate because thermal cycling is functionally central, and LCE is coherent with a mineral-dominated façade insert intended for crack resistance, UV shielding, and abrasive dust exposure.<br>• Technological feasibility (non-prescriptive)<br>A thermally responsive porous ceramic bilayer with embedded bacterial niches is feasible in principle using scaffold-based shaping plus mineralization; modular inserts are construction-compatible, replaceable, and field-repairable. | • Biological plausibility<br>Non-pathogenic biomineralizing bacteria such as Sporosarcina pasteurii or spore-forming Bacillus subtilis are compatible with dormant survival during dry periods and reactivation when graywater flows through the module.<br>• Structural / material plausibility<br>A 3D living ceramic is coherent because porous ceramic blocks provide filtration pathways, mechanical stability, thermal resistance, and a mineral-friendly environment for precipitation-based crack sealing.<br>• Technological feasibility (non-prescriptive)<br>Service-wall cartridges, replaceable filter bricks, or distributed utility modules are realistic integration routes; the concept scales better than exposed living façades and fits desert buildings where water recycling and maintainability are critical. |
| **ELM candidate 2** | Name | Sealed archaeal solar-modulating living film. | Archaeal living-ceramic greywater reactor cartridge. | Cross-phylum fungal composite insulation panel. |
| | Gap-defining 5-slot combination | ARC–HYB–ARA–2D–LPO | ARC–HYB–ARA–3D–LCE | EUK–HYB–FUN–CP–LCO |
| | Full 7-slot hypothesis | ARC–HYB–ARA–2D–LPO–SLS–BTI+OPM | ARC–HYB–ARA–3D–LCE–SLB+BRE–CMM | EUK–HYB–FUN–CP–LCO–SLB–FBT+CMM |
| | Concept description | A thin, sealed archaeal living film inside a transparent polymer laminate that modulates incoming light and thermal load on façades or louvers under extreme heat and radiation. | A hot, saline greywater-treatment cartridge built from porous ceramic modules colonized by nonpathogenic haloarchaea, intended for water polishing and nutrient cycling in desert buildings. | A multi-scale mycelium–fiber–mineral composite panel in which fungal growth binds a fibrous scaffold and porous core, providing passive cooling, thermal insulation, and local crack-bridging capacity. |
| | Justification | • Biological plausibility<br>Halophilic archaea are among the most naturally compatible cells for high salt, high irradiation, and high temperature environments, especially when maintained in contained saline microchambers.<br>• Structural / material plausibility<br>The 2D phylum fits a continuous interfacial film, and LPO is coherent with a polymeric laminate or membrane architecture that also | • Biological plausibility<br>Haloarchaea are unusually compatible with high salt, heat, and low water activity, making them stronger candidates than most conventional cells for contained service-core treatment modules handling concentrated or brackish water streams.<br>• Structural / material plausibility<br>3D matches a volumetric reactor cartridge, and LCE fits a ceramic host that tolerates heat, fouling control, and repeated wet–dry operation better than fragile hydrated matrices. | • Biological plausibility<br>Fungi already used in mycelium materials, such as Pleurotus ostreatus, can colonize fibrous porous substrates efficiently; in desert use, the biologically active region can remain protected inside the panel and operate intermittently rather than continuously.<br>• Structural / material plausibility<br>CP is justified because the system combines 1D hyphal networks with a 3D porous composite body; LCO is also coherent because the panel can combine |
| | | provides biological containment and dust shielding.<br>● Technological feasibility and practical implications<br>The concept is realistic as a replaceable façade insert or shutter cartridge rather than an exposed coating; this reduces abrasion risk, simplifies containment, and supports low-maintenance building integration. | ● Technological feasibility (non-prescriptive)<br>Porous ceramic bioreactor cartridges are realistic in principle as modular building-service components; they are scalable, swappable, and easier to contain biologically than open environmental systems. | organic fibers, mineral fillers, and fungal extracellular binding phases.<br>● Technological feasibility (non-prescriptive)<br>Panel-scale manufacturing is compatible with existing insulation and sandwich-panel logic; the concept is field-repairable through modular replacement and more scalable than free-form living masonry. |
| **ELM candidate 3** | Name | Cross-domain fiber cartridge for dust capture and local $CO_2$ conditioning. | Cross-domain lichen-like living-ceramic dust and $CO_2$ skin. | Cross-kingdom bacterial–fungal self-healing render for sand stabilization. |
| | Gap-defining 5-slot combination | CRD–HYB–CRK–1D–LPO | CRD–HYB–CRK–FR–LCE | CRD–HYB–CRK–CP–LCO |
| | Full 7-slot hypothesis | CRD–HYB–CRK–1D–LPO–SLB–FBT+OPM | CRD–HYB–CRK–FR–LCE–SLB+ENV–SBP+BMN | CRD–HYB–CRK–CP–LCO–SLB–CMM+BMN |
| | Concept description | A replaceable cross-domain fiber cartridge, for example a fungal–cyanobacterial or fungal–microalgal capillary network, that traps dust, buffers humidity locally, and contributes to localized $CO_2$ control in air-intake or courtyard ventilation systems. | A lichen-inspired ramified cyanobacteria–fungal mineralized skin grown within a porous ceramic screen to intercept dust, locally polish $CO_2$, and slowly consolidate eroding building surfaces. | A protected bacterial–fungal living render in which bacteria mineralize and fungi reinforce the matrix, enabling self-healing, dust suppression, and stabilization of sand-exposed exterior surfaces. |
| | Justification | ● Biological plausibility<br>Cross-domain consortia are plausible where one partner provides structural colonization and the other contributes gas exchange or nutrient cycling, provided the system is immobilized and physically contained in a sheltered cartridge.<br>● Structural / material plausibility<br>The 1D phylum fits capillary fibers and woven filter media, while LPO is coherent with hydrogel-coated or polymer-supported fibrous architectures designed for air handling rather than load-bearing structure.<br>● Technological feasibility and practical implications<br>This is much more credible as a serviceable building subsystem than as an exposed façade skin: it can be swapped like a filter, isolated from occupants, protected from wind abrasion, and integrated into low-pressure ventilation hardware. | ● Biological plausibility<br>A cyanobacteria–fungus consortium is cell-centric and biologically credible for a cross-domain ELM; both partners can tolerate repeated desiccation better than many hydrated cell systems when held in shaded pore networks and supplied only intermittently.<br>● Structural / material plausibility<br>FR fits a ramified, branching colonization pattern, and LCE fits a mineral-rich crust or ceramic-supported shell intended to face wind, UV, and abrasive particles.<br>● Technological feasibility (non-prescriptive)<br>Patterned inoculation of porous ceramic screens and subsequent mineral consolidation are feasible in principle; the format is scalable as replaceable façade or wind-screen modules rather than as free-growing environmental release. | ● Biological plausibility<br>A cross-kingdom pairing such as Sporosarcina pasteurii with Aureobasidium pullulans is cell-centrically coherent: the bacterium contributes mineral precipitation, while the fungus contributes surface adhesion, extracellular binding, and crack-spanning growth.<br>● Structural / material plausibility<br>CP–LCO is appropriate because the render is inherently multi-scale: bacterial microcolonies, fungal filament networks, and a bulk mineral–polymer composite matrix act together rather than as a simple single-phase material.<br>● Technological feasibility (non-prescriptive)<br>The concept is realistic only as a contained or semi-contained render layer with controlled nutrient access; that constraint is acceptable in desert buildings, where serviceable façade skins and sacrificial outer layers are already practical design strategies. |

Table 8. Use case 2 Pipeline E output: full instantiation of gap-derived ELM candidates for different runs.

| *ELM candidate code* | *Component* | *Iteration 1* | *Iteration 2* | *Iteration 3* |
|---|---|---|---|---|
| **BAC–BIO–EUB–FR–LCE–SLB–BFF+BMN** | One-line ELM identity | Spore-forming bacterial living ceramic that biomineralizes branched crack and sand-pore networks to seal defects and stabilize desert building interfaces. | Spore-forming eubacterial living ceramic biofilm that mineralizes branched crack and sand-pore networks to seal defects and stabilize desert building perimeters. | Spore-forming eubacterial calcite living ceramic that mineralizes branched crack and sand networks for defect sealing and near-foundation stabilization in desert buildings. |
| | Biological system | Sporosarcina pasteurii spores as the primary mineralizing eubacterial system; alternative candidate: Bacillus subtilis mineralizing biofilm-forming strain. | Sporosarcina pasteurii as the recommended mineralizing eubacterium; alternative: Bacillus subtilis mineralizing/biofilm-forming strain. | Sporosarcina pasteurii deployed primarily as dormant spores; alternative: mineralizing biofilm-forming Bacillus subtilis strain. |
| | Functional role | Moisture-triggered germination, biofilm formation, $CaCO_3$ biomineralization, crack bridging, and grain-to-grain biocementation. | Biofilm formation, inorganic particle anchoring/interfacing, and biomineralization-driven generation of a ceramic extracellular sealing phase. | Moisture-triggered reactivation, interfacial biofilm formation, extracellular $CaCO_3$ biomineralization, crack bridging, and grain-to-grain biocementation. |
| | Material system | Calcite-dominant living ceramic composed of bacterially precipitated $CaCO_3$ integrated with bacterial ECM, optionally incorporating native silica-rich fines from the treated substrate. | Predominantly calcium carbonate living ceramic extracellular matrix; alternative calcium phosphate/apatite-like mineral phase. | Calcite-dominant living ceramic composed of bacterially precipitated $CaCO_3$ integrated with bacterial ECM, with optional trapping/bridging of native silica-rich fines in sanded or dusty mineral substrates. |
| | Structural / geometric instantiation | Fractal multiscale network filling irregular crack branches and near-surface sand pores, with mineral bridges spanning ~20–200 µm per activation cycle across defect pathways. | Fractal ramified biofilm-mineral network occupying 10–800 µm branched crack paths and 100–500 µm sand-grain pore networks, with 10–80 µm biofilm coatings and localized mineral bridges. | Fractal multiscale network coating crack walls and near-surface pore throats, with hierarchical mineral bridges spanning micron-to-submillimetric gaps across irregular defect branches. |
| | Target substrate / environment | Cracked earthen or mineral building envelopes, façade defects, and wind-erodible desert perimeter sands near foundations. | Near-surface cracks in earthen, mortar, or cementitious desert buildings and adjacent wind-erodible sand around foundations or walls. | Cracked earthen or mineral building envelopes, porous façade defects, mortar/render discontinuities, and wind-erodible desert perimeter sands near foundations. |
| | Application / deployment scenario | Dry powder, slurry, or infiltrant applied to defected building/sand interfaces and activated by brief wetting or rare moisture events for autonomous ceramic repair and stabilization. | Dry powder patch, injectable mineralizing grout, or surface/perimeter stabilization treatment activated by intermittent wetting events. | Dry powder, aqueous infiltrant, or brushable mineralizing treatment applied to mineral defects and activated by brief wetting cycles using low-volume water delivery. |
| | Production mechanism / workflow | Spore fermentation, dry blending with mineralization salts, field infiltration onto inorganic substrates, biofilm anchoring, and in situ biomineralization during wet-dry activation cycles. | Fermentation of spore-forming bacteria, dry formulation with mineralizing precursors, field application into defect networks, wetting-triggered biofilm establishment, followed by repeated in situ biomineralization cycles. | Spore fermentation and stabilization, low-volume substrate prewetting, bacterial attachment/biofilm establishment on native inorganic surfaces, followed by repeated calcium/mineralization pulses and controlled drying cycles. |
| | Main functional outcome / performance objective | Reduced crack connectivity and permeability, increased near-surface cohesion and erosion resistance, and persistent low-maintenance repair under intermittent desert moisture. | Reduction of connected crack permeability, increased substrate cohesion, enhanced sand-grain cementation, and lower wind-erosion-induced material loss. | Persistent crack sealing, reduced permeability, increased sand cohesion and erosion resistance, and formation of a self-grown mineral repair layer without synthetic scaffolds. |
| **ARC–HYB–ARA–3D–LCE–SLB+BRE–CMM** | One-line ELM identity | Haloarchaeal biofilm–colonized porous ceramic (3D) cartridge for hot, saline building greywater polishing and bioremediation via stable archaeal metabolism and retention within a swappable reactor module. | Haloarchaeal biofilm–colonized porous ceramic composite monolith cartridge for saline greywater polishing as a replaceable smart-building bioremediation module. | Haloarchaeal biofilm–colonized porous living-ceramic (3D) cartridge that reduces soluble organics in hot saline greywater as a swappable smart-building bioremediation module. |
| | Biological system | Primary: Haloferax volcanii perfusion-colonized biofilm within ceramic porosity. Alternative: Halobacterium salinarum for extreme-brine robustness. | Haloferax volcanii (recommended). Alternative: Halobacterium salinarum immobilized as internal biofilms within ceramic porosity. | Primary: Haloferax volcanii biofilm within ceramic porosity. Alternative: Halobacterium salinarum for higher brine robustness. |
| | Functional role | Cell-centric removal of soluble organics (COD fraction) via archaeal uptake/metabolism; EPS/biofilm formation for retention and particulate/microdroplet capture within pores. | Aerobic saline-tolerant heterotrophic polishing of dissolved organics (DOC/COD/TOC proxies), nutrient assimilation into biomass with controlled biomass turnover; stable biofilm-mediated retention in porous ceramic. | Cell-centric COD reduction via archaeal uptake/metabolism; EPS/biofilm retention for biomass stability and microdroplet/particulate capture in pores. |
| | Material system | Living ceramics (LCE): porous $Al_2O_3$ monolith/foam (baseline) or SiC open-cell foam (premium thermal/abrasion robustness). | Porous alumina ($\alpha$-$Al_2O_3$) ceramic composite monolith (preferred); alternative: cordierite ($2MgO\cdot2Al_2O_3\cdot5SiO_2$) honeycomb with porous walls. | Living ceramic scaffold: porous $\alpha$-$Al_2O_3$ honeycomb/foam (baseline) or open-cell SiC foam (premium thermal/abrasion robustness). |
| | Structural / geometric instantiation | 3D flow-through cartridge: honeycomb channels 0.8–2.0 mm + interconnected colonization pores 50–300 µm (optionally graded: 200–500 µm upstream; 50–150 µm downstream). | 3D stacked monolith cartridge: macro-channels ~0.8–1.5 mm; wall open porosity ~35–55%; pore throats ~10–80 µm enabling protected internal colonization with flow-through transport. | 3D flow-through cartridge with 0.8–2.0 mm channels + interconnected 50–300 µm colonization pores; target open porosity 60–80% (optionally graded upstream-to-downstream). |
| | Target substrate / environment | Hot-to-warm, saline/brackish greywater streams in desert-building service cores; repeated wet–dry and salinity fluctuations managed by buffered recirculation. | Hot, saline/brackish greywater stream in desert-building service cores (wet–dry duty cycles, high fouling potential). | Hot-to-warm, saline/brackish building greywater streams (service-core plumbing loop). |
| | Application / deployment scenario | SLB+BRE: swappable building-service greywater polishing cartridge installed after coarse filtration; sealed housing with downstream retention barrier. | Replaceable building-service cartridge downstream of coarse filtration; conductivity-controlled operation, periodic backflush and swap schedule; sealed housing with outlet microfiltration containment. | Replaceable cartridge in smart-building greywater line after coarse screening; conductivity/temperature monitored; periodic brine backflush and scheduled swap-out. |
| | Production mechanism / workflow | CMM: composite-style porous/graded ceramic scaffold fabrication (preform + slurry infiltration + sintering) followed by perfusion seeding and staged commissioning under controlled brine conditions. | CMM: ceramic-fiber preform + slurry infiltration (functionally graded porosity) → burnout/sinter → scaffold sterilization → controlled archaeal seeding/recirculation colonization → cartridge assembly + QC. | CMM-consistent: ceramic fiber preform formation → ceramic slurry infiltration + fugitive porogens → debind/partial sinter → sterilize → archaeal seeding in recirculating brine bioreactor → sealed cartridge packaging. |
| | Main functional outcome / performance objective | Stable, contained archaeal colonization enabling sustained organic polishing with controlled pressure drop and minimal cell escape over a defined service interval. | Stable, contained greywater polishing at service-core flow with measurable organic-load reduction while maintaining acceptable pressure drop and preventing viable-cell release. | Stable, contained archaeal biomass enabling measurable COD reduction with controlled pressure drop and minimal cell release to effluent. |
| **CRD–HYB–CRK–CP–LCO–SLB–CMM+BMN** | One-line ELM identity | Cross-domain, cross-kingdom bacterial–fungal living composite façade render that self-heals by biomineral crack sealing and suppresses dust/sand erosion in smart-building envelopes. | Encapsulated bacterial–fungal hybrid living composite render that episodically biomineralizes and binds sand at façade microcracks, enabling self-healing and dust suppression for sand-exposed smart building skins. | Cross-domain bacterial–fungal hybrid living composite façade render that seals microcracks via localized biomineralization and suppresses dust/erosion through fungal EPS-mediated surface binding in smart-building envelopes. |
| | Biological system | Recommended: Sporosarcina pasteurii (biomineralizing bacterium) + Aureobasidium pullulans (adhesive/EPS-producing fungus). Alternative: Lysinibacillus sphaericus + Trichoderma harzianum. | Recommended: Sporosarcina pasteurii (spores, encapsulated) + Aureobasidium pullulans (EPS/adhesion, constrained microvoid colonization). Alternative: Bacillus subtilis (spores) + Trichoderma reesei (filamentous reinforcement, higher containment burden). | Candidate 1 (recommended): Sporosarcina pasteurii (spores; mineralizing bacterium) + Aureobasidium pullulans (EPS/adhesive fungus). Alternative: Lysinibacillus sphaericus (spore-forming, mineral-associated) + Trichoderma harzianum (filamentous colonizer). |
| | Functional role | Bacteria: $CaCO_3$ precipitation for crack sealing + sand-grain cementation. Fungi: EPS-driven dust binding + filament/hyphal crack spanning + adhesion to composite microtexture. | Bacterial ureolysis-driven $CaCO_3$ biomineralization for crack sealing + sand-grain cementation; fungal EPS/biomass for adhesion, retention, and microcrack bridging, stabilizing fines and supporting mineral nucleation. | Bacteria: moisture-triggered activity → mineral precipitation at crack tips/pores (sealing/stiffening). Fungi: EPS-driven dust binding, adhesion to microtexture, and filament/hyphal spanning that stabilizes repair zones and reduces erosion-driven particle loss. |
| | Material system | Polymer-modified mineral render composite: lime/$CaCO_3$–silica fillers + waterborne polymer modifier + microfibers; designed as a porous colonizable scaffold (living composite). | Breathable siloxane–acrylic polymer–mineral composite render: silica sand + silica fume + reinforcement microfibers + polymer–silica microcapsules carrying spores and finite nutrient/Ca reservoirs; optional mineral-rich geopolymer–latex variant. | Candidate 1 (recommended): polymer-modified mineral render composite (lime/$CaCO_3$ + silica filler + acrylic/SBR latex + microfibers + buffered hydrogel microcapsules). Alternative: cementitious–polymer composite with micro-silica + cellulose microfibers + porous lightweight fillers as protected cell refugia. |
| | Structural / geometric instantiation | CP multi-scale: 2–5 mm 2D coating; 10–25% interconnected 3D porosity (20–200 µm throats); 0D spore microcapsules (50–300 µm); 1D microfiber + fungal hyphal reinforcement; surface microtexture 50–500 µm. | CP multi-scale architecture: 2D render skin (3–6 mm) + 3D pore network (15–30% porosity; 20–150 µm throats) + 0D capsules (50–300 µm) as living microreactors + 1D reinforcement (basalt microfibers + constrained fungal biomass in microvoids/cracks). | CP multi-scale: 2–5 mm 2D coating + 3D interconnected porosity (10–25%, throats ~20–200 µm) + 0D microcapsule reservoirs (100–500 µm) + 1D microfibers (3–12 mm) enabling crack deflection, hyphal anchoring, and localized mineral bridging. |
| | Target substrate / environment | Desert-building exterior façades and sand-exposed perimeter surfaces (thermal cycling, abrasion, episodic humidity). | Sand-exposed exterior façades (masonry/earthen envelopes) under wind abrasion, dust loading, thermal cycling, and episodic moisture (desert/peri-desert buildings). | Exterior building envelopes and façade renders exposed to sand/dust abrasion and wet–dry cycling (desert/perimeter conditions). |
| | Application / deployment scenario | Serviceable façade “living render skin” with condition-based activation (brief misting cycles) to | Serviceable living façade render skin applied by trowel/spray; activated by natural moisture events or | Serviceable façade “living skin” activated by periodic controlled wetting; aims to maintain barrier integrity (reduced permeability) and reduce dust release/erosion |
| | | trigger mineral sealing and maintain low dust emission and surface cohesion. | scheduled misting; functions as sacrificial outer layer to reduce dust release and self-heal microcracking. | while remaining compatible with building maintenance practices (SLB). |
| | Production mechanism / workflow | CMM: composite scaffold manufacture + curing → post-colonization by spray-inoculation and humid anchoring → drying to dormancy. BMN: periodic low-dose mineralization feed to precipitate $CaCO_3$ in cracks/pores. | CMM+BMN: composite formulation with graded fillers + low-shear incorporation of living microcapsules; on-wall curing; episodic moisture-triggered biomineralization for localized crack/sand binding. | CMM+BMN: composite render manufacturing + embedding buffered living microcapsules → coat/cure → post-cure priming colonization → controlled biomineralization activation cycles → breathable protective finish; QC on porosity, adhesion, viability, and crack-healing performance. |
| | Main functional outcome / performance objective | Reduce dust emission and wind erosion while achieving repeatable microcrack closure (50–300 µm) and measurable recovery of surface cohesion/abrasion resistance after activation cycles. | Reduced wind-driven erosion and dust emission + repeated microcrack sealing and sand-grain cementation over wetting cycles, maintaining façade integrity and surface cohesion. | Repeatable microcrack sealing (permeability reduction), erosion/dust suppression under abrasion, and retention of façade mechanical/adhesive integrity across multiple wet–dry activation cycles. |

Table 9. Use case 2 prompt F output: generated images for ELM final candidates using different styles.

| ***Iterations*** | ***ELM candidate 1: BAC–BIO–EUB–FR–LCE–SLB–BFF+BMN***<br>*Style: Futuristic hyper-real 3D (general / sci-fi)* | ***ELM candidate 2: ARC–HYB–ARA–3D–LCE–SLB+BRE–CMM***<br>*Style: Scientific schematic* | ***ELM candidate 3: CRD–HYB–CRK–CP–LCO–SLB–CMM+BMN***<br>*Style: Realistic photoreal 3D* |
|---|---|---|---|
| **1** | 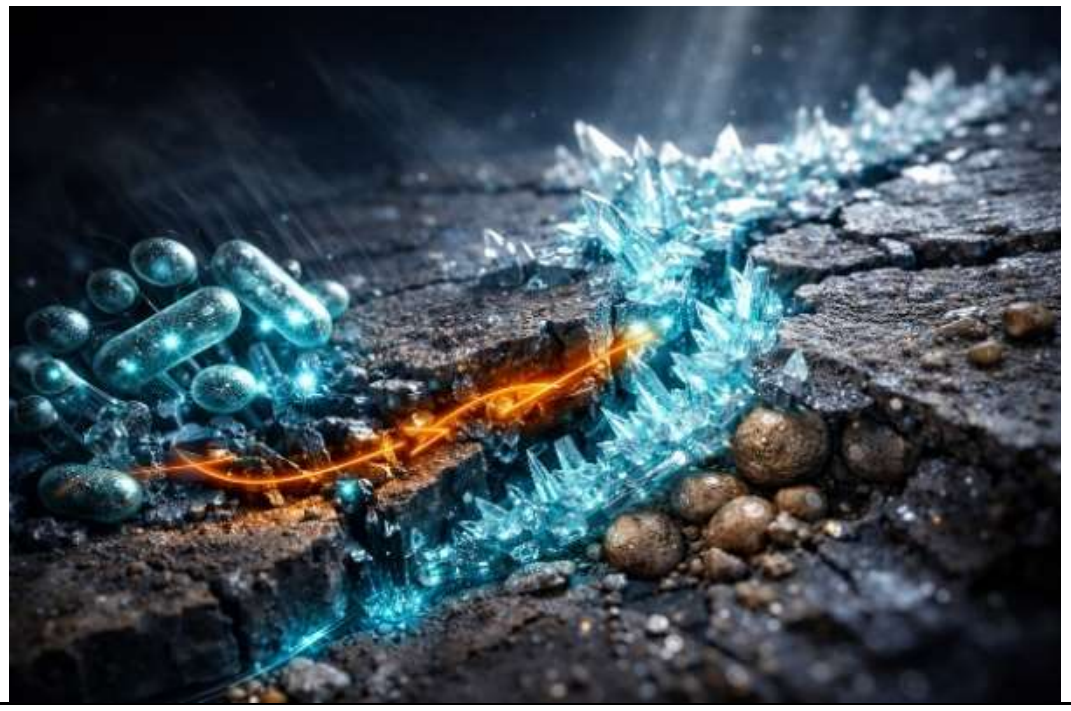 | 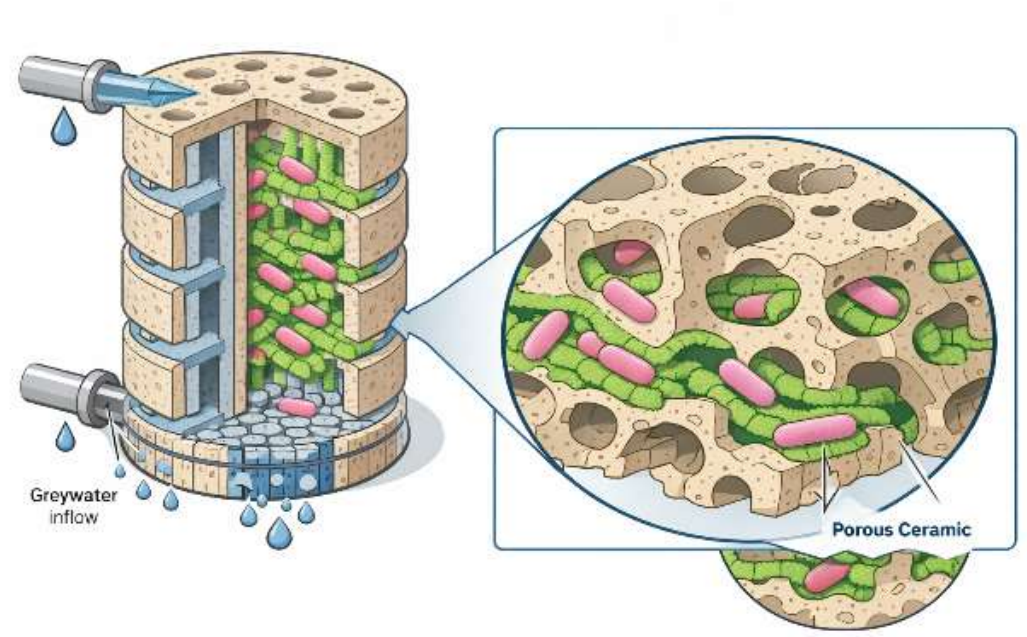 | 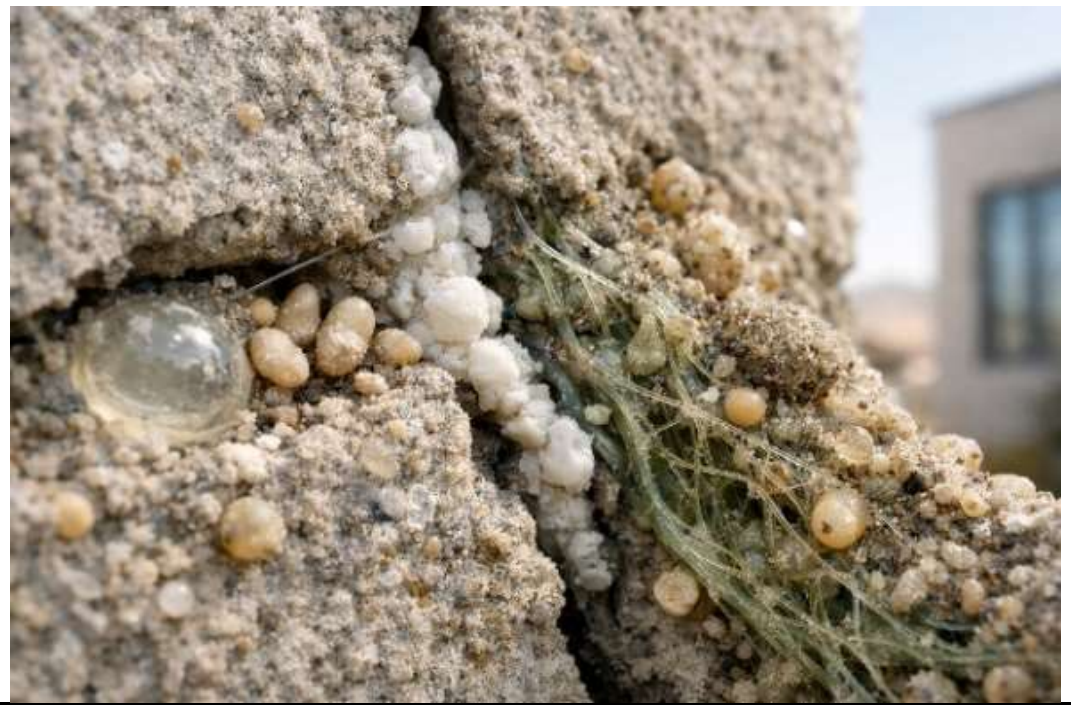 |
| **2** | 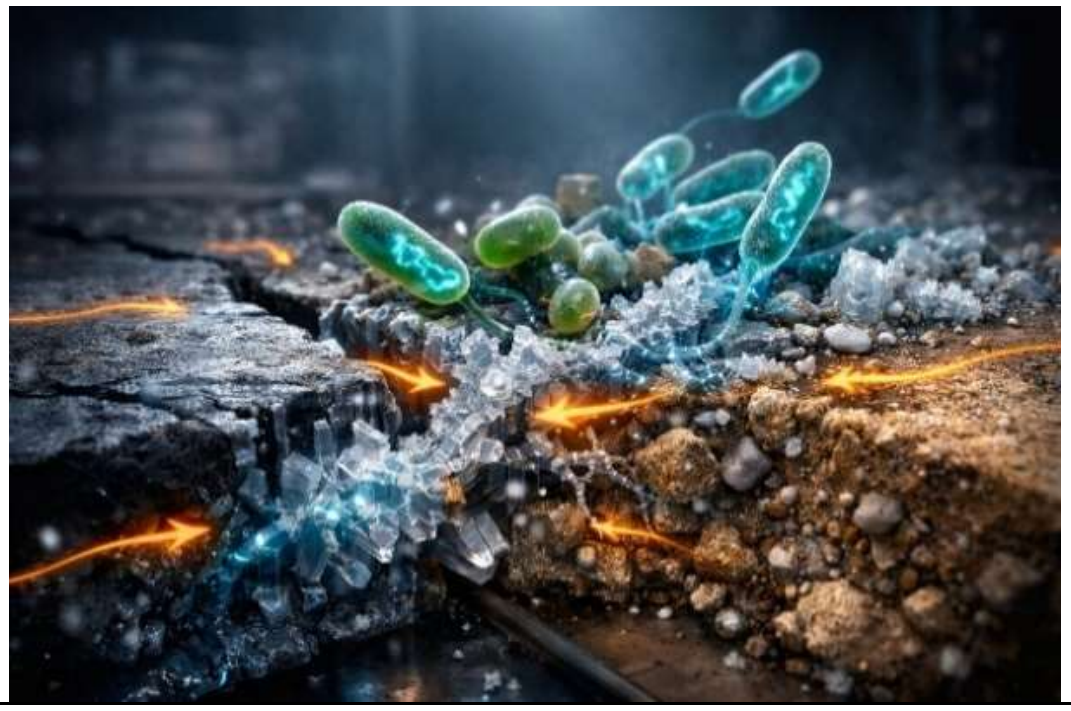 | 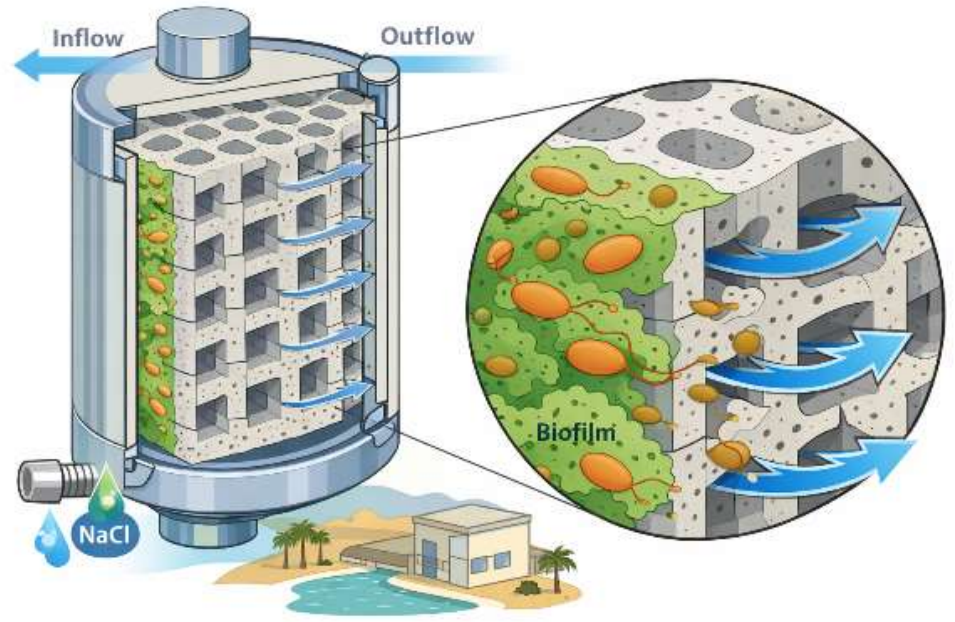 | 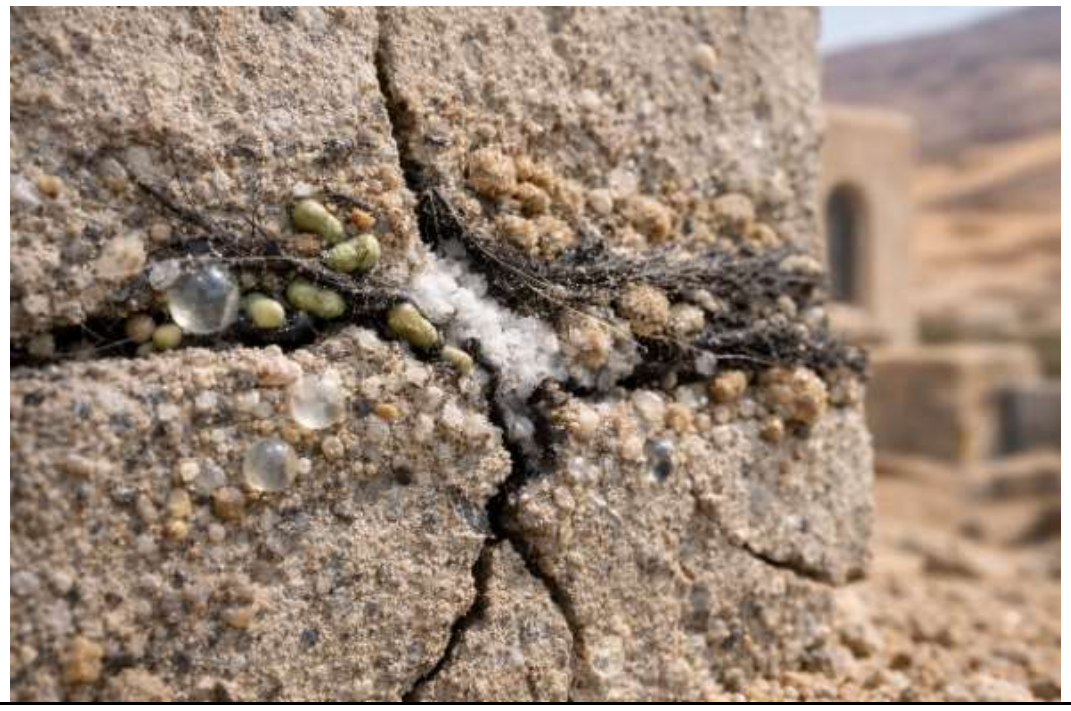 |
| **3** | 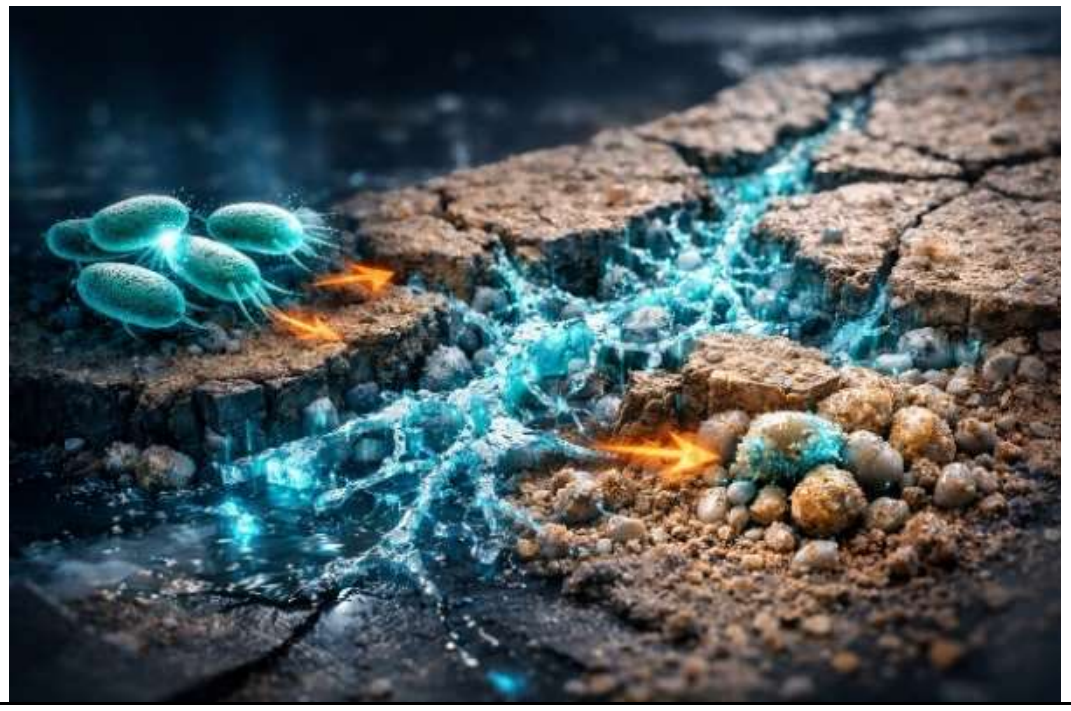 | 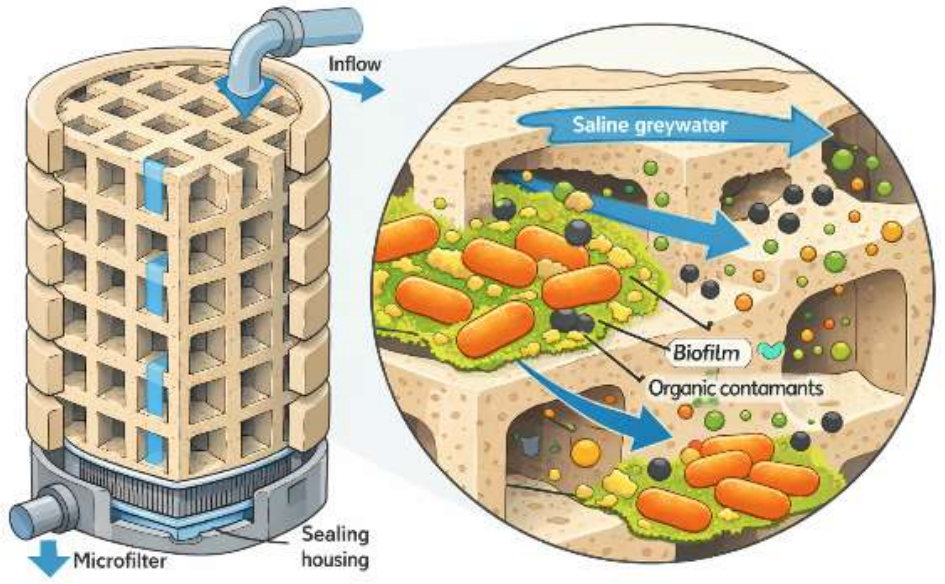 | 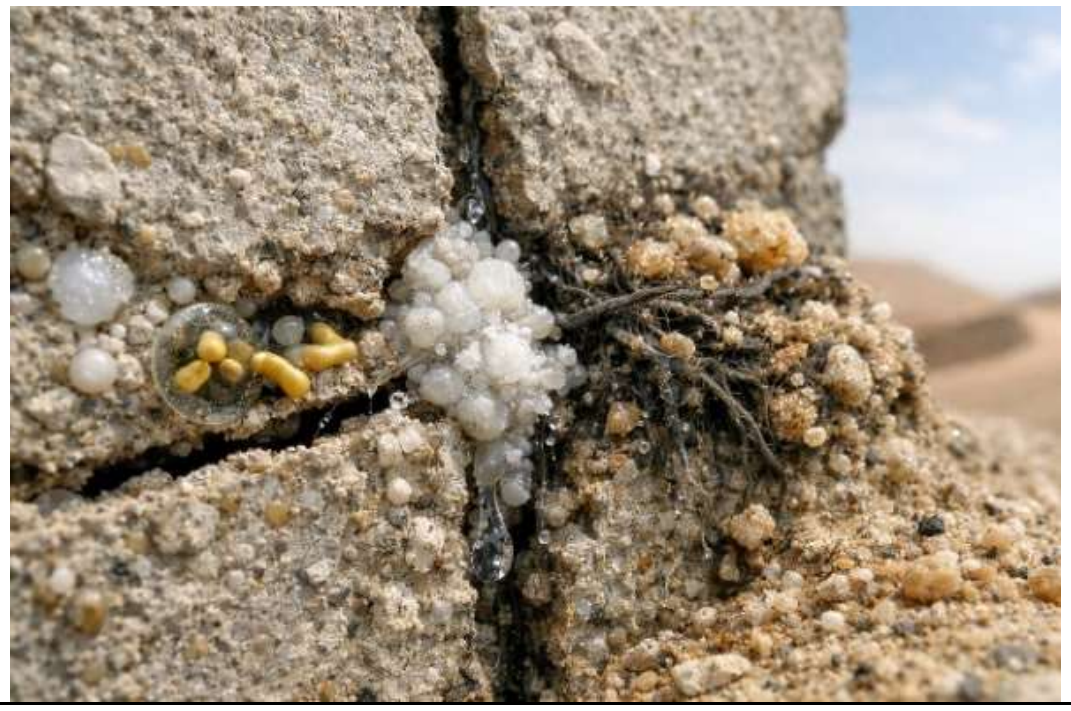 |

### 4.3. Use case 3: ELM design for target application domain 2 via workflow comparison

Use case 3 addresses ELM design for target application domain 2 (long-term colonization of Titan) by executing workflow D, which consists of an application-conditioned retrieval from the database followed by downstream design generation. In this case, candidate ELMs already represented in the database are selected according to their relevance to the boundary conditions and functional requirements of the Titan application domain (Pipeline D), and their corresponding 7-slot ontological codes are then passed to Pipeline E for full specification. Workflow D operates on the initial dataset from bibliography annotations.

For use case 3, the same cascaded multi-iteration strategy adopted in use case 2 was followed. First, Pipeline D was executed in three independent iterations. Based on the outcomes of these runs, three ELM candidates were selected, with one candidate chosen from each iteration (in this case the ELM candidate 1 of every iteration). Subsequently, Pipeline E was applied to each of the three selected ELMs, again using three independent iterations per ELM input. From these outputs, the first iteration of each candidate was selected for the final stage. Prompt F was conducted on the chosen final candidates also under three independent iterations to support the final generation and comparative assessment of the resulting designs (Table 12). This sequential implementation ensures methodological consistency with use case 2 while enabling the exploration of robustness, convergence, and diversity of outcomes in use case 3.

Table 10 and Table 11 summarize the generative AI responses for use case 3, while the complete model outputs are provided in Supplementary Material S6 (S6_1 providing the generative AI output following prompt D and S6_2 giving the generative AI response to prompt E). As output of workflow D, three ELMs fully instantiated are proposed for target application domain 2:

- ELM candidate 1: BAC–HYB–EUB–3D–LPO–CAT+BTP–OPM
- ELM candidate 2: BAC–HYB–EUB–3D–LCE–SPE+SLB–BMN
- ELM candidate3: BAC–HYB–EUB–2D–LCA–SPE+ENP+BTP+CAT–BFF

Further assessment is intended to examine whether application-conditioned selection from the database, when combined with downstream design generation, yields plausible and informative candidate solutions under the imposed boundary conditions. In any case, again, the generative AI is able to propose highly innovative engineered living materials (a total of 9 in the first stage if we consider the 3 iterations for the 3 candidates of Table 10), even if working upon and already existing type of ELM from the database in this third use case. Also in this case, a justification of their biological and structural plausibility is provided and indications towards their technological materialization are given. In the second stage (results shown in Table 11), for the selected candidates, descriptions with a remarkable level of detail and justifying reliability are provided, toward the practical implementation of these ELMs. Both qualitative, as well as extremely precise quantitative indications are provided for their engineering design, materialization and in-service operation. The accuracy of the descriptions and the well-justified design decisions demonstrate that the generative AI, after being trained with our described ontology and knowledge graph and guided with our collection of prompts, constitutes a remarkable companion, not only for the discovery and co-creation of ELMs, but also for their reinvention or reengineering towards practical application in new environments and aiming at increased societal impacts. The detailed information of these breakthroughs is provided in Tables 10 and 11, in supplementary materials S6_1 and S6_2, and the resulting ELMs schematically illustrated in Table 12.

Table 10. Use case 3 Pipeline D output: app-conditioned retrieved ELM candidates across different runs.

| ***ELM candidate*** | ***Component*** | ***Iteration 1*** | ***Iteration 2*** | ***Iteration 3*** |
|---|---|---|---|---|
| **ELM candidate 1** | Name | Organic-solvent biocatalyst hydrogel for hydrocarbon ISRU chemistry. | Self-healing bacterial concrete for habitat structures. | Bioelectrochemical graphene living carbon interfaces (energy + electrosynthesis support). |
| | Origin & selection rationale (dataset-derived) | Graph candidate explicitly targets biocatalysis in pure organic solvents, which is unusually aligned with Titan's hydrocarbon-dominant environment, provided operation occurs in a shielded, temperature-controlled cartridge. | Selected because it directly supports low-maintenance, long-duration infrastructure integrity under remote operation constraints by enabling crack self-healing inside a controlled habitat envelope. | Graphene biofilms with electroactive bacteria for microbial fuel cells / electrosynthesis map cleanly onto Titan needs: very low solar energy pushes toward electrochemically driven, closed-loop systems for waste-to-energy and carbon recycling. |
| | Completed ontology code | BAC–HYB–EUB–3D–LPO–CAT+BTP–OPM | BAC–HYB–EUB–3D–LCE–SPE+SLB–BMN | BAC–HYB–EUB–2D–LCA–SPE+ENP+BTP+CAT–BFF |
| | Application-oriented concept | A sealed alginate hydrogel biocatalyst module hosting genetically modified Escherichia coli to catalyze targeted conversions of hydrocarbon-derived feedstocks into useful precursors for habitat/ISRU chemical loops under Titan-relevant handling constraints. | A sealed, habitat-integrated self-healing concrete module where calcite-precipitating bacteria are activated only during repair cycles, minimizing water/energy use while extending structural lifetime. | A modular, aqueous bioelectrochemical surface (graphene + bacterial biofilm) used inside habitat processing loops to support energy recovery and electrically assisted bioconversion. |
| | Justification | • Biological suitability: E. coli is stated as genetically modified and embedded in alginate hydrogel, enabling controlled biocatalysis while physically confining cells.<br>• Titan boundary conditions: Direct compatibility with hydrocarbon media (entry states "pure organic solvents"); operation still assumed inside a thermally managed unit (cryogenic exterior otherwise incompatible with active metabolism).<br>• Water constraint: Hydrogel provides a microaqueous niche while the external process stream can be hydrocarbon-rich; this supports Titan deployment where bulk liquid water is not available at the surface.<br>• Low-maintenance integration: Cartridge-style, replaceable biocatalyst bed fits remote operation + limited resupply (swap modules vs. servicing open cultures).<br>• Containment note: Genetic safeguard specifics are Not present in the ontology/graph; only physical encapsulation is directly supported by the entry. | • Boundary conditions (cryogenic, hydrocarbons, no surface water): operates inside shielded/controlled subsystems (habitat walls / internal structural layers), not exposed as an open surface biota.<br>• Low-maintenance / autonomy: healing can be event-triggered (crack detection → activation), matching severe resupply constraints.<br>• Material suitability: living ceramic (LCE) + 3D structural integration aligns with load-bearing infrastructure.<br>• Resource efficiency: uses intermittent, localized activation, compatible with recycled water loops (no reliance on surface water).<br>• Bio-containment: can be implemented as encapsulated micro-reservoirs within the concrete matrix (specific containment strategy details: Not present in the ontology/graph). | • Biological suitability: biofilm-based operation (BFF) enables immobilization and containment, improving robustness and low-maintenance behavior.<br>• Functional relevance: supports energy handling (ENP) and catalysis/bioprocessing (CAT/BTP) in closed-loop systems—useful where solar is scarce.<br>• Titan boundary conditions: placed in shielded, temperature-controlled wet loops; does not require Titan surface water.<br>• Technological feasibility: graphene provides a defined living carbon (LCA) conductive scaffold; biofilm interfaces are compatible with modular reactors and replacement cartridges.<br>• Explicit identifiers: material (graphene) is explicit; electroactive bacterial species is not present in the ontology/graph. |
| **ELM candidate 2** | Name | Expandable yeast–hydrogel living polymer for internal insulation/filler + recycling linkage. | Archaeal intracellular thermoplastic synthesis for ISRU-compatible parts. | Waste-to-material expansion biocomposite (Saccharomyces cerevisiae in acrylic hydrogel). |
| | Origin & selection rationale (dataset-derived) | Graph candidate explicitly describes embedded Saccharomyces cerevisiae in a synthetic acrylic hydrogel that yields a dramatic dry mass and volume increase, which can be re-used as expandable filler/packing/insulation inside controlled habitat layers. | Selected because it supports in-situ biomanufacturing of thermoplastics (PHA/PHB)—valuable for repairs, consumables, and modular components under resupply constraints. | Embedding Saccharomyces cerevisiae in an acrylic hydrogel matrix grown using bread-waste–derived media is directly aligned with Titan's resource efficiency and waste cycling constraints, while producing usable bulk biocomposite mass. |
| | Completed ontology code | EUK–HYB–FUN–3D–LPO–MAP+BRE–CMM | ARC–BIO–ARA–3D–LPO–SPE+MAP–ICS | EUK–HYB–FUN–3D–LPO–SPE+MAP+BRE+BTP–CMM |
| | Application-oriented concept | A sealed growth-and-lock insulation/filler cartridge where S. cerevisiae proliferates within an acrylic hydrogel to generate bulk living polymer material for interior panel infill, vibration damping, or void-filling in modular habitat components. | A contained archaeal bioprocessing module producing biopolymer thermoplastics for fabrication/repair workflows within Titan habitats and robotic outposts | A contained biomanufacturing panel/cassette that converts habitat organic waste streams into expanded yeast–hydrogel biocomposite material for low-criticality parts, packing, or insulation layers inside the habitat. |
| | Justification | • Biological suitability: Saccharomyces cerevisiae and the synthetic acrylic hydrogel matrix are explicitly stated; hydrogel format supports physical containment and process control.<br>• Titan boundary conditions: Growth must occur in a controlled, warmed compartment; the produced bulk material can then be integrated into multi-layer panels that face cryogenic gradients.<br>• Closed-loop relevance: The entry's growth is linked to "media derived from bread waste"; Titan-specific feedstock substitution is Not present in the ontology/graph, but the waste-to-material framing motivates BRE for habitat waste conversion pathways.<br>• Resource efficiency: If coupled to internal waste streams, can reduce imported polymers (details of conversion efficiency are Not present in ontology/graph).<br>• Maintenance: Modular panels/cartridges match remote operation constraints. | • Cryogenic exterior: run as a thermally managed bioreactor inside habitat/ISRU bays; no claim of direct Titan-surface activity.<br>• Long-duration robustness: archaea can be selected for tolerance to operational stressors (specific extremophiles: Not present in the ontology/graph).<br>• Material suitability: direct route to living polymer (LPO) outputs (thermoplastic feedstock for printing/molding downstream).<br>• Resource efficiency: can, in principle, couple to recycled carbon streams (exact substrates: Not present in the ontology/graph).<br>• Containment: intracellular production supports closed harvesting/processing with reduced risk of environmental release. | • Biological suitability: Saccharomyces cerevisiae is explicitly identified and well-suited to controlled, contained cultivation.<br>• Functional relevance: combines waste treatment (BRE) with materials production (MAP), supporting long-duration autonomy.<br>• Titan boundary conditions: operates in internal controlled subsystems; avoids dependence on Titan's external environment and mitigates maintenance via modular cartridges.<br>• Technological feasibility: matches CMM (composite manufacturing / colonization in a synthetic matrix); scalable as low-complexity reactors with predictable inputs/outputs.<br>• Practical integration: best for non-load-critical materials and internal habitat use where bio-containment is straightforward. |
| **ELM candidate 3** | Name | Yeast-in-porous sponge biocatalyst cartridges for closed-loop synthesis. | Spatially patterned bacterial cellulose matrices for membranes + liners. | Flexible bacterial-cellulose ELM platform (symbiotic bacteria + yeast) for membranes/sensing/biocatalysis. |
| | Origin & selection rationale (dataset-derived) | Graph candidate explicitly provides a platform material (chitosan sponges) hosting metabolically active S. cerevisiae "with diverse applications as biocatalysts," aligning with Titan's need for compact, modular bioprocess units inside habitats. | Selected because it enables on-demand fabrication of structured polymer matrices (membranes, gaskets, liners, packing materials) that reduce imported consumables. | A synergic symbiotic culture of bacteria and yeast producing bacterial cellulose-based ELMs is compelling for Titan because cellulose sheets can serve as low-mass membranes, carriers, and functional interfaces in closed-loop processing units. |
| | Completed ontology code | EUK–HYB–CRK–3D–LPO–BTP+CAT–PSP | BAC–BIO–EUB–3D–LPO–SPE+MAP–SBP | CRD–HYB–CRK–3D–LPO–SPE+MAP+BSE+CAT–BFF |
| | Application-oriented concept | A replaceable porous biocatalyst cartridge (chitosan sponge) hosting S. cerevisiae to perform selected biosynthesis/bioconversion steps that support habitat logistics (e.g., chemical intermediates, processing aids) under sealed operation. | A contained bacterial patterning process producing cellulose strands/matrices with controlled geometry for functional membranes and structural polymer inserts in habitat subsystems. | A contained production module that grows bacterial-cellulose living polymer sheets used as functional membranes/interfaces for biosensing and biocatalysis inside habitat/ISRU process skids. |
| | Justification | • Biological suitability: S. cerevisiae explicitly stated; porous sponge scaffold supports high surface area and immobilization—good for stable bioprocessing. | • Low solar & cryogenic outside: production occurs in indoor controlled units, then materials are deployed as abiotic components. | • Biological suitability: biofilm-mediated growth (BFF) supports immobilization and sheet formation while improving containment and stability. |
| | | • Titan boundary conditions: Requires a shielded, aqueous/temperate bioprocess bay; not intended for direct exposure to Titan surface conditions.<br>• Operational feasibility: Sponge-cartridge format supports low-maintenance swaps, consistent with limited resupply.<br>• Containment: Physical immobilization in a sponge aids biological containment; genetic containment is Not present in the ontology/graph.<br>• Chemistry note: Specific target reactions and Titan feedstocks are Not present in the ontology/graph; the entry supports the platform capability. | • Manufacturability: spatial patterning supports repeatable geometries (important for sealing/filtration parts).<br>• Resource efficiency: cellulose-like matrices can reduce dependence on imported polymers (exact yields/feeds: Not present in the ontology/graph).<br>• Bio-containment: production is compatible with closed culture + post-processing, limiting biological release into habitat spaces.<br>• Integration: outputs can serve as interfaces between mechanical systems and fluid loops (water, air, process gases). | • Functional relevance: cellulose ELMs can be deployed as membranes/coatings/carriers for biosensing (BSE) and biocatalysis (CAT) in closed loops.<br>• Titan boundary conditions: designed for internal, water-managed operation; supports remote maintenance through replaceable membrane cartridges.<br>• Technological feasibility: bacterial cellulose is a well-defined living-polymer matrix compatible with modular fabrication and functionalization workflows.<br>• Explicit identifiers: product matrix (bacterial cellulose) is explicit; the specific bacterial and yeast species are not present in the ontology/graph. |

Table 11. Use case 3 Pipeline E output: full instantiation of retrieved ELM candidates across different runs

| *ELM candidate code* | *Component* | *Iteration 1* | *Iteration 2* | *Iteration 3* |
|---|---|---|---|---|
| **BAC–HYB–EUB–3D–LPO–CAT+BTP–OPM** | One-line ELM identity | Sealed 3D hydrogel-packed hybrid living polymer cartridge embedding engineered eubacteria as a replaceable whole-cell biocatalyst module for solvent-phase ISRU chemical conversions. | 3D packed-bed alginate hydrogel cartridge containing encapsulated eubacterial whole-cell biocatalysts for sealed, replaceable catalytic bioprocessing modules that convert hydrocarbon-derived intermediates into ISRU-relevant chemical precursors. | Sealed 3D polymer-hydrogel hybrid packed-bed cartridge hosting immobilized eubacteria as a replaceable whole-cell biocatalyst module for catalytic conversion steps in hydrocarbon process loops. |
| | Biological system | Primary: genetically modified Escherichia coli (K-12 derivative) as immobilized resting-cell biocatalyst; Alternative: Pseudomonas putida KT2440 as solvent-tolerant chassis. | Genetically modified Escherichia coli whole-cell biocatalyst (primary); alternative: engineered Pseudomonas putida KT2440 for higher solvent-stress tolerance. | Recommended: engineered E. coli (K-12 derivative) as immobilized resting-cell biocatalyst; Alternative: solvent-tolerant Pseudomonas putida KT2440. |
| | Functional role | Selective bioconversion of hydrocarbon-derived feedstocks to targeted chemical intermediates under flow-through operation; activity maintained in microaqueous hydrogel niches. | Enzyme-module expression + intracellular cofactor regeneration to drive selective biotransformations of pre-activated hydrocarbon-derived intermediates in a sealed bioprocess step. | Whole-cell catalysis of a defined biotransformation step (redox/functionalization) on organic-phase substrates with aqueous microdomain cofactor support. |
| | Material system | Calcium alginate hydrogel beads (primary); PEGDA hydrogel beads/monolith (alternative) within solvent-compatible housing and a cell-retentive membrane barrier. | $Ca^{2+}$-crosslinked alginate hydrogel (2–3 wt%) as living-polymer matrix; optional stabilizing outer coat to limit leakage/abrasion. | Primary: $Ca^{2+}$-crosslinked alginate hydrogel beads; Alternative: PEGDA photocrosslinked hydrogel monolith. |
| | Structural / geometric instantiation | 3D packed-bed cartridge: 0.8–1.2 mm hydrogel beads; diffusion path ≤200–300 µm to active cells; modular cylindrical reactor (lab-to-industrial by numbering-up). | 3D packed-bed of 0.8–1.5 mm hydrogel beads in a 0.1–2 L replaceable cartridge with flow distributors and <200 µm retainers. | 3D packed-bed cartridge (beads 0.8–1.2 mm; diffusion length target ≤200–300 µm) with outlet cell-retentive membrane; alternative channeled monolith reactor. |
| | Target substrate / environment | Hydrocarbon-rich solvent process stream handled in a sealed, temperature-controlled ISRU unit (external Titan environment isolated). | Hydrocarbon-dominant ISRU process streams handled in a thermally controlled, sealed unit; microaqueous niche maintained inside hydrogel. | Hydrocarbon-rich organic process stream handled in a sealed, temperature-controlled ISRU/habitat unit with internal aqueous microdomains. |
| | Application / deployment scenario | Replaceable biocatalyst cartridge in a Titan habitat/ISRU chemistry skid enabling catalysis + bioprocessing in hydrocarbon media with internal aqueous microenvironments. | Swap-in biocatalyst cartridge integrated into an ISRU chemical skid to produce functionalized/oxygenated precursors for habitat chemical loops (CAT+BTP). | Replaceable biocatalyst cartridge in a Titan habitat/ISRU chemical skid enabling catalytic conversion + bioprocessing steps under containment. |
| | Production mechanism / workflow | OPM: strain build → culture/resting-cell prep → alginate encapsulation → bead curing/rinsing → cartridge packing + membrane integration → conditioning + QC release testing. | OPM encapsulation + cartridge assembly: strain banking → culture → alginate mixing → $Ca^{2+}$ bead gelation → packing → conditioning → QC (activity, leakage, hydraulics). | OPM: culture → resting-cell prep → hydrogel encapsulation → cartridge assembly + membrane → conditioning → QC (activity/leakage/pressure drop) → numbering-up scale-out. |
| | Main functional outcome / performance objective | High-selectivity conversion with low-maintenance cartridge swap, quantified by conversion/selectivity, space–time yield, operational half-life, and zero-leakage biocontainment performance. | High selectivity conversion with controlled containment (minimal cell escape) and predictable cartridge lifetime under continuous or cyclic operation. | Stable conversion/selectivity over extended flow-through operation with zero/near-zero cell leakage and cartridge-level swap maintenance. |
| **BAC–HYB–EUB–3D–LCE–SPE+SLB–BMN** | One-line ELM identity | Spore-forming eubacterial biomineralization microcapsules embedded in a 3D silicate-rich living-ceramic concrete that seals structural cracks on-demand in space-habitat building modules. | Spore-based eubacterial hybrid living ceramic–concrete that seals structural microcracks in sealed habitat building modules via trigger-on biomineralized $CaCO_3$ deposition for long-duration space infrastructure. | Spore-based eubacterial hybrid living ceramic concrete that executes on-demand biomineralization inside 3D habitat wall panels to seal cracks and restore barrier/mechanical performance in space-colonization smart-building infrastructure. |
| | Biological system | Primary: Sporosarcina pasteurii dormant spores (ureolytic $CaCO_3$ precipitation); Alternative: Bacillus subtilis spores (ECM-assisted mineral nucleation). | Primary: Bacillus subtilis spores (carbonic-anhydrase–driven or bicarbonate-assisted $CaCO_3$ nucleation); Alternative: Sporosarcina pasteurii spores (ureolytic MICP). | Bacillus pseudofirmus spores (recommended) as mineralizing eubacterial chassis; alternative: Sporosarcina pasteurii resting cells/spores for high-rate carbonate precipitation. |
| | Functional role | Event-triggered germination during repair cycles → localized $CaCO_3$ precipitation, crack-face | Dormant-by-default, crack-triggered biomineralization: $CaCO_3$ precipitation to bridge cracks, reduce permeability, and partially restore stiffness/strength | Event-triggered germination/resting-cell activation in crack-connected micro-niches → $CaCO_3$ precipitation, crystal nucleation on |
| | | bridging, permeability reduction, partial stiffness recovery. | | cell/EPS surfaces, crack bridging, permeability reduction |
| | Material system | Primary: cementitious silicate matrix (C–S–H–type binder + mineral aggregates) with silica/aluminosilicate brittle-shell microcapsules. Alternative: metakaolin-based aluminosilicate geopolymer with buffered microcapsules. | Ceramic-dominant cementitious concrete matrix (Portland cement + silica fume + basalt/silica aggregates; optional basalt microfibers) with embedded buffered hydrogel microcapsules containing spores + Ca-source. | Cementitious living ceramic matrix (C–S–H / silicate-dominant) with embedded porous aluminosilicate/silica ceramic micro-reservoirs carrying spores + dry mineral precursors; alternative matrix: alkali-activated geopolymer aluminosilicate. |
| | Structural / geometric instantiation | 3D load-bearing module with distributed microcapsules (150–500 µm; 1–3 vol%; 5–15 mm spacing) designed to intercept and heal 50–500 µm cracks. | 3D structural panel/layer (50–150 mm thick) with distributed 200–800 µm healing microcapsules (0.5–2 vol%, 1–3 mm spacing) + optional porous ceramic depot granules (1–3 mm) for repeat activation. | 3D precast wall panel layer (20–60 mm) containing 200–800 µm ceramic micro-reservoirs at 1–3 vol%, tuned for 0.05–0.5 mm crack sealing with ≤0.2–0.5 mm diffusion distance from crack fluid to reservoirs. |
| | Target substrate / environment | Internal, controlled habitat structural layer (shielded from external cryogenic/hydrocarbon conditions); alkaline crack microenvironment with controlled repair-fluid dosing. | Cracked cementitious/ceramic concrete inside a sealed habitat structural envelope; activation only when controlled moisture is delivered to crack zones. | Pressurized, controlled habitat structural elements (internal load-bearing ceramic/cementitious layers), operated with closed-loop water management and limited activation dosing. |
| | Application / deployment scenario | Smart/living building element for space colonization habitats: crack detection → localized fluid activation → BMN sealing → dry-down dormancy; designed for low-maintenance remote operation. | Smart/living habitat walls or internal structural layers for space colonization: event-driven self-healing under remote/low-maintenance operation constraints. | Space exploration/colonization habitat construction: smart-living building panels with autonomous crack repair cycles triggered by damage detection, minimizing maintenance and resupply. |
| | Production mechanism / workflow | Microencapsulate spores + precursors in brittle ceramic shells → mix into cementitious matrix with low-shear handling → cast/cure modules → activate BMN only upon crack event via minimal repair-fluid delivery. | Spore production → buffered $Ca^{2+}$-loaded hydrogel microencapsulation → low-shear incorporation into concrete → controlled curing → in-service moisture-triggered BMN crack sealing. | Spore production → loading into porous ceramic micro-reservoirs → mixing/casting into living ceramic concrete → controlled curing → crack-triggered activation fluid delivery → BMN-driven $CaCO_3$ deposition → verification and re-dormancy. |
| | Main functional outcome / performance objective | Repeatable on-demand crack sealing (permeability ↓, crack aperture ↓, stiffness/strength partially restored) with minimal water/energy and strong biocontainment by physical/chemical gating. | ≥10× permeability reduction and ≥50–80% partial mechanical recovery after crack events, with repeatable, localized healing cycles while maintaining baseline structural capacity. | Repeated crack closure and barrier restoration (permeability reduction + mechanical stiffness recovery) with low water/nutrient duty cycle and long dormant stability. |
| **BAC–HYB–EUB–2D–LCA–SPE+ENP+BTP+CAT–BFF** | One-line ELM identity | Planar eubacterial electroactive biofilm grown on a graphene-family carbon electrode to provide a sealed bioelectrochemical interface for energy recovery and electrically assisted bioprocess/catalysis in space-habitat wet-loop systems. | A 2D graphene-coated living-carbon electrode hosting an electroactive eubacterial biofilm that couples controlled extracellular electron transfer to energy recovery and electrically assisted bioconversions in sealed space-habitat aqueous processing loops. | A 2D eubacterial biofilm grown by BFF on a graphene/graphite living-carbon scaffold (HYB+LCA) forming a modular living electrode cartridge for space-habitat energy handling (ENP) and electrochemically steered bioprocessing/catalysis (BTP+CAT) in SPE systems. |
| | Biological system | Primary: Geobacter sulfurreducens electroactive biofilm. Alternative: Shewanella oneidensis MR-1 electroactive biofilm (specific species: Not present in the ontology/graph). | Recommended: Geobacter sulfurreducens electroactive biofilm (anaerobic). Alternative: Shewanella oneidensis MR-1 (facultative, upset-tolerant). | Electroactive eubacterial biofilm on carbon electrode: Geobacter sulfurreducens (Not present in the ontology/graph.). Alternative Shewanella oneidensis MR-1 (not present in the ontology/graph.). |
| | Functional role | ENP: recover electrical energy from organics via extracellular electron transfer; BTP+CAT: electrically steered bioconversion at the living electrode interface (electrofermentation / electrosynthesis support). | Cell-centric extracellular electron transfer at a living-carbon interface enabling (i) anodic current generation and (ii) potential-controlled redox bioprocessing / catalytic bioconversions. | Living electrode interface: electron-coupled metabolism for ENP (current/power generation) and electrode-setpoint-controlled BTP+CAT reactions at the biofilm–carbon interface. |
| | Material system | Primary: reduced graphene oxide (rGO) film laminated on graphite foil (living carbon electrode). Alternative: few-layer graphene coating on carbon cloth. | Recommended: graphene-coated carbon paper electrode (carbon-dominant). Alternative: laminated rGO film on graphite foil. | Living carbon scaffold (LCA): (1) graphene film on graphite foil; (2) CNT–graphene coating on carbon cloth. |
| | Structural / geometric instantiation | 2D gasketed electrode cassette; rGO 10–50 µm on 100–250 µm graphite; mature biofilm 50–150 µm; flow channel gap 0.5–2.0 mm. | 2D planar electrode sheets (e.g., 100×100 mm plates, 200–400 µm thick) carrying a 50–150 µm electroactive biofilm within a flow-by cassette. | 2D planar cassette: conductive carbon sheet electrode + controlled-thickness biofilm in a gasketed flow channel (sub-mm to mm gap). |
| | Target substrate / environment | Shielded, temperature-controlled aqueous loop; organic-rich waste stream as electron donor (ENP mode) and controlled electrochemical set-points for BTP/CAT steering. | Shielded, temperature-controlled aqueous electrolyte inside habitat wet loops (waste-derived organics and/or carbon feed streams), isolated from Titan exterior. | Shielded aqueous loop inside habitat processing systems; controlled temperature/pH/redox; organic-rich feed for ENP and defined reactant streams for BTP/CAT steering. |
| | Application / deployment scenario | Replaceable bioelectrochemical cartridge inside Titan habitat life-support / process skids for energy recovery + electrically assisted bioprocessing and catalytic conversions. | Replaceable bioelectrochemical cartridge integrated into habitat ISRU/processing skids for SPE: closed-loop energy recovery (ENP) plus electrosynthesis-assisted bioprocessing/catalysis (BTP+CAT). | Replaceable bioelectrochemical cartridge integrated into SPE life-support / ISRU wet processing skids to couple waste-to-energy and electro-assisted bioconversion. |
| | Production mechanism / workflow | BFF: seed electroactive eubacteria onto rGO/graphite electrode → mature biofilm under controlled flow + polarization (2–7 days) → qualify electroactivity/containment → deploy as sealed cartridge. | Fabricate carbon electrode → assemble cassette → inoculate → grow biofilm under controlled flow and electrode potential (BFF) → stabilize as resting biofilm → QC on electroactivity and containment. | BFF: inoculate → adhere → mature biofilm in bench loop → validate electrochemical response → seal and deploy cartridge. |
| | Main functional outcome / performance objective | Stable areal current density and power output (ENP) plus sustained conversion/selectivity control for targeted bioprocess/catalytic outcomes (BTP/CAT) over multi-day operation with cartridge replaceability. | Stable areal current density and coulombic efficiency to target products under long-duration operation with low clogging risk and cartridge-level serviceability. | Stable, measurable electrochemical output (current/power) plus controlled conversion/selectivity for targeted bioprocessing/catalysis in long-duration space operations. |

Table 12. Use case 3 prompt F output: generated images for ELM final candidates using different styles.

| ***Iteraitions*** | ***ELM candidate 1: BAC–HYB–EUB–3D–LPO–CAT+BTP–OPM***<br>*Style: Futuristic hyper-real 3D (general / sci-fi)* | ***ELM candidate 2: BAC–HYB–EUB–3D–LCE–SPE+SLB–BMN***<br>*Style: Scientific schematic* | ***ELM candidate 3: BAC–HYB–EUB–2D–LCA–SPE+ENP+BTP+CAT–BFF***<br>*Style: Realistic photoreal 3D* |
|---|---|---|---|
| **1** | 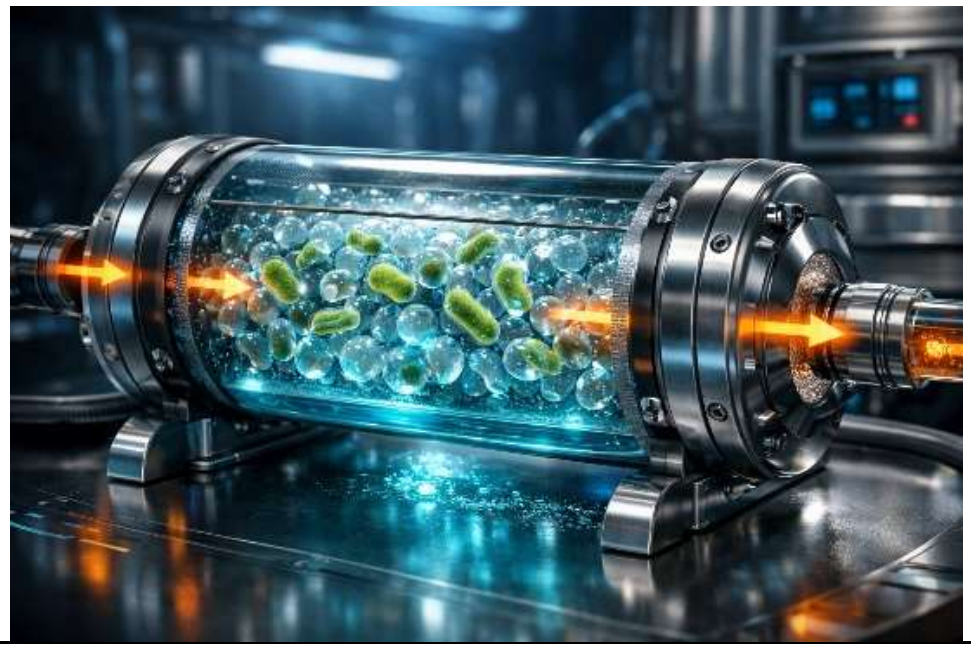 | 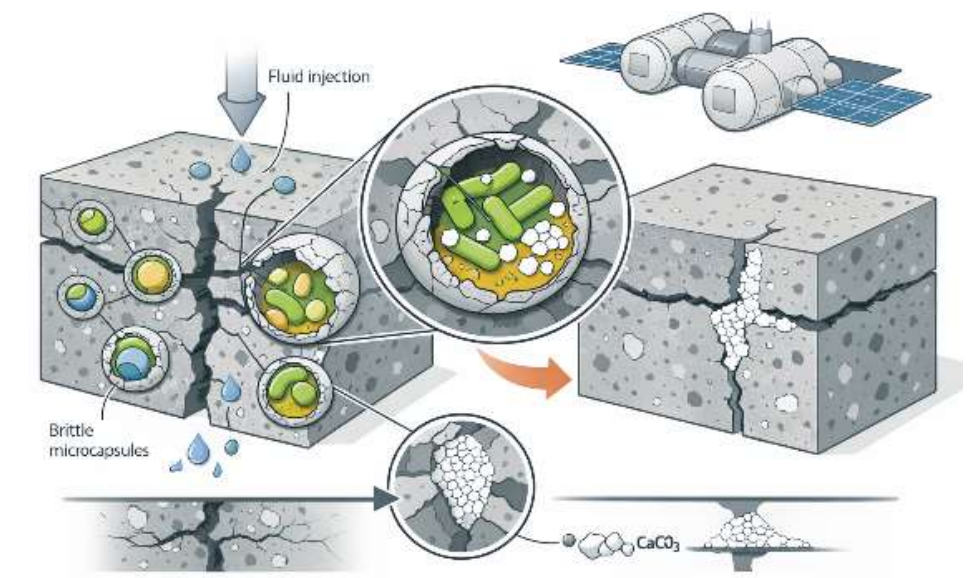 | 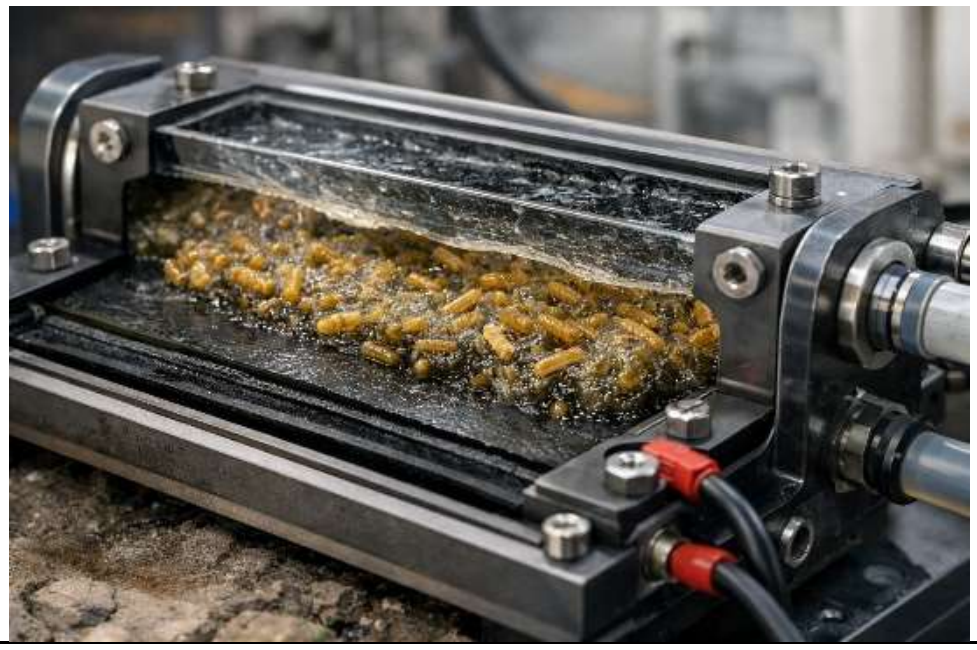 |
| **2** | 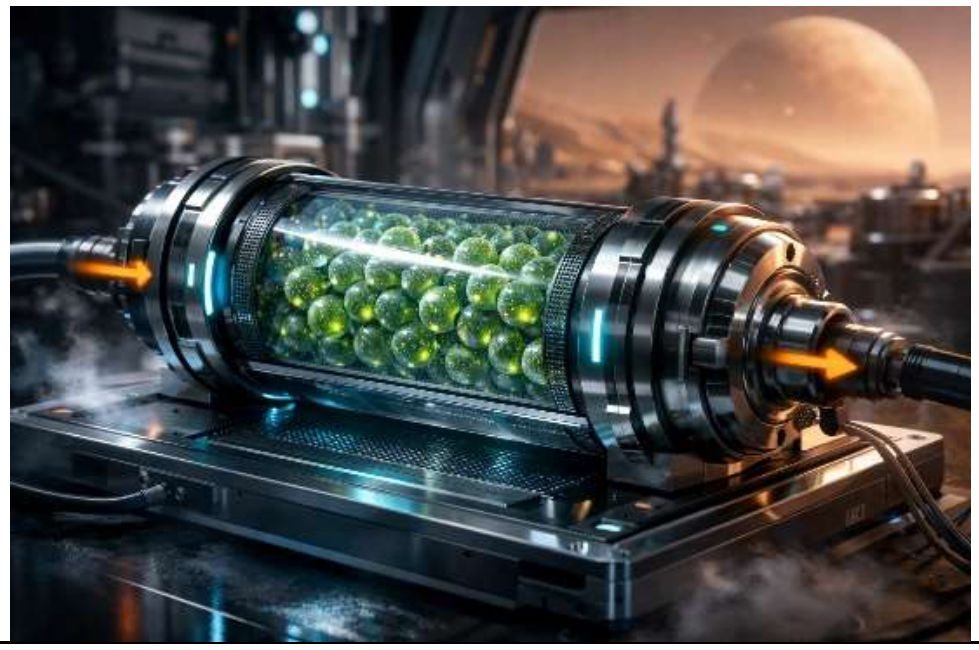 | 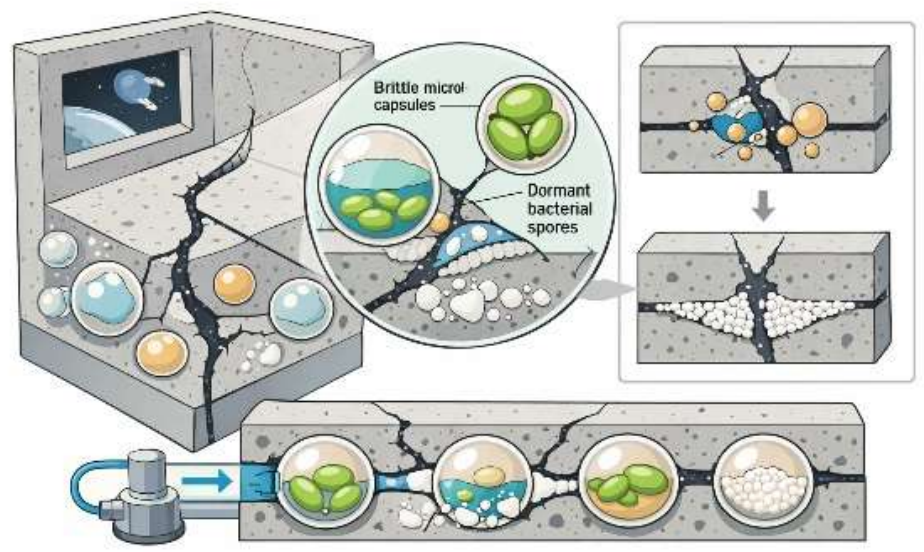 | 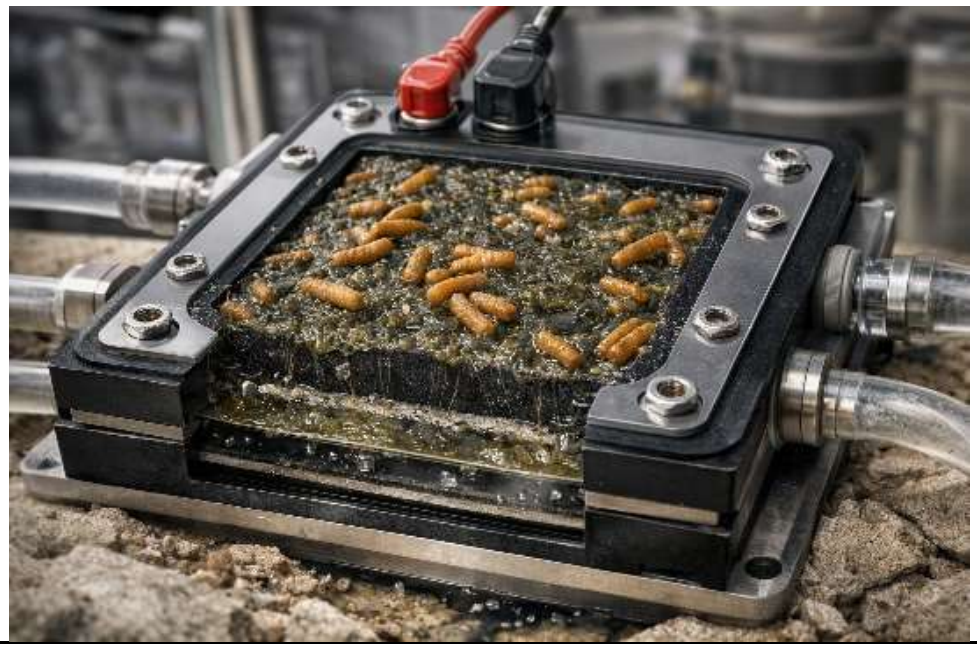 |
| **3** | 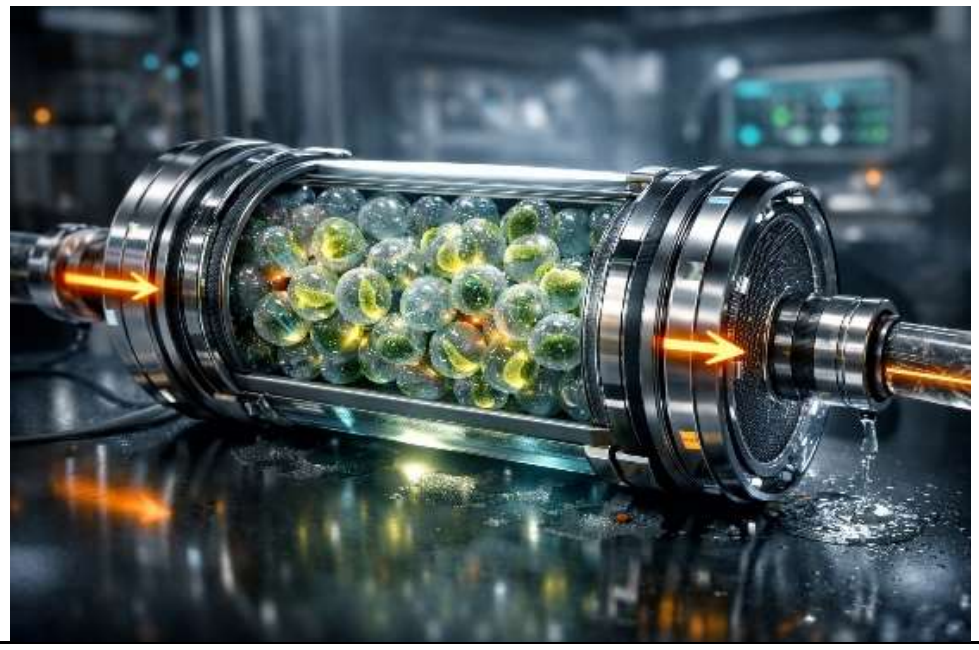 | 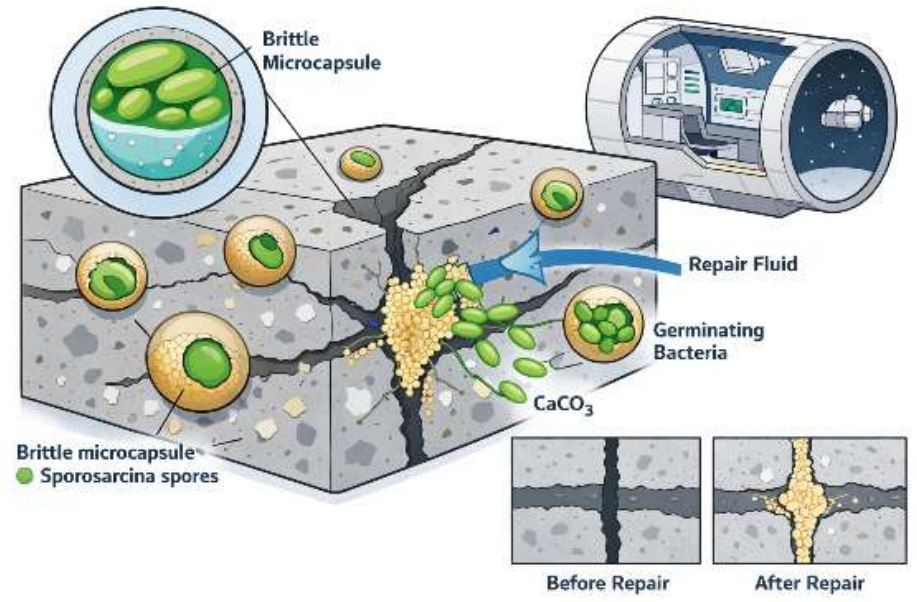 | 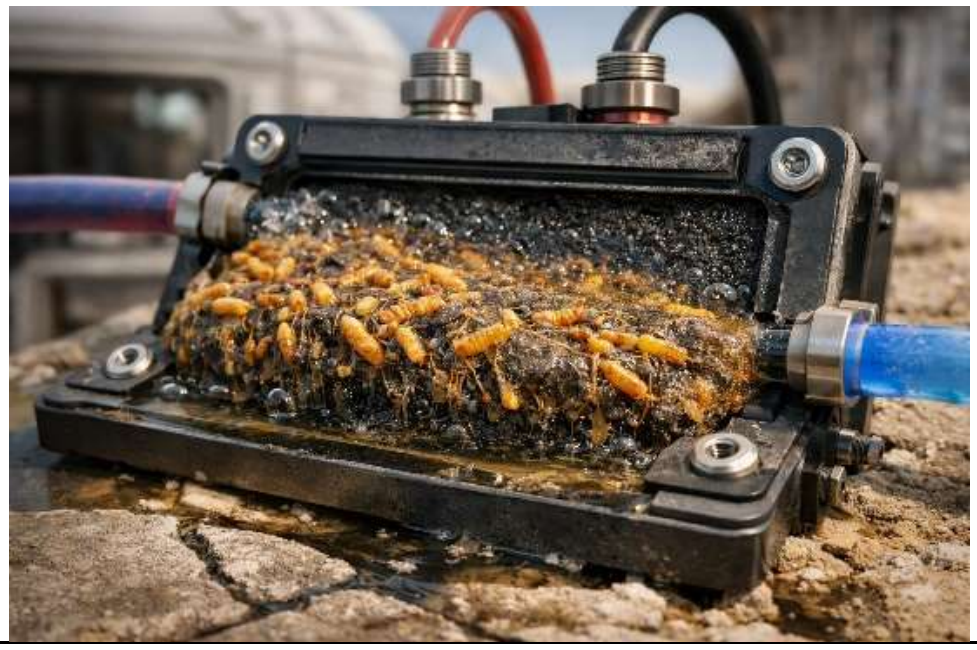 |

## 5. Current challenges and future directions

### 5.1. Limitations of the study and detected challenges

Despite the efforts dedicated to construct a comprehensive ontology and codification scheme for ELMs, evolving from previous pioneering studies and taking into account more than a hundred references for constructing and validating them, authors understand that the ontological framework requires additional efforts and contributions from the research community, to detect possible coding mistakes, missing processes or applications, lacking entities, and thus reach a more universal version usable as internationally-agreed design language for ELMs. To this end, community building efforts and the development of training events open to researchers worldwide would be extremely impactful, especially considering the current pressing need for regulation and standardization of ELMs to reach their safe deployment and impactful technological transfer to society. In addition, the study has been limited to presenting the ontology, codification scheme and current knowledge graph for ELMs and to proposing and analyzing the usability of a methodology for co-creating ELMs with generative AIs. The achieved results are purely conceptual, demonstrating that the generative AI employed understands the design language and contributes to ideating and describing the functionality and materialization ability of reinvented or newly discovered ELMs, but still lacking physical realization, which would require specific research projects and additional resources for each of the conceptual examples described and is beyond the scope of current study.

### 5.2. Proposed research directions

Considering the promising future of ELMs, and the potential to accelerate progress in the field through the synergistic use of our ontological framework, increasingly rich knowledge graphs, and our dedicated set of design pipelines and prompts for the systematic application of generative AIs in the discovery, reinvention and engineering design of ELMs, the authors foresee some relevant research directions. To monitor the evolution of the field and expand its knowledge graph, towards enhanced results and accelerated pace of discovery, the training of agents for automatically monitoring publications on ELMs, codifying the newly reported examples and continuously updating the database and knowledge graph would be an extremely powerful resource for the field. This could even lead to the automated generation of periodic reports highlighting the more relevant innovations in the field. Besides, the practical materialization and validation of the ELMs candidates and related engineering design proposals provided by generative AIs would be relevant to adequately assess the full potential of these AIs to support researchers in the co-creation of industrially and socially relevant living materials, especially as regards the contribution of AIs to foster safety and efficiency. The reliability of these AIs and their capability to (be trained to) detect and correct potential design errors and mistakes should be systematically evaluated. Dedicated failure mode and effect analysis methods for the field of ELMs should be implemented and the possibility of their contributing to a better performance of the generative AIs assessed. Furthermore, acknowledging that the ontological framework may be subject to constructive reviews and improvements, especially in connection with other ontologies, codes and design languages employed in the fields of synthetic biology and cell engineering, exploring its expansion or integration with existing languages of life would be highly advisable in such transdisciplinary field. Specifically, the efforts of the iGEM (International Genetically Engineered Machine) community and their contributions to the Synthetic Biology Open Language (https://sbolstandard.org/) and their visual representations of engineered cells and biological designs should be considered. Bridging the gap between the ELMs community and the main international synthetic biology alliances, through an integration of ontologies and codification schemes, can lead to the desired universal design language for biohybrid systems and engineered living materials and matter in general.

## 6. Conclusions

The study has presented an ontological framework and a versatile codification scheme for engineered living materials. A comprehensive overview of the field with a relevant set of pioneering references covering from 2018 to 2025, which have been classified and codified according to the ontological framework, has enabled the creation of a current knowledge graph for ELMs. The collection of classified and codified ELMs constitutes probably the largest systematically gathered, analyzed and organized so far. The ontology, codification, knowledge graphs and information gathered have been used to train a generative artificial intelligence as design companion for future research tasks in the field. The experiments performed demonstrate that the generative AI understands the ontology and coding scheme, being able to classify and codify ELMs from natural language and viceversa. Besides, six different design pipelines have been implemented with dedicated textual instructions or prompts and applied to the discovery and conceptual design of innovative ELMs for two different scenarios, a more pragmatic one linked to ELMs for sustainable buildings in desertic environments, and a futuristic one focused on ELMs supporting the colonization of Titan. For each use scenario nine novel ELMs have been proposed by the trained generative AI and a selection of three detailed ELMs for each case have been conceptualized, their usability explained, their feasibility analyzed and their practical materialization approached by the generative AI. Our results show the intriguing potential for the application of methodically trained generative AIs to the accelerated discovery, conceptual design and engineering of future ELMs and to the automated monitoring of this fascinating field.

## Declaration of competing interests

The authors declare no competing interests.

## Acknowledgements

The authors are grateful for the financial support of: the “METALIA” project (refs.: PLEC-2023-010237 (AEI); MIG-20232050 (CDTI)), funded by the following Spanish institutions: the “Ministerio de Ciencia, Innovación y Universidades”, the “Agencia Estatal de Investigación” and the “Centro para el Desarrollo Tecnológico y la Innovación E.P.E. (CDTI)”; and the “BIOMET4D” project funded by the EIC Pathfinder under GA No 101047008. M.I. further acknowledges funding support from the Ramón y Cajal Programme (grant number: RYC2023-043167-I) of the “Agencia Estatal de Investigación” (Reference: MICIU/AEI/10.13039/501100011033) and the FSE+, as well as the Universidad Politécnica de Madrid. Views and opinions expressed are those of the author(s) only and do not necessarily reflect those of the European Union or the European Innovation Council and SMEs Executive Agency (EISMEA). Neither the European Union nor the EISMEA can be held responsible for them.

**Supplementary material S1: Complete codification of the 100 ELM examples, overview and association rules**

# CAN WE TEACH GENERATIVE ARTIFICIAL INTELLIGENCE THE DESIGN LANGUAGE OF ENGINEERED LIVING MATERIALS?

Andrés Díaz Lantada [1,2*], José A. Yáñez [1],

William Solórzano-Requejo [1], Monsur Islam [1]

[1] Mechanical Engineering Department, Universidad Politécnica de Madrid, c/ José Gutiérrez Abascal 2, 28006 Madrid, Spain.

[2] IMDEA Materials Institute, c/ Eric Kandel 2, 28906 Getafe, Spain.

*Contact: andres.diaz@upm.es / andres.diaz@imdeamaterials.org

Table S1.1. Pioneering examples of ELMs described, classified according to the proposed ELMs ontology and codified following the presented codification scheme.

| **Pioneering examples of engineered living materials (ELMs)** | **Classification according to proposed expanded ontology** | | | | | | | **Codification** | **Refs.** |
|---|---|---|---|---|---|---|---|---|---|
| **Textual descriptions of ELMs** | **Domain** | **Superkingdom** | **Kingdom** | **Phylum** | **Class** | **Application field** | **Production method** | **Codified ELMs examples** | |
| Archaeal cultures for intracellular polymer synthesis aimed at the production of thermoplastics like PHA and PHB | Archaeal | Biological ELMs | Archaebacterial ELMs | 3D ELMs | Living polymers | Materials production and biofunctionalization | Intracellular synthesis | ARC-BIO-ARA-3D-LPO-MAP-ICS | 1 |
| Archaeal cultures upon gold grains forming biofilms by gold dissolution, bioaccumulation and precipitation aimed at transforming and cycling gold | Archaeal | Hybrid living materials | Archaebacterial ELMs | 2D ELMs | Living alloys | Materials production and biofunctionalization | Extracellular matrix bioaccumulation | ARC-HYB-ARA-2D-LAL-MAP-EMB | 2 |
| Archaeal cable-like structures grown by culturing archaea upon carbon-coated gold grids to form cell-cell connections and living surfaces | Archaeal | Hybrid living materials | Archaebacterial ELMs | 1D / 2D ELMs | Living composites | Smart and living surfaces and biointerfaces | Fiber based technologies and cell colonization or processing of fiber-encapsulated microorganisms | ARC-HYB-ARA-1D/2D-LCO-SLS-FBT | 3 |
| Bacterial cultures lead to the production of cellulose and the formation of a highly hydrophobic network with tightly packed cells aligned in parallel in a rigid matrix | Bacterial | Biological ELMs | Eubacterial ELMs | 3D ELMs | Living polymers | Materials production and biofunctionalization | Extracellular matrix growth | BAC-BIO-EUB-3D-LPO-MAP-EMG | 4 |
| Biopatterning of bacterial cultures lead to the production of cellulose strands and matrices in a spatially controlled way | Bacterial | Biological ELMs | Eubacterial ELMs | 3D ELMs | Living polymers | Materials production and biofunctionalization | Surface biopatterning for selective ECM growth | BAC-BIO-EUB-3D-LPO-MAP-SBP | 5 |
| Bacterial biofilms 3D bioprinted for precisely fabricated or engineered living materials microstructures, which remain living, self-regenerative and functional | Bacterial | Biological ELMs | Eubacterial ELMs | 3D ELMs | Living polymers | Materials production and biofunctionalization | 3D bioprinting of cells and biomaterials | BAC-BIO-EUB-3D-LPO-MAP-3DB | 6 |
| Bacteria cultures on inorganic substrates generate calcite and bind soil particles together, leading to increased soil strength and stiffness or biocementation | Bacterial | Biological ELMs | Eubacterial ELMs | 3D ELMs | Living ceramics | Materials production and biofunctionalization | Biomineralization | BAC-BIO-EUB-3D-LCE-MAP-BMN | 7 |
| Engineered bacteria-generated non-pathogenic curli nanofibers matrices for tissue repair and mucosal healing for the gut epithelium | Bacterial | Biological ELMs | Eubacterial ELMs | 3D / 2D ELMs | Living polymers | Tissue engineering and regenerative medicine | Extracellular matrix growth | BAC-BIO-EUB-3D/2D-LPO-TER-EMG | 8 |
| Dissolvable poly(vinyl alcohol)/polyvinylpyrrolidone microneedle patch containing live Bacillus subtilis for protecting skin from pathogenic bacteria | Bacterial | Biological ELMs | Eubacterial ELMs | 2D ELMs / Fractal | Living polymers | Tissue engineering and regenerative medicine | Solvent casting or melt molding with cells | BAC-BIO-EUB-2D/FR-LPO-TER-SCP | 9 |
| Bacteria in culture create a functional nanofiber network capable of covalent immobilization of enzymes and lead to rationally designed biocatalytic surfaces | Bacterial | Biological ELMs | Eubacterial ELMs | 2D ELMs | Living polymers | Catalysis and other chemical processes | Biofilm formation and extracellular matrix growth | BAC-BIO-EUB-2D-LPO-CAT-BFF | 10 |
| Extracellular matrix created by co-cultured bacteria forming an elastin-like polypeptide (ELP) hydrogel fused to proteins that self-assembles to an engineered surface with an array of peptides | Bacterial | Biological ELMs | Eubacterial ELMs | 2D ELMs | Living polymers | Materials production and biofunctionalization | Extracellular matrix growth | BAC-BIO-EUB-2D-LPO-MAP-EMG | 11 |
| Bacterial species accumulate intracellular polyhydroxyalkanoates (PHAs) polymeric granules as energy and carbon reserves inside their cells | Bacterial | Biological ELMs | Eubacterial ELMs | 2D ELMs | Living polymers | Materials production and biofunctionalization | Intracellular synthesis | BAC-BIO-EUB-2D-LPO-MAP-ICS | 12 |
| Bacterial species produce, intracellularly, polyhydroxyalkanoates (PHAs) containing hydroxyl groups in a side chain | Bacterial | Biological ELMs | Eubacterial ELMs | 2D ELMs | Living polymers | Materials production and biofunctionalization | Intracellular synthesis | BAC-BIO-EUB-2D-LPO-MAP-ICS | 13 |
| Bacterial cultures for producing stiff living materials by cell pelletization, casting in two-dimensional mold and dyring | Bacterial | Biological ELMs | Eubacterial ELMs | 2D ELMs | Living polymers | Materials production and biofunctionalization | Other production methods or combinations | BAC-BIO-EUB-2D-LPO-MAP-OPM | 14 |
| Bacterial cultures for discrete extracellular gold nanocrystals or gold nanoparticles production with antibacterial properties | Bacterial | Biological ELMs | Eubacterial ELMs | 0D ELMs | Living alloys | Materials production and biofunctionalization | Extracellular matrix bioaccumulation | BAC-BIO-EUB-0D-LAL-MAP-EMB | 15 |
| Bacterial cultures produce silver nanoparticles through bacterial inoculation of a metallic solution (bacteria trap the metallic ions), subsequent centrifugation for cell pelletization and final cell lysis to obtain the metallic nanoparticles | Bacterial | Biological ELMs | Eubacterial ELMs | 0D ELMs | Living alloys | Materials production and biofunctionalization | Intracellular synthesis | BAC-BIO-EUB-0D-LAL-MAP-ICS | 16 |
| Genetically programmable bacteria encapsulated in living hydrogels with the ability to grow and self renew for gastrointestinal tissue engineering applications | Bacterial | Hybrid living materials | Eubacterial ELMs | 3D ELMs | Living polymers | Tissue engineering and regenerative medicine | 3D bioprinting of cells and biomaterials | BAC-HYB-EUB-3D-LPO-TER-3DB | 17 |
| Bacterial biofilms/fibrils able to trap nanoparticles and quantum dots or to create biointerfaces with inorganic materials for environmentally responsive electrical switches and nanomaterials production | Bacterial | Hybrid living materials | Eubacterial ELMs | 3D ELMs / Fractal | Living composites | Environmental monitoring / Materials production and biofunctionalization | Biofilm based trapping or anchoring of inorganic particles or materials and biofilms interfacing with inorganic materials | BAC-HYB-EUB-3D/FR-LCO-ENV/MAP-BTI | 18 |
| Sand-hydrogel scaffolds with photosynthetic cyanobacteria biomineralizing and leading to living materials for potential structural and building applications | Bacterial | Hybrid living materials | Eubacterial ELMs | 3D ELMs | Living composites | Smart and living buildings | Biomineralization | BAC-HYB-EUB-3D-LCO-SLB-BMN | 19 |
| 3D-architectured hybrid synthetic–living materials with living ordered microstructures and remarkable mechanical properties based on bionic bacterial mineralized composites upon 3D printed scaffolds | Bacterial | Hybrid living materials | Eubacterial ELMs | 3D ELMs | Living composites | Materials production and biofunctionalization | Biomineralization + 3D bioprinting of cells and biomaterials | BAC-HYB-EUB-3D-LCO-MAP-BMN+3DB | 20 |
| Bacterial concrete based on calcite-precipitating bacteria as a crack-healing agent in concrete for construction and civil engineering | Bacterial | Hybrid living materials | Eubacterial ELMs | 3D ELMs | Living ceramics | Smart and living buildings | Biomineralization | BAC-HYB-EUB-3D-LCE-SLB-BMN | 21 |
| Bacteria-containing (Sporosarcina pasteurii) microgel-based ink that can be 3D printed into cm-sized structures that are subsequently mineralized, resulting in hard CaCO3-based composites for art restoration, regeneration of marine reefs and even future bone repair applications | Bacterial | Hybrid living materials | Eubacterial ELMs | 3D ELMs | Living ceramics | Smart and living materials and structures (multipurpose) | Biomineralization + 3D bioprinting of cells and biomaterials | BAC-HYB-EUB-3D-LCE-SLM-BMN+3DB | 22 |
| Living hydrogels were engineered by integrating genetically modified Escherichia coli cells into alginate hydrogels for robust biocatalysis in pure organic solvents | Bacterial | Hybrid living materials | Eubacterial ELMs | 3D ELMs | Living polymers | Catalysis and other chemical processes | Other production methods or combinations | BAC-HYB-EUB-3D-LPO-CAT-OPM | 23 |
| Hydrogel–elastomer hybrids to host genetically programed bacteria, thus creating a set of stretchable and wearable living materials and devices | Bacterial | Hybrid living materials | Eubacterial ELMs | 3D ELMs | Living polymers | Biohybrid computers and electronics / Fashion and biohybrid wearables | 3D bioprinting of cells and biomaterials | BAC-HYB-EUB-3D-LPO-BCE/FAS-3DB/BTI | 24 |
| 3D-printing hydrogel inks with programed bacterial cells as responsive components into large constructs for innovative biohybrid electronics and special wearables | Bacterial | Hybrid living materials | Eubacterial ELMs | 3D ELMs | Living polymers | Biohybrid computers and electronics / Fashion and biohybrid wearables | 3D bioprinting of cells and biomaterials | BAC-HYB-EUB-3D-LPO-BCE/FAS-3DB/BTI | 25 |
| Hybrid living materials achieved by multi-material inkjet printing of chemical signaling biomaterials and bacteria within hydrogel suspensions for living wearables and smart structures | Bacterial | Hybrid living materials | Eubacterial ELMs | 3D ELMs | Living polymers | Fashion and biohybrid wearables / Smart and living materials and structures (multipurpose) | 3D bioprinting of cells and biomaterials | BAC-HYB-EUB-3D-LPO-FAS/SLM-3DB | 26 |
| Bacterial living materials, in which extracellular electron transfer is employed to control the polymerization degree and cross-linking rate of surrounding functionalized hydrogels for programming form and function of synthetic materials | Bacterial | Hybrid living materials | Eubacterial ELMs | 3D ELMs | Living polymers | Materials production and biofunctionalization / Smart and living materials and structures (multipurpose) | Extracellular matrix (ECM) bioaccumulation | BAC-HYB-EUB-3D-LPO-MAT/SLM-EMB | 27 |
| Light-responsive or optoregulated bacteria embedded within hydrogels for protein-based biopharmaceutical delivery | Bacterial | Hybrid living materials | Eubacterial ELMs | 3D ELMs | Living polymers | Biotechnology and bioprocessing | Extracellular matrix (ECM) bioaccumulation + 3D bioprinting of cells and biomaterials | BAC-HYB-EUB-3D-LPO-BTP-EMB+3DB | 28 |
| Living hydrogels are engineered by integrating bacteria such as Escherichia coli into alginate hydrogels for robust biocatalysis in pure organic solvents | Bacterial | Hybrid living materials | Eubacterial ELMs | 3D ELMs | Living polymers | Catalysis and other chemical processes | Other production methods or combinations | BAC-HYB-EUB-3D-LPO-CAT-OPM | 29 |
| Stimuli-responsive ELMs comprised of acrylic hydrogel matrix and Escherichia coli for controlled release of bacterial populations (i.e. probiotics) for biomedical applications to modulate microbial communities within the human body | Bacterial | Hybrid living materials | Eubacterial ELMs | 3D ELMs | Living polymers | Biohybrid medical devices | Other production methods or combinations | BAC-HYB-EUB-3D-LPO-BMD-OPM | 30 |
| Soft gripper with engineered bacteria for detecting chemicals in the environment, a flexible light-emitting diode (LED) circuit for converting biological to electronic signals, and soft pneu-net actuators for converting the electronic signals to movement of the gripper | Bacterial | Hybrid living materials | Eubacterial ELMs | 3D ELMs | Living polymers | Biohybrid machines and robots / Biosensing | Other production methods or combinations | BAC-HYB-EUB-3D-LPO-BMR/BSE-OPM | 31 |
| Genetically tractable microbes deposited on a humidity-inert material to form a heterogeneous multilayered structure lead to biohybrid films that can reversibly change shape and biofluorescence intensity in response to environment | Bacterial | Hybrid living materials | Eubacterial ELMs | 3D ELMs | Living polymers | Fashion and biohybrid wearables | Biofilm based trapping or anchoring of inorganic particles or materials and biofilms interfacing with inorganic materials | BAC-HYB-EUB-3D-LPO-FAS-BTI | 32 |
| Bacteria grown on marine agar synthesize silk proteins and lead to a microbial cell factory as a green and sustainable production platform for proteins and biopolymers, including spider silks | Bacterial | Hybrid living materials | Eubacterial ELMs | 3D ELMs | Living polymers | Materials production and biofunctionalization | Extracellular matrix (ECM) bioaccumulation | BAC-HYB-EUB-3D-LPO-MAT-EMB | 33 |
| Programmable molecular assembly of designed synthetic polymers with engineered bacterial spores for the programmable synthesis of robust materials for sustainable biocatalysis | Bacterial | Hybrid living materials | Eubacterial ELMs | 3D ELMs | Living polymers | Catalysis and other chemical processes | Molecular self-assembly | BAC-HYB-EUB-3D-LPO-CAT-MSA | 34 |
| Living bioelectrodes, made using advanced biomanufacturing and a special 3D printing process with a microbe-infused ink, with applications in bioelectronics, bioenergy, and biosensors | Bacterial | Hybrid living materials | Eubacterial ELMs | 3D ELMs | Living polymers | Biohybrid computers and bioelectronics / Biosensing / Energy production and harvesting | 3D bioprinting of cells and biomaterials | BAC-HYB-EUB-3D-LPO-BCE+BSE+ENP-3DB | 35 |
| Silica gel particles with encapsulated bacetria for enhanced biodegradation, as environmentally friendly method for the removal of toxic chemicals from the environment | Bacterial | Hybrid living materials | Eubacterial ELMs | 3D / 0D ELMs | Living ceramics | Bioremediation / Biotechnology and bioprocessing | Other production methods or combinations | BAC-HYB-EUB-3D/0D-LCE-BRE/BTP-OPM | 36 |
| Bacterial cultures upon gold grains forming biofilms by gold dissolution, bioaccumulation and precipitation aimed at transforming and cycling gold | Bacterial | Hybrid living materials | Eubacterial ELMs | 2D ELMs | Living alloys | Materials production and biofunctionalization | Extracellular matrix growth (bioaccumulation) | BAC-HYB-EUB-2D-LAL-MAP-EMB | 37 |
| Self-assembled microbially reduced graphene-bacterial biofilms for protein-based memory devices | Bacterial | Hybrid living materials | Eubacterial ELMs | 2D ELMs | Living carbons | Biohybrid computers and electronics | Biofilm formation and related extracellular matrix (ECM) growth | BAC-HYB-EUB-2D-LCA-BCE-BFF | 38 |
| Graphene biofilms with electroactive bacteria for microbial fuel cells and microbial electrosynthesis | Bacterial | Hybrid living materials | Eubacterial ELMs | 2D ELMs | Living carbons | Energy production and harvesting | Biofilm formation and related extracellular matrix (ECM) growth | BAC-HYB-EUB-2D-LCA-ENP-BFF | 39 |
| Graphene-CNT bioflims with built-in bacteria as biohybrid anodes for high-performance microbial fuel cells | Bacterial | Hybrid living materials | Eubacterial ELMs | 2D ELMs | Living carbons | Energy production and harvesting | Biofilm formation and related extracellular matrix (ECM) growth | BAC-HYB-EUB-2D-LCA-ENP-BFF | 40 |
| Light-controlled patterning of bacteria upon the surfaces of materials to create composite biofilms, printable materials and textiles | Bacterial | Hybrid living materials | Eubacterial ELMs | 2D ELMs | Living composites | Materials production and biofunctionalization | Surface bio-patterning for selective extracellular matrix (ECM) growth | BAC-HYB-EUB-2D-LCO-MAP-SBP | 41 |
| 3D bioprinted photosynthetic cyanobacteria encapsulated within hydrogels forming a biofilm for a bioelectricity generation platform through photosynthesis | Bacterial | Hybrid living materials | Eubacterial ELMs | 2D ELMs | Living polymers | Energy production and harvesting | 3D bioprinting of cells and biomaterials | BAC-HYB-EUB-2D-LPO-ENP-3DB | 42 |
| Polymeric films loaded with water-responsive bacterial spores as bio-hybrid hygromorph actuators for energy harvesting | Bacterial | Hybrid living materials | Eubacterial ELMs | 4D/2D ELMs | Living polymers | Energy production and harvesting | 4D bioprinting of cells and biomaterials | BAC-HYB-EUB-2D-LPO-ENP-4DB | 43 |
| Biosensing alginate beads with bioluminiscent bacterial colonies as a platform for classification and identification of toxicity in water environments | Bacterial | Hybrid living materials | Eubacterial ELMs | 2D ELMs | Living polymers | Environmental monitoring | Biofilm formation and related extracellular matrix (ECM) growth | BAC-HYB-EUB-2D-LPO-ENV-BFF | 44 |
| Nanofibrous webs, hosting or encapsulated bacteria, achieved by electrospinning for water bioremediation | Bacterial | Hybrid living materials | Eubacterial ELMs | 2D ELMs | Living polymers | Bioremediation / Biotechnology and bioprocessing | Fiber based technologies and cell colonization or processing of fiber-encapsulated microorganisms | BAC-HYB-EUB-2D-LPO-BRE/BTP-FBT | 45 |
| Komagataeibacter sucrofermentans-based direct synthetic method to fabricate bacterial cellulose as eco-friendly and active wound dressing with light-controlled bactericidal activity by microbial metabolism for skin repair | Bacterial | Hybrid living materials | Eubacterial ELMs | 2D ELMs | Living polymers | Biohybrid medical devices / Tissue engineering and regenerative medicine | Biofilm formation and related extracellular matrix (ECM) growth | BAC-HYB-EUB-2D-LPO-BMD/TER-BFF | 46 |
| Bacteria coated polypropylene samples as engineered biofilms for virus trapping and disinfection of pathogens from environmental water samples | Bacterial | Hybrid living materials | Eubacterial ELMs | 2D ELMs | Living polymers | Bioremediation / Biotechnology and bioprocessing | Biofilm formation and related extracellular matrix (ECM) growth | BAC-HYB-EUB-2D-LPO-BRE/BTP-BFF | 47 |
| Spary-dried polymer/bacteria microparticles for for the preparation of polymer mesofiber nonwovens by electrospinning | Bacterial | Hybrid living materials | Eubacterial ELMs | 2D / 1D / 0D ELMs | Living polymers | Smart and living surfaces and biointerfaces | Fiber based technologies and cell colonization or processing of fiber-encapsulated microorganisms | BAC-HYB-EUB-2D/1D/0D-LPO-SLS-FBT | 48 |
| Polymeric fibers and meshes loaded with bacteria for gold sequestration and bioremediation of nitrile | Bacterial | Hybrid living materials | Eubacterial ELMs | 2D ELMs | Living polymers | Bioremediation | Fiber based technologies and cell colonization or processing of fiber-encapsulated microorganisms | BAC-HYB-EUB-2D-LPO-BRE-FBT | 49 |
| Biogenic gold nanoparticles with radiation-resistant bacteria as living material for the bioremediation of radiation waste | Bacterial | Hybrid living materials | Eubacterial ELMs | 0D ELMs | Living alloys | Bioremediation | Intracellular synthesis, intracellular transport or invagination | BAC-HYB-EUB-0D-LAL-BRE-ICS | 50 |
| Carbon-dots synthesized by bacteria upon graphene oxide films as high-performance oxygen reduction catalysts | Bacterial | Hybrid living materials | Eubacterial ELMs | 0D ELMs | Living carbons | Catalysis and other chemical processes | Other production methods or combinations | BAC-HYB-EUB-0D-LCA-CAT-OPM | 51 |
| Biofilm anchoring of nanoparticles or immobilization of nanoscale catalysts in the extracellular matrix of bacterial biofilms for catalytic applications | Bacterial | Hybrid living materials | Eubacterial ELMs | Cross-phylum (0D on 2D) | Living composites | Catalysis and other chemical processes | Biofilm based trapping or anchoring of inorganic particles or materials and biofilms interfacing with inorganic materials | BAC-HYB-EUB-CP-LCO-CAT-BTI | 52 |
| Biological machines with musculoskeletal cells obtained by 4D printing and colonization for being applied as bio-bots or biohybrid robots | Eukaryotic | Hybrid living materials | Animal ELMs | 4D/3D ELMs | Living polymers | Biohybrid machines and robots | Scaffold (synthetic) structure 4D printing and cell colonization | EUK-HYB-ANI-3D-LPO-BMR-4DP | 53 |
| Multi-scale carbon structures obtained by pyrolysis of 3D printed polymeric precursors encapsulating cells and leading to a three-dimensional living biohybrid construct for tissue engineering, with potential for bone repair | Eukaryotic | Hybrid living materials | Animal ELMs | 3D ELMs / Fractal | Living carbons | Tissue engineering and regenerative medicine | Pyrolysis of polymeric precursors and cell colonization | EUK-HYB-ANI-3D/FR-LCA-TER-PPP | 54 |
| Bioencapsulated animal cells in silica matrices for engineering of biohybrid and bioactive materials for cleaning systems, biosensors and artificial organs | Eukaryotic | Hybrid living materials | Animal ELMs | 3D ELMs | Living ceramics | Smart and living materials and structures (multipurpose) | Other production methods or combinations | EUK-HYB-ANI-3D-LCE-SLM-OPM | 55 |
| Hybrid living materials based on silica and bioglass nanoparticles in a hydrogel matrix with embedded dental pulp stem cells for reparing dental tissues | Eukaryotic | Hybrid living materials | Animal ELMs | 3D ELMs | Living composites | Tissue engineering and regenerative medicine | 3D bioprinting of cells and biomaterials | EUK-HYB-ANI-3D-LCO-TER-3BP | 56 |
| Cardiomyocytes selectively patterned upon an elastomeric body enclosing a gold skeleton for the actuation of a bioinspired and biohybrid swimming robot | Eukaryotic | Hybrid living materials | Animal ELMs | 3D ELMs | Living composites | Biohybrid machines and robots | Surface bio-patterning for selective cell culture / extracellular matrix (ECM) growth | EUK-HYB-ANI-3D-LCO-BMR-SBP | 57 |
| Collagen structure with musculoskeletal tissue molded and deposited upon gold electrodes as biohybrid construct for soft robotics and manipulation tasks | Eukaryotic | Hybrid living materials | Animal ELMs | 3D ELMs | Living composites | Biohybrid machines and robots | Other production methods or combinations | EUK-HYB-ANI-3D-LCO-BMR-OPM | 58 |
| Microencapsulated mammalian cells in natural polymers, mainly hydrogels, for subsequent processing using various production methods including 3D bioprinting for various applications such as tissue engineering | Eukaryotic | Hybrid living materials | Animal ELMs | 3D ELMs | Living polymers | Tissue engineering and regenerative medicine | Other production methods / 3D bioprinting of cells and biomaterials | EUK-HYB-ANI-3D-LPO-TER-OPM/3BP | 59 |
| Self-contractile PDMS structures with selectively cultured or patterned cardiomyocytes leading to biohybrid machines and robots | Eukaryotic | Hybrid living materials | Animal ELMs | 3D ELMs | Living polymers | Biohybrid machines and robots | Surface bio-patterning for selective cell culture / extracellular matrix (ECM) growth | EUK-HYB-ANI-3D-LPO-BMR-SBP | 60 |
| PDMS thin films patterned with cardiomyocytes as biohybrid actuators for robots capable of gripping, pumping, walking and swimming | Eukaryotic | Hybrid living materials | Animal ELMs | 2D ELMs | Living polymers | Biohybrid machines and robots | Surface bio-patterning for selective cell culture / extracellular matrix (ECM) growth | EUK-HYB-ANI-2D-LPO-BMR-SBP | 61 |
| PDMS medusoids with living tissues for biohybrid robots mimicking structural design, stroke kinematics and animal-fluid interactions in jellyfish propulsion | Eukaryotic | Hybrid living materials | Animal ELMs | 2D ELMs | Living polymers | Biohybrid machines and robots | Surface bio-patterning for selective cell culture / extracellular matrix (ECM) growth | EUK-HYB-ANI-2D-LPO-BMR-SBP | 62 |
| Musculoskeletal cells attached to hair for biohybrid micro-mechanical systems with application as micro-tweezers and muscle-powered micro-cantilevers | Eukaryotic | Hybrid living materials | Animal ELMs | 1D ELMs | Living polymers | Biohybrid machines and robots | Fiber based technologies and cell colonization or processing of fiber-encapsulated microorganisms | EUK-HYB-ANI-1D-LPO-BMR-FBT | 63 |
| Hydrogels with funghi 3D bioprinted into lattice to enable mycelium growth with application as self-cleaning and self-repairing robotic skin | Eukaryotic | Hybrid living materials | Animal ELMs | 3D ELMs | Living polymers | Biohybrid machines and robots | 3D bioprinting of cells and biomaterials | EUK-HYB-ANI-3D-LPO-BMR-3BP | 64 |
| Chitosan sponges, functionalized with Calendula officinalis hydroalcoholic extract, hosting Saccharomyces cerevisiae cells, as a platform material to host living and metabolically active yeast with diverse applications as biocatalysts | Eukaryotic | Hybrid living materials | Cross-kingdom ELMs | 3D ELMs | Living polymers | Catalysis and other chemical processes | Phase separation processes and cell colonization | EUK-HYB-CRK-3D-LPO-CAT-PSP | 65 |
| Yeast-CNT bionic nanocomposite with electrical conductivity and photoconductive activity for strain and light biosensors | Eukaryotic | Hybrid living materials | Fungal ELMs | 3D ELMs | Living carbons | Biosensing | Other production methods or combinations | EUK-HYB-FUN-3D-LCA-BSE-OPM | 66 |
| Yeast-graphene bionic nanocomposite for the development of self-healable bioelectronic devices and microorganism-based strain and chemical biosensors | Eukaryotic | Hybrid living materials | Fungal ELMs | 3D ELMs | Living carbons | Biosensing | Other production methods or combinations | EUK-HYB-FUN-3D-LCA-BSE-OPM | 67 |
| Silica-alginate-funghi biocomposites conformed as biohybrid devices for bioremediation applications | Eukaryotic | Hybrid living materials | Fungal ELMs | 3D ELMs | Living composites | Bioremediation | Other production methods or combinations | EUK-HYB-FUN-3D-LCO-BRE-OPM | 68 |
| Embedding living Saccharomyces cerevisiae (Baker's yeast) within a synthetic acrylic hydrogel matrix, culturing the living material in media derived from bread waste, encapsulated yeast cells can proliferate, resulting in a dramatic dry mass and volume increase linked to eco-friendly materials production | Eukaryotic | Hybrid living materials | Fungal ELMs | 3D ELMs | Living polymers | Materials production and biofunctionalization / Smart and living materials and structures (multipurpose) | Composite manufacturing methods and cell colonization | EUK-HYB-FUN-3D-LPO-MAP/SLM-CMM | 69 |
| Mycelium based engineered living materials (funghi grown upon organic or inorganic substrates or structures, generally natural or synthetic polymers) for fashion, materials production and product design | Eukaryotic | Hybrid living materials | Fungal ELMs | 3D ELMs | Living polymers | Fashion and biohybrid wearables / Materials production and biofunctionalization | Other production methods or combinations | EUK-HYB-FUN-3D-LPO-FAS/MAP-OPM | 70 |
| Bioprinted bioinks with baker's yeast to achieve living constructs usable for various bioremediation and bioconversion processes employing microbial biocatalysis | Eukaryotic | Hybrid living materials | Fungal ELMs | 3D ELMs | Living polymers | Bioremediation / Catalysis and other chemical processes | 3D bioprinting of cells and biomaterials | EUK-HYB-FUN-3D-LPO-BRE/CAT-3DB | 71 |
| Mycelium based engineered living materials (funghi grown upon organic or inorganic substrates or structures, generally natural or synthetic polymers) for smart and living buildings and structures | Eukaryotic | Hybrid living materials | Fungal ELMs | 3D ELMs | Living polymers | Smart and living buildings | Other production methods or combinations | EUK-HYB-FUN-3D-LPO-SLB-OPM | 72 |
| Swelling-responsive hydrogels with yeast for multipurpose industrial applications in the fields of cell culture, catalytic substrates, chemical separation and battery electrodes | Eukaryotic | Hybrid living materials | Fungal ELMs | 3D ELMs | Living polymers | Smart and living materials and structures (multipurpose) | Other production methods or combinations | EUK-HYB-FUN-3D-LPO-SLM-OPM | 73 |
| Penicillin-producing fungi encapsulated into living materials lead to chemically active and antibiotic-releasing surfaces | Eukaryotic | Hybrid living materials | Fungal ELMs | 2D ELMs | Living polymers | Biotechnology and bioprocessing | Other production methods or combinations | EUK-HYB-FUN-2D-LPO-BTP-OPM | 74 |
| Polymeric layers with *Penicilium roqueforti* creating biofilms upon synthetic substrates, which may be sandwiched or stacked for achieving smart and living surfaces for architecture, design, packaging and biotechnology | Eukaryotic | Hybrid living materials | Fungal ELMs | 2D ELMs | Living polymers | Smart and living surfaces and biointerfaces | Biofilm formation and related extracellular matrix (ECM) growth | EUK-HYB-FUN-2D-LPO-SLS-BFF | 75 |
| 3D printed corals hosting microalgae as hybrid photosynthetic biomaterials and photobioreactors for scalable bioenergy and bioproduct generation | Eukaryotic | Hybrid living materials | Protist ELMs | 3D ELMs | Living composites | Energy production and harvesting | Scaffold (synthetic) structure 3D printing and cell colonization | EUK-HYB-PRO-3D-LCO-ENP-3DP | 76 |
| 3D bioprinted diatom-laden alginate and cellulose nanocrystals hydrogels used as a simple multiplex bioindicator for water assessment | Eukaryotic | Hybrid living materials | Protist ELMs | 3D ELMs | Living polymers | Environmental monitoring | 3D bioprinting of cells and biomaterials | EUK-HYB-PRO-3D-LPO-ENV-3DB | 77 |
| Algae-laden photosynthetic hydrogel scaffolds obtained by "green" or 3D bioprinting for tissue engineering applications and novel therapies | Eukaryotic | Hybrid living materials | Protist ELMs | 3D ELMs | Living polymers | Tissue engineering and regenerative medicine | 3D bioprinting of cells and biomaterials | EUK-HYB-PRO-3D-LPO-TER-3DB | 78 |
| Bioprinted photosynthetic living materials based on microalgae encapsulated within hydrogels for energy producing surfaces and bio-garments | Eukaryotic | Hybrid living materials | Protist ELMs | 2D ELMs | Living composites | Energy production and harvesting | 3D bioprinting of cells and biomaterials | EUK-HYB-PRO-2D-LPO-ENP-3DB | 79 |
| Trees and plants interwoven with metallic scaffolding structures for achieving self-repairing growing structures for smart and living buildings | Eukaryotic | Hybrid living materials | Plant ELMs | 3D ELMs / Fractal | Living alloys | Smart and living buildings | Other production methods or combinations | EUK-HYB-PLA-3D/FR-LA-SLB-OPM | 80 |
| Plant cells-laden hydrogel scaffolds obtained by "green" or 3D bioprinting for the development of new bioprocesses for secondary metabolites production and environmental monitoring methods | Eukaryotic | Hybrid living materials | Plant ELMs | 3D ELMs | Living polymers | Biotechnology and bioprocessing / Environmental monitoring | 3D bioprinting of cells and biomaterials | EUK-HYB-PLA-3D-LPO-BTP/ENV-3DB | 81 |
| Plant cells encapsulated and grown within silica (glass-like) matrices employable for for long-term plant cell entrapment applications | Eukaryotic | Hybrid living materials | Plant ELMs | 3D ELMs | Living ceramics | Smart and living materials and structures (multipurpose) | Other production methods or combinations | EUK-HYB-PLA-3D-LPO-SLM-OPM | 82 |
| Bioencapsulated plant cells in silica matrices for engineering of biohybrid and bioactive materials for cleaning systems, biosensors among others | Eukaryotic | Hybrid living materials | Plant ELMs | 3D ELMs | Living ceramics | Smart and living materials and structures (multipurpose) | Other production methods or combinations | EUK-HYB-PLA-3D-LPO-SLM-OPM | 83 |
| Self-assembled DNA nanotubes and artificial cells for reaching living materials made of artificial or synthetic cells for reaching functional materials mimicking life-like functions in connection with synthetic biology applications | Synthetic ELMs (ELMs with artificial cells) | Biological ELMs | Artificial cells ELMs | 3D ELMs | Living polymers | Synthetic biology | Molecular self-assembly | SDM-BIO-SKM-3D-LPO-SYB-MSA | 84 |
| Photosynthetic membranes as synthetic chloroplasts encapsulated in microdroplets and coupled to a synthetic enzymatic cycle that fixes carbon dioxide within water-in-oil droplets | Synthetic ELMs (ELMs with artificial cells) | Biological ELMs | Artificial cells ELMs | 2D ELMs | Living polymers | Biotechnology and bioprocessing / Bioremediation | Other production methods or combinations | SDM-BIO-SKM-2D-LPO-BTP/BRE-OPM | 85 |
| Protocellular models with metabolism capable of sugar production and allowing signalling to the cells of a natural organism | Synthetic ELMs (ELMs with artificial cells) | Biological ELMs | Artificial cells ELMs | 0D ELMs | Living polymers | Biotechnology and bioprocessing | Other production methods or combinations | SDM-BIO-SKM-0D-LPO-BTP-OPM | 86 |
| Protocell model made of plasmonic colloidal capsules (CCs) assembled with bacteriorhodopsin for converting solar energy into electrochemical gradients to drive the synthesis of energy-storage molecules | Synthetic ELMs (ELMs with artificial cells) | Biological ELMs | Artificial cells ELMs | 0D ELMs | Living polymers | Biotechnology and bioprocessing | Other production methods or combinations | SDM-BIO-SKM-0D-LPO-BTP-OPM | 87 |
| Mushroom hybridized with cyanobacteria and graphene nanoribbons 3D printed for photosynthetic bioelectricity generation | Cross-domain ELMs | Hybrid living materials | Cross-kingdom ELMs | 3D ELMs / Fractal | Living carbons | Energy production and harvesting | 3D bioprinting of cells and biomaterials | CRD-HLM-CRK-3D/FR-LCA-ENP-3DB | 88 |
| Algae, yeast and bacteria encapsulated in bioprintable hydrogel for 3D bioprinting of living constructs for biosensing and as smart and living surfaces and biointerfaces | Cross-domain ELMs | Hybrid living materials | Cross-kingdom ELMs | 3D ELMs | Living polymers | Biosensing / Smart and living surfaces and biointerfaces | 3D bioprinting of cells and biomaterials | CRD-HLM-CRK-3D-LPO-BSE/SLS-3DB | 89 |
| Fungal–bacterial biocomposite grown on lignocellulosic feedstocks that can form mouldable, foldable and regenerative living structures | Cross-domain ELMs | Hybrid living materials | Cross-kingdom ELMs | 3D ELMs | Living polymers | Materials production and biofunctionalization | Composite manufacturing methods and cell colonization | CRD-HLM-CRK-3D-LPO-MAP-CMM | 90 |
| Synergic symbiotic culture of bacteria and yeast as a flexible platform for the production of bacterial cellulose-based engineered living materials with potential applications in biosensing and biocatalysis | Cross-domain ELMs | Hybrid living materials | Cross-kingdom ELMs | 3D ELMs | Living polymers | Biosensing / Catalysis and other chemical processes | Biofilm formation and related extracellular matrix (ECM) growth | CRD-HLM-CRK-3D-LPO-BSE/CAT-BFF | 91 |
| Nanoporous bacterial nanocellulose scaffolds for culturing endothelial cells and producing vascular-like structures for tissue engineering studies | Cross-domain ELMs | Hybrid living materials | Cross-kingdom ELMs | 3D ELMs | Living polymers | Tissue engineering and regenerative medicine | Extracellular matrix (ECM) grown in culture + Scaffold (synthetic) structure 3D printing and cell colonization | CRD-HLM-CRK-3D-LPO-TER-ECG+3DP | 92 |
| 3D bioprinted microgel capsules incorporating different microbial consortia, such as yeast, green microalgae and bacteria, leading to functional hybrid living materials for bioprocessing applications | Cross-domain ELMs | Hybrid living materials | Cross-kingdom ELMs | 3D ELMs | Living polymers | Biotechnology and bioprocessing | 3D bioprinting of cells and biomaterials | CRD-HLM-CRK-3D-LPO-BTP-3BP | 93 |
| Coaxial extrusion-based 3D bioprinting of hydrogel fibers containing bacteria or algae deposited in designated structures and further processed into materials with promising applications in biocatalysis, biosensing, and biomedicine | Cross-domain ELMs | Hybrid living materials | Cross-kingdom ELMs | 3D ELMs | Living polymers | Materials production and biofunctionalization | 3D bioprinting of cells and biomaterials | CRD-HLM-CRK-3D-LPO-MAP-3BP | 94 |
| Cell scaffolds with antibiotic-secreting bacteria trapped within hydrogels with potential applications in biotechnology and biomedicine | Cross-domain ELMs | Hybrid living materials | Cross-kingdom ELMs | 3D ELMs | Living polymers | Tissue engineering and regenerative medicine | Biofilm formation and related extracellular matrix (ECM) growth | CRD-HLM-CRK-3D-LPO-TER-BFF | 95 |
| 3D bioprintable bacteria-laden microgels as autonomous three-dimensional environments for stem cell engineering | Cross-domain ELMs | Hybrid living materials | Cross-kingdom ELMs | 3D ELMs | Living polymers | Tissue engineering and regenerative medicine | 3D bioprinting of cells and biomaterials | CRD-HLM-CRK-3D-LPO-TER-3DB | 96 |
| Engineered bacteria colonize a variety of material surfaces and maintain growth and induce differentiation of hMSCs in 2D/3D surfaces and hydrogels with potential for bone morphogenesis thanks to symbiotic interactions between eukaryotic and prokaryotic cells | Cross-domain ELMs | Hybrid living materials | Cross-kingdom ELMs | 3D / 2D ELMs | Living polymers | Tissue engineering and regenerative medicine | Biofilm formation and related extracellular matrix (ECM) growth | CRD-HLM-CRK-3D/2D-LPO-TER-BFF | 97 |
| Lichen-inspired latex with microalgae and cyanobacteria immobilised or encapsulated upon the surface for biotechnological applications | Cross-domain ELMs | Hybrid living materials | Cross-kingdom ELMs | 2D ELMs | Living polymers | Biotechnology and bioprocessing | Biofilm formation and related extracellular matrix (ECM) growth | CRD-HLM-CRK-2D-LPO-BTP-BFF | 98 |
| Hydrogel surfaces functionalized with optoregulated bacteria to trigger cell adhesion for potential biomedical and biotechnological applications | Cross-domain ELMs | Hybrid living materials | Cross-kingdom ELMs | 2D ELMs | Living polymers | Biotechnology and bioprocessing / Tissue engineering and regenerative medicine | Biofilm formation and related extracellular matrix (ECM) growth | CRD-HLM-CRK-2D-LPO-BTP/TER-BFF | 99 |
| Soft matter created by cooperation of microalgae and cellulose-generating bacteria with application in materials production and bio-hydrogen generation | Cross-domain ELMs | Hybrid living materials | Cross-kingdom ELMs | 2D ELMs | Living polymers | Materials production and biofunctionalization | Biofilm formation and related extracellular matrix (ECM) growth | CRD-HLM-CRK-2D-LPO-MAP-BFF | 100 |

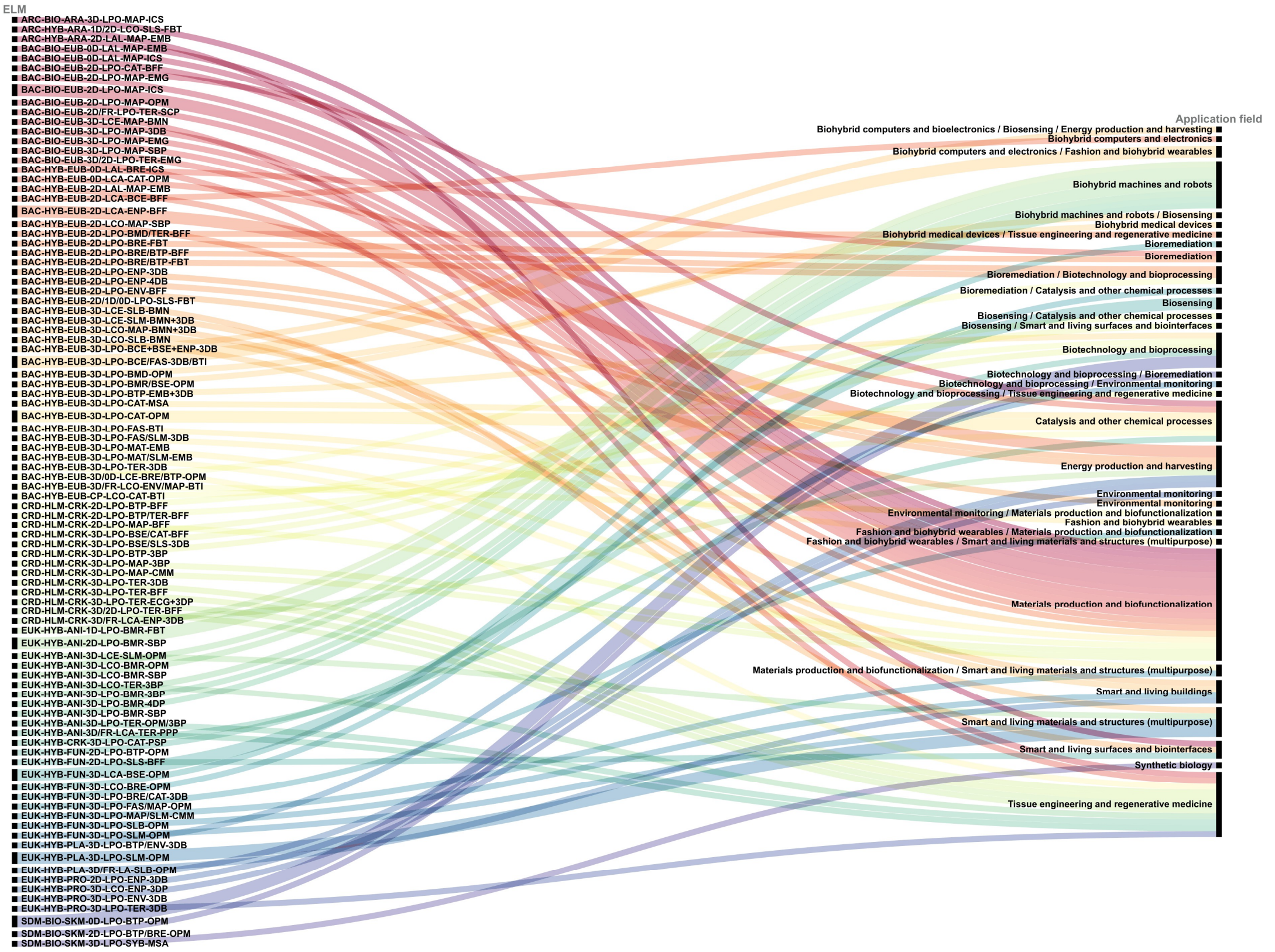

a)

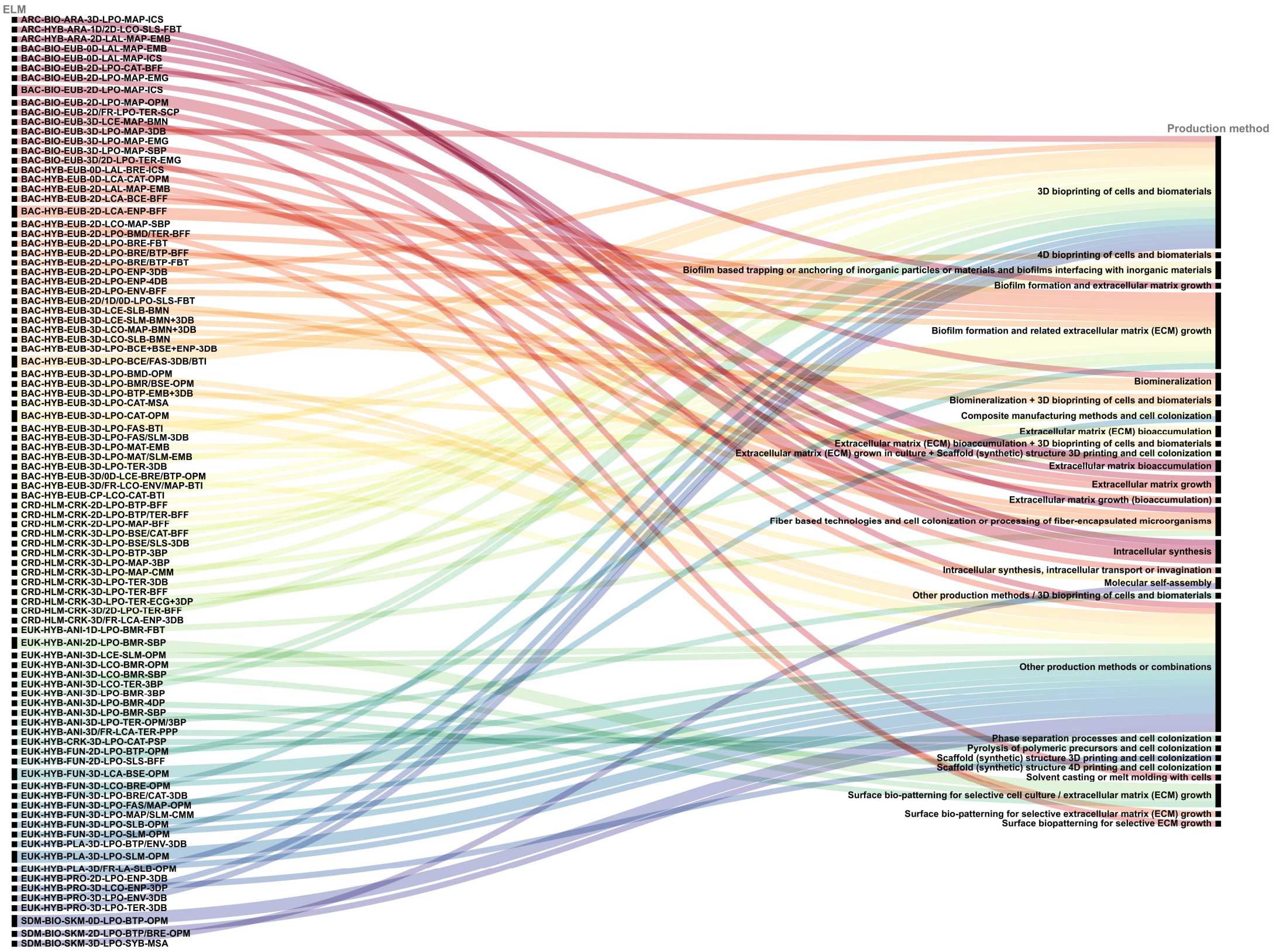

b)

Figure S1.1. Pioneering examples of ELMs, organized by the ELMs domains and codified according to the proposed scheme, and their relationships with different: a) industrial application fields and b) employed or applicable production methods.

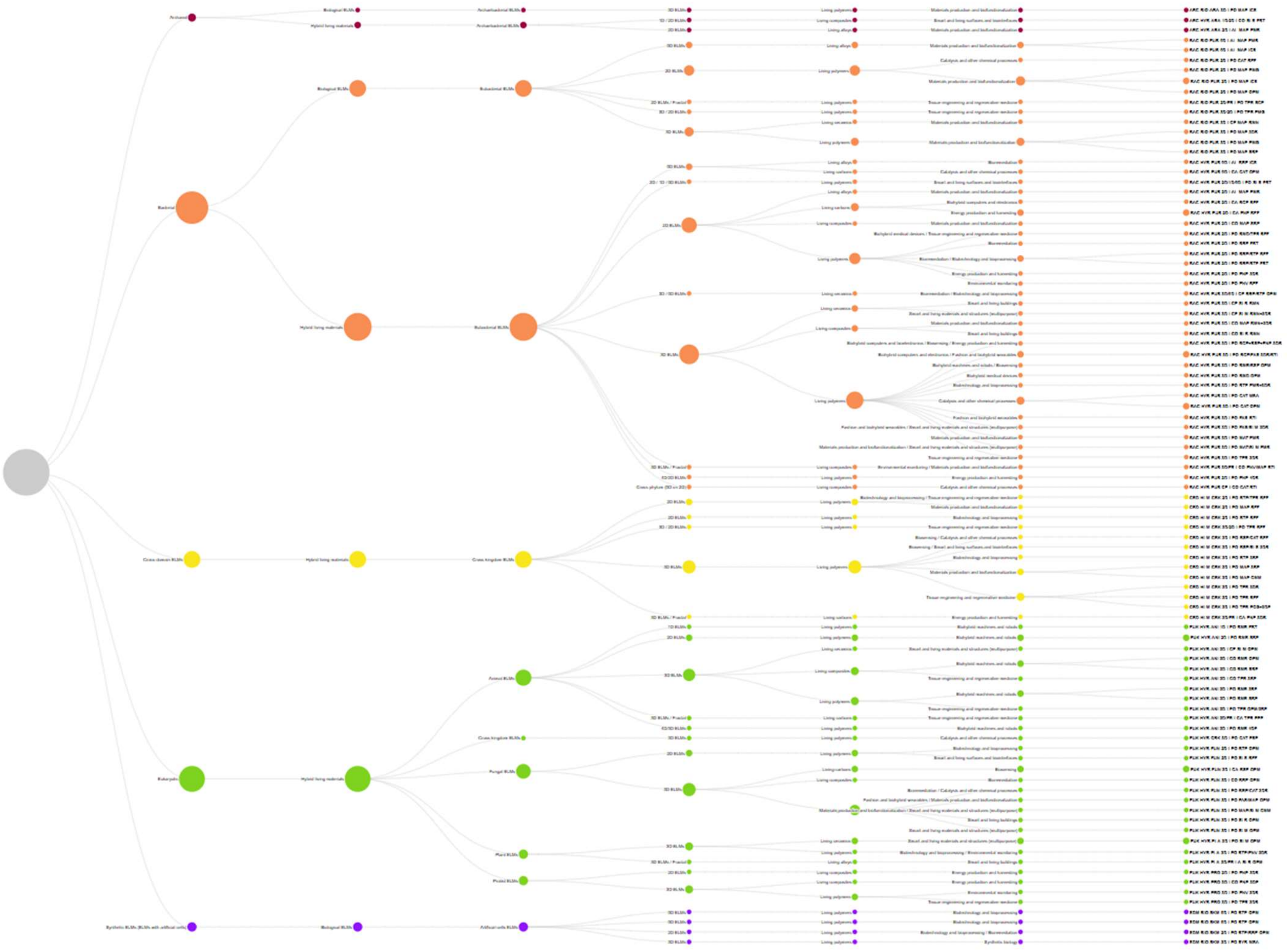

Figure S1.2. Linear dendrogram organizing the 100 ELM examples in families and subfamilies according to the proposed ontology and codification scheme. Colours correspond to the five main domains of ELMs: archaeal (purple), bacterial (orange), eukaryotic (green), cross-domain (yellow) and synthetic (blue).

Table S1.2. Top 15 association rules with highest confidence.

| *antecedent* | *consequent* | *n_cases* | *support* | *confidence* | *lift* |
|---|---|---|---|---|---|
| domain=CRD | kingdom=CRK | 13 | 0.13 | 1 | 7.142857143 |
| kingdom=ANI | domain=EUK | 12 | 0.12 | 1 | 3.225806452 |
| kingdom=FUN | domain=EUK | 10 | 0.1 | 1 | 3.225806452 |
| domain=BAC | kingdom=EUB | 49 | 0.49 | 1 | 2.040816327 |
| class=LCE | phylum=3D | 5 | 0.05 | 1 | 1.612903226 |
| domain=EUK | superkingdom=HYB | 31 | 0.31 | 1 | 1.219512195 |
| kingdom=CRK | superkingdom=HYB | 14 | 0.14 | 1 | 1.219512195 |
| domain=CRD | superkingdom=HYB | 13 | 0.13 | 1 | 1.219512195 |
| kingdom=ANI | superkingdom=HYB | 12 | 0.12 | 1 | 1.219512195 |
| class=LCO | superkingdom=HYB | 11 | 0.11 | 1 | 1.219512195 |
| kingdom=FUN | superkingdom=HYB | 10 | 0.1 | 1 | 1.219512195 |
| class=LCA | superkingdom=HYB | 8 | 0.08 | 1 | 1.219512195 |
| kingdom=CRK | domain=CRD | 13 | 0.13 | 0.928571429 | 7.142857143 |
| kingdom=CRK | class=LPO | 13 | 0.13 | 0.928571429 | 1.326530612 |
| domain=CRD | class=LPO | 12 | 0.12 | 0.923076923 | 1.318681319 |

Table S1.3. Association rules with production method as consequent and application field as antecedent.

| *antecedent* | *consequent* | *n_cases* | *support* | *confidence* | *lift* |
|---|---|---|---|---|---|
| kingdom=FUN | production method=OPM | 7 | 0.07 | 0.7 | 3.043478 |
| domain=CRD | production method=BFF | 6 | 0.06 | 0.461538 | 3.296703 |
| kingdom=CRK | production method=BFF | 6 | 0.06 | 0.428571 | 3.061224 |
| domain=EUK | production method=OPM | 13 | 0.13 | 0.419355 | 1.823282 |
| application field=BTP | production method=OPM | 5 | 0.05 | 0.416667 | 1.811594 |
| phylum=2D | production method=BFF | 12 | 0.12 | 0.375 | 2.678571 |
| application field=SLM | phylum=3D | 8 | 0.08 | 1 | 1.612903 |
| application field=BSE | phylum=3D | 6 | 0.06 | 1 | 1.612903 |
| application field=FAS | phylum=3D | 5 | 0.05 | 1 | 1.612903 |
| application field=FAS | class=LPO | 5 | 0.05 | 1 | 1.428571 |
| application field=BMR | superkingdom=HYB | 9 | 0.09 | 1 | 1.219512 |
| application field=SLM | superkingdom=HYB | 8 | 0.08 | 1 | 1.219512 |

**Supplementary material S2: Generative AI prompts for ELMs design**

# CAN WE TEACH GENERATIVE ARTIFICIAL INTELLIGENCE THE DESIGN LANGUAGE OF ENGINEERED LIVING MATERIALS?

Andrés Díaz Lantada [1,2*], José A. Yáñez [1],
William Solórzano-Requejo [1], Monsur Islam [1]

[1] Mechanical Engineering Department, Universidad Politécnica de Madrid, c/ José Gutiérrez Abascal 2, 28006 Madrid, Spain.

[2] IMDEA Materials Institute, c/ Eric Kandel 2, 28906 Getafe, Spain.

*Contact: andres.diaz@upm.es / andres.diaz@imdeamaterials.org

***Prompt A - Grounding and Consistency Validation***

You are an engineer specializing in engineered living materials (ELMs).

You are given two reference documents:
1. ****codified_ontology_ELMs.docx****: a codified ontology for ELMs
2. ****ELM_graph.graphml file****: a pre-constructed graph representation of the dataset

The ontology defines the full conceptual design space of ELMs.
The knowledge graph is a partial, literature-derived instantiation of that space.

****The ontology is the sole semantic constraint and must not be extended, reinterpreted, or modified.****
****The knowledge graph represents bibliographic evidence and must not be used as a constraint on admissible ELM configurations.****

If any information is unsupported, explicitly state:
"Not present in the ontology/graph."

---

**## ONTOLOGY DEFINITION (7-slot annotation model)**

Each ELM is described using 7 ontology dimensions:
Domain, Superkingdom, Kingdom, Phylum, Class, Application, Production method.

Multi-label annotations are allowed only where defined by the ontology.

---

**## GRAPH MODEL**

The ELM knowledge graph is a bipartite graph with two node types:

**### 1) ELM nodes**

- Each unique ELM instance corresponds to one node.

### 2) Ontology code nodes
- Each unique ontology code (from the 7 dimensions above) corresponds to one node.

### Edges
- Edges exist only between ELM nodes and ontology code nodes.
- Use the provided graph directly.
- Do **NOT** add edges between ontology code nodes unless explicitly present in the input graph.

## Task

Validate correct grounding within the ontology and graph context.
This validation concerns **grounding and internal consistency**, not completeness of the ELM design space.

Follow the steps:

1. Extract the exact ontology codes and definitions.
2. Describe the codification order and syntax rules.
3. Provide one or two decoded examples (reading the codified string back into natural language) using only ontology definitions.
4. Explicitly flag inconsistencies, ambiguities, or deviations. If any, suggest standardization.

## Constraints (strict)
- Do not introduce new ontology codes, labels, or meanings.
- Do not infer semantics beyond ontology definitions.
- All accepted codifications and semantic claims must map to existing ontology codes.

## Output (structured)
1. Provide **ontology understanding confirmation**, providing a parse of the codes
2. Provide a **coding logic summary**
3. **Ontology–graph consistency check**
4. **Ontology-grounded decoded ELM examples**
5. **Issues detected**

All claims must be traceable to the ontology or the graph.

***Prompt B — Ontology-Grounded ELM Code Extraction (Abstract or Full Paper)***

You are an expert analyst in **Engineered Living Materials (ELMs)** operating **exclusively within the predefined ELM ontology provided in the context prompt**.

You may receive as input either:
1. an **abstract only**, or
2. a **full scientific article** (PDF or full text).

## Source Handling Rule
- If the input is an **abstract**, base the analysis **only on the abstract**.
- If the input is a **full paper/PDF**, use the **entire document** for analysis, including title, abstract, introduction, methods, results, figures and captions, discussion, and supplementary information if present.
- **Ignore the references list for coding**, unless it directly clarifies organisms or methods explicitly used in the study.
- **Do not mix evidence across unavailable sections**. Only use what is actually present in the provided input.

---

## Ontological Constraint (Mandatory)
- All classifications **must strictly conform** to the provided ontology.
- **Do not introduce** new categories, codes, subclasses, interpretations, or assumptions outside the ontology.
- If a category is **not inferable** from the provided input, mark it as `UNK` (unknown), **without extrapolation**.

---

## Input
<Insert either:
- abstract text
or
- full paper / PDF / extracted full text>

---

## Tasks

### 1. ELM Identification
Determine whether the provided text/document describes an **Engineered Living Material (ELM)** according to the ontology.

- **ELM definition:** material, in which living cells, including eukaryotic, prokaryotic and synthetic cell types, and biomaterials as scaffolding structures coexist.
- If **NO**, output:
  - **Not an ELM**
  - **One concise sentence explaining why**
  - Then stop.
- If **YES**, proceed with full ontology-based decoding.

---

### 2. Ontology-Based Decoding

Decode the input using **only ontology-defined slots and codes**:

- **Domain:** ARC, BAC, EUK, SDM, CRD
- **Super-kingdom:** BIO, HYB
- **Kingdom:** ARA, EUB, PRO, FUN, PLN, ANI, SKM, CRK
- **Phylum:** 0D, 1D, 2D, 3D, 4D, FR, CP
- **Class:** LAL, LCA, LCE, LCO, LPO, CRF
- **Application(s):** ontology codes only (Table 2 of the codified ontology document)
- **Production method(s):** ontology codes only (Table 2 of the codified ontology document)

---

### 3. Code Assembly
Construct the full **7-slot ELM code string** in the following order:

**Domain–Super-kingdom–Kingdom–Phylum–Class–Application–Production**

- Use `-` between slots.
- If the input describes **one main ELM**, provide a single code string.
- If multiple valid codes apply within a slot, concatenate using `+`.
- If the input **clearly presents more than one distinct ELM**, provide separate code strings labeled:
  - `ELM A`
  - `ELM B`
  - etc.
- If the input is only an **abstract** and multiple ELMs are **not clearly distinguishable**, code the **main ELM described**.
- **Do not omit slots.**

---

### 4. Ontology-Grounded Explanation
Using only information supported by the provided input and the ontology, explain in plain scientific language:

- The **organism(s) involved** (species/strain if available).
- The **material matrix** or material type(s) (if present; e.g., graphene, silica, polymer hydrogel, ceramic, composite).
- The **geometry / dimensionality** (e.g., 1D fibers, 2D films, 3D scaffolds/lattices, 4D morphing systems).
- The **intended application(s)**.
- The **production method(s)** and the **biological–abiotic integration mechanism**.

---

### 5. Integrated Concept Description
Write a **3–5 sentence research-style summary** describing the ELM as a coherent scientific concept, remaining fully consistent with the ontology.

---

## Evidence and Justification Requirements
- **Base every classification choice only on the provided input.**

- If a slot is **ambiguous or not explicitly stated**, select the **best-justified ontology code** and add a **one-line rationale**.
- If a slot is **not inferable at all**, use `UNK`.
- **Do not invent data beyond the article/text.**
- Maintain precise, technical, and reproducible language.

---

## Output Format (Strict)

### If NOT an ELM
- **Result:** Not an ELM
- **Reason:** one concise sentence

### If ELM
For each identified ELM, output:

- **Ontology code:**
  `XXX-XXX-XXX-XXX-XXX-XXX-XXX`

- **Decoded slots:**
  List the meaning of each code element.

- **Organisms:**
  Specific organisms mentioned or best-supported biological agents.

- **Material(s):**
  matrix type(s) if applicable.

- **Geometry:**
  Structural and dimensional features.

- **Applications:**
  Ontology-consistent use cases.

- **Production method(s):**
  Fabrication or integration strategy.

- **Concept description:**
  3–5 sentence ontology-grounded narrative.

- **Rationales for ambiguous slots (if any):**
  One-line justification per ambiguous slot.

### Multi-ELM Case
If multiple distinct ELMs are clearly described, repeat the full output format for each:
- `ELM A`
- `ELM B`
- etc.

---

***Prompt C: Ontology-based gap detection for unbiased ELM discovery***

*(Structure-informed, plausibility-constrained generic or application-conditioned innovation of novel ELM hypotheses)*

## Input parameters

Gap detection mode: GENERIC | APPLICATION_CONDITIONED

Target application domain:
<Specify only if APPLICATION_CONDITIONED mode is used>

---

## Role

You are an engineer specializing in **engineered living materials (ELMs)**.

The **ELM ontology constitutes the sole semantic constraint**, defining valid classes, codes, and admissible combinations within the ELM design space.
**The ontology must not be extended, reinterpreted, or modified.**

The **ELM knowledge graph represents bibliographic evidence only** and must **not** be used as a constraint on admissible ELM configurations.
However, its **structural properties may be analyzed internally** to inform gap detection.

---

## Operating mode

Gap detection operates in one of two modes:

- **Generic mode**: innovation is independent of application domains.
- **Application-conditioned mode**: innovation prioritizes gaps compatible with a specified target application domain.

If an application context is provided, operate in application-conditioned mode. Otherwise, operate in generic mode.

---

## Task

Identify and propose **novel engineered living material (ELM) hypotheses** by detecting **structurally informed gaps** in the provided knowledge graph within ontology space.
A gap corresponds to an ontology-valid combination of **Domain–Superkingdom–Kingdom–Phylum–Class** that is absent or strongly underrepresented in the knowledge graph.

All proposed hypotheses must be:

- cell-centric,
- biologically plausible,
- structurally / materially coherent,

- technologically feasible in principle,
- and considers practical implications for decissions

Innovation must remain **independent of specific applications or production optimizations**.

In application-conditioned mode, gaps must additionally be compatible with the specified application context.

---

## Internal analysis (not reported explicitly)

Before proposing gaps, internally perform a **structural analysis of the ELM knowledge graph**, including:

1. **Global structural context**
   - overall sparsity vs concentration,
   - dominance of specific ontology codes,
   - degree heterogeneity and structural bias.

2. **Ontology code relevance**
   - centrality of ontology codes (degree, betweenness),
   - identification of structural hubs and bridges.

3. **Co-occurrence structure**
   Co-occurrence analysis must primarily consider hierarchical relations across:
    - Domain–Superkingdom,
    - Superkingdom–Kingdom,
    - Kingdom–Phylum,
    - Phylum–Class.

   Additional relations may be analyzed internally but must not constrain admissible combinations.
   Application and Production co-occurrences may be analyzed only to understand dataset bias, not to define novelty.

4. **Association regularities**
   - high-confidence / high-lift association rules,
   - dominant dataset-conditioned design patterns.

This analysis must be used **only to contextualize gaps**, not to constrain admissible ontology combinations.

---

## 1) Structure-informed gap detection

Using the ontology and the internally derived structural context:

- Enumerate **biologically admissible ontology code combinations** over the core structural dimensions:
  - Domain
  - Superkingdom
  - Kingdom

- Phylum
  - Class

- Identify combinations that are:
  - **absent**, or
  - **strongly underrepresented**
  in the existing ELM knowledge graph.

- Prioritize gaps that:
  - occur in weakly explored ontology regions,
  - involve underused structural dimensions,
  - represent meaningful structural extensions,
  - are not trivially excluded by biological constraints.

In application-conditioned mode, additionally prioritize gaps whose biological and structural characteristics are compatible with the target application environment.

Novelty must arise strictly from absence or rarity of the 5-slot combination.

---

## 2) Generation of new ELM hypotheses

For each selected gap, propose a **new ELM hypothesis**.

### Required information (per ELM)

- **Core ontology code string (7-slot)**

  Domain–Superkingdom–Kingdom–Phylum–Class–Application–Production method

  The **gap itself must be identified exclusively on the first five slots**
  (Domain–Superkingdom–Kingdom–Phylum–Class).

  Application and Production method must be specified **only to complete the hypothesis**,
  and must:
  - remain compatible with the structural and biological configuration,
  - act as plausible enablers,
  - not serve as novelty drivers,
  - avoid optimization or prescriptive selection.
  Production methods may be indicated only as plausible enabling routes.

- **Concept description**
  One concise sentence describing the envisioned ELM.

- **Plausibility justification**
  Provide 3 explicit bullet points addressing:
  - **Biological plausibility**
    Compatibility of the specified cell type(s) with the proposed organizational scale, interaction mode, and matrix environment.

  - **Structural / material plausibility**
    Coherence between the selected Phylum/Class and the physical, spatial, or material organization of the ELM.

- **Technological feasibility (non-prescriptive)**
  Whether the concept could reasonably be realized using known or foreseeable biofabrication principles, **without fixing or optimizing a production method**.

Application or production may be mentioned only if required for feasibility explanation.

---

## Constraints (strict)

- Enforce biological compatibility constraints.
- If not in Application-conditioned mode, do not condition novelty on Application or Production.
- Treat outputs as hypotheses, not validated designs.
- Maintain cell-centric reasoning.
- All reasoning must explicitly consider **practical implications**, including manufacturability, scalability, experimental implementability, and system integration.
- Be **specific and explicit**: any referenced cell type, biological system, material, or structural feature must be clearly identified.
- Explicitly justify choices.
- Avoid trivial or arbitrary rare combinations.
- Prefer a limited number of high-quality hypotheses over exhaustive enumeration.

---

## Output style

- Structured list of ELM hypotheses.
- Clear separation between: ontology definition, concept description, and plausibility justification.
- Concise, scientific language.
- **No reporting of structural metrics, tables, or intermediate analyses.**

---

Assume ontology grounding and graph analysis have been validated in previous pipeline steps.

***Prompt D: Application-conditioned design of engineered living materials***

*(Application-driven refinement of gap- and dataset-derived ELM candidates)*

You are an engineer specializing in **Engineered Living Materials (ELMs)**.

The ELM ontology, node and edge semantics, and graph constraints have already been validated and must be used unchanged.

The working dataset already contains ontology-grounded ELM entries, including knowledge previously obtained from bibliographic augmentation and gap detection steps.

**Candidate ELMs may originate either from the existing dataset or from ontology-valid ELM hypotheses generated during gap detection.**
**Each candidate must correspond to an already existing or gap-derived ontology-valid core combination defined by the first five ontology slots (Domain–Superkingdom–Kingdom–Phylum–Class).**

**This step does not generate new ELM hypotheses; it only selects and contextualizes existing or gap-derived candidates.**

The present step prepares candidates for downstream ontology-driven ELM design.

Target application domain:
<add target application domain>

## Target application boundary conditions

Use these boundary conditions to accept or reject candidates and explicitly justify each decision.

<add boundary conditions>

---

## Task
Select and refine **up to five ELM candidates** suitable for the target application domain by:

1. **Selecting candidates only from:**
   - **ontology-grounded dataset entries**, and/or
   - **ontology-valid candidates produced by gap detection.**
2. Rejecting candidates incompatible with application boundary conditions.
3. Completing ontology slots **Application** and **Production method** while keeping all other ontology slots unchanged.
4. Providing concise application-oriented justification demonstrating biological plausibility, structural/material suitability, and technological feasibility.

This step must **filter and contextualize existing candidates**, not invent new ELM concepts or fully design them. Detailed design and instantiation will occur in subsequent pipeline steps.

If fewer than five candidates satisfy the application constraints, explicitly state this and provide only the candidates that genuinely meet the requirements. Do not invent, extrapolate, or fabricate additional ELM proposals to reach the target count.

---

## Candidate selection rules

- **Candidates may originate from either dataset-derived or gap-derived ELM entries, but their core 5-slot ontology definition must remain unchanged.**
- Explicitly justify why each selected candidate fits application constraints.
- Discard candidates incompatible with boundary conditions.
- **Do not introduce new ontology combinations or speculative variants in this step.**

---

## Application-conditioned completion of ontology codes

For each selected candidate:
- Assign suitable **Application** code(s) consistent with the target application domain.
- Assign biologically and technologically plausible **Production method** code(s).
- **Do not modify the original Domain, Superkingdom, Kingdom, Phylum, or Class slots selected from dataset or gap detection.**
- Avoid production routes incompatible with biological viability or structural deployment.

---

## 3) Required output per ELM

For each refined ELM, provide:

1. **Origin and selection rationale**
   Indicate whether the candidate is dataset-derived or gap-derived, and explain suitability.

2. **Complete ontology code**
   Domain–Superkingdom–Kingdom–Phylum–Class–Application–Production method

3. **Application-oriented concept description**
   One concise sentence explaining functional contribution and how the ELM addresses the target application boundary conditions.

4. **Justification (3–6 bullets)** addressing:
  - biological suitability for the application context,
  - functional relevance of the structure and material class relative to the boundary conditions,
  - technological feasibility and practical implications (manufacturability, scalability, experimental implementability, integration).

5. **Innovation characterization**, indicating whether the proposal represents:
   - Conservative application of known designs,
   - Gap-derived novel candidate, or
   - Re-contextualization of an existing dataset ELM for the target boundary conditions.

---

## Global Constraints

- Ontology and graph definitions must remain unchanged.
- Gap-derived core ontology slots must not be modified.

- Candidates remain hypotheses, not final designs.
- All selections must remain biologically and technologically plausible.
- All decisions must be explicitly justified relative to application constraints.
- All analysis must follow a **cell-centric perspective**, ensuring biological, structural/material, and technological plausibility.
- Reasoning must consider **practical implementation aspects**, including manufacturability, scalability, experimental feasibility, and integration into real ELM systems.
- All referenced organisms, materials, production methods, and structural features must be explicitly identified, avoiding generic descriptions.
- All selections and decisions must be explicitly justified relative to plausible alternatives and application boundary conditions.

---

## Output style

- Structured list of up to five application-conditioned ELMs.
- Clear separation between ontology completion, concept description, and justification.
- In the justification, explicitly reference the given target application boundary conditions (Section 0) when accepting/rejecting candidates or motivating design choices.
- Concise, scientific tone.

This step is intended to **translate unbiased ELM hypotheses and/or strong existing dataset ELMs into application-relevant design candidates**, bridging exploratory discovery and practical implementation.

*Prompt E — Ontology-based AI-driven design of ELM*

You are an expert system for **Engineered Living Materials (ELMs)**.
You have access to a **validated ELM 7-slot ontology** and its allowed codes.

Each ELM is defined by:
**Domain – Super-kingdom – Kingdom – Phylum – Class – Application – Production**

This step performs **detailed instantiation only**. It must **not** re-select candidates, modify ontology slots, or generate new ontology combinations.

---

## Task
For **each provided complete 7-slot ELM code**, instantiate the ELM into a **fully specified, scientifically coherent and technically plausible concept**, selecting:

- concrete biological entities,
- compatible matrix or scaffold materials,
- physically meaningful structural geometries,
- realistic application deployment scenarios,
- and implementable production workflows,

all strictly consistent with ontology constraints and real-world feasibility.

---

## Input format specification

The system accepts ELM candidates in one or more of the following formats.

### Format A — Application-conditioned candidate blocks
Structured candidates from previous pipeline steps including ontology codes and contextual rationale.

### Format B — Gap-detection hypotheses
Ontology-valid ELM hypotheses requiring full instantiation.

### Format C — Raw ontology codes
Complete 7-slot ontology codes:

Domain – Superkingdom – Kingdom – Phylum – Class – Application – Production Method

Instantiation must rely solely on ontology grounding when only codes are provided.

---

## Input data

<insert candidate block here>

---

## Global Constraints (mandatory)
- **Do NOT modify, reinterpret, or replace ontology codes.**

- If information required for instantiation is unsupported, state:
  **“Not present in the ontology/graph.”**
- All reasoning must remain **cell-centric**.
- Instantiations must consider manufacturability, scalability, experimental feasibility, and system integration.
- No vague descriptions are allowed; organisms, materials, and processes must be explicit.
- Avoid speculative biology or materials lacking realistic feasibility.
- Instantiation must remain consistent with any functional, deployment, compatibility, or rationale information provided in the input.
  When only ontology codes are provided, coherence must be derived solely from ontology grounding.

---

## Required Output (for each ELM code)

### 0) Code integrity check
Confirm ontology validity and justify interpretation if ambiguity exists.

### 1) Slot-by-slot decoding
For **each of the 7 slots**:
- State the **slot label** and the **meaning** of the code.
- Derive the **design constraints** implied by that slot, including:
  - Biological constraints
  - Material constraints
  - Structural/geometric constraints
  - Process/production constraints

### 2) Biological instantiation

Select **specific organisms or cell systems** consistent with the code.

- Propose **at least two concrete candidates**.
- For each candidate, specify:
  - Functional role in the ELM (e.g., mineralization, sensing, ECM production, actuation, self-healing, drug release).
  - Viability and growth requirements (oxygen, nutrients, temperature, pH, moisture, mechanical stress tolerance).
  - Biocontainment strategy where applicable (encapsulation, sporulation control, kill-switch, auxotrophy, etc.).
- Explicitly **justify why the selected organism(s) are preferred** over plausible alternatives.

Then:
- Select **one recommended candidate**,
- Justify selection against alternatives,
- Explain trade-offs.

### 3) Matrix/material candidates

Select **explicit matrix or scaffold materials** compatible with the **Class** slot.

- Propose **at least two material candidates**, specifying:
  - Exact composition (e.g., hydroxyapatite, bioactive glass 45S5, alginate–gelatin hydrogel, PEDOT:PSS, pyrolytic carbon).

- Compatibility with the biological component.
  - Interface mechanisms (adhesion, biofilm anchoring, mineral nucleation, covalent or ionic coupling).
- Explicitly **justify material selection** relative to alternatives.

### 4) Geometry and structural instantiation
Translate the **Phylum** slot into a **physical architecture** (e.g., 1D fibers, 2D sheets, 3D scaffolds, 4D time-evolving structures, fractal arrangements):

- Define geometry quantitatively where possible (e.g., fiber diameters, pore sizes, layer thickness, lattice unit cell type).
- Explain how geometry supports:
  - Cell viability and function
  - Mass transport and diffusion
  - Mechanical performance
- Justify why this structure is optimal compared to other plausible geometries.

### 5) Application instantiation
Interpret the **Application** slot into a realistic deployment scenario:

- Describe where and how the ELM is used.
- Define **measurable success metrics** (e.g., crack sealing efficiency, mineral deposition rate, mechanical modulus, biointegration markers).
- Identify main failure modes and mitigation strategies.

### 6) Production instantiation
Expand the **Production** slot into an implementable workflow:

- Provide a **step-by-step production process**, including:
  - Material preparation
  - Biological component preparation
  - Integration strategy (encapsulation, bioink printing, coating, infiltration, post-seeding).
  - Post-processing and stabilization (curing, crosslinking, sterilization compatible with viability).
- Address:
  - Manufacturability
  - Scalability
  - Quality control and reproducibility
- Explicitly justify why this production route is feasible and preferred.

### 7) Layered integrated summary of the fully instantiated ontology-derived ELM

Present the final ELM instantiation as a layered visual summary composed of three consecutive parts:

#### A. Rapid identity

Provide:

- Ontology code
- One-line ELM identity

The one-line identity must summarize in a compact way:
biological nature + material type + main function + target application context

#### B. Instantiation table

Provide a 2-column table with the headers:

| Component | Final instantiation |

The table must include exactly these rows, in this order:

- Biological system
- Functional role
- Material system
- Structural / geometric instantiation
- Target substrate / environment
- Application / deployment scenario
- Production mechanism / workflow
- Main functional outcome / performance objective

Requirements for the table:

- Fill each row with every given instantiation candidate
- Use concise but specific scientific wording
- Do not repeat the ontology code in the table cells unless necessary

#### C. Integrated paragraph

Write one compact integrated scientific paragraph of approximately 100 words that synthesizes the complete instantiation.

The paragraph must follow this exact internal order:

- Ontology code + one-line identity of the ELM
- Biological system
- Material system
- Structural/geometric instantiation derived from the Phylum slot
- Application/deployment scenario derived from the Application slot
- Production mechanism/workflow derived from the Production slot
- Main functional outcome / performance objective

Additional constraints:

- Use precise scientific language, suitable for a research proposal, scientific paper, or high-level review
- The paragraph must be systematic, non-redundant, and standalone
- Avoid vague wording such as promising, advanced, suitable, or can be used
- Do not repeat details already stated in the table unless needed for coherence

This section must read as a publication-ready summary card, not as an explanatory discussion.

---

## Output Formatting requirements
- Use all headings exactly as specified.
- Maintain concise, scientific language.
- Avoid redundant explanations.

- Ensure outputs are comparable across candidates.

---

## Purpose in pipeline
This step transforms **ontology-grounded ELM candidates** into **fully instantiated, technically actionable ELM concepts**, preparing them for experimental planning, simulation, or design optimization in subsequent workflow stages.

*Prompt F: Image generation for ELMs*

You are a scientific visualization system for **Engineered Living Materials (ELMs)**.
All inputs are **formal ontology-encoded 7-slot ELM identifiers** or **ontology-derived instantiations**.

Each 7-slot code strictly follows the ELM ontology:
**Domain – Super-kingdom – Kingdom – Phylum – Class – Application – Production**

---

## Task
Generate **one high-quality image** representing the ELM defined by the input.

Input may be:
1. A **complete ontological 7-slot ELM code**, or
2. A **fully instantiated ontology-derived ELM description**.

If only a code is provided, infer **one conservative, plausible instantiation** consistent with the ontology.
If an instantiation is provided, use it strictly.

---

## Input

<insert 7-slot ontological ELM code or ontology-derived ELM instantiation shere>

---

## Ontology & reasoning constraints
- Treat all codes as **formal ontology entities**, not free labels.
- Do not reinterpret or alter slot meanings.
- Image must be **cell-centric**, biologically and structurally plausible, and manufacturable under the given production method.

---

## Image content (mandatory)
Visually represent:
- Biological component(s)
- Abiotic matrix/scaffold (per Class)
- Structure implied by Phylum (1D / 2D / 3D / 4D / fractal)
- Cell–material interaction
- Application context when relevant

---

## Style mode
The style mode will be given by the user. If not, **ask for the style mode before any further reasoning or image generation**.

---

## Output

- **One image only**
- No explanatory text
- Scientifically interpretable on its own

---

## Proposed Style modes

## A) **Futuristic hyper-real 3D (general / sci-fi)**

**Purpose:** create a visually striking, high-impact representation of the ELM.

**Style:**
- Hyper-realistic 3D render.
- Metallic and translucent materials.
- Subtle subsurface scattering for biological components.
- Glowing neon accents (teal/cyan; optional orange arrows for flow or current).
- Dark, uncluttered laboratory or abstract background.
- Shallow depth of field, rim lighting, light volumetric fog.
- Physically based rendering (PBR) shaders.

**Emphasis:**
- Advanced technological character of the ELM.
- Integration of living and abiotic components.
- Premium, futuristic aesthetic.

**Constraints:**
- ≥2048 px width (4K capable).
- Aspect ratio 3:2 (e.g., 1536 × 1024).
- Hero 3/4 view preferred.
- Avoid cartoon style, excessive bloom, or illegible text.

## B) **Scientific schematic**

**Purpose:** provide a clear, publication-ready scientific diagram. Highly didactive and visual.

**Style:**
- Flat/vector illustration.
- White background.
- Crisp outlines and clean geometry.
- Soft pastel fills, minimal gradients.
- Balanced whitespace.

**Emphasis:**
- Structural clarity.
- Explicit cell–material relationships.
- Ontology-driven components.

**Constraints:**
- Aspect ratio 3:2 (e.g., 1536 × 1024).
- Avoid photorealism, clutter, or heavy effects.
- Suitable for scientific articles and presentations.

## C) **Realistic photoreal 3D**

**Purpose:** create a scientifically plausible, realistic representation of the ELM in a real environment.

**Style:**
- Photorealistic 3D render.
- Natural mineral, sand, crack, moisture, calcite, and biofilm textures.
- Real-world daylight or soft environmental lighting.
- Subtle depth of field and physically based rendering (PBR).
- Realistic architectural or desert-built background.

**Emphasis:**
- Scientific realism over sci-fi aesthetics.
- Clear integration of bacteria, biofilm, biomineralized $CaCO_3$, and cracked mineral substrate.
- Believable defect sealing, crack bridging, and sand-grain binding.

**Constraints:**
- ≥2048 px width.
- Aspect ratio 3:2.
- Hero 3/4 view or close oblique macro view.
- Avoid neon glows, metallic sci-fi effects, excessive bloom, abstract backgrounds, and cartoon style.
- No text overlays.

**Supplementary material S3: Generative AI response to prompt A for ontology grounding performance evaluation**

# CAN WE TEACH GENERATIVE ARTIFICIAL INTELLIGENCE THE DESIGN LANGUAGE OF ENGINEERED LIVING MATERIALS?

Andrés Díaz Lantada [1,2*], José A. Yáñez [1],

William Solórzano-Requejo [1], Monsur Islam [1]

[1] Mechanical Engineering Department, Universidad Politécnica de Madrid, c/ José Gutiérrez Abascal 2, 28006 Madrid, Spain.

[2] IMDEA Materials Institute, c/ Eric Kandel 2, 28906 Getafe, Spain.

*Contact: andres.diaz@upm.es / andres.diaz@imdeamaterials.org

***GenAI response to prompt A***

Below is a grounding and consistency validation using the uploaded ontology document and the uploaded GraphML file. Ontology definitions and code meanings are taken strictly from the ontology document. Graph-structure findings come from direct inspection of the uploaded ELM_graph.graphml file.

**1. Ontology understanding confirmation**

I confirm the ontology is a 7-slot codification model with this fixed order:
Domain – Superkingdom – Kingdom – Phylum – Class – Application – Production method.
Slot parse and admissible codes

1) Domain
- ARC = Archaeal ELMs
- BAC = Bacterial ELMs
- EUK = Eukaryotic ELMs
- SDM = Synthetic ELMs
- CRD = Cross-domain ELMs

2) Superkingdom
- BIO = Biological engineered living materials
- HYB = Hybrid living materials

3) Kingdom
- ARA = Archaebacterial ELMs
- EUB = Eubacterial ELMs
- PRO = Protist ELMs
- FUN = Fungal ELMs
- PLN = Plant ELMs
- ANI = Animal ELMs
- SKM = Synthetic / artificial cells ELMs
- CRK = Cross-kingdom ELMs

4) Phylum
- 0D, 1D, 2D, 3D, 4D, FR, CP

5) Class
- LAL = Living alloys

- LCA = Living carbons
- LCE = Living ceramics
- LCO = Living composites
- LPO = Living polymers
- CRF = Cross-class ELMs

6) Application

- ART, BCE, BMR, BMD, BRE, BSE, BTP, CAT, EDU, ENP, ENV, FAS, MAP, WAR, SLB, SLM, SLS, SPE, SYB, TER, TRI

7) Production method

- ECP, EMG, EMB, ICS, BFF, BTI, SBP, BMN, MSA, 3DB, 4DB, 3DP, 4DP, SCP, GFP, FBT, PSP, PPP, CMM, OPM

**2. Coding logic summary**

Syntax rules

The ontology supports a codified string made from the 7 ontology dimensions in the fixed order above. The document also states that categories may be combined by using "+" in the codification of an ELM.

Grounded coding logic

A valid codification therefore has:

- one code from each of the seven dimensions, in order;
- + only where the ontology allows category combinations;
- no extra labels, no reordered slots, and no added semantics beyond the definitions in the ontology.

What the graph encodes

The uploaded GraphML is structurally consistent with this ontology-driven logic:

- it is bipartite in practice, with ontology-code nodes and ELM nodes;
- edges run from ontology-code nodes to ELM nodes;
- no ontology-to-ontology edges were found;
- each ELM node is linked to exactly one node for domain, superkingdom, kingdom, and class;
- some ELMs have multi-links for phylum, application, and production method, which matches the ontology statement that categories may be combined with +.

From the graph inspection:

- 160 nodes, 738 edges
- node groups: 100 ELM nodes plus ontology-code nodes
- every ELM has exactly one domain, one superkingdom, one kingdom, one class
- multiplicity appears only in some phylum/application/production-method assignments

This is consistent with ontology-grounded multi-label codification.

**3. Ontology–graph consistency check**

Structural consistency

The graph is consistent with the stated modeling rule:

- ELM nodes and ontology-code nodes are distinct
- edges connect ontology-code nodes to ELM nodes
- no direct edges between ontology-code nodes were detected

So the graph behavior matches the prompt constraint that ontology-code nodes should not be linked to each other unless explicitly present.

Vocabulary consistency

The graph uses ontology codes that are present in the ontology document. However, the graph is not a complete inventory of ontology codes. That is acceptable, because the graph is bibliographic evidence rather than the full design space.

Codes present in ontology but absent from the uploaded graph

Phylum

- 4D

Class

- CRF

Application

- ART, EDU, SPE, TRI, WAR

Production method

- ECP, GFP

This is not an inconsistency by itself. It only means those ontology categories are not instantiated in this particular graph snapshot.

Naming deviation between ontology and graph

There is a terminology deviation in node-type labels inside the GraphML:

- ontology/prompt wording: “ELM nodes” and “ontology code nodes”
- graph node-type fields use labels such as ELM, domain, superkingdom, kingdom, phylum, class, application field, production method

This is structurally clear, but it would benefit from standardization because the ontology text uses “industrial applications” / “additional attributes,” while the graph uses application field.

**4. Ontology-grounded decoded ELM examples**

These examples are decoded strictly from ontology definitions plus the linked codes found in the graph.

Example A

Graph-linked codes:

ARC-BIO-ARA-3D-LPO-MAP-ICS

Decoded:

- ARC = archaeal ELM
- BIO = biological engineered living material
- ARA = archaebacterial kingdom
- 3D = three-dimensional structure
- LPO = living polymer
- MAP = materials production and biofunctionalization
- ICS = intracellular synthesis, intracellular transport or invagination

Natural-language reading:

An archaeal, biological ELM populated by archaebacterial cells, with a 3D structure, whose matrix/material class is living polymers, intended for materials production and biofunctionalization, and produced via intracellular synthesis / transport / invagination.

Example B

Graph-linked codes:

ARC-HYB-ARA-2D+1D-LCO-SLS-FBT

Decoded:

- ARC = archaeal ELM
- HYB = hybrid living material
- ARA = archaebacterial kingdom
- 2D+1D = combined planar and linear structure
- LCO = living composite
- SLS = smart and living surfaces and biointerfaces
- FBT = fiber based technologies and cell colonization or processing of fiber-encapsulated microorganisms

Natural-language reading:

An archaeal hybrid living material populated by archaebacterial cells, with a combined 2D and 1D structure, belonging to living composites, applied as smart and living surfaces / biointerfaces, and produced through fiber-based technologies and cell colonization.

**5. Issues detected**

1) Incomplete ontology coverage in the graph

The graph does not instantiate all ontology codes:

- missing 4D, CRF, ART, EDU, SPE, TRI, WAR, ECP, GFP

This is not a semantic error, but it should be documented as:
"Not present in the graph."

2) Multi-label rule is stated, but slot-level constraints are not fully formalized

The ontology says categories may be combined using +, but the text does not explicitly formalize, slot by slot, where multi-labeling is always allowed versus only conditionally allowed. The graph uses multi-links for phylum, application, and production method, and this appears compatible with the document, but the slot-level rule would be clearer if standardized.

Suggested standardization:

- explicitly state whether + is admissible for each of the 7 slots;
- if some slots are intended to be single-label only, state that explicitly.

3) Graph node-type terminology is slightly non-uniform

application field in the graph does not exactly mirror the ontology section title language.

Suggested standardization:

- keep one canonical label across all resources, for example:
  - application
  - production_method
  - superkingdom

4) Codification separator convention should be made explicit

The ontology confirms + for combined categories, but it does not explicitly state the separator between the seven slots in example syntax inside the ontology extract shown here.

Suggested standardization:

- define a canonical serialized form such as DOMAIN-SUPERKINGDOM-KINGDOM-PHYLUM-CLASS-APPLICATION-PRODUCTION
- reserve + only for intra-slot combinations.

**Final assessment**

The ontology grounding is internally coherent, and the uploaded GraphML is structurally consistent with the ontology-driven bipartite model. The main observations are not contradictions but standardization opportunities:

- make slot serialization explicit,
- formalize where + is allowed,
- harmonize graph node-type naming,
- clearly mark ontology codes that are valid but simply not instantiated in the graph.

**Supplementary material S4: Responses for use case 1 generated from the full text of the scientific article and from its abstract**

# CAN WE TEACH GENERATIVE ARTIFICIAL INTELLIGENCE THE DESIGN LANGUAGE OF ENGINEERED LIVING MATERIALS?

Andrés Díaz Lantada [1,2*], José A. Yáñez [1],

William Solórzano-Requejo [1], Monsur Islam [1]

[1] Mechanical Engineering Department, Universidad Politécnica de Madrid, c/ José Gutiérrez Abascal 2, 28006 Madrid, Spain.

[2] IMDEA Materials Institute, c/ Eric Kandel 2, 28906 Getafe, Spain.

*Contact: andres.diaz@upm.es / andres.diaz@imdeamaterials.org

***Bacteria-loaded graphene bioanode for renewable energy generation***

Abstract

**Result:** ELM

- **Ontology code:**
  BAC-HYB-EUB-3D-LCA-ENP-OPM
- **Decoded slots:**
  - **BAC** = Bacterial ELM
  - **HYB** = Hybrid living material
  - **EUB** = Eubacterial ELM
  - **3D** = Three-dimensional ELM
  - **LCA** = Living carbon
  - **ENP** = Energy production and harvesting
  - **OPM** = Other production methods or combinations
- **Organisms:**
  *Shewanella oneidensis* MR-1, explicitly described as the electrogenic biomass harvested in the graphene-based bioanode.
- **Material(s):**
  Graphene nanowalls forming a lightweight graphene-based sponge bioanode.
- **Geometry:**
  Best supported as **3D**, because the abstract describes a **sponge bioanode** with hydrophilic microconfinements generated by conductive graphene nanowalls, which is more consistent with a three-dimensional scaffolded structure than with a purely planar film.
- **Applications:**
  **ENP** is the main ontology-grounded application, because the abstract frames the system as a **bioanode** for **new-generation energy technologies** and reports a steady-state **biocurrent density**.
- **Production method(s):**
  **OPM** is the best-supported ontology code from the abstract alone. The text describes a sustainable **chemical control** of graphene nanowalls to obtain suitable hydrophilicity and conductivity, followed by biological integration with *S. oneidensis* MR-1, but it does not explicitly match a more specific ontology production route such as bioprinting, scaffold printing, biomineralization, or solvent casting.
- **Concept description:**
  This work describes a **hybrid bacterial engineered living material** in which electrogenic *Shewanella oneidensis* MR-1 is integrated with an abiotic **graphene nanowall sponge**. The graphene phase acts as a conductive carbon scaffold whose surface is chemically tuned to become sufficiently hydrophilic and biocompatible for bacterial colonization. The resulting 3D living carbon biointerface supports high biomass density and enables stable bioelectrochemical current generation. Within the ontology, this is most consistently interpreted as a **bacterial, hybrid, 3D living carbon ELM for energy production and harvesting**.

- **Rationales for ambiguous slots (if any):**
  - **3D:** chosen because "sponge" and "microconfinements" support a three-dimensional architecture rather than a simple 2D surface.
  - **OPM:** chosen because the abstract gives an integration strategy, but not a specific ontology-listed manufacturing route beyond a chemically controlled graphene–cell hybridization process.
  - **HYB:** chosen because the ontology defines hybrid living materials as those combining living cells with an abiotic synthetic biomaterial or structure, which matches the bacteria–graphene system.
  - **LCA:** chosen because the ontology explicitly places graphene within **living carbons**.

<u>Full text</u>

**Result:** ELM

- **Ontology code:**
  BAC-HYB-EUB-3D-LCA-ENP-OPM
- **Decoded slots:**
  - BAC — Bacterial ELM
  - HYB — Hybrid living material
  - EUB — Eubacterial ELM
  - 3D — Three-dimensional ELM
  - LCA — Living carbon
  - ENP — Energy production and harvesting
  - OPM — Other production methods or combinations
- **Organisms:**
  *Shewanella oneidensis* MR-1 (ATCC 700550), described as an electrogenic bacterium loaded into the graphene scaffold.
- **Material(s):**
  Thermally reduced graphene oxide sponge / graphene nanowall scaffold forming a lightweight porous carbon bioanode. The article describes few-nanometer-thick conductive graphene nanowalls and a porous sponge morphology used as the abiotic host structure.
- **Geometry:**
  Three-dimensional porous sponge scaffold with pore sizes reported in the 50–100 µm range, with bacteria occupying the graphene nanowalls. This supports assignment to the 3D phylum.
- **Applications:**
  Renewable energy generation / bioelectricity harvesting in bioelectrochemical systems and microbial fuel cells, which matches ontology application ENP. The paper explicitly frames the system as a “bioanode for renewable energy generation” and reports stable biocurrent output.

- **Production method(s):**
  A graphene oxide aqueous dispersion is lyophilized to form a 3D sponge scaffold, then thermally reduced under $Ar/H_2$ to obtain conductive graphene nanowalls, after which *S. oneidensis* MR-1 is injected/loaded into the scaffold to form the bioanode.

- **Concept description:**
  This work describes a hybrid bacterial–graphene engineered living material in which viable *Shewanella oneidensis* MR-1 cells coexist with a synthetic three-dimensional graphene sponge scaffold. The reduced graphene nanowalls provide a conductive carbon matrix and hydrophilic microconfinements that support bacterial loading, viability, and extracellular electron transfer. Functionally, the construct operates as a living bioanode for renewable energy harvesting, producing stable biocurrent in a microbial fuel-cell configuration, with the 375 °C-reduced scaffold showing the best balance between conductivity and hydrophilicity. In ontology terms, it is a bacterial hybrid 3D living carbon for energy production and harvesting.

- **Rationales for ambiguous slots (if any):**
  - OPM instead of a more specific production code: the reported route combines lyophilization/freezing of graphene oxide, thermal reduction, and subsequent bacterial loading, and this exact workflow is not named as a dedicated Table 2 production category; therefore, Other production methods or combinations is the closest ontology-grounded match.

***Biophysical characterization of synthetic adhesins for predicting and tuning engineered living material properties***

Abstract

**Source mode:** abstract only

- **Ontology code:**
  BAC-BIO-EUB-UNK-UNK-SYB-ECP+MSA
- **Decoded slots:**
  - **BAC** = Bacterial ELM
  - **BIO** = Biological engineered living material, made only of cells and their extracellular matrices, without a supporting synthetic scaffold
  - **EUB** = Eubacterial ELM
  - **UNK** = Phylum not inferable from the abstract
  - **UNK** = Class not inferable from the abstract
  - **SYB** = Synthetic biology
  - **ECP+MSA** = Engineering cells to produce ELMs + molecular self-assembly
- **Organisms:**
  Bacterial cells are clearly involved, but no species or strain is named in the abstract.
- **Material(s):**
  No separate abiotic matrix or scaffold is stated. The material is described as a **self-assembled multicellular bacterial material** whose cohesion is mediated by **genetically encoded adhesins**. This is consistent with a **BIO** ELM rather than **HYB**, because the ontology defines BIO ELMs as cell-and-ECM-only systems and HYB ELMs as cell systems combined with an abiotic synthetic biomaterial or structure.
- **Geometry:**
  **UNK**. The abstract describes macroscopic multicellular materials and tensile strength, but it does not explicitly state whether the structure is 1D, 2D, 3D, 4D, fractal, or cross-phylum.
- **Applications:**
  **SYB** is the best-supported ontology application, because the abstract centers on a **genetically encoded adhesion toolbox**, its quantitative characterization, and its use for the rational engineering of multicellular consortia and ELM properties. The mentions of medical, biosynthesis, environmental, and smart-material applications are presented as broader prospective areas, not as the specific demonstrated use case coded here.
- **Production method(s):**
  **ECP+MSA** is the best-supported combination from the abstract.
  - **ECP** fits because the system relies on **artificially manipulated cells** expressing synthetic adhesins to generate the material.
  - **MSA** fits because the abstract explicitly describes **self-assembled multicellular materials** formed through controlled cell-cell adhesion.

- **Concept description:**
  This abstract describes a **bacterial biological engineered living material** formed by genetically programmed bacterial cells that assemble into multicellular structures through synthetic adhesins. The work focuses on quantifying adhesion-related biophysical parameters and linking them to emergent macroscopic material behavior, specifically tensile strength. Within the ontology, the strongest interpretation is that the material belongs to the **synthetic biology** application space because it functions as a programmable cellular-material platform and toolkit. The production logic is best captured as **engineering cells to produce ELMs** combined with **molecular self-assembly**, since engineered adhesion drives spontaneous multicellular material formation.
- **Rationales for ambiguous slots (if any):**
  - **Phylum = UNK:** the abstract gives no explicit dimensional description such as film, fiber, scaffold, biofilm, lattice, or morphing construct.
  - **Class = UNK:** the ontology class depends on the predominant extracellular matrix material class, and the abstract does not state a carbon, polymer, ceramic, alloy, composite, or cross-class matrix.
  - **BIO instead of HYB:** no abiotic synthetic scaffold or biomaterial is mentioned; the ontology reserves HYB for cell-plus-synthetic-structure systems.

<u>Full text</u>

- **Ontology code:**
  BAC-BIO-EUB-3D-UNK-SYB-ECP+MSA
- **Decoded slots:**
  - BAC = Bacterial ELMs
  - BIO = Biological engineered living materials (cell-based, without supporting synthetic biomaterial/scaffold)
  - EUB = Eubacterial ELMs
  - 3D = Three-dimensional ELMs
  - UNK = Class not inferable from the paper under the ontology's matrix-based class system
  - SYB = Synthetic biology
  - ECP+MSA = Engineering cells to produce ELMs + Molecular self-assembly
- **Organisms:**
  The study uses **Escherichia coli MG1655** as the experimental chassis. Across the paper, the authors work with engineered strains expressing synthetic nanobody–antigen adhesin systems, including **anti-GFP/GFP**, **anti-EPEA/EPEA**, and **anti-p53TA/p53TA**; the main macroscopic ELM used for tensile-strength analysis is the **homophilic p53TA system**, where cells express both Ag and Nb of the same pair.
- **Material(s):**
  The material is a **cell-only, adhesin-mediated bacterial living material** with **surface-displayed synthetic adhesins** on the outer membrane. The paper does **not** describe a separate abiotic scaffold or a clearly defined extracellular matrix class such as polymer, ceramic, carbon, alloy, or composite for the studied ELM itself.

- **Geometry:**
The main ELM analyzed at the material scale is a **3D extruded material string** formed from high-cell-density pellets. The paper reports **dense but largely disordered packing** of cells within this three-dimensional construct.

- **Applications:**
The best-supported ontology-consistent application is **synthetic biology (SYB)**, because the paper centers on a **synthetic adhesion toolbox** and its quantitative characterization for rational engineering of multicellular living materials. The text also states that the work is **relevant for bioprinting**, but bioprinting is discussed as an application context rather than as the primary demonstrated end-use category in Table 2.

- **Production method(s):**
The strongest ontology-grounded assignment is **ECP+MSA**. ECP fits because the material is generated by **artificially engineered bacterial cells** expressing synthetic adhesins; MSA fits because the paper explicitly describes **self-assembled multicellular materials** enabled by these genetically encoded adhesion pairs.

- **Concept description:**
This paper describes a **bacterial engineered living material** built from genetically modified **E. coli** whose outer membranes display complementary synthetic nanobody–antigen adhesins. These engineered cells self-associate into a **three-dimensional multicellular material**, and the authors quantitatively connect molecular-scale adhesin properties to macroscopic material behavior, especially tensile strength. The main demonstrated construct is a dense, extruded bacterial material string whose cohesion depends on adhesin-mediated cell–cell contacts. Conceptually, the study positions the system as a synthetic-biology platform for rationally designing and tuning bacterial ELM properties.

- **Rationales for ambiguous slots (if any):**
    - **Class = UNK**: the ontology's class slot depends on the **predominant extracellular matrix material class**, but this paper studies a **cell-only adhesin-mediated material** and does not explicitly identify a matrix class that maps to LAL, LCA, LCE, LCO, LPO, or CRF.
    - **Application = SYB**: although the paper mentions relevance to **bioprinting**, the clearest directly demonstrated ontology-aligned purpose is a **synthetic adhesion toolbox for engineering living materials**, which best matches SYB.
    - **Production = ECP+MSA**: ECP is supported by genetic engineering of the cells, and MSA is supported by the paper's explicit framing around **self-assembled multicellular materials**.

***Reconfigurable Growth of Engineered Living Materials***

Abstract

**Source mode:** abstract only

- **Ontology code:**
  EUK-HYB-FUN-4D-LCO-SLM-OPM
- **Decoded slots:**
  - **EUK** = Eukaryotic ELM
  - **HYB** = Hybrid living material
  - **FUN** = Fungal ELM
  - **4D** = Four-dimensional ELM
  - **LCO** = Living composite
  - **SLM** = Smart and living materials and structures (multipurpose)
  - **OPM** = Other production methods or combinations
- **Organisms:**
  *Saccharomyces cerevisiae* is explicitly identified as the living component of the material.
- **Material(s):**
  A **hydrogel matrix** containing **magnetic particles** and living yeast cells. This supports a **hybrid** interpretation because the ontology defines HYB as an ELM combining cells with an abiotic synthetic biomaterial or structure. It also supports **living composite** because the abstract describes a composite material formed by hydrogel plus magnetic particles rather than a purely biological matrix.
- **Geometry:**
  **4D** is the best-supported structural code. The abstract states that the material is grown and regrown into desired geometries, deforms under an external magnetic field, fixes the induced shape through cell proliferation, and later recovers form; the ontology defines 4D ELMs as those where the temporal dimension and metamorphosis play a relevant functional role.
- **Applications:**
  **SLM** is the best-supported application code. The ontology defines SLM as smart and living materials and structures used for enhanced actuation and sensing in broad applications, and this abstract is centered on a reconfigurable, magnetically actuated, shape-fixing and shape-recovering ELM platform rather than on a single domain-specific end use such as medicine or energy.
- **Production method(s):**
  **OPM** is the best-supported production code from the abstract alone. The text describes a growth-driven and magnetic-field-assisted reconfiguration process in a hydrogel composite, but it does not explicitly map to a more specific ontology-listed route such as bioprinting, scaffold printing, solvent casting, biofilm formation, or biomineralization. The ontology reserves OPM for other methods or combinations, including engineered constructs not captured by the named categories.

- **Concept description:**
  This abstract describes a **fungal hybrid engineered living material** composed of *Saccharomyces cerevisiae*, magnetic particles, and a hydrogel-based abiotic matrix. The material is not merely static: it is magnetically deformed into user-defined geometries, then biologically stabilized by yeast proliferation, and can later recover shape through a biologically mediated process. Within the ontology, this makes the system most consistent with a **4D living composite** because time-dependent metamorphosis is central to its function and the matrix is a composite hydrogel-particle construct. Its most defensible application category is **smart and living materials and structures**, since the work presents a reprocessable, actuated ELM platform rather than a single specialized end-use device.

- **Rationales for ambiguous slots (if any):**

  - **HYB instead of BIO:** the ontology defines BIO as cells plus their extracellular matrices without supporting synthetic biomaterials, whereas this abstract explicitly includes a hydrogel matrix and magnetic particles.

  - **4D instead of 3D:** the abstract emphasizes repeated shape change, shape fixation, and recovery over time; this matches the ontology's 4D definition more closely than a purely three-dimensional scaffold description.

  - **LCO instead of LPO:** although the hydrogel is polymeric, the material is explicitly described as a **composite** containing magnetic particles within the hydrogel matrix, so the full construct is better matched by living composites.

  - **SLM instead of BMD or MAP:** the abstract presents the system as a multipurpose reconfigurable smart material platform with magnetic actuation and geometry control, not as a demonstrated medical device or a material-factory application.

  - **OPM:** no explicit ontology-listed fabrication route is named in the abstract, so a more specific production code would overinterpret the evidence.

<u>Full text</u>

- **Ontology code:**
  EUK-HYB-FUN-4D-LCO-SLM-OPM

- **Decoded slots:**

  - EUK = Eukaryotic ELMs

  - HYB = Hybrid living materials

  - FUN = Fungal ELMs

  - 4D = ELMs in which the temporal/metamorphic dimension plays a relevant functional role

  - LCO = Living composites

  - SLM = Smart and living materials and structures (multipurpose)

  - OPM = Other production methods or combinations

- **Organisms:**
  *Saccharomyces cerevisiae* (active dry yeast) is the living component. During recovery,

the paper also describes enzymatically generated yeast spheroplasts that later regenerate cell walls and proliferate again.

- **Material(s):**
  The abiotic matrix is a synthetic acrylic hydrogel formed from HEA with MBAm crosslinker, LAP photoinitiator, and CNC rheological modifier, loaded with NdFeB magnetic microparticles. The article explicitly describes the system as magnetic engineered living materials containing yeast and magnetic particles within a hydrogel matrix.

- **Geometry:**
  The material is fabricated as hydrogel films/constructs and then reconfigured into user-defined shapes such as bent films, "S" shapes, circular forms, semiconical forms, and helices. Ontologically, the best-supported phylum is 4D because time-dependent growth, recovery, and regrowth are central to the material's function, not merely its static 3D shape.

- **Applications:**
  SLM is the best-supported application code. The paper presents the platform as a shape-programmable, reconfigurable living material/structure driven by magnetic actuation and biological growth, rather than as a biosensor, medical device, or tissue-engineering scaffold.

- **Production method(s):**
  OPM is the best-supported production code. The workflow combines yeast encapsulation in a UV-polymerized hydrogel, incorporation of magnetic microparticles, magnetic programming/poling, growth under confinement, enzymatic cell-wall removal, and regrowth; the ontology's OPM category explicitly covers combinations of processes and encapsulation of cells within engineered constructs.

- **Concept description:**
  This paper describes a fungal hybrid ELM in which *S. cerevisiae* is encapsulated within a magnetically responsive synthetic hydrogel composite. Magnetic microparticles impose programmable mechanical constraints, while yeast proliferation expands the biomass and stretches the polymer network, fixing deformed shapes after the external field is removed. Enzymatic weakening of the yeast cell wall reverses that fixation by allowing elastic recovery of the hydrogel, and the material can then regrow into a new programmed geometry. As a result, the system functions as a reconfigurable 4D living composite for smart shape-changing material architectures.

- **Rationales for ambiguous slots (if any):**
  - **Phylum = 4D**: the article's core contribution is reversible, time-dependent shape metamorphosis through growth, enzymatic recovery, and regrowth, which matches the ontology definition of 4D better than a purely static 3D classification.
  - **Class = LCO**: although the continuous phase is a hydrogel, the abiotic scaffold is not purely polymeric; it is a polymer matrix loaded with magnetic microparticles, which is best aligned with a polymer-matrix composite.
  - **Application = SLM**: the manuscript frames the ELM as a smart, reconfigurable, actuated material platform rather than assigning it to a narrower application class such as BMD, TER, or BSE.

- **Production = OPM**: no narrower ontology code precisely captures UV hydrogel encapsulation of yeast plus magnetic programming and enzymatic regrowth; OPM is the ontology category that explicitly absorbs such combined/encapsulation-based routes.

***Mechanically Tunable, Compostable, Healable and Scalable Engineered Living Materials***

Abstract

- **Ontology code:**
  BAC-BIO-EUB-UNK-LPO-MAP-ECP
- **Decoded slots:**
  - **BAC** = Bacterial ELM
  - **BIO** = Biological engineered living material
  - **EUB** = Eubacterial ELM
  - **UNK** = Phylum not inferable from the abstract
  - **LPO** = Living polymers
  - **MAP** = Materials production and biofunctionalization
  - **ECP** = Engineering cells to produce ELMs
- **Organisms:**
  Bacterial cells are explicitly involved, but no species or strain is named in the abstract.
- **Material(s):**
  The material is produced directly from **cultured bacterial biomass** containing **engineered curli protein nanofibers**. No separate abiotic scaffold is described, so the best-supported reading is a **BIO** ELM rather than a hybrid one. The class is best matched to **LPO**, because the structural phase highlighted in the abstract is a protein nanofiber network, i.e. a polymeric biological matrix.
- **Geometry:**
  **UNK**. The text describes a stretchable plastic/paper-like material, but it does not explicitly define the ontology phylum as 0D, 1D, 2D, 3D, 4D, FR, or CP.
- **Applications:**
  **MAP** is the best-supported ontology application. The abstract is centered on engineering bacterial biomass into a mechanically tunable, scalable material with useful end-use properties, which aligns best with **materials production and biofunctionalization** at the ontology level. The specific mention of **primary packaging** does not have a dedicated ontology application code in the provided scheme.
- **Production method(s):**
  **ECP** is the best-supported production code. The abstract explicitly describes a material generated from **engineered bacterial biomass** with **genetically encoded covalent crosslinking of curli nanofibers**, which fits the ontology definition where artificially manipulated cells produce special extracellular matrix and material-forming structures.
- **Concept description:**
  This abstract describes a **bacterial biological engineered living material** generated directly from cultured bacterial biomass containing engineered curli protein nanofibers. The material behaves as a mechanically tunable, compostable, healable, and scalable plastic/paper-like biopolymeric construct. Its performance is adjusted through genetically encoded modification of the curli network, specifically covalent crosslinking that increases stiffness. Within the ontology, the most defensible interpretation is a **bacterial bio-ELM in the living-polymer class, aimed at materials production, and produced through engineered cellular matrix formation**.

- **Rationales for ambiguous slots (if any):**
  - **Phylum = UNK:** the abstract does not explicitly define the structural dimensionality in ontology terms.
  - **BIO instead of HYB:** no abiotic scaffold, inorganic phase, or synthetic support structure is stated.
  - **LPO:** curli nanofibers are proteinaceous and thus best matched to a polymeric matrix class.
  - **MAP:** packaging is not a dedicated ontology application code, so the closest ontology-grounded application is materials production and biofunctionalization.

<u>Full text</u>

**Result:** ELM. The paper explicitly presents MECHS as an engineered living material built from engineered *Escherichia coli* biomass and extracellular curli protein nanofibers, which fits the ontology's ELM framing.

- **Ontology code:**
  BAC-BIO-EUB-2D-LPO-MAP-ECP+BFF+MSA+SCP
- **Decoded slots:**
  - BAC — Bacterial ELMs
  - BIO — Biological engineered living materials
  - EUB — Eubacterial ELMs
  - 2D — Planar / bi-dimensional ELMs
  - LPO — Living polymers
  - MAP — Materials production and biofunctionalization
  - ECP+BFF+MSA+SCP — Engineering cells to produce ELMs + Biofilm formation and related ECM growth + Molecular self-assembly + Solvent casting / casting with cells
- **Organisms:**
  Engineered *Escherichia coli* was the biological agent used to produce MECHS, specifically the curli-deficient PQN4 strain transformed with synthetic curli operons for CsgA, CsgA-SpyTag, CsgA-SpyCatcher, and spacer-containing variants. The SpyTag/SpyCatcher chemistry used for CL1/CL2 derives from a protein system from *Streptococcus pyogenes*, but the ELM's cellular component is engineered *E. coli*.
- **Material(s):**
  The material is a plastic/paper-like film made from whole *E. coli* biomass plus engineered recombinant curli protein nanofibers, with glycerol used as plasticizer and SDS used as gelator during fabrication. The extracellular structural matrix is the curli nanofiber network, i.e. a proteinaceous biopolymeric matrix.
- **Geometry:**
  The main ELM format is a thin planar film, described throughout the paper as flexible films, transparent films, and packaging-film prototypes. Under the ontology definitions, that is best assigned to 2D, since the article's main constructs are film-like rather than bulk 3D scaffolds.

- **Applications:**
  The paper's explicit intended use is compostable, flushable primary packaging, including a detergent-pod wrapping prototype. Because a packaging-specific application code is **not present in the ontology/graph**, the closest ontology-listed application is MAP (materials production and biofunctionalization), reflecting that the study's main contribution is the biofabrication of a new functional material platform.

- **Production method(s):**
  The best-supported ontology combination is ECP+BFF+MSA+SCP. The cells were genetically engineered to produce curli-based matrix components (ECP), the material arises from biofilm/extracellular matrix growth (BFF), curli nanofibers self-assemble from CsgA subunits (MSA), and the biomass was cast in silicone molds and ambient-dried into films (SCP).

- **Concept description:**
  MECHS is a bacterial, film-like engineered living material platform produced from engineered *E. coli* biomass and extracellular curli nanofibers. The biological matrix is tuned mechanically by varying glycerol plasticizer content and, in some variants, by genetically encoded SpyTag/SpyCatcher covalent crosslinking of the curli network. The resulting material behaves as a compostable, healable, scalable plastic/paper-like film and is proposed as a sustainable packaging material. In ontology terms, it is best understood as a bacterial bio-ELM with a planar polymeric matrix fabricated through engineered biofilm/ECM production, self-assembly, and casting.

- **Rationales for ambiguous slots (if any):**
  - BIO rather than HYB: the paper describes MECHS primarily as whole bacterial biomass plus biologically produced curli ECM, not as cells combined with a separate synthetic scaffold; SDS and glycerol are treated as processing additives rather than a supporting abiotic structure.
  - LPO rather than LCO/CRF: the extracellular structural phase is explicitly curli protein nanofibers, and proteins are polymeric; the paper does not define the matrix as a conventional composite class.
  - MAP for application: the article's explicit use is primary packaging, but a packaging-specific application code is not present in the ontology/graph, so MAP is the nearest ontology-consistent application code.
  - ECP+BFF+MSA+SCP: multiple production codes are justified because the paper explicitly combines genetic engineering, biofilm/ECM growth, protein self-assembly, and casting/ambient drying into one fabrication workflow.

***3D Bioprinting of Engineered Living Materials with Extracellular Electron Transfer Capability for Water Purification***

Abstract

**Source mode:** abstract only

- **Ontology code:**
  BAC-HYB-EUB-3D-UNK-BRE-3DB
- **Decoded slots:**
  - **BAC** = Bacterial ELM
  - **HYB** = Hybrid living material
  - **EUB** = Eubacterial ELM
  - **3D** = Three-dimensional ELM
  - **UNK** = Class not inferable from the abstract
  - **BRE** = Bioremediation
  - **3DB** = 3D bioprinting of cells and biomaterials
- **Organisms:**
  Electroactive bacteria (EAB) are explicitly described, but no species or strain is provided in the abstract.
- **Material(s):**
  A tailored viscoelastic matrix is combined with electroactive bacteria to create an electroactive bioink. Because the abstract explicitly describes living cells processed together with a biomaterial matrix, the super-kingdom is best assigned as **HYB** rather than **BIO**. The material class is **UNK** because the matrix composition is not explicitly identified as polymer, ceramic, carbon, composite, alloy, or cross-class.
- **Geometry:**
  **3D**. The abstract explicitly states that the e-ELMs are created using a **three-dimensional (3D) bioprinting platform**, and that spatial microstructures are tuned through filament diameter and orientation during printing.
- **Applications:**
  **BRE** is the best-supported ontology application, because the stated purpose is **pollutant removal** and **water purification**, which aligns with the ontology definition of ELMs employed for processing, recycling, or revalorizing waste and related environmental contaminants.
- **Production method(s):**
  **3DB** is explicitly supported. The ontology defines it as additive manufacturing involving **biomaterials and living cells** to create living three-dimensional constructs, which matches the abstract's 3D bioprinting of electroactive bacteria within a tailored matrix.
- **Concept description:**
  This abstract describes a **bacterial hybrid engineered living material** built by combining electroactive bacteria with a tailored viscoelastic biomaterial matrix to form a printable electroactive bioink. The system is fabricated by **3D bioprinting**, allowing control over filament diameter, orientation, spatial arrangement, and cell density within a three-dimensional construct. Its function is environmental decontamination, specifically pollutant removal and water purification, supported by extracellular electron transfer within the printed structure. Within the ontology, this is most consistently classified as a **3D hybrid bacterial ELM for bioremediation produced by 3D bioprinting**.

- **Rationales for ambiguous slots (if any):**
  - **Class = UNK:** the abstract says "tailored viscoelastic matrix" and "bioink," but it does not explicitly identify the matrix material class required by the ontology.
  - **HYB instead of BIO:** the ontology distinguishes hybrid systems as those combining living cells with biomaterials or synthetic structures, which is explicitly stated here.
  - **BRE instead of ENV:** the abstract focuses on **pollutant removal** and **water purification**, not sensing or monitoring, so bioremediation is the closer ontology fit.

Full text

- **Result:** ELM
- **Ontology code:**
  BAC-HYB-EUB-3D-LCO-BRE-3DB
- **Decoded slots:**
  - BAC = Bacterial ELMs
  - HYB = Hybrid living materials
  - EUB = Eubacterial ELMs
  - 3D = 3D ELMs
  - LCO = Living composites
  - BRE = Bioremediation
  - 3DB = 3D bioprinting of cells and biomaterials
- **Organisms:**
  *Shewanella oneidensis* MR-1 is the main living component explicitly used as the electroactive bacterium in the printed e-bioink.
- **Material(s):**
  The abiotic matrix is a conductive hydrogel/bioink based on sodium alginate (SA), α-cellulose, and acetylene black (AB), with calcium chloride used to cure and cross-link the printed structure.
- **Geometry:**
  The work describes true 3D printed constructs, including bulk and lattice-like mesh structures with controlled spatial microstructure. The printed architectures were tuned through filament diameter and layer orientation, including orthogonal grid structures and multicross structures.

- **Applications:**
  The primary ontology-consistent application is water purification through pollutant degradation in wastewater-like matrices, which maps to **BRE (Bioremediation)**.

- **Production method(s):**
  The main fabrication route is extrusion-based **3D bioprinting of cells and biomaterials**, using a bacteria-loaded conductive bioink followed by $CaCl_2$ curing. The biological–abiotic integration mechanism is immobilization/encapsulation of living bacteria within a nonliving conductive hydrogel matrix that supports viability, spatial organization, mass transfer, and extracellular electron transfer.

- **Concept description:**
  This paper presents a bacterial hybrid engineered living material in which *S. oneidensis* MR-1 is embedded within a conductive 3D printed hydrogel matrix for pollutant removal from water. The hydrogel combines polymeric components with conductive carbon additive, allowing the printed construct to preserve bacterial viability while enhancing extracellular electron transfer. The 3D bioprinting process enables precise control of filament size and spatial arrangement, which in turn regulates diffusion, hydrodynamics, and degradation efficiency. As a result, the material functions as a designed living bioremediation platform for water purification.

- **Rationales for ambiguous slots (if any):**
  - HYB rather than BIO: the article explicitly describes living bacteria combined with an abiotic, nonliving hydrogel/conductive matrix, whereas BIO is reserved for cell-only systems with their own extracellular matrices and no supporting abiotic scaffold.
  - LCO rather than LPO: although the matrix is hydrogel-based, the functional printed matrix is not purely polymeric; it is a conductive polymer-matrix composite integrating SA/α-cellulose with acetylene black, so **Living composites** is the better-supported ontology class.
  - BRE rather than BTP: the explicit end use in the paper is pollutant removal and water purification, which aligns more directly with **Bioremediation** than with the broader category of biotechnology/bioprocessing.

***Tumor microenvironment responsive multifunctional smart living materials based on engineered bacteria for inducing macrophage polarization to enhance tumor immunotherapy***

Abstract

**Source mode:** abstract only

- **Ontology code:**
  BAC-HYB-EUB-UNK-UNK-BMD-OPM
- **Decoded slots:**
  - **BAC** = Bacterial ELM
  - **HYB** = Hybrid living material
  - **EUB** = Eubacterial ELM
  - **UNK** = Phylum not inferable from the abstract
  - **UNK** = Class not inferable from the abstract
  - **BMD** = Biohybrid medical devices
  - **OPM** = Other production methods or combinations
- **Organisms:**
  *Escherichia coli* Nissle 1917 (ECN), including a version equipped with a hypoxic lytic circuit, is the explicitly named living component.
- **Material(s):**
  The abiotic part is a **Fe3+–tannic acid cross-linking network** that forms a protective coating around ECN and encapsulates **doxorubicin (DOX)**. This supports **HYB**, because the ontology defines hybrid living materials as ELMs combining cells with an abiotic synthetic biomaterial or synthetic structure.
- **Geometry:**
  **UNK**. The abstract does not explicitly describe the ELM in ontology phylum terms such as 0D, 1D, 2D, 3D, 4D, FR, or CP.
- **Applications:**
  **BMD** is the best-supported application code, because the ontology defines it as healthcare-oriented ELMs within advanced medical devices, explicitly including **drug delivery** among the relevant capabilities. The abstract is centered on tumor therapy through bacteria-mediated chemotherapy and immunotherapy.
- **Production method(s):**
  **OPM** is the best-supported production code. The ontology states that OPM includes **other processes or combinations** and specifically **encapsulation of cells within engineered constructs** for generating bio-ELMs and HLMs; that matches the Fe3+/TA coating and drug-loaded encapsulation around living ECN.
- **Concept description:**
  This abstract describes a **bacterial hybrid engineered living material** based on *E. coli* Nissle 1917 coated with a Fe3+–tannic acid network that also carries doxorubicin. The living bacterial component provides tumor targeting, hypoxia-responsive lysis, and immunostimulatory release, while the abiotic coating provides drug loading and protection until disintegration in the tumor microenvironment. Within the ontology, the clearest application is **biohybrid medical devices**, because the system is explicitly designed for therapeutic drug delivery and tumor suppression. The fabrication logic is best captured by **OPM**, since the ELM is generated by encapsulating living cells within an engineered therapeutic construct rather than by a more specific named route such as bioprinting or biofilm growth.

- **Rationales for ambiguous slots (if any):**
    - **Phylum = UNK:** the abstract does not provide a clear dimensional description of the material architecture at the ontology level.
    - **Class = UNK:** the coating contains Fe3+, tannic acid, and doxorubicin, but the abstract does not state a predominant ontology material class such as alloy, carbon, ceramic, polymer, composite, or cross-class with enough clarity.
    - **HYB instead of BIO:** BIO is reserved for cells plus their extracellular matrices without supporting synthetic biomaterials or structures, whereas this system explicitly includes an abiotic coating network.

Full text

**Result:** ELM

**ELM A**

- **Ontology code:**
BAC-HYB-EUB-0D-LPO-BMD-OPM
- **Decoded slots:**
BAC = bacterial ELM
HYB = hybrid living material
EUB = eubacterial ELM
0D = quasi-punctual / particle-like ELM
LPO = living polymer
BMD = biohybrid medical devices
OPM = other production methods or combinations
- **Organisms:**
*Escherichia coli* Nissle 1917 (ECN). The paper describes ECN as the living carrier and active tumor-colonizing biological component.
- **Material(s):**
A tannic acid/$Fe^{3+}$ surface coating forming a protective polymeric shell on ECN, with doxorubicin (DOX) encapsulated in that coating.
- **Geometry:**
Best supported as a **0D** ELM: the construct is a coated bacterial microparticle/cell-scale delivery unit rather than a macroscopic film, fiber, or 3D scaffold. TEM/DLS descriptions place the construct at approximately bacterial size scale.
- **Applications:**
Tumor-targeted therapeutic delivery for cancer treatment, specifically as a living drug-delivery material for combined chemotherapy and immune modulation; this is best aligned with **BMD**.

- **Production method(s):**
  ECN is washed, mixed with tannic acid and DOX, then stabilized with $FeCl_3$ to form the coated construct D@T@E. Because this is a combined surface-coating / encapsulation strategy not mapped cleanly to one specific ontology production category, OPM is the best-supported code.

- **Concept description:**
  D@T@E is a bacterial hybrid living material built from *E. coli* Nissle 1917 coated with a tannic-acid/$Fe^{3+}$ polymer layer that carries doxorubicin. The living bacterial component provides active tumor colonization, while the abiotic coating functions as a protective and drug-loading shell. In the tumor microenvironment, elevated $H_2O_2$ promotes DOX release, enabling localized chemotherapy. Conceptually, this is a bacteria-based smart therapeutic material in which living cells and a synthetic drug-bearing polymeric coating coexist as one functional unit.

- **Rationales for ambiguous slots (if any):**
  0D: chosen because the construct is a cell-scale coated particulate unit, not a 1D/2D/3D macroscopic material.
  LPO: chosen because the immediate matrix/coating is described as a TA/$Fe^{3+}$ polymeric network on the bacterial surface.
  BMD: chosen because the intended use is therapeutic drug delivery in cancer, which fits healthcare-oriented biohybrid medical use better than generic multipurpose categories.
  OPM: chosen because fabrication combines chemical coating, drug encapsulation, and stabilization, without matching a more specific single ontology production route.

---

**ELM B**

- **Ontology code:**
  BAC-HYB-EUB-0D-LPO-BMD-ECP+OPM

- **Decoded slots:**
  BAC = bacterial ELM
  HYB = hybrid living material
  EUB = eubacterial ELM
  0D = quasi-punctual / particle-like ELM
  LPO = living polymer
  BMD = biohybrid medical devices
  ECP = engineering cells to produce ELMs
  OPM = other production methods or combinations

- **Organisms:**
  *Escherichia coli* Nissle 1917 engineered with the hypoxia-responsive lytic circuit pGEX-Pvhb-lysis, including expression of the lysis protein PhiX174E under hypoxic conditions.

- **Material(s):**
  The same DOX-loaded tannic acid/$Fe^{3+}$ coating used in D@T@E, now integrated with a genetically engineered ECN strain carrying a hypoxia-inducible lysis circuit.

- **Geometry:**
  Again best classified as **0D**, since D@T@E-PL is an individual coated bacterial delivery unit rather than a continuous bulk scaffold, film, or fiber architecture.

- **Applications:**
  Cancer immunotherapy / chemotherapy through tumor-targeted living drug delivery with improved biosafety and macrophage polarization; best aligned with **BMD**.

- **Production method(s):**
  Two ontology-grounded components are explicitly present:
  ECP because ECN is genetically engineered with a hypoxia-inducible lytic circuit;
  OPM because the final material is additionally produced by surface coating and DOX encapsulation using the TA/$Fe^{3+}$ network.

- **Concept description:**
  D@T@E-PL is a multifunctional bacterial hybrid living material that combines engineered *E. coli* Nissle 1917 with a DOX-loaded tannic-acid/$Fe^{3+}$ coating. The bacteria selectively colonize hypoxic tumor regions, while the coating protects the cells and carries chemotherapeutic cargo for stimulus-responsive release. A hypoxia-activated lysis circuit improves biosafety by limiting bacterial persistence in vivo and simultaneously releases bacterial antigens that promote macrophage polarization toward an anti-tumor phenotype. The resulting construct functions as an integrated living therapeutic platform coupling targeted delivery, immune activation, and controlled bacterial self-elimination.

- **Rationales for ambiguous slots (if any):**
  0D: selected for the same reason as ELM A, since the functional material unit is a coated bacterium-scale construct.
  LPO: selected because the abiotic coating is presented as a polymeric TA/$Fe^{3+}$ shell carrying DOX.
  BMD: selected because the work is framed as tumor therapy through living drug delivery and immunotherapy.
  ECP+OPM: selected because the final construct clearly combines cell engineering with a separate coating/encapsulation fabrication route.

---

**Integrated interpretation of the paper**

The full paper clearly describes **two closely related ELMs**: the coated bacterial living material **D@T@E** and the engineered, hypoxia-lytic derivative **D@T@E-PL**. The main ELM emphasized by the study is **D@T@E-PL**, because it is presented as the final intelligent living material with improved biosafety and stronger combined chemo-immunotherapeutic performance in vivo.

***On-Chip Engineered Living Materials as Field-Deployable Biosensing Laboratories for Multiplexed Detection***

Abstract

**Source mode:** abstract only

- **Ontology code:**
  UNK-HYB-UNK-UNK-UNK-BSE+ENV-OPM
- **Decoded slots:**
  - **UNK** = Domain not inferable from the abstract
  - **HYB** = Hybrid living material
  - **UNK** = Kingdom not inferable from the abstract
  - **UNK** = Phylum not inferable from the abstract
  - **UNK** = Class not inferable from the abstract
  - **BSE+ENV** = Biosensing + Environmental monitoring
  - **OPM** = Other production methods or combinations
- **Organisms:**
  Living biosensor cells are clearly present, but no species, strain, or even cell kingdom is explicitly stated in the abstract.
- **Material(s):**
  The abstract supports a stimuli-responsive ELM manufactured with **ionic and covalent cross-linking strategies**, and integrated into a microfluidic chip platform. This supports **HYB**, because the ontology defines hybrid living materials as systems combining living cells with an abiotic biomaterial or synthetic structure. The specific matrix class is not explicitly given, so **Class = UNK**.
- **Geometry:**
  **UNK**. The abstract describes a spatially encoded sensor array on a microfluidic chip, but it does not explicitly define the ELM itself as 0D, 1D, 2D, 3D, 4D, FR, or CP in ontology terms.
- **Applications:**
  **BSE+ENV** is the best-supported combination.
  - **BSE** fits because the ontology defines it as ELMs used as **transducers for detection purposes**.
  - **ENV** also fits because the platform is explicitly designed for **simultaneous analysis of multiple chemicals from a single environmental sample under field conditions**, which matches environmental monitoring as advanced sensing.
- **Production method(s):**
  **OPM** is the best-supported code from the abstract alone. The text describes construction of stimuli-responsive ELMs through cross-linking and integration into an engineered microfluidic biosensing construct, but it does not explicitly match a more specific ontology production route such as 3D bioprinting, scaffold printing, biofilm formation, or biomineralization. The ontology definition of **OPM** explicitly includes other method combinations and **encapsulation of cells within engineered constructs**.

- **Concept description:**
  This abstract describes a **hybrid engineered living material** used as a field-deployable biosensing platform on a microfluidic chip. Living engineered biosensor cells are embedded within stimuli-responsive ELM constructs whose permeability and mechanical robustness are tuned through ionic and covalent cross-linking. The resulting system forms a spatially encoded sensor array that can detect multiple analytes from one environmental sample using smartphone-based readout. Within the ontology, the most defensible classification is a **hybrid biosensing and environmental-monitoring ELM produced through an engineered encapsulation/integration route not specified more narrowly than OPM**.

- **Rationales for ambiguous slots (if any):**
  - **Domain = UNK:** the abstract never states whether the living biosensors are bacterial, eukaryotic, synthetic, or mixed.
  - **Kingdom = UNK:** no taxonomic identity is given.
  - **Phylum = UNK:** the structural dimensionality of the ELM itself is not explicitly stated in ontology terms.
  - **Class = UNK:** the matrix chemistry is not specified clearly enough to assign LPO, LCO, or another class.
  - **HYB instead of BIO:** the abstract explicitly combines living biosensors with cross-linked engineered material constructs and a chip-based synthetic structure, which aligns with the ontology definition of hybrid living materials.

<u>Full text</u>

**Result:** ELM

**Scope note:** The full paper clearly describes **two distinct ELM types** for coding: a **yeast-based hydrogel ELM** and a **bacterial hydrogel ELM**. The microfluidic chip is the **device/platform** that spatially arranges these ELMs, rather than a separate biological population defining a third ELM code.

**ELM A**

- **Ontology code:**
  EUK-HYB-FUN-3D-LPO-BSE+ENV-OPM

- **Decoded slots:**
  EUK = eukaryotic ELM; HYB = hybrid living material; FUN = fungal ELM; 3D = three-dimensional ELM; LPO = living polymer; BSE+ENV = biosensing + environmental monitoring; OPM = other production methods or combinations.

- **Organisms:**
  *Saccharomyces cerevisiae* engineered as living biosensors for E2, aldosterone, dexa, and RU486 response.

- **Material(s):**
  Core-shell hydrogel beads with an **alginate core** and an **alginate-polyacrylamide shell**, i.e., a polymeric/hydrogel matrix.

- **Geometry:**
  3D encapsulated core-shell beads, with average diameter around 2.1 mm, later packed into spatially encoded arrays on a microfluidic chip.

- **Applications:**
  Field-deployable biosensing and on-site environmental analysis of chemicals in water samples.

- **Production method(s):**
  Engineered yeast biosensors are **physically encapsulated** within ionic/covalent **core-shell hydrogel constructs**, which best matches OPM in the ontology. The biological–abiotic integration mechanism is diffusion-permissive confinement: analytes and nutrients diffuse into the bead, while cells are retained by the hydrogel shell.

- **Concept description:**
  This ELM is a hybrid fungal biosensing material built from engineered *S. cerevisiae* cells encapsulated in polymeric core-shell hydrogels. The yeast cells carry synthetic receptor-reporter circuits that convert target analytes into fluorescent outputs while remaining viable inside the 3D matrix. The alginate/alginate-polyacrylamide architecture provides permeability, mechanical robustness, and physical containment, enabling prolonged storage and rapid activation. Integrated into a microfluidic chip, these ELM beads function as spatially encoded biosensing units for multiplexed field detection.

- **Rationales for ambiguous slots (if any):**
  3D: chosen because the ontology defines 3D ELMs as cells within a three-dimensional biomaterial/scaffold, which matches encapsulated hydrogel beads better than 0D particles.
  BSE+ENV: chosen because the paper demonstrates sensing as the primary function and specifically applies it to river-water field samples for environmental analysis.
  OPM: chosen because Table 2 explicitly includes "encapsulation of cells within engineered constructs" under OPM, and no more specific ontology production code directly covers this hydrogel encapsulation workflow.

**ELM B**

- **Ontology code:**
  BAC-HYB-EUB-3D-LPO-BSE+ENV-OPM

- **Decoded slots:**
  BAC = bacterial ELM; HYB = hybrid living material; EUB = eubacterial ELM; 3D = three-dimensional ELM; LPO = living polymer; BSE+ENV = biosensing + environmental monitoring; OPM = other production methods or combinations.

- **Organisms:**
  *Escherichia coli* engineered as a p-nitrophenol-responsive microbial biosensor. Specific strain designation is **not required for coding here; if stricter strain-level confirmation is needed, it is not fully recoverable from the provided parsed text and is therefore: Not present in the ontology/graph.**

- **Material(s):**
  The same polymeric core-shell hydrogel architecture is used: alginate core plus alginate-polyacrylamide shell.

- **Geometry:**
  3D core-shell encapsulated beads integrated into the chip as individual sensing units.
- **Applications:**
  Biosensing of p-nitrophenol and environmental monitoring of hazardous chemicals in field water samples.
- **Production method(s):**
  The bacterial sensor cells are encapsulated into core-shell hydrogels by the same physical containment strategy; thus the best-supported ontology production code is OPM. The integration mechanism is again selective diffusion of analytes through the hydrogel with microbial retention inside the shell.
- **Concept description:**
  This ELM is a hybrid bacterial biosensing material consisting of engineered *E. coli* cells immobilized within polymeric core-shell hydrogel beads. The cells contain a p-nitrophenol-responsive sensing circuit that produces fluorescence after analyte exposure, while the hydrogel matrix preserves viability, prevents leakage, and supports analyte diffusion. The resulting 3D ELM operates as a portable biosensing unit that can be arranged spatially in a microfluidic array. In the integrated platform, it contributes to multiplexed environmental detection from real water samples.
- **Rationales for ambiguous slots (if any):**
  3D: selected for the same reason as ELM A: cells are embedded within a 3D hydrogel scaffold, not merely attached to discrete particles.
  BSE+ENV: selected because the p-NP sensor is presented as a biosensor for an environmental contaminant and the platform is validated in field water analysis.
  OPM: selected because hydrogel encapsulation/combination workflows are covered by the ontology's residual "other methods or combinations" category, which explicitly includes encapsulation within engineered constructs.

**Integrated concept description**

The paper describes a family of hybrid engineered living materials in which genetically engineered microbial biosensors are embedded inside mechanically robust, diffusion-permissive polymeric core-shell hydrogels. These ELMs are three-dimensional living sensing units rather than free-cell assays, and they are designed for rapid, portable, multiplexed chemical detection. The fungal and bacterial variants are combined at the platform level through spatial encoding in a microfluidic chip, where each bead population acts as a discrete analyte-responsive module. Ontologically, the work is best represented as separate **fungal** and **bacterial** hybrid polymeric 3D ELMs dedicated to **biosensing** and **environmental monitoring**.

***Engineered Living Biofilm with Enhanced Metal Binding Ability for Corrosion Protection in Seawater***

Abstract

**Source mode:** abstract only

- **Ontology code:**
  BAC-HYB-EUB-2D-LCE-SLS+MAP-ECP+BTI+BMN
- **Decoded slots:**
  - **BAC** = Bacterial ELM
  - **HYB** = Hybrid living material
  - **EUB** = Eubacterial ELM
  - **2D** = Two-dimensional ELM
  - **LCE** = Living ceramic
  - **SLS+MAP** = Smart and living surfaces and biointerfaces + Materials production and biofunctionalization
  - **ECP+BTI+BMN** = Engineering cells to produce ELMs + Biofilm-based interfacing with inorganic materials + Biomineralization
- **Organisms:**
  *Escherichia coli* is explicitly named; the biofilm is genetically engineered by appending a metal-binding domain to extracellular amyloids.
- **Material(s):**
  The living component is an engineered *E. coli* biofilm. The abiotic context is the metallic substrate, and the abstract also reports a biofilm-induced mineralization layer mainly composed of **calcite** on the steel surface. This supports a **HYB** interpretation because the ontology defines hybrid ELMs as systems combining cells with an abiotic structure, and the best-supported class is **LCE** because the protective barrier is described as mainly calcite, i.e. a ceramic mineral.
- **Geometry:**
  **2D**. The ontology defines 2D ELMs as planar structures, usually biofilms or cells generating a planar extracellular matrix, which matches a corrosion-protective biofilm grown on a metal surface.
- **Applications:**
  **SLS+MAP** is the best-supported application combination.
  - **SLS** fits because the ontology defines it as smart/living **surfaces and biointerfaces**, and this work is explicitly a living protective surface on steel.
  - **MAP** also fits because the biofilm is used to **functionalize** the metallic material by improving corrosion resistance.
- **Production method(s):**
  **ECP+BTI+BMN** is the best-supported combination.

- **ECP** fits because the cells are genetically engineered to produce modified extracellular amyloids.
- **BTI** fits because the ontology explicitly includes biofilms interfacing with inorganic materials, which matches the steel-bound biofilm.
- **BMN** fits because the abstract explicitly reports a mineralization layer mainly composed of calcite.

- **Concept description:**
This abstract describes a **bacterial hybrid engineered living material** consisting of an engineered *E. coli* biofilm grown as a planar living coating on metallic substrates. Genetic introduction of a metal-binding domain strengthens attachment to the metal surface, while biofilm-induced calcite formation generates a stable ceramic corrosion barrier. The material therefore acts as a living surface/biointerface that functionally upgrades steel by reducing corrosion in water environments. Within the ontology, it is most consistently interpreted as a **2D hybrid bacterial ELM with a ceramic-dominant protective layer, aimed at living-surface functionality and material biofunctionalization**.

- **Rationales for ambiguous slots (if any):**
  - **HYB instead of BIO:** the ELM is not only cells plus their own ECM; it is explicitly functioning in direct combination with an abiotic metallic structure.
  - **LCE instead of LCO:** the abstract states that the stable corrosion barrier is mainly composed of **calcite**; that makes a ceramic classification better supported than a generic composite one, although the organic biofilm remains part of the construct.
  - **SLS+MAP:** the work is simultaneously a living surface/interface on steel and a material-functionalization strategy for corrosion protection.

<u>Full text</u>

**Result:** ELM

The paper describes **one main engineered ELM**: an engineered *Escherichia coli* biofilm with a genetically appended metal-binding domain (MBD) on extracellular amyloid fibers, used as a living anti-corrosion surface on steel in seawater-like conditions. The article explicitly frames this as an engineered living biofilm/ELM and shows coexistence of living cells, extracellular matrix, and an abiotic interfacing surface, with additional calcite biomineralization during operation.

**ELM A**

- **Ontology code:**
BAC-HYB-EUB-2D-LPO+LCE-SLS-ECP+BFF+BTI+BMN

- **Decoded slots:**
BAC = bacterial ELM
HYB = hybrid living material
EUB = eubacterial ELM
2D = planar / biofilm-like ELM
LPO+LCE = living polymer + living ceramic
SLS = smart and living surfaces and biointerfaces

ECP+BFF+BTI+BMN = engineering cells to produce ELMs + biofilm formation and ECM growth + biofilm interfacing with inorganic materials + biomineralization

- **Organisms:**
  Engineered *Escherichia coli* MG1655 derivatives, specifically the OmpR234 background with ΔcsgA, transformed with pCA24N-CsgA-MBD to produce the MBD-fused curli subunit. The MBD sequence is derived from the type IV pilus receptor-binding region of *Pseudomonas aeruginosa*, but the living ELM population itself is engineered *E. coli*.

- **Material(s):**
  The living matrix is a **polymeric extracellular amyloid biofilm** based on curli/CsgA-derived fibers and associated EPS. During seawater exposure, the system additionally forms a **mineralized calcite calcium-carbonate layer**, giving the mature protective interface both organic/polymeric and ceramic/mineral character.

- **Geometry:**
  A **2D surface-associated biofilm/coating** formed on planar metal coupons, specifically X70 carbon steel and 304 stainless steel. The ontology definition for 2D explicitly includes biofilms and planar extracellular matrices.

- **Applications:**
  The document's explicit end use is **corrosion protection of metallic surfaces in seawater/simulated seawater**. Within the provided ontology, the closest supported application code is SLS, because the construct functions as a **living surface/biointerface** that improves interaction with the surrounding environment by protecting the metal substrate. A dedicated corrosion-protection application code is **not present in the ontology/graph**.

- **Production method(s):**
  ECP: the cells are genetically engineered to alter curli production and display the MBD peptide.
  BFF: the material is produced by deliberate **biofilm formation** and extracellular matrix growth.
  BTI: the engineered biofilm is intentionally grown as an interface on **inorganic metallic surfaces**.
  BMN: the protective system is strengthened by **biofilm-induced calcite biomineralization** during immersion.

- **Concept description:**
  This ELM is an engineered bacterial living coating in which genetically modified *E. coli* forms a curli-based biofilm with enhanced affinity for metal surfaces through an appended metal-binding peptide. Once established on steel, the living biofilm acts as a planar biointerface that suppresses corrosion and also promotes formation of a calcite-rich mineral barrier. The resulting construct therefore combines a living polymeric matrix with an inorganic mineralized layer at the metal interface. In functional terms, it behaves as a hybrid living surface for anti-corrosion protection in seawater environments, reducing corrosion current and improving electrochemical resistance on both X70 carbon steel and 304 stainless steel.

- **Rationales for ambiguous slots (if any):**
  **HYB instead of BIO:** selected because the engineered biofilm is deployed as a living coating **interfacing directly with an abiotic metallic structure**, which matches the ontology logic for hybrid living materials and the BTI production category.

**LPO+LCE instead of only LPO or only LCE:** selected because the paper explicitly supports both a **polymeric amyloid/EPS biofilm** and a **calcite mineral layer**; assigning only one class would discard directly stated material evidence.
**SLS application:** selected because anti-corrosion is the clear end use, but a dedicated anti-corrosion application code is **not present in the ontology/graph**; SLS is the closest ontology-supported fit for an active living surface/biointerface on metal.
**BTI in the production slot:** included because the ontology explicitly covers **biofilms interfacing with inorganic materials**, and that is a central operational feature of this system on steel and stainless steel.

***Engineered Bacillus subtilis Biofilm@Biochar living materials for in-situ sensing and bioremediation of heavy metal ions pollution***

Abstract

**Source mode:** abstract only

**ELM A**

- **Ontology code:**
  BAC-BIO-EUB-2D-LPO-BSE+BRE-ECP+BFF
- **Decoded slots:**
  - **BAC** = Bacterial ELM
  - **BIO** = Biological engineered living material
  - **EUB** = Eubacterial ELM
  - **2D** = Two-dimensional ELM
  - **LPO** = Living polymers
  - **BSE+BRE** = Biosensing + Bioremediation
  - **ECP+BFF** = Engineering cells to produce ELMs + Biofilm formation and related extracellular matrix growth
- **Organisms:**
  *Bacillus subtilis* is explicitly identified as the chassis organism.
- **Material(s):**
  An engineered *B. subtilis* **TasA-metallothionein biofilm**. No separate abiotic scaffold is described for this first material, so the best-supported super-kingdom is **BIO**, which the ontology defines as cells plus their extracellular matrices without supporting synthetic biomaterials or structures. The class is best matched to **LPO** because the functional matrix is a proteinaceous amyloid biofilm.
- **Geometry:**
  **2D** is the best-supported phylum because the ontology explicitly describes 2D ELMs as planar or bidimensional structures, usually **biofilms**.
- **Applications:**
  **BSE+BRE** is the best-supported combination. The abstract explicitly states that the engineered system both **senses** $Pb^{2+}$, $Cu^{2+}$, and $Hg^{2+}$ in real time and **adsorbs/removes** these heavy metals from the environment. In ontology terms, that aligns with **biosensing** and **bioremediation**.
- **Production method(s):**
  **ECP+BFF** is the best-supported production combination.
  - **ECP** fits because genetically engineered circuits and engineered TasA fusion functionality are used.
  - **BFF** fits because the material is explicitly an engineered **biofilm** with extracellular matrix growth.

- **Concept description:**
  This abstract describes a **bacterial biological engineered living material** based on an engineered *Bacillus subtilis* biofilm. The cells are genetically programmed for heavy-metal sensing and adsorption, while the TasA-based extracellular matrix provides the material-forming biofilm framework. Functionally, this bio-ELM acts both as a detection transducer and as a remediation platform for $Pb^{2+}$, $Hg^{2+}$, and $Cu^{2+}$. Within the ontology, it is most consistently classified as a **2D living-polymer biofilm ELM for biosensing and bioremediation produced through cell engineering and biofilm growth**.

- **Rationales for ambiguous slots (if any):**
  - **LPO:** chosen because the named matrix-forming component is **TasA**, a proteinaceous amyloid extracellular matrix component, which is best matched to a polymeric class.
  - **BSE+BRE rather than BSE+ENV+BRE:** the abstract directly demonstrates sensing and removal/remediation; environmental monitoring is implicit but not needed to capture the main demonstrated dual function.

**ELM B**

- **Ontology code:**
  BAC-HYB-EUB-UNK-LCA-BRE-ECP+BTI

- **Decoded slots:**
  - **BAC** = Bacterial ELM
  - **HYB** = Hybrid living material
  - **EUB** = Eubacterial ELM
  - **UNK** = Phylum not inferable from the abstract
  - **LCA** = Living carbons
  - **BRE** = Bioremediation
  - **ECP+BTI** = Engineering cells to produce ELMs + Biofilm-based trapping/anchoring of materials and biofilms interfacing with materials

- **Organisms:**
  *Bacillus subtilis* is again the living component, now used in the hybrid **biofilm@biochar** material.

- **Material(s):**
  The hybrid material is formed by incubating the engineered strain with **biochar** to create **biofilm@biochar (BBC)**. This supports **HYB**, because the ontology defines hybrid living materials as systems combining cells with an abiotic biomaterial or synthetic structure. The best-supported class is **LCA**, because the abiotic support is **biochar**, a carbonaceous material, so the hybrid matrix is most plausibly carbon-dominant.

- **Geometry:**
  **UNK**. The abstract does not explicitly define the BBC material in ontology phylum terms such as 0D, 1D, 2D, 3D, 4D, FR, or CP.

- **Applications:**
  **BRE** is the best-supported application code for the BBC material. The abstract specifically states that the hybrid biofilm@biochar material is applied in the **bioremediation** of heavy metal ions in soil and reduces $Pb^{2+}$ accumulation in maize plants.

- **Production method(s):**
  **ECP+BTI** is the best-supported production combination.

  - **ECP** remains justified because the living component is still genetically engineered.

  - **BTI** is the closest ontology route because the hybrid is formed by a **biofilm interfacing with an abiotic material support**.

- **Concept description:**
  This abstract also describes a second, distinct ELM: a **hybrid bacterial biofilm@biochar material** built from the engineered *B. subtilis* strain and a carbonaceous support. In this case, the living biofilm is combined with biochar to generate a field-applicable remediation material for heavy-metal-contaminated soil. The hybrid construct reduces exchangeable $Pb^{2+}$ in soil, lowers $Pb^{2+}$ accumulation in maize, and improves plant growth. Within the ontology, this is most consistently interpreted as a **bacterial hybrid living-carbon ELM for bioremediation produced through engineered-cell biofilm formation on an abiotic support**.

- **Rationales for ambiguous slots (if any):**

  - **Phylum = UNK:** the abstract does not explicitly describe the BBC architecture in ontology dimensional terms.

  - **LCA instead of LCO:** chosen because the named abiotic support is **biochar**, which is carbonaceous; however, a composite interpretation would also be arguable.

  - **BTI:** selected as the closest ontology production code because the hybrid is created by forming a biofilm in direct association with an abiotic support material.

Full text

**Result:** ELM

The full paper clearly describes **two distinct ELMs**:
**ELM A**, an engineered *Bacillus subtilis* biofilm used as a sensing and adsorption material, and **ELM B**, a hybrid **biofilm@biochar (BBC)** material formed by combining that engineered biofilm with biochar for soil remediation. The article explicitly presents both the engineered biofilm and the hybrid BBC as engineered living materials for heavy-metal sensing/remediation.

**ELM A**

- **Ontology code:**
  BAC-BIO-EUB-2D-LPO-BSE+BRE+ENV-ECP+BFF

- **Decoded slots:**
  BAC = bacterial ELM; BIO = biological engineered living material made only of cells and their extracellular matrices; EUB = eubacterial; 2D = planar/biofilm-type ELM; LPO = living polymer; BSE+BRE+ENV = biosensing + bioremediation +

environmental monitoring; ECP+BFF = engineering cells to produce the ELM + biofilm formation and extracellular-matrix growth.

- **Organisms:**
The living chassis is engineered *Bacillus subtilis*, specifically derivatives of strain 1935 carrying the heavy-metal sensing circuit and the TasA-metallothionein adsorption construct; the metallothionein sequence is reported as coming from *Pongo abelii*, but the living material itself is populated by *B. subtilis* cells.

- **Material(s):**
The matrix is a **biofilm extracellular matrix** built around engineered **TasA amyloid fibers** displaying **metallothionein (TasA-MT)** for heavy-metal binding; no separate synthetic scaffold is used for this ELM.

- **Geometry:**
This ELM is best classified as a **biofilm-type 2D structure**. The paper repeatedly describes it as an engineered **biofilm (BF)**, even though the matrix contains fibrous TasA structures.

- **Applications:**
The biofilm is used for **biosensing** $Pb^{2+}$, $Cu^{2+}$, and $Hg^{2+}$, and for **bioremediation** by adsorbing those ions from water; because the sensing target is environmental heavy-metal pollution, **environmental monitoring** is also supported.

- **Production method(s):**
The ELM is produced by **genetic engineering of cells** to install promoter-reporter and adsorption circuits, followed by **biofilm formation** in culture to grow the functional extracellular matrix. The sensing circuit uses metal-responsive promoters controlling fluorescent reporters, while the adsorption function is achieved by TasA-MT display on biofilm fibers.

- **Concept description:**
This ELM is an engineered *Bacillus subtilis* biofilm that integrates heavy-metal detection and heavy-metal capture in a single living material platform. The cells carry metal-responsive genetic circuits for $Pb^{2+}$, $Cu^{2+}$, and $Hg^{2+}$ sensing, while the extracellular biofilm architecture presents TasA-metallothionein fibers that bind those ions. Functionally, the material behaves as both a whole-cell biosensor and a remediation matrix in aqueous environments. In ontology terms, it is a bacterial, biofilm-based, polymeric bio-ELM with sensing and remediation roles.

- **Rationales for ambiguous slots (if any):**
2D: selected because the ontology explicitly associates **biofilms** with the 2D phylum, and the paper consistently frames this construct as a biofilm, despite the presence of fibrous substructures.

**ELM B**

- **Ontology code:**
BAC-HYB-EUB-3D-LCA-BRE-ECP+BFF+BTI

- **Decoded slots:**
BAC = bacterial ELM; HYB = hybrid living material combining cells with an abiotic scaffold/structure; EUB = eubacterial; 3D = three-dimensional ELM; LCA = living carbon; BRE = bioremediation; ECP+BFF+BTI = engineering cells to produce the living phase + biofilm formation + biofilm-based trapping/anchoring/interfacing with inorganic material.

- **Organisms:**
  The living component is the engineered *Bacillus subtilis* HJFbgrTM biofilm-forming strain. No second living organism is incorporated into the material; maize is only the downstream test plant in the soil-remediation assay, not a constituent of the ELM.

- **Material(s):**
  The hybrid matrix combines engineered **biofilm** with **maize-stover biochar (BC)**. The biochar provides the porous carbonaceous scaffold, and the engineered *B. subtilis* cells adhere to its surface and form a thick biofilm on it.

- **Geometry:**
  This ELM is best described as a **3D hybrid material**, because the living biofilm colonizes a rough, porous **three-dimensional biochar scaffold**. The SEM description in the paper states that the BC surface is rough and porous and that the engineered strain adhered well and formed a thick biofilm on the BC surface.

- **Applications:**
  The explicitly demonstrated use is **bioremediation** of $Pb^{2+}$-contaminated soil, including reduction of exchangeable $Pb^{2+}$ in soil and lower Pb accumulation in maize tissues.

- **Production method(s):**
  The living phase is first created by **engineering *B. subtilis*** and allowing it to form a functional **biofilm**, after which the biofilm is **combined with biochar** to generate a **biofilm@biochar** hybrid. Ontologically, the key integration step is the **interfacing/anchoring of biofilm with an inorganic material**, which matches the hybrid HLM production logic.

- **Concept description:**
  This ELM is a hybrid biofilm@biochar material in which an engineered *Bacillus subtilis* biofilm is immobilized on a porous biochar scaffold. The resulting construct functions as a living remediation material for $Pb^{2+}$-contaminated soil, lowering exchangeable lead in soil and reducing lead accumulation in maize. Scientifically, it combines the biological specificity and self-renewing biofilm functionality of the engineered bacterium with the structural and adsorptive support of a carbonaceous abiotic substrate. In ontology terms, it is a bacterial hybrid ELM organized as a three-dimensional carbon-based remediation material.

- **Rationales for ambiguous slots (if any):**
  LCA: selected because the hybrid's dominant abiotic scaffold is explicitly **biochar**, a carbonaceous material; this makes **living carbon** the best-supported class, although the living phase also contributes a polymeric biofilm.

If you want, I can next convert this into a **compact dataset-ready table** with one row per ELM and one column per ontology slot.

***Living Cellulose Materials with Tunable Viscoelasticity through Probiotic Proliferation***

Abstract

**Source mode:** abstract only

- **Ontology code:**
  UNK-BIO-UNK-4D-LPO-MAP-OPM
- **Decoded slots:**
  - **UNK** = Domain not inferable from the abstract
  - **BIO** = Biological engineered living material
  - **UNK** = Kingdom not inferable from the abstract
  - **4D** = Four-dimensional ELM
  - **LPO** = Living polymers
  - **MAP** = Materials production and biofunctionalization
  - **OPM** = Other production methods or combinations
- **Organisms:**
  Probiotics are explicitly the living component, but no species, strain, domain, or kingdom is stated in the abstract.
- **Material(s):**
  The matrix is **bacterial cellulose**, and the material is described as probiotics integrated into that cellulose matrix. The class is best assigned as **LPO** because the ontology defines living polymers as ELMs whose extracellular matrices are made predominantly of polymers, and cellulose is a polymeric biomaterial. The super-kingdom is best assigned as **BIO** because the abstract describes living cells within a biological cellulose matrix and does not mention any supporting synthetic biomaterial or synthetic scaffold.
- **Geometry:**
  **4D** is the best-supported phylum. The abstract explicitly states that **time** is the input that tunes the viscoelastic response, driving a transition from a more fluid-like to a more solid-like biomaterial; this matches the ontology definition of 4D ELMs, where the temporal dimension and metamorphosis play a relevant functional role.
- **Applications:**
  **MAP** is the best-supported application code. The abstract presents the system primarily as a route to produce a **living bio-ink with tunable viscoelastic response**, which fits the ontology category of **materials production and biofunctionalization** better than sensing, medical-device, or environmental-monitoring uses.
- **Production method(s):**
  **OPM** is the best-supported production code from the abstract alone. The text states that probiotics are **integrated into bacterial cellulose**, but it does not explicitly describe a more specific ontology-listed route such as ECM growth in culture, biofilm formation, or bioprinting; the ontology defines **OPM** as other methods or combinations, including encapsulation/integration of cells within engineered constructs.
- **Concept description:**
  This abstract describes an engineered living material in which probiotic cells are integrated into a bacterial-cellulose matrix to create a living biomaterial whose viscoelasticity evolves over time. The key functional mechanism is cell proliferation inside the matrix, which progressively modulates the mechanical response from more fluid-like to more solid-like behavior. Within the ontology, the construct is most consistently interpreted as a **biological, polymer-based, 4D ELM** because the matrix is cellulose and the relevant functional transformation is time-dependent. Its main ontology-grounded use is as a **materials-production platform**, specifically a living bio-ink with tunable rheological properties.

- **Rationales for ambiguous slots (if any):**
  - **Domain = UNK:** the abstract says "probiotics" but does not explicitly identify whether the living cells are bacterial, fungal, or mixed.
  - **Kingdom = UNK:** no taxonomic identity is given for the probiotics.
  - **BIO instead of HYB:** no supporting synthetic biomaterial or synthetic scaffold is stated; the matrix is bacterial cellulose, which is a biological matrix.
  - **4D instead of 3D/UNK:** the abstract explicitly makes **time** the controlling input for functional metamorphosis of the material, which matches the ontology's 4D criterion.
  - **OPM:** the integration route is described functionally but not with enough manufacturing detail to justify a narrower ontology production code.

<u>Full text</u>

**Result:** ELM. The paper explicitly describes "probiotic cellulose" as a living material composed of living probiotics integrated within a bacterial cellulose matrix, which fits the ontology's definition of an ELM.

**ELM A**

- **Ontology code:**
  BAC-BIO-EUB-3D-LPO-BMD-EMG
- **Decoded slots:**
  BAC = bacterial ELM; BIO = biological engineered living material; EUB = eubacterial ELM; 3D = three-dimensional ELM; LPO = living polymer; BMD = biohybrid medical devices; EMG = extracellular matrix grown in culture.
- **Organisms:**
  *Komagataeibacter xylinus* (Kx), the cellulose-producing bacterium, and *Lactobacillus fermentum* (Lf), the probiotic that proliferates through the cellulose matrix. The paper states that Kx produces the bacterial cellulose film and that Lf progressively invades the matrix during incubation.
- **Material(s):**
  The matrix is bacterial cellulose, described as a nonliving cellulose matrix formed by bacterial biosynthesis. Within the ontology, cellulose is best classified under living polymers (LPO).
- **Geometry:**
  The material is best read as a 3D cellulose membrane/pellicle: the article describes a thick gel-like membrane, reports a thickness of about 1.5 ± 0.2 mm, and shows by cross-sectional FESEM on page 3 that probiotics progressively invade the full interior of the matrix rather than remaining only on the surface.

- **Applications:**
  The best-supported ontology application is BMD because the paper presents the material as a tunable living bio-ink of interest for 3D printing and states that it could be used to obtain health devices; it also highlights antimicrobial potential relevant to biomedical use.

- **Production method(s):**
  EMG is the best-supported production code because the cellulose matrix is grown in culture by Kx and the final living material is obtained by co-culture and subsequent probiotic proliferation within that biologically produced extracellular matrix.

- **Concept description:**
  This ELM is a bacterial cellulose-based living polymer in which a biologically produced cellulose matrix hosts living probiotic bacteria. Kx generates the cellulose pellicle, and subsequent proliferation of Lf inside that matrix tunes the viscoelastic response from more fluid-like to more solid-like behavior. The FESEM series on page 3 shows increasing volumetric colonization of the matrix, while the photographs on page 5 illustrate the corresponding macroscopic shift from a transparent gel-like pellicle to a more opaque solid-like material. In ontology terms, it is a bacterial, biological, three-dimensional living polymer intended toward biomedical-device-oriented use as a tunable living bio-ink.

- **Rationales for ambiguous slots (if any):**
  - BIO rather than HYB: the matrix is bacterial cellulose grown by cells, not a synthetic scaffold or synthetic abiotic support.
  - 3D rather than 2D: although the material is membrane-like, the paper analyzes it as a thick volumetric pellicle with internal bacterial invasion across the cross-section.
  - BMD: this is assigned from the authors' stated intended use in health devices and bio-inks, not from a demonstrated finished medical device in the paper.
  - EMG: this is the closest ontology-listed production route because the key structural matrix is extracellular cellulose grown in culture; the article does not map more specifically to another listed manufacturing category.

***Employing photocurable biopolymers to engineer photosynthetic 3D-printed living materials for production of chemicals***

Abstract

**Source mode:** abstract only

**ELM A**

- **Ontology code:**
  BAC-HYB-EUB-3D-LPO-CAT+BTP-3DB
- **Decoded slots:**
  - **BAC** = Bacterial ELM
  - **HYB** = Hybrid living material
  - **EUB** = Eubacterial ELM
  - **3D** = Three-dimensional ELM
  - **LPO** = Living polymers
  - **CAT+BTP** = Catalysis and other chemical processes + Biotechnology and bioprocessing
  - **3DB** = 3D bioprinting of cells and biomaterials
- **Organisms:**
  Specifically engineered **Synechocystis** cell factories.
- **Material(s):**
  A photocurable bioink made of **alginate**, **galactoglucomannan-methacrylate**, and living photosynthetic cells. This supports **HYB**, because the ontology defines hybrid living materials as systems combining cells with an abiotic biomaterial or scaffold, and **LPO**, because the matrix is a hydrogel/polymeric biomaterial.
- **Geometry:**
  **3D**. The abstract explicitly frames the platform as **3D-printing** complex, tunable **3D architectures**, and the ontology defines 3D ELMs as cells within a three-dimensional biomaterial or scaffold.
- **Applications:**
  **CAT+BTP** is the best-supported combination. The construct is described as a **green biocatalyst** for **solar-chemical production** of ethylene, which aligns with **catalysis and other chemical processes**, and it also operates as a living production platform within a **biotechnological process**.
- **Production method(s):**
  **3DB** is explicitly supported. The ontology defines it as additive manufacturing involving biomaterials and living cells for creating living three-dimensional constructs, which matches the photocurable bioink and 3D-printing workflow described here.
- **Concept description:**
  This abstract describes a **bacterial hybrid engineered living material** based on engineered *Synechocystis* cells immobilized in a photocurable polymeric bioink. The material is fabricated by 3D bioprinting into hydrogel-based constructs that maintain photosynthetic activity and enable solar-driven chemical production. Functionally, the ELM acts as a living biocatalyst for ethylene production within a scalable bioprocessing platform. Within the ontology, it is best classified as a **3D hybrid living-polymer ELM for catalysis and bioprocessing produced by 3D bioprinting**.

- **Rationales for ambiguous slots (if any):**
    - **CAT+BTP:** both are directly supported because the abstract presents the construct both as a **biocatalyst** and as a production platform in the **chemical industry**.
    - **3D instead of 2D:** although thin hydrogel layers are mentioned, the abstract consistently emphasizes **3D-printing** and **3D architectures** as the defining structural framework.

---

**ELM B**

- **Ontology code:**
EUK-HYB-PRO-3D-LPO-CAT+BTP-3DB
- **Decoded slots:**
    - **EUK** = Eukaryotic ELM
    - **HYB** = Hybrid living material
    - **PRO** = Protist ELM
    - **3D** = Three-dimensional ELM
    - **LPO** = Living polymers
    - **CAT+BTP** = Catalysis and other chemical processes + Biotechnology and bioprocessing
    - **3DB** = 3D bioprinting of cells and biomaterials
- **Organisms:**
Specifically engineered **Chlamydomonas** cell factories.
- **Material(s):**
The same photocurable bioink platform: **alginate** plus **galactoglucomannan-methacrylate** with living photosynthetic cells. This again supports **HYB** and **LPO** under the ontology definitions for hybrid systems and polymeric matrices/hydrogels.
- **Geometry:**
**3D**. The material is described as a 3D-bioprinted living construct, and the ontology definition of 3D ELMs matches cells entrapped in a three-dimensional biomaterial.
- **Applications:**
**CAT+BTP** is again the best-supported combination. Here the living construct performs the **biotransformation of cyclohexanone to ε-caprolactone**, which is a catalytic chemical process embedded in a biotechnology/bioprocessing context.
- **Production method(s):**
**3DB**, for the same reason as above: the abstract explicitly describes a bioink containing biomaterials and living cells used for **3D printing** living constructs.

- **Concept description:**
  This abstract also describes a second, distinct ELM built from engineered *Chlamydomonas* cells immobilized in the same photocurable hydrogel bioink. The living construct is fabricated by 3D bioprinting and functions as a photosynthetic biocatalyst for the conversion of cyclohexanone to ε-caprolactone. The abiotic hydrogel matrix provides structural support and improved mechanical tolerance, while the entrapped eukaryotic cells supply the catalytic function. Within the ontology, this is best interpreted as a **3D hybrid living-polymer protist ELM for catalysis and bioprocessing produced via 3D bioprinting**.

- **Rationales for ambiguous slots (if any):**

  - **PRO instead of PLN:** the ontology defines **PRO** for eukaryotic cells that are not animal, plant, or fungal cells; based on the abstract alone, *Chlamydomonas* is best treated as a microalgal protist rather than assigned to plantae.

  - **CAT+BTP:** justified because the abstract describes a specific **chemical biotransformation** and frames the platform as relevant to **industrial chemical production**.

<u>Full text</u>

**Result:** ELM. The paper describes living cells intentionally immobilized within a 3D-printed polymeric scaffold, which matches the ontology definition of an engineered living material.

**ELM A**

- **Ontology code:**
  BAC-HYB-EUB-3D-LPO-CAT-3DB

- **Decoded slots:**
  BAC = bacterial ELM
  HYB = hybrid living material
  EUB = eubacterial ELM
  3D = three-dimensional ELM
  LPO = living polymer
  CAT = catalysis and other chemical processes
  3DB = 3D bioprinting of cells and biomaterials

- **Organisms:**
  Engineered *Synechocystis* sp. PCC 6803 expressing the ethylene-forming enzyme (Efe).

- **Material(s):**
  A photocurable hydrogel/bioink composed of alginate, galactoglucomannan-methacrylate (GGMMA), and the photoinitiator LAP, printed on TMPSM-treated glass support. The matrix is polymeric, so the ontology class is living polymers.

- **Geometry:**
  3D-printed thin hydrogel films, including approximately 200 μm-thick layers, attached to glass substrates. Although the constructs are thin films, the paper consistently frames them as 3D-printed living materials and three-dimensional printed constructs/scaffolds.

- **Applications:**
  Photosynthetic ethylene production for chemical manufacturing. The paper presents the construct as a green biocatalyst for solar-chemical production, with ethylene identified as an industrially relevant chemical and polymer building block.

- **Production method(s):**
Cells were mixed directly into the photocurable bioink, extrusion-printed onto pre-treated glass, and then photocured with 405 nm light to form a covalently crosslinked hydrogel. The biological–abiotic integration mechanism is physical entrapment of living cyanobacterial cells inside the polymer matrix, with matrix crosslinking and glass attachment mediated by methacrylate chemistry.

- **Concept description:**
This ELM is a hybrid photosynthetic biocatalyst built by embedding engineered *Synechocystis* cells into a 3D-printed photocurable polymer matrix. The alginate/GGMMA/LAP scaffold provides a mechanically stable hydrogel environment that retains living cells while permitting light-driven metabolism and gas exchange. Within this construct, the entrapped cyanobacteria convert inorganic carbon into ethylene under illumination. The material therefore functions as a solid-state living catalytic platform for renewable chemical production.

- **Rationales for ambiguous slots (if any):**
HYB: selected because the living cells coexist with an abiotic processed scaffold/matrix rather than constituting a cells-only material.
3D: selected because the article explicitly describes 3D-printing of living materials and three-dimensional printed constructs, even though the printed format is a thin layer.
CAT: selected because the ELM is used as a biocatalyst for ethylene production in a chemical-production process; this is more directly supported than broader application labels.

**ELM B**

- **Ontology code:**
EUK-HYB-PRO-3D-LPO-CAT-3DB

- **Decoded slots:**
EUK = eukaryotic ELM
HYB = hybrid living material
PRO = protist ELM
3D = three-dimensional ELM
LPO = living polymer
CAT = catalysis and other chemical processes
3DB = 3D bioprinting of cells and biomaterials

- **Organisms:**
Engineered *Chlamydomonas reinhardtii* UVM11-CW/CHMO_PSAD clone 10 expressing heterologous cyclohexanone monooxygenase (CHMO).

- **Material(s):**
The same photocurable polymeric hydrogel system: alginate + GGMMA + LAP, printed on glass support. This is ontology-consistent with living polymers.

- **Geometry:**
3D-printed thin hydrogel films on glass, used as immobilized living biocatalytic layers.

- **Applications:**
Whole-cell biotransformation of cyclohexanone to ε-caprolactone, an industrially relevant chemical used in polymer production. The use case is explicitly framed as chemical production by a living biocatalyst.

- **Production method(s):**
  *Chlamydomonas* cells were incorporated into the photocurable bioink, 3D-printed, and photocured under 405 nm light. The integration mechanism is entrapment of viable algal cells within a covalently crosslinked polymer matrix that allows substrate diffusion inward and product release outward.
- **Concept description:**
  This ELM is a eukaryotic hybrid living material in which engineered *Chlamydomonas* cells are immobilized inside a 3D-printed photocurable hydrogel. The polymer matrix creates a stable structural environment while preserving photosynthetic viability after recovery from printing stress. Functionally, the construct operates as a whole-cell catalytic material that converts cyclohexanone into ε-caprolactone. In ontology terms, it is a protist-based living polymer designed for chemical biotransformation.
- **Rationales for ambiguous slots (if any):**
  PRO: selected because *Chlamydomonas reinhardtii* is a eukaryotic microalga and the ontology places eukaryotic organisms that are not animal, plant, or fungal cells under protists.
  HYB: selected because the algae are immobilized in a non-cellular scaffold/material rather than forming a cells-only matrix.
  CAT: selected because the demonstrated function is a catalytic chemical conversion, cyclohexanone to ε-caprolactone.

**Integrated assessment**

The full paper clearly describes **two distinct ELMs**, not just one platform, because it applies the same 3D-printed photocurable polymer scaffold to two different engineered living systems with different taxa and different catalytic outputs: a bacterial cyanobacterial ethylene-producing material and a eukaryotic algal biotransformation material. Both satisfy the ontology-level ELM definition because they combine living cells with a scaffold-like biomaterial in a structured printed construct.

***Robust myco-composites: a biocomposite platform for versatile hybrid-living materials***

Abstract

**Source mode:** abstract only

- **Ontology code:**
  EUK-HYB-FUN-3D-LCO-MAP-3DP
- **Decoded slots:**
  - **EUK** = Eukaryotic ELM
  - **HYB** = Hybrid living material
  - **FUN** = Fungal ELM
  - **3D** = Three-dimensional ELM
  - **LCO** = Living composites
  - **MAP** = Materials production and biofunctionalization
  - **3DP** = Scaffold structure 3D printing and cell colonization
- **Organisms:**
  Fungal **mycelium** is the living component. No fungal species or strain is explicitly named in the abstract.
- **Material(s):**
  The abstract describes **hybrid-living mycelium composites** / **myco-composites** based on mycelium growing on **lignocellulosic waste** and processed through **biocomposite additive manufacturing**. This supports **HYB**, because the ontology defines hybrid living materials as systems combining living cells with a biomaterial or scaffold, and **LCO**, because the matrix is explicitly framed as a **composite** rather than a single polymer class.
- **Geometry:**
  **3D** is the best-supported phylum. The ontology defines 3D ELMs as cells within a three-dimensional biomaterial or scaffold, and the abstract explicitly emphasizes **complex geometries**, **structural myco-composites**, and **additive manufacturing** with subsequent mycelium colonization.
- **Applications:**
  **MAP** is the best-supported application code. The abstract is centered on a manufacturing platform for creating scalable, tunable structural myco-composites with improved mechanical and surface properties, which aligns best with the ontology definition of ELMs employed as **factories of materials or as support to materials functionalization**. The mentions of foldable containers and flexible textiles are presented as illustrative examples rather than as the main ontology-driving application.
- **Production method(s):**
  **3DP** is the best-supported production code. The ontology defines it as **scaffold structure 3D printing and cell colonization**, and the abstract explicitly describes a platform combining **high-resolution biocomposite additive manufacturing** with **robust mycelium colonization** through indirect inoculation.

- **Concept description:**
This abstract describes a **fungal hybrid engineered living material** in which mycelium colonizes an additively manufactured biocomposite scaffold derived from lignocellulosic material. The resulting construct is a structural myco-composite with improved mechanical performance, tunable properties, and compatibility with complex geometries. Within the ontology, the material is most consistently classified as a **3D living composite**, because the living fungal network develops inside a three-dimensional composite scaffold rather than forming a purely biological matrix on its own. Its primary ontology-grounded role is **materials production and biofunctionalization**, enabled by a **3D-printed scaffold plus cell-colonization** manufacturing route.

- **Rationales for ambiguous slots (if any):**
    - **LCO instead of LPO:** the abstract repeatedly uses **myco-composites / biocomposite additive manufacturing**, so a composite class is more directly supported than a generic polymer class.
    - **MAP instead of FAS or SLS:** containers and textiles are examples, but the abstract's main contribution is a platform for **manufacturing structural myco-composites**.
    - **3DP instead of CMM or OPM:** the text explicitly highlights **additive manufacturing** followed by **mycelium colonization**, which matches the ontology description of **3DP** more closely than broader composite-manufacturing categories.

<u>Full text</u>

**ELM A**

- **Ontology code:**
EUK-HYB-FUN-3D-LPO-SLM-OPM

- **Decoded slots:**
EUK = eukaryotic ELM
HYB = hybrid living material
FUN = fungal ELM
3D = three-dimensional ELM
LPO = living polymer
SLM = smart and living materials and structures (multipurpose)
OPM = other production methods or combinations

- **Organisms:**
Fungal mycelium is the living component. The optimized biocomposite was reported as compatible with *Trametes versicolor*, *Pleurotus ostreatus*, and an Ecovative fungal strain, while the subsequent main experiments used Ecovative mycelium for indirect inoculation and colonization.

- **Material(s):**
The matrix is a 3D-printed biocomposite made from chitosan, cellulose fibers, and spent coffee grounds; the paper describes chitosan as the binding matrix, cellulose as the structural reinforcement, and coffee grounds as the nutrient source that supports fungal growth. This is best aligned with a polymer-dominant matrix in the ontology.

- **Geometry:**
  The main ELM is a three-dimensional printed biocomposite colonized by fungal mycelium. The paper explicitly describes high-resolution 3D printing of complex geometries, full colonization of 10 cm tall structures, and demonstration objects including bio-welded containers and flexible textile-like prints, but the core platform remains a 3D myco-composite system.

- **Applications:**
  The strongest ontology-consistent application is multipurpose smart/living materials and structures, because the paper frames the platform as enabling advanced structural mycelium applications and demonstrates functional behaviors such as hydrophobicity, self-sealing, flexible hinge formation, and textile-like deformation. Packaging-like use is discussed, but "packaging" is not an ontology application code in the provided taxonomy.

- **Production method(s):**
  The paper uses high-resolution extrusion-based additive manufacturing of a biocomposite scaffold, followed by fungal indirect inoculation and colonization from a mycelium bed. Because the ontology options do not exactly name this combined workflow, the best-supported production slot is OPM rather than a narrower code.

- **Concept description:**
  This paper describes a hybrid fungal living material in which living mycelium colonizes a pre-formed, high-resolution 3D-printed biocomposite made of chitosan, cellulose, and coffee grounds. The printed scaffold provides structural form and mechanical robustness, while the mycelium contributes colonization, bridging, hydrophobic skin formation, and regenerative sealing behavior. The resulting myco-composite is presented as a versatile structural platform with improved mechanical performance relative to previously reported mycelium composites, alongside demonstrators such as self-sealing containers and flexible textile-like assemblies. As an ontology-grounded concept, it is best interpreted as a eukaryotic hybrid fungal 3D living polymer for multipurpose smart/living material functions, fabricated through a combined additive-manufacturing-plus-colonization workflow.

- **Rationales for ambiguous slots (if any):**
  **Application = SLM:** The article demonstrates several functional structural behaviors and multiple end uses, but no single narrower ontology application code is stated as the definitive main one; SLM is the best-supported umbrella category.
  **Production = OPM:** The workflow is specifically "3D-printed biocomposite + indirect inoculation," which does not map exactly to a single named ontology production category; OPM is therefore the most reproducible choice.

A stricter alternative coding, if you want a more manufacturing-specific interpretation, would be to test EUK-HYB-FUN-3D-LPO-SLM-3DP, but I consider OPM better justified by the exact wording of the paper and the ontology boundaries.

***Engineered algae microrobots as photosynthetic living materials promote T cells' anti-tumor immunity***

Abstract

**Source mode:** abstract only

- **Ontology code:**
  EUK-HYB-PRO-0D-UNK-BMD-OPM
- **Decoded slots:**
  - **EUK** = Eukaryotic ELM
  - **HYB** = Hybrid living material
  - **PRO** = Protist ELM
  - **0D** = Quasi-punctual ELM, usually cells coupled to micro-/nano-particles
  - **UNK** = Class not inferable from the abstract
  - **BMD** = Biohybrid medical devices
  - **OPM** = Other production methods or combinations
- **Organisms:**
  *Chlorella vulgaris* is the explicitly named living component. The dendritic-cell-derived part is described as a **genetically engineered DC membrane** / **cell membrane nanovesicles**, not as a second living cell population.
- **Material(s):**
  The construct combines living *Chlorella vulgaris* with engineered dendritic-cell membrane nanovesicles displaying OX40L, 4-1BBL, and CD70. This supports **HYB**, because the ontology defines hybrid living materials as systems combining cells with an abiotic biomaterial or structure rather than only cells plus their own extracellular matrix. **Class = UNK** is the safest assignment, because the abstract does not identify a predominant ontology matrix class such as polymer, ceramic, carbon, alloy, or composite.
- **Geometry:**
  **0D** is the best-supported phylum. The abstract describes a **photosynthetic micron robot**, i.e. a particle-like microscale construct assembled from a living cell and nanoscale membrane components, which is closest to the ontology definition of a quasi-punctual ELM with cells coupled to micro-/nano-objects.
- **Applications:**
  **BMD** is the best-supported application code. The ontology defines this category for healthcare-oriented biohybrid medical devices, including improved biointeraction and drug-delivery-type therapeutic functionality, and the abstract is explicitly centered on tumor immunotherapy and anti-relapse treatment.
- **Production method(s):**
  **OPM** is the best-supported production code. The abstract describes an assembled therapeutic construct made by combining a living microalga with engineered membrane nanovesicles, but it does not match a more specific ontology route such as biofilm growth, biomineralization, or bioprinting. The ontology definition of **OPM** explicitly covers other combinations and engineered constructs used to generate bio-ELMs or HLMs.

- **Concept description:**
This abstract describes a **eukaryotic hybrid engineered living material** in which living *Chlorella vulgaris* is integrated with engineered dendritic-cell membrane nanovesicles to form a photosynthetic therapeutic micron-scale construct. The living algal component generates oxygen in situ to alleviate tumor hypoxia, while the membrane-displayed TNF-ligand proteins stimulate T-cell activation and proliferation. Functionally, the system is a biohybrid therapeutic platform for improving anti-tumor immune response and reducing relapse and metastasis. Within the ontology, it is most consistently interpreted as a **0D hybrid medical ELM produced through a specialized assembly route best captured by OPM**.

- **Rationales for ambiguous slots (if any):**
    - **PRO instead of PLN:** the abstract identifies *Chlorella vulgaris* only as a **microalga**; under the ontology, **PRO** is the safer assignment when the eukaryotic cell is not explicitly stated as animal, plant, or fungal.
    - **0D instead of UNK:** "photosynthetic micron robot" most strongly suggests a particle-like microscale construct rather than a film, fiber, scaffold, or morphing architecture.
    - **HYB instead of BIO:** the construct is not just cells plus their own extracellular matrix; it includes engineered nonliving membrane nanovesicles as part of the functional material.
    - **Class = UNK:** the abstract does not provide enough evidence to assign a predominant matrix class from the ontology.

<u>Full text</u>

- **Ontology code:**
EUK-BIO-PRO-0D-UNK-BMD+BMR-ECP+OPM

- **Decoded slots:**
    - EUK — Eukaryotic ELM
    - BIO — Biological engineered living material
    - PRO — Protist ELM
    - 0D — Quasi-punctual / particle-like ELM
    - UNK — Class not inferable from the provided input
    - BMD+BMR — Biohybrid medical device + biohybrid machine/robot
    - ECP+OPM — Engineering cells to produce ELMs + other production methods or combinations

- **Organisms:**
The living organism constituting the coded ELM is **Chlorella vulgaris**. The paper also uses a **genetically engineered mouse dendritic cell line DC2.4** as the source of the membrane coating that displays OX40L, 4-1BBL, and CD70, but those dendritic cells are not the living population in the final coated microrobot itself. T cells and B16F10 melanoma cells are experimental target/context cells rather than constitutive elements of the ELM.

- **Material(s):**
The material is a **membrane-coated living microalga**: photosynthetic **C. vulgaris** coated with an **engineered dendritic-cell membrane**. The paper does not describe an abiotic synthetic scaffold as part of the intrinsic "photosynthetic micron robot"; the HP hydrogel appears only as a delivery vehicle in one in vivo treatment setup, not as the core coded ELM structure.

- **Geometry:**
The ELM is best classified as **0D** because it is presented as a population of discrete, quasi-spherical micrometer-scale "micron robots," with **C. vulgaris** described as "green spheres" of about **2.5 µm** average diameter, later coated with engineered membrane.

- **Applications:**
The intended use is **anti-tumor immunotherapy**, specifically oxygenation of hypoxic tumor microenvironments and enhancement of effector and memory T-cell responses to reduce tumor relapse and metastasis in mice. Within the ontology, this supports **BMD** because the system is used for healthcare as an advanced therapeutic biohybrid device, and **BMR** because the paper explicitly frames it as a **"photosynthetic micron robot."**

- **Production method(s):**
The paper first **genetically engineers DC2.4 cells** using lentiviral constructs so they overexpress OX40L, 4-1BBL, and CD70, then isolates their membranes, and finally coats **C. vulgaris** by a membrane-processing/mixing sequence including extraction, ultrasound smoothing, magnetic stirring, and collection of the coated algae. This fits **ECP** because artificially manipulated cells are used to produce the functional ELM-enabling biological interface, and **OPM** because the membrane-coating/combination route is a hybrid assembly workflow not explicitly enumerated elsewhere in Table 2.

- **Concept description:**
This paper describes a biological engineered living material built from living **Chlorella vulgaris** cells whose surfaces are coated with genetically engineered dendritic-cell membrane displaying OX40L, 4-1BBL, and CD70. Under red-light irradiation, the living microalgae generate oxygen photosynthetically, alleviating tumor hypoxia, while the membrane-presented ligands stimulate T-cell activation and proliferation. Functionally, the construct behaves as a microrobotic therapeutic biointerface that couples oxygen generation with immune co-stimulation. In mouse models, especially when combined with anti-PD-1, it improves anti-tumor immunity and reduces relapse and metastasis.

- **Rationales for ambiguous slots (if any):**
    - BIO rather than HYB: the coded micron robot itself is described as living algae plus a **biologically derived** engineered cell membrane, not as cells combined with an **abiotic synthetic scaffold**.
    - PRO rather than PLN: the paper explicitly identifies **C. vulgaris** as a **microalga / freshwater green alga**, but it does not explicitly place it in **Plantae**; under the ontology, the safer directly supported kingdom assignment is PRO.
    - UNK for class: the ontology class slot is defined by the **predominant extracellular-matrix material class** (alloy, carbon, ceramic, composite, polymer, cross-class), whereas this paper describes a **membrane-coated living**

**cell construct** rather than a clearly classifiable extracellular matrix. **Not present in the ontology/graph.**

- BMD+BMR: BMD is justified by the therapeutic anti-cancer use, and BMR is justified because the article repeatedly frames the system as a **micron robot**.
- ECP+OPM: ECP captures the genetically engineered DC source cells; OPM captures the subsequent membrane-isolation/coating combination process.

***Light-Regulated Pro-Angiogenic Engineered Living Materials***

Abstract

**Source mode:** abstract only

- **Ontology code:**
  BAC-HYB-EUB-3D-LPO-TER-OPM
- **Decoded slots:**
  - **BAC** = Bacterial ELM
  - **HYB** = Hybrid living material
  - **EUB** = Eubacterial ELM
  - **3D** = Three-dimensional ELM
  - **LPO** = Living polymers
  - **TER** = Tissue engineering and regenerative medicine
  - **OPM** = Other production methods or combinations
- **Organisms:**
  Optogenetically engineered bacteria are the living component, but no species or strain is explicitly named in the abstract.
- **Material(s):**
  The abiotic matrix is a **bilayer hydrogel construct** that encapsulates the bacteria. This supports **HYB**, because the ontology defines hybrid living materials as ELMs that combine cells with an abiotic biomaterial or scaffold, and **LPO**, because hydrogels fall under the ontology's living-polymer class.
- **Geometry:**
  **3D** is the best-supported phylum. The ontology defines 3D ELMs as cells within a three-dimensional biomaterial or scaffold, which matches bacteria encapsulated in bilayer hydrogel constructs.
- **Applications:**
  **TER** is the best-supported application code. The ontology defines this category for advanced healthcare solutions connected to **tissue engineering and regenerative medicine**, and the abstract is explicitly framed in regenerative medicine, with light-controlled release of a pro-angiogenic factor that promotes endothelial network formation.
- **Production method(s):**
  **OPM** is the safest ontology-grounded production code. The ontology explicitly includes under OPM the **encapsulation of cells within engineered constructs** for generating bio-ELMs and HLMs, which matches secure bacterial encapsulation in bilayer hydrogel constructs.
- **Concept description:**
  This abstract describes a **bacterial hybrid engineered living material** in which optogenetically engineered bacteria are confined within a bilayer hydrogel scaffold. The living cells synthesize and secrete a collagen-binding VEGF peptidomimetic under light control, enabling tunable local delivery of a pro-angiogenic cue. The hydrogel provides containment and structural support while preserving bacterial activity and preventing escape. Within the ontology, this is most consistently interpreted as a **3D living-polymer hybrid ELM for tissue engineering and regenerative medicine**, produced through an encapsulation-based route captured by **OPM**.

- **Rationales for ambiguous slots (if any):**
  - **HYB instead of BIO:** the ontology reserves **BIO** for cells plus their own extracellular matrices without supporting synthetic biomaterials; here the bacteria are encapsulated in a hydrogel scaffold.
  - **TER instead of BMD:** the abstract is framed around **regenerative medicine** and angiogenesis rather than a general medical-device or drug-delivery platform, so **TER** is the closer primary application.
  - **OPM instead of ECP:** although the bacteria are engineered, the ontology definition of **ECP** is centered on engineered cells producing matrix/material, whereas the clearest explicit fabrication step here is **encapsulation in an engineered hydrogel construct**, which is directly covered by **OPM**.

<u>Full text</u>

- **Ontology code:**
  BAC-HYB-EUB-3D-LPO-TER-ECP
- **Decoded slots:**
  - BAC = Bacterial ELM
  - HYB = Hybrid living material
  - EUB = Eubacterial ELM
  - 3D = Three-dimensional ELM
  - LPO = Living polymer
  - TER = Tissue engineering and regenerative medicine
  - ECP = Engineering cells to produce ELMs
- **Organisms:**
  *ClearColi* BL21(DE3), described in the paper as an endotoxin-free variant of *E. coli*, optogenetically engineered to synthesize and secrete the collagen-binding QK fusion protein YCQ.
- **Material(s):**
  Bilayer acrylate-modified Pluronic F127 / Pluronic F127-diacrylate (PluDA) hydrogel constructs. The secreted protein is a collagen-binding VEGF peptidomimetic, but the scaffold material of the ELM itself is polymeric hydrogel.
- **Geometry:**
  3D bilayer hydrogel discs with a bacterial core and a protective shell layer. The text explicitly describes the constructs as bilayer hydrogel discs with defined core and shell dimensions.
- **Applications:**
  Tissue engineering and regenerative medicine, specifically pro-angiogenic / regenerative use through light-regulated release of a VEGF-mimetic that promotes angiogenic network formation in endothelial cells. This matches the ontology category TER.

- **Production method(s):**
  ECP is the best-supported production code, because the study centers on optogenetically engineering bacteria to produce and secrete the active therapeutic fusion protein inside the material. The hydrogel is the encapsulating scaffold, but the ontology-defined production emphasis best matched by the paper is engineered cells producing the functional material output.

- **Concept description:**
  This paper describes a bona fide engineered living material in which genetically engineered bacterial cells are combined with a synthetic hydrogel scaffold to create a remotely controllable therapeutic construct. The bacteria are programmed with a light-responsive circuit to secrete a collagen-binding VEGF peptidomimetic (YCQ), while the bilayer PluDA hydrogel maintains bacterial viability and prevents escape. The construct operates as a 3D hybrid bacterial–polymeric ELM that enables tunable light-regulated release of the pro-angiogenic factor over several days. The released protein binds collagen-containing matrices and promotes angiogenic network formation in HUVEC cultures, supporting its regenerative-medicine function.

- **Rationales for ambiguous slots (if any):**
  - **Application = TER rather than BMD:** the paper frames the system primarily as a regenerative, pro-angiogenic material for healing and vascular regeneration, not as an already configured medical device. TER is therefore the closest ontology-grounded fit.
  - **Production = ECP rather than OPM:** although the hydrogel encapsulation is important, the defining functional mechanism is that engineered bacteria produce and secrete the therapeutic agent in response to light; this aligns most directly with Engineering cells to produce ELMs.
  - **Class = LPO rather than LCO:** the matrix is described as Pluronic-based hydrogel, which is polymeric; no evidence is given that the ELM matrix is predominantly composite in the ontology sense.

Result: **ELM**. The article explicitly describes an engineered living material composed of live engineered bacteria securely encapsulated in synthetic bilayer hydrogel constructs for light-controlled therapeutic factor release.

***Gene expression dynamics in input-responsive engineered living materials programmed for bioproduction***

Abstract

**Source mode:** abstract only

**ELM A**

- **Ontology code:**
  BAC-HYB-EUB-3D-LPO-BTP-OPM
- **Decoded slots:**
  - **BAC** = Bacterial ELM
  - **HYB** = Hybrid living material
  - **EUB** = Eubacterial ELM
  - **3D** = Three-dimensional ELM
  - **LPO** = Living polymers
  - **BTP** = Biotechnology and bioprocessing
  - **OPM** = Other production methods or combinations
- **Organisms:**
  Engineered *Escherichia coli*.
- **Material(s):**
  F127-bisurethane methacrylate (**F127-BUM**) hydrogel with encapsulated engineered bacteria.
- **Geometry:**
  **3D**. The ontology defines 3D ELMs as cells within a three-dimensional biomaterial or scaffold, which matches microbes encapsulated in a hydrogel construct.
- **Applications:**
  **BTP** is the best-supported application code because the abstract frames these ELMs as **bioreactors for sustained bioproduction** and as platforms for long-term, input-responsive metabolic production. The ontology defines BTP as ELMs employed within biotechnological processes.
- **Production method(s):**
  **OPM** is the best-supported production code. The abstract explicitly states that the ELMs are fabricated by **encapsulating microbes in hydrogels**, and the ontology definition of OPM explicitly includes **encapsulation of cells within engineered constructs** for generating bio-ELMs and HLMs.
- **Concept description:**
  This abstract describes a **bacterial hybrid engineered living material** based on engineered *E. coli* encapsulated in a polymeric hydrogel matrix. The construct supports repeated induction of gene expression over multiple weeks and enables cyclic bioproduction through inducible genetic programs, including CRISPRa-regulated pathway control. Within the ontology, the system is most consistently interpreted as a **3D living-polymer hybrid ELM** because the bacteria are confined inside a three-dimensional hydrogel biomaterial. Its clearest ontology-grounded role is **biotechnology and bioprocessing**, with fabrication best captured as **encapsulation-based OPM**.

- **Rationales for ambiguous slots (if any):**
  - **HYB instead of BIO:** the ontology reserves BIO for cells plus their own extracellular matrices, whereas this system uses an external hydrogel biomaterial.
  - **BTP instead of CAT or MAP:** the abstract is centered on sustained **bioproduction** and long-term gene-expression control in a **bioreactor-like** setting, which aligns more directly with biotechnology and bioprocessing.

---

**ELM B**

- **Ontology code:**
  EUK-HYB-FUN-3D-LPO-BTP-OPM
- **Decoded slots:**
  - **EUK** = Eukaryotic ELM
  - **HYB** = Hybrid living material
  - **FUN** = Fungal ELM
  - **3D** = Three-dimensional ELM
  - **LPO** = Living polymers
  - **BTP** = Biotechnology and bioprocessing
  - **OPM** = Other production methods or combinations
- **Organisms:**
  Engineered *Saccharomyces cerevisiae*.
- **Material(s):**
  Bovine serum albumin (**BSA**)–polyethylene glycol diacrylate (**PEGDA**) hydrogel with encapsulated engineered yeast.
- **Geometry:**
  **3D**. As above, the ontology definition of 3D ELMs matches cells embedded within a three-dimensional hydrogel biomaterial.
- **Applications:**
  **BTP** is again the best-supported application code because the yeast ELM is programmed for scheduled switching between two bioproduction outputs over 27 days under continuous cultivation. This is most directly a biotechnological process platform.
- **Production method(s):**
  **OPM** is the best-supported production code because the defining fabrication step is again **microbial encapsulation within an engineered hydrogel construct**, which the ontology places under OPM.
- **Concept description:**
  This abstract also describes a second, distinct ELM built from engineered *S. cerevisiae*

cells encapsulated in a BSA-PEGDA hydrogel. The material supports chemically controlled switching between production of betaxanthin pigments and proteinase A over multi-week cultivation. Within the ontology, it is best classified as a **3D hybrid living-polymer fungal ELM**, since the yeast cells operate inside a three-dimensional polymeric hydrogel matrix. Its primary ontology-grounded function is **biotechnology and bioprocessing**, and its fabrication route is best captured by **OPM** because it is defined by encapsulation rather than a more specific named manufacturing route.

- **Rationales for ambiguous slots (if any):**
  - **FUN:** *Saccharomyces cerevisiae* is a yeast, so within the ontology it belongs to the fungal kingdom.
  - **OPM instead of ECP:** although the cells are engineered, the explicit fabrication step described in the abstract is **encapsulation in hydrogels**, which is directly covered by OPM.

<u>Full text</u>

**ELM A**

- **Ontology code:**
  BAC-HYB-EUB-3D-LPO-BTP-OPM
- **Decoded slots:**
  - BAC = Bacterial ELM
  - HYB = Hybrid living material
  - EUB = Eubacterial ELM
  - 3D = Three-dimensional ELM
  - LPO = Living polymer
  - BTP = Biotechnology and bioprocessing
  - OPM = Other production methods or combinations
- **Organisms:**
  Engineered *Escherichia coli* MG1655.
- **Material(s):**
  Pluronic F127-bisurethane methacrylate (F127-BUM) hydrogel; this is a polymeric hydrogel matrix.
- **Geometry:**
  3D cylindrical hydrogel constructs, cast in silicone molds with approximately 4 mm diameter and 2 mm height.
- **Applications:**
  Sustained, programmable bioproduction, including inducible gene expression and pteridine-pathway bioproduction. This fits BTP rather than medical-device or sensing categories in the demonstrated experiments.
- **Production method(s):**
  Engineered *E. coli* were encapsulated directly in F127-BUM hydrogel, transferred into a mold, and photocured with UV light. The best ontology match is OPM because the paper describes encapsulation within an engineered hydrogel construct, but not a named ontology process such as 3D bioprinting, scaffold printing followed by colonization, or solvent casting/leaching.

- **Concept description:**
  This ELM is a bacterial hybrid living material formed by embedding engineered *E. coli* in a 3D polymeric F127-BUM hydrogel. The encapsulated cells retain inducible CRISPRa-based gene expression over multi-week culture and can repeatedly switch bioproduction ON and OFF. The demonstrated output includes pteridine-pathway activation for pyruvoyl tetrahydropterin production, showing that the construct functions as a programmable microbial bioreactor. In ontology terms, it is a 3D hybrid bacterial living polymer directed toward biotechnology and bioprocessing.
- **Rationales for ambiguous slots (if any):**
  - LPO: selected because the ontology explicitly places hydrogels under living polymers, and F127-BUM is described as a hydrogel matrix.
  - BTP: selected because the experimentally demonstrated purpose is programmable bioproduction under continuous culture.
  - OPM: selected because direct cell encapsulation plus molding/photocuring is described, but no narrower ontology production category is explicitly matched by the paper.

---

**ELM B**

- **Ontology code:**
  EUK-HYB-FUN-3D-LPO-BTP-3DB
- **Decoded slots:**
  - EUK = Eukaryotic ELM
  - HYB = Hybrid living material
  - FUN = Fungal ELM
  - 3D = Three-dimensional ELM
  - LPO = Living polymer
  - BTP = Biotechnology and bioprocessing
  - 3DB = 3D bioprinting of cells and biomaterials
- **Organisms:**
  Engineered *Saccharomyces cerevisiae*; the printed strain is reported as *S. cerevisiae* spk05, derived from BY4741-based engineering.
- **Material(s):**
  BSA-PEGDA hydrogel, described as bovine serum albumin–polyethylene glycol diacrylate bioconjugates.
- **Geometry:**
  3D printed hydrogel constructs fabricated by SLA printing; therefore the geometry is clearly 3D.

- **Applications:**
  Dynamic multi-product bioproduction, specifically scheduled switching between betaxanthin pigment production and proteinase A production over 27 days. This aligns with BTP.

- **Production method(s):**
  The yeast-laden BSA-PEGDA resin was printed with a stereolithographic apparatus (SLA) 3D printer and photocured. This directly matches ontology code 3DB for 3D bioprinting of cells and biomaterials.

- **Concept description:**
  This ELM is a fungal hybrid living material built by embedding engineered *S. cerevisiae* in a 3D-printed BSA-PEGDA hydrogel. The construct contains two inducible bioproduction modules, enabling scheduled switching between betaxanthin synthesis and secreted proteinase A production. The work demonstrates that multi-input genetic control can be maintained over weeks in a printed yeast-laden hydrogel. Ontologically, it is a 3D fungal living polymer for biotechnology and bioprocessing produced by 3D bioprinting.

- **Rationales for ambiguous slots (if any):**
  - LPO: selected because the matrix is a hydrogel bioconjugate centered on PEGDA/BSA and the ontology places hydrogels under living polymers.
  - BTP: selected because the demonstrated use is controlled production of biomolecules, not tissue engineering, sensing, or medical-device deployment.

**Result:** ELM

The paper clearly describes **two distinct ELMs**, not just one: a bacterial F127-BUM hydrogel system based on *E. coli* and a fungal BSA-PEGDA printed hydrogel system based on *S. cerevisiae*. Both satisfy the ontology's ELM definition because living cells coexist with an abiotic scaffold/hydrogel matrix.

***Engineered living materials for the conversion of a low-cost food-grade precursor to a high-value flavonoid***

Abstract

**Source mode:** abstract only

- **Ontology code:**
  BAC-HYB-EUB-3D-LPO-BMD+BTP-OPM
- **Decoded slots:**
  - **BAC** = Bacterial ELM
  - **HYB** = Hybrid living material
  - **EUB** = Eubacterial ELM
  - **3D** = Three-dimensional ELM
  - **LPO** = Living polymers
  - **BMD+BTP** = Biohybrid medical devices + Biotechnology and bioprocessing
  - **OPM** = Other production methods or combinations
- **Organisms:**
  Engineered probiotic *Escherichia coli* Nissle 1917.
- **Material(s):**
  Poly(vinyl alcohol)-based hydrogels encapsulating the engineered probiotic bacteria. This supports **HYB**, because the ontology defines hybrid living materials as systems combining living cells with an abiotic biomaterial or scaffold, and **LPO**, because hydrogels are explicitly included under living polymers.
- **Geometry:**
  **3D** is the best-supported phylum. The ontology defines 3D ELMs as cells within a three-dimensional biomaterial or scaffold, which matches bacteria encapsulated inside a hydrogel matrix.
- **Applications:**
  **BMD+BTP** is the best-supported combination.
  - **BTP** fits because the abstract describes the system as a sustained **microbial biofactory / bioreactor** for bioproduction.
  - **BMD** also fits because the stated alternative strategy is **direct delivery of the microbial biofactories to the body**, with emphasis on **biocontainment in the body** and controlled in-body production, which aligns with the ontology definition of biohybrid medical devices including drug-delivery-like healthcare functions.
- **Production method(s):**
  **OPM** is the best-supported production code. The ontology explicitly includes under OPM the **encapsulation of cells within engineered constructs** for generating bio-ELMs and HLMs, which matches probiotic *E. coli* Nissle 1917 encapsulated in PVA-based hydrogels.

- **Concept description:**
  This abstract describes a **bacterial hybrid engineered living material** composed of engineered probiotic *E. coli* Nissle 1917 encapsulated within a biocompatible polymeric hydrogel. The living cells function as contained microbial biofactories that convert cinnamic acid into pinocembrin while remaining metabolically active for about one month. The hydrogel provides biocontainment and also regulates bacterial growth, thereby helping control production levels inside the material. Within the ontology, this is most consistently interpreted as a **3D living-polymer hybrid ELM** positioned between **biotechnology/bioprocessing** and **biohybrid medical-device** use, produced through an **encapsulation-based OPM route**.

- **Rationales for ambiguous slots (if any):**

  - **BMD+BTP instead of only BTP:** the abstract does not stop at ex vivo production; it explicitly frames **direct delivery to the body** as the target strategy and addresses **in-body biocontainment**, which supports a medical-device interpretation alongside bioprocessing.

  - **HYB instead of BIO:** the ontology reserves **BIO** for cells plus their own extracellular matrices, whereas this system uses an external **PVA hydrogel** scaffold.

  - **OPM instead of ECP:** although the bacteria are engineered, the explicit fabrication step described is **encapsulation in hydrogels**, and that is directly covered by **OPM** in the ontology.

<u>Full text</u>

- **Ontology code:**
  BAC-HYB-EUB-3D-LPO-BMD+BTP-OPM

- **Decoded slots:**

  - **BAC** = Bacterial ELMs
  - **HYB** = Hybrid living materials
  - **EUB** = Eubacterial ELMs
  - **3D** = Three-dimensional ELMs
  - **LPO** = Living polymers
  - **BMD+BTP** = Biohybrid medical devices + Biotechnology and bioprocessing
  - **OPM** = Other production methods or combinations

- **Organisms:**
  The main living component is **engineered probiotic *Escherichia coli* Nissle 1917**. The engineered biosynthetic cascade includes genes encoding **4-coumaroyl-CoA ligase from *Nicotiana tabacum*** and **chalcone synthase from *Arabidopsis thaliana***, but the actual cellular chassis inside the material is *E. coli* Nissle 1917.

- **Material(s):**
  The abiotic matrix is a **poly(vinyl alcohol) (PVA)-based hydrogel**, specifically **vinyl sulfone-functionalized PVA (PVA-VS)**, chemically crosslinked into a stable hydrogel network. This supports classification as a **living polymer** and as a **hybrid** ELM because living bacteria coexist with a synthetic polymer scaffold.

- **Geometry:**
  The ELM is described as **hydrogel films**, including a **bilayer film** format in which a bacteria-containing hydrogel layer is enclosed by a bacteria-free outer hydrogel layer. Although film-shaped, the bacteria are encapsulated **within a volumetric hydrogel matrix**, which best matches the ontology's **3D** category.

- **Applications:**
  The study presents the material as a platform for **conversion of cinnamic acid into pinocembrin**, which supports **biotechnology and bioprocessing (BTP)**. The paper also explicitly frames the system as a prospective **living drug-delivery / living medical device** for production of a bioactive flavonoid in the body, supporting **biohybrid medical devices (BMD)**.

- **Production method(s):**
  The bacteria were genetically engineered and then **encapsulated in photocrosslinked PVA-VS hydrogels**; bilayer films were fabricated by sequential deposition and crosslinking of a bacteria-containing inner layer and a bacteria-free enveloping layer. The ontology does not provide a more specific dedicated code for this exact hydrogel encapsulation workflow, so **OPM** is the best-supported production-method code.

- **Concept description:**
  This work describes an engineered living material composed of probiotic *E. coli* Nissle 1917 encapsulated within a synthetic PVA-based hydrogel matrix, forming a hybrid bacterial living polymer. The engineered bacteria catalyze the conversion of the low-cost precursor cinnamic acid into the higher-value flavonoid pinocembrin while remaining spatially contained in bilayer hydrogel films. The material therefore functions both as a microbial bioprocessing platform and as a prototype biohybrid therapeutic device concept. Its central design principle is the integration of metabolically active engineered bacteria with a polymeric containment matrix that permits nutrient and product diffusion while restricting bacterial escape.

- **Rationales for ambiguous slots (if any):**
  - **3D:** The construct is a film, but the cells are embedded throughout a crosslinked hydrogel volume rather than on a purely planar surface, so **3D** is better supported than 2D.
  - **BMD+BTP:** The paper demonstrates biochemical conversion in vitro (**BTP**) and explicitly motivates the system as a future in-body drug-delivery / living medical-device platform (**BMD**).
  - **OPM:** The fabrication uses hydrogel encapsulation and bilayer photocrosslinking, which is clearly described in the paper but not more specifically codified elsewhere in the ontology tables provided.

***Fungal Engineered Living Materials: The Viability of Pure Mycelium Materials with Self-Healing Functionalities***

Abstract

**Source mode:** abstract only

- **Ontology code:**
  EUK-BIO-FUN-2D-LPO-FAS-EMG
- **Decoded slots:**
  - **EUK** = Eukaryotic ELM
  - **BIO** = Biological engineered living material
  - **FUN** = Fungal ELM
  - **2D** = Two-dimensional ELM
  - **LPO** = Living polymers
  - **FAS** = Fashion and biohybrid wearables
  - **EMG** = Extracellular matrix (ECM) grown in culture
- **Organisms:**
  Fungal mycelium is the living component. No fungal species or strain is explicitly named in the abstract.
- **Material(s):**
  The abstract describes **pure mycelium materials** composed entirely of fungal cells, with no supporting synthetic scaffold or abiotic matrix. That supports **BIO**, because the ontology defines bio-ELMs as materials made only of cells and their extracellular matrices, without synthetic biomaterials or structures. The best-supported class is **LPO**, because the material is a pure fungal mycelial matrix rather than a ceramic, carbon, alloy, or explicit composite system, and the ontology places predominantly polymeric matrices under living polymers.
- **Geometry:**
  **2D** is the best-supported phylum. The ontology defines 2D ELMs as planar or bi-dimensional structures, and this abstract specifically frames the material as a **leather substitute**, which is most consistently interpreted as a planar material form.
- **Applications:**
  **FAS** is the best-supported application code. The ontology defines FAS for ELMs used as innovative materials for **fashion**, and the abstract explicitly states that the pure mycelium materials are fabricated for use as **leather substitutes**.
- **Production method(s):**
  **EMG** is the best-supported production code. The ontology defines EMG as cases where cultured cells grow an extracellular matrix, applying to both hybrid and biological ELMs. This is the closest ontology match for a **pure mycelium material** grown from fungal cells without any stated synthetic scaffold, printing route, or encapsulation step.
- **Concept description:**
  This abstract describes a **fungal biological engineered living material** formed entirely from mycelium and developed as a leather-substitute material. The living fungal network remains viable under dry and oligotrophic conditions and can self-heal after damage, with chlamydospores implicated as a key mechanism in recovery. Within the ontology, the construct is most consistently interpreted as a **planar fungal living-polymer ELM** because it is a pure mycelial material aimed at a leather-like sheet application. Its primary ontology-grounded role is **fashion material development**, and its fabrication is best captured as **matrix growth in culture** rather than a scaffold-based hybrid route.

- **Rationales for ambiguous slots (if any):**
    - **2D instead of UNK or 3D:** chosen because the intended product is a **leather substitute**, which most strongly implies a planar material format under the ontology's 2D definition.
    - **LPO:** chosen because the abstract describes a **pure mycelium** material and does not indicate a carbon, ceramic, alloy, or explicit composite matrix; under the ontology, the closest supported class is living polymers.
    - **EMG instead of OPM:** chosen because the abstract supports growth of a pure fungal material in culture more directly than any specific encapsulation, printing, or scaffold-colonization route.

<u>Full text</u>

**Result:** ELM. The paper explicitly describes a **fungal engineered living material** made from **pure mycelium**, where the fungal organism provides both the structural material and the self-healing function through viable chlamydospores embedded in the dried matrix.

- **Ontology code:**
EUK-BIO-FUN-2D-LPO-FAS+TRI-EMG+BFF
- **Decoded slots:**
**EUK** = Eukaryotic ELMs
**BIO** = Biological engineered living materials (bio-ELMs)
**FUN** = Fungal ELMs
**2D** = Planar / bi-dimensional ELMs
**LPO** = Living polymers
**FAS+TRI** = Fashion and biohybrid wearables + Transport and related infrastructures
**EMG+BFF** = Extracellular matrix grown in culture + Biofilm formation and related extracellular matrix growth.
- **Organisms:**
The main ELM is based on **Ganoderma lucidum**, specifically the **GL-M9726** strain used as the model organism in the experiments. Other fungal species and strains were tested comparatively for revivability, but they are not the main engineered material described.
- **Material(s):**
The material is a **pure mycelium skin/pellicle** with no synthetic scaffold, making it a bio-ELM rather than a hybrid material. The paper describes fungal tissue with structural contribution from fungal cell walls, notes the relevance of **chitin** and **hydrophobins**, and also mentions a possible **polysaccharide coating** at the liquid interface; after growth, the skins are plasticized with **20% glycerol** and dried.

- **Geometry:**
  The geometry is best described as a **2D sheet-like membrane / pellicle / skin**. The material is grown at the **liquid–air interface** as a floating mycelium skin and used as a leather-like sheet, even though it has measurable thickness.

- **Applications:**
  The paper frames the material as a **leather substitute** and points to **fashionwear** and **automotive seats** as explicit future product applications. A furniture use is also mentioned in the paper, but a furniture-specific application code is **not present in the ontology/graph**.

- **Production method(s):**
  The material is produced by **liquid-state culture / fermentation**, followed by homogenization of fungal biomass, transfer to wells, **static growth into a planar skin at the liquid–air interface**, then glycerol plasticization and drying. In ontology terms, the closest supported production codes are **EMG+BFF**.

- **Concept description:**
  This ELM is a **living fungal mycelium skin** made from *Ganoderma lucidum* and designed as a **self-healing leather-like material**. Its key functional element is the presence of **chlamydospores** embedded in the dried mycelium matrix; after hydration and nutrient exposure, these spores germinate into new hyphae that refill punctures and cuts and reconnect with the existing network. The material remains viable after drying at **40 °C** and plasticization with **20% glycerol**, and healed regions recover mechanical performance to the point that samples often fail outside the repaired zone. As an ontology-grounded concept, it is a **2D fungal bio-ELM with a polymeric matrix**, targeted primarily toward wearable / leather-like product applications and secondarily toward vehicle-interior uses.

- **Rationales for ambiguous slots (if any):**
  **2D:** Chosen because the material is a planar **skin/pellicle/sheet** grown at the liquid–air interface; although it has thickness, the ontology's 2D definition best matches the reported geometry.
  **LPO:** Chosen because the reported material is a pure, flexible fungal tissue with polymeric biological structure; the paper does not describe a predominantly metallic, ceramic, carbon, or cross-class matrix.
  **FAS+TRI:** Chosen because the paper explicitly mentions **fashionwear** and **automotive seats** as target applications; furniture is mentioned, but no furniture-specific ontology code is present in the ontology/graph.
  **EMG+BFF:** Chosen because the ontology has no dedicated "liquid-state mycelium pellicle fermentation" code; the closest supported match is **ECM grown in culture** plus **biofilm-related ECM growth** for a cultured planar skin formed at the interface.

***Additive Manufacturing of Engineered Living Materials with Bio-Augmented Mechanical Properties and Resistance to Degradation***

Abstract

**Source mode:** abstract only

- **Ontology code:**
  UNK-HYB-UNK-3D-LPO-BMR+MAP-3DB
- **Decoded slots:**
  - **UNK** = Domain not inferable from the abstract
  - **HYB** = Hybrid living material
  - **UNK** = Kingdom not inferable from the abstract
  - **3D** = Three-dimensional ELM
  - **LPO** = Living polymers
  - **BMR+MAP** = Biohybrid machines and robots + Materials production and biofunctionalization
  - **3DB** = 3D bioprinting of cells and biomaterials
- **Organisms:**
  Engineered microbes are explicitly involved, but no species, strain, domain, or kingdom is identified in the abstract.
- **Material(s):**
  The matrix is an aqueous **BSA-PEGDA hydrogel resin**, i.e. a polymeric hydrogel containing engineered microbes. This supports **HYB**, because the ontology defines hybrid living materials as systems combining living cells with abiotic biomaterials or structures, and **LPO**, because hydrogels are explicitly included in the ontology's living-polymer class.
- **Geometry:**
  **3D** is the best-supported phylum. The abstract explicitly describes vat photopolymerization to create objects with a wide range of **3D form factors**, and the ontology defines 3D ELMs as cells within a three-dimensional biomaterial or scaffold.
- **Applications:**
  **BMR+MAP** is the best-supported combination.
  - **MAP** fits because the abstract explicitly shows that metabolites produced by the encapsulated cells **bioaugment the matrix properties**, which aligns with ELMs used as support to **materials functionalization**.
  - **BMR** also fits because the printed ELM parts are assembled into **mechanically functional bolts and gears** to demonstrate potential for **synthetic living machines**, which matches the ontology definition of constituents of biohybrid machines and robots and their subsystems.
- **Production method(s):**
  **3DB** is the best-supported production code. The ontology defines it as additive manufacturing involving **biomaterials and living cells** to create living three-dimensional constructs, which matches vat photopolymerization of aqueous BSA-PEGDA resins containing engineered microbes.

- **Concept description:**
  This abstract describes a **hybrid 3D engineered living material** in which engineered microbes are directly encapsulated in a BSA-PEGDA hydrogel during additive manufacturing. The living cells continuously produce metabolites such as l-DOPA, naringenin, and betaxanthins, and these metabolites actively modify the surrounding matrix by increasing stiffness or resistance to degradation. The printed ELM components are then assembled into mechanically functional parts such as bolts and gears, showing that the living constructs can operate as functional elements of synthetic living machines. Within the ontology, this is most consistently interpreted as a **3D living-polymer hybrid ELM for materials biofunctionalization and biohybrid-machine subsystems, produced by 3D bioprinting**.

- **Rationales for ambiguous slots (if any):**
  - **Domain = UNK:** the abstract only states "engineered microbes" and does not explicitly identify whether they are bacterial, fungal, or mixed.
  - **Kingdom = UNK:** no taxonomic identity is given for the encapsulated microbes.
  - **BMR+MAP instead of only MAP:** matrix enhancement is central, but the abstract also explicitly demonstrates assembly into **bolts and gears** for **synthetic living machines**, which is directly covered by **BMR**.
  - **3DB instead of 3DP:** the cells are processed **within** the printable resin during fabrication, not colonizing a pre-printed scaffold afterward, so the ontology definition of **3D bioprinting of cells and biomaterials** is the closer match.

<u>Full text</u>

**Result:** ELM
**Finding:** The paper clearly describes multiple distinct ELMs, specifically **ELM-EC-LDOPA**, **ELM-EC-NGN**, and **ELM-SC-BXN**, all based on engineered microorganisms encapsulated in a BSA-PEGDA hydrogel and fabricated by vat photopolymerization / 3D printing.

**ELM A — ELM-EC-LDOPA**

- **Ontology code:**
  BAC-HYB-EUB-3D-LPO-MAP+BMR-3DB

- **Decoded slots:**
  **BAC** = Bacterial ELM; **HYB** = Hybrid living material; **EUB** = Eubacterial; **3D** = three-dimensional ELM; **LPO** = living polymer; **MAP** = materials production and biofunctionalization; **BMR** = biohybrid machines and robots; **3DB** = 3D bioprinting of cells and biomaterials. The ontology defines HYB for ELMs combining cells with abiotic synthetic biomaterials, 3D for cell-containing 3D structures, LPO for polymer-dominant matrices including hydrogels, MAP for biofunctionalization of materials, BMR for machine/robot constituents, and 3DB for additive manufacturing where cells and biomaterials are processed together.

- **Organisms:**
  Engineered **Escherichia coli** were used. The l-DOPA-generating strain was created by transforming a tyrosine-overproducing strain, and the culture conditions identify the working strain as **E. coli eBL0430D** for l-DOPA production.

- **Material(s):**
The matrix is a **BSA-PEGDA hydrogel**, formed from bovine serum albumin and poly(ethylene glycol) diacrylate, with Ru(bpy)3Cl2 and sodium persulfate used for photopolymerization. In this ELM, the encapsulated bacteria produce **l-DOPA**, which then interacts with BSA and contributes to matrix stiffening.

- **Geometry:**
This is a **3D hydrogel ELM**. The paper reports 3D printed cylinders, lattices, tubular stents, and specifically a **nut-and-bolt assembly** fabricated from the l-DOPA-producing resin.

- **Applications:**
Best-supported applications are **materials production and biofunctionalization (MAP)**, because in situ l-DOPA production enhances matrix stiffness, and **biohybrid machines and robots (BMR)**, because the printed material is demonstrated as mechanically functional nut-and-bolt hardware for prospective synthetic living machines.

- **Production method(s):**
The best-supported production code is **3DB**, because the living cells and the BSA-PEGDA biomaterial were combined in the same aqueous resin and then directly processed by vat photopolymerization / SLA 3D printing into living constructs.

- **Concept description:**
ELM-EC-LDOPA is a bacterial hybrid living material in which engineered **E. coli** are encapsulated within a **BSA-PEGDA 3D hydrogel** fabricated by vat photopolymerization. The embedded cells continuously produce **l-DOPA**, and the paper shows that this metabolite binds to the surrounding BSA-containing matrix and increases compressive stiffness relative to the cell-free control. The construct therefore functions as a bio-augmented living polymer whose biological component actively improves material performance rather than merely being passively enclosed. Its printing into mechanically functional **nuts and bolts** further supports its interpretation as a candidate material platform for **synthetic living machines**.

- **Rationales for ambiguous slots (if any):**
**Application = MAP+BMR:** MAP was selected because the central demonstrated effect is matrix biofunctionalization by in situ l-DOPA production; BMR was also selected because the same resin was used to print mechanically functional bolt-and-nut components for living-machine proof-of-concept.
**Production = 3DB:** The ontology does not name "vat photopolymerization" separately, so it is best mapped to 3DB because cells and biomaterial are printed together, not colonized afterward on a previously printed scaffold.

**ELM B — ELM-EC-NGN**

- **Ontology code:**
BAC-HYB-EUB-3D-LPO-BTP-3DB

- **Decoded slots:**
**BAC** = Bacterial ELM; **HYB** = Hybrid living material; **EUB** = Eubacterial; **3D** = three-dimensional ELM; **LPO** = living polymer; **BTP** = biotechnology and bioprocessing; **3DB** = 3D bioprinting of cells and biomaterials. This mapping follows the ontology definitions for bacterial cell-containing hydrogels processed as 3D living constructs and used for metabolite production.

- **Organisms:**
  The naringenin-producing system uses engineered **Escherichia coli BL21(DE3)** carrying multiple plasmids for naringenin biosynthesis.

- **Material(s):**
  The matrix is again the **BSA-PEGDA hydrogel**. The biological output is **naringenin**, a flavonoid therapeutic compound produced in situ by the encapsulated engineered bacteria.

- **Geometry:**
  This ELM is a **3D hydrogel construct** fabricated by the same vat-photopolymerization route used throughout the paper, with representative printed geometries including cylinders and other 3D forms.

- **Applications:**
  The best-supported ontology application is **BTP**, because the paper presents this ELM primarily as a construct for **continuous in situ production of naringenin** rather than as a finalized medical device.

- **Production method(s):**
  The best-supported production method is **3DB**, since engineered bacteria and biomaterial resin were processed together in a 3D printing workflow.

- **Concept description:**
  ELM-EC-NGN is a bacterial hybrid living polymer in which engineered **E. coli BL21(DE3)** are embedded in a **BSA-PEGDA 3D hydrogel** and printed by vat photopolymerization. Its defining function is the **continuous production of naringenin**, which the paper uses as a comparative example of bio-augmentation through metabolite synthesis. Although naringenin interacts with BSA, the paper reports that it does **not** induce the same matrix-structure change or stiffness increase observed for l-DOPA. This makes ELM-EC-NGN best interpreted as a **bioprocessing-oriented ELM** rather than a mechanically reinforced one.

- **Rationales for ambiguous slots (if any):**
  **Application = BTP:** Although naringenin is described as a therapeutic compound, the paper does not present this construct as a concrete medical device; the clearest supported use is metabolite production within a living material, which fits biotechnology and bioprocessing.
  **Production = 3DB:** The same rationale applies as above: cells and biomaterial are printed together in one additive-manufacturing step.

**ELM C — ELM-SC-BXN**

- **Ontology code:**
  EUK-HYB-FUN-3D-LPO-MAP+BMR-3DB

- **Decoded slots:**
  **EUK** = Eukaryotic ELM; **HYB** = Hybrid living material; **FUN** = Fungal ELM; **3D** = three-dimensional ELM; **LPO** = living polymer; **MAP** = materials production and biofunctionalization; **BMR** = biohybrid machines and robots; **3DB** = 3D bioprinting of cells and biomaterials. The organism is fungal yeast, the matrix is a hydrogel, the construct is 3D, the function biofunctionalizes the material by improving degradation resistance, and the same resin is demonstrated as a gear system.

- **Organisms:**
  The living component is engineered **Saccharomyces cerevisiae**. The betaxanthin-producing strain was created from **BY4741 yeast**, and the culturing section names **S. cerevisiae BY01** as the engineered strain.

- **Material(s):**
  The matrix is the same **BSA-PEGDA hydrogel**. The yeast produce **betaxanthins**, which bind to BSA and are associated with resistance to microbial / enzymatic degradation of the surrounding matrix.

- **Geometry:**
  This is also a **3D hydrogel ELM**. The paper reports cylinders and lattices for characterization, and specifically prints **gear components** from the betaxanthin-producing resin.

- **Applications:**
  The best-supported applications are **MAP**, because betaxanthin production biofunctionalizes the material by improving degradation resistance, and **BMR**, because the paper demonstrates an assembled **gear system** made from this ELM as a mechanically functional subsystem for prospective living machines.

- **Production method(s):**
  The best-supported production method is **3DB**, because the engineered yeast were directly entrapped in the printable resin and processed together with the biomaterial during additive manufacturing.

- **Concept description:**
  ELM-SC-BXN is a fungal hybrid living polymer in which engineered **S. cerevisiae** are encapsulated within a **BSA-PEGDA 3D hydrogel** fabricated by vat photopolymerization. The embedded yeast continuously produce **betaxanthins**, and the paper shows that this metabolite alters interactions with the protein matrix and is associated with a strong increase in resistance to microbial degradation. Unlike a passive yeast-loaded scaffold, this ELM actively stabilizes its own surrounding material through ongoing biochemical production. Its fabrication into working **gear assemblies** further supports interpretation as both a **biofunctionalized material** and a **biohybrid machine component**.

- **Rationales for ambiguous slots (if any):**
  **Application = MAP+BMR:** MAP was selected because the principal demonstrated effect is material biofunctionalization via enhanced degradation resistance; BMR was also selected because the same living material was printed into assembled gears with transmitted rotational motion.
  **Production = 3DB:** As with the bacterial constructs, the cells were included in the printable biomaterial resin before 3D fabrication, so 3DB is a closer ontology match than scaffold-printing-plus-colonization.

If you want, I can turn this into a reusable extraction template for the next ELM paper.

***Phenotypically complex living materials containing engineered cyanobacteria***

Abstract

**Source mode:** abstract only

- **Ontology code:**
  BAC-HYB-EUB-3D-LPO-BRE-3DB
- **Decoded slots:**
  - **BAC** = Bacterial ELM
  - **HYB** = Hybrid living material
  - **EUB** = Eubacterial ELM
  - **3D** = Three-dimensional ELM
  - **LPO** = Living polymers
  - **BRE** = Bioremediation
  - **3DB** = 3D bioprinting of cells and biomaterials
- **Organisms:**
  *Synechococcus elongatus* PCC 7942 is the explicitly named living component. The abstract also describes engineered cyanobacterial strains for reporter expression, laccase production, and inducible cell death, but all remain within the same cyanobacterial platform.
- **Material(s):**
  The living cells are printed **within a hydrogel matrix**, and the ontology defines **HYB** for ELMs combining cells with an abiotic scaffold or biomaterial, while **LPO** covers matrices made predominantly of polymers, including **hydrogels**.
- **Geometry:**
  **3D** is the best-supported phylum because the abstract explicitly states **3D printing**, **volumetric 3D-printed designs**, and a hydrogel-embedded biocomposite, which matches the ontology definition of cells within a three-dimensional biomaterial or scaffold.
- **Applications:**
  **BRE** is the best-supported application code. The strongest demonstrated functional use in the abstract is **decolorization of the textile dye pollutant indigo carmine**, explicitly framed as a tool for **environmental bioremediation**, which aligns directly with the ontology description of ELMs used for processing, recycling, or revalorizing waste and debris.
- **Production method(s):**
  **3DB** is the best-supported production code because the ontology defines it as additive manufacturing involving **living cells and biomaterials** to create living three-dimensional constructs, which matches 3D printing of engineered cyanobacteria directly within a hydrogel matrix.
- **Concept description:**
  This abstract describes a **bacterial hybrid engineered living material** composed of engineered *Synechococcus elongatus* cells embedded in a polymeric hydrogel and fabricated by 3D printing. The cyanobacteria are programmed to generate different stimulus-responsive outputs, including reporter expression, pollutant-degrading laccase production, and inducible cell death for biocontainment. The printed construct is therefore both structurally shaped by additive manufacturing and functionally programmed by synthetic biology. Within the ontology, the clearest classification is a **3D living-polymer hybrid bacterial ELM for bioremediation produced by 3D bioprinting**.

- **Rationales for ambiguous slots (if any):**
  - **BRE instead of BSE:** the YFP riboswitch is presented as an initial proof-of-concept, but the main explicit end-use demonstrated in the abstract is **pollutant decolorization / environmental bioremediation**.
  - **HYB instead of BIO:** the ontology reserves **BIO** for cell-plus-ECM-only systems, whereas this material explicitly uses a **hydrogel matrix** as the surrounding abiotic component.
  - **3DB instead of 3DP:** the cells are processed **within** the printable hydrogel during additive manufacturing, rather than colonizing a separately printed scaffold afterward.

<u>Full text</u>

**ELM A**

- **Ontology code:**
  BAC-HYB-EUB-3D-LPO-BRE-3DB
- **Decoded slots:**
  - BAC = Bacterial ELM
  - HYB = Hybrid living material
  - EUB = Eubacterial ELM
  - 3D = Three-dimensional ELM
  - LPO = Living polymers
  - BRE = Bioremediation
  - 3DB = 3D bioprinting of cells and biomaterials
- **Organisms:**
  Engineered *Synechococcus elongatus* PCC 7942 cyanobacteria. The main functional construct in the paper is the laccase-producing strain with inducible lysis, denoted Laccase+-riboF-Lysis+; the article also includes reporter/control variants built on the same alginate-hydrogel platform.
- **Material(s):**
  The abiotic matrix is an alginate hydrogel, described as a natural polymer/polysaccharide scaffold, crosslinked with calcium salts and used as a cell-laden printable bioink.
- **Geometry:**
  The material is a volumetric 3D-printed hydrogel scaffold produced by direct-ink-writing. The paper emphasizes grid-type printed geometries with high surface-area-to-volume ratio and explicitly frames the construct as a three-dimensional hydrogel composite, not a 2D film or a 4D morphing system.

- **Applications:**
  The clearest ontology-supported application is **bioremediation**: the engineered laccase-producing hydrogel decolorizes the textile dye indigo carmine. The inducible cell-death module is a biocontainment feature, but the paper's stated end-use application is remediation of chemical pollutants.
- **Production method(s):**
  The best-supported production code is **3DB**, because the study uses direct-ink-writing / 3D printing of a **cell-laden alginate bioink**, meaning cells and biomaterial are processed together during fabrication. The paper does not describe a separately printed empty scaffold followed by later colonization as the primary route.
- **Concept description:**
  This paper describes a bona fide engineered living material because genetically engineered cyanobacterial cells coexist with an abiotic alginate hydrogel scaffold and produce regulated functional outputs from within that material. The platform uses 3D bioprinting to shape porous, photosynthetic hydrogel constructs containing *S. elongatus* PCC 7942, enabling external chemical control through a theophylline-responsive riboswitch. In its main functional form, the material expresses CotA laccase to oxidize and decolorize indigo carmine, establishing a clear bioremediation use case. The same platform also integrates an inducible lysis circuit to reduce environmental contamination by escaped cells, making the ELM both functional and partially self-limiting.
- **Rationales for ambiguous slots (if any):**
  - **HYB rather than BIO:** the cells are embedded in an abiotic alginate hydrogel scaffold, so the construct is scaffold-based rather than cell/ECM-only.
  - **LPO rather than LCO:** although the paper uses the word "biocomposite," the matrix itself is predominantly a polymeric hydrogel, and the ontology explicitly places hydrogels under living polymers.
  - **BRE only:** bioremediation is the only explicit end-use application clearly claimed and experimentally demonstrated at the material level. The YFP and lysis modules support regulation and containment but do not define a separate primary application slot more strongly than BRE.
  - **3DB rather than 3DP:** the study prints a cell-containing bioink directly, which matches "3D bioprinting of cells and biomaterials" better than "printed scaffold followed by cell colonization."

**Identification decision:** ELM. The article explicitly describes genetically engineered cyanobacteria integrated into an alginate hydrogel scaffold to form a stimuli-responsive functional material, which matches the ontology's ELM definition.

**Scope note:** The paper includes several engineered strain variants within the same alginate-hydrogel platform. I coded the main application-bearing construct (Laccase+-riboF-Lysis+) because it is the clearest fully specified ELM with a defined ontology-supported end use; the reporter/control variants are better interpreted as validation variants of the same platform rather than separate end-use ELM classes.

***Engineered living materials grown from programmable Aspergillus niger mycelial pellets***

Abstract

**Source mode:** abstract only

- **Ontology code:**
  EUK-BIO-FUN-2D-LPO-BSE-ECP+EMG
- **Decoded slots:**
  - **EUK** = Eukaryotic ELM
  - **BIO** = Biological engineered living material
  - **FUN** = Fungal ELM
  - **2D** = Two-dimensional ELM
  - **LPO** = Living polymers
  - **BSE** = Biosensing
  - **ECP+EMG** = Engineering cells to produce ELMs + Extracellular matrix grown in culture
- **Organisms:**
  *Aspergillus niger* mycelial pellets are the explicitly named living component.
- **Material(s):**
  The abstract describes **self-supporting living membrane materials** grown directly from programmable fungal mycelial pellets. No abiotic scaffold, hydrogel, or synthetic support is stated, so the best-supported super-kingdom is **BIO**, which the ontology defines for ELMs made only of cells and their extracellular matrices, without supporting synthetic biomaterials or structures. The class is best assigned as **LPO**, because the material is a pure fungal/mycelial matrix rather than an explicitly carbon, ceramic, alloy, or composite construct.
- **Geometry:**
  **2D** is the best-supported phylum. The ontology defines 2D ELMs as planar or bi-dimensional structures, and the abstract explicitly describes **large-area self-supporting** and **living membrane materials**, which are most consistent with a planar membrane format.
- **Applications:**
  **BSE** is the best-supported application code. The strongest demonstrated function in the abstract is tunable color response to environmental xylose concentration, proposed as a **biosensor** for detecting xylose levels in industrial wastewater. The other mentions of fabrics and packaging are future opportunities rather than the main validated use in the abstract.
- **Production method(s):**
  **ECP+EMG** is the best-supported combination.
  - **ECP** fits because the abstract describes **programmable** *A. niger* and inducible expression tuning of melanin-biosynthesis genes.

- **EMG** fits because the material is grown from fungal cells into a self-supporting matrix in culture, without a separate synthetic scaffold.

- **Concept description:**
This abstract describes a **fungal biological engineered living material** formed directly from programmable *Aspergillus niger* mycelial pellets into self-supporting membrane-like sheets. The living fungal network remains viable, self-regenerative, and functional over long storage, and its visible output can be tuned through inducible melanin biosynthesis. Within the ontology, the material is most consistently interpreted as a **2D fungal living-polymer bio-ELM**, because it is a planar membrane grown entirely from fungal biomass without an abiotic scaffold. Its clearest validated function is **biosensing**, specifically color-based detection of xylose levels relevant to industrial wastewater monitoring.

- **Rationales for ambiguous slots (if any):**
  - **BIO instead of HYB:** no supporting synthetic biomaterial or synthetic structure is described in the abstract.
  - **2D instead of 3D:** "living membrane materials" and "large-area self-supporting structures" are more consistent with a planar material than a bulk 3D scaffold.
  - **BSE instead of BSE+ENV:** the abstract explicitly supports **detection** of xylose; environmental monitoring is plausible from the wastewater context, but the demonstrated ontology fit is more directly **biosensing**.
  - **ECP+EMG:** the fabrication includes both programmed gene control and growth of the fungal material itself; a single narrower production code would miss one of those two explicitly supported aspects.

<u>Full text</u>

**Result:** ELM. The paper explicitly presents a fungi-derived engineered living material grown from programmable *Aspergillus niger* mycelial pellets, forming self-supporting living membrane materials that remain alive, self-regenerative, and functional.

- **Ontology code:**
EUK-BIO-FUN-2D-LPO-BSE+ENV-ECP+EMG

- **Decoded slots:**
  - EUK = Eukaryotic ELMs
  - BIO = Biological engineered living materials (cells plus their own extracellular matrices, without synthetic scaffold)
  - FUN = Fungal ELMs
  - 2D = Planar / membrane-like ELMs
  - LPO = Living polymers
  - BSE+ENV = Biosensing + Environmental monitoring
  - ECP+EMG = Engineering cells to produce ELMs + Extracellular matrix grown in culture

- **Organisms:**
The biological chassis is *Aspergillus niger*, specifically wild-type ATCC1015 and engineered derivatives including the melanin-auxotrophic SΔpptA strain and the xylose-inducible Spxyn-pptA strain.

- **Material(s):**
  The material is built from dense *A. niger* mycelial pellets / hyphae as the sole raw material. The paper states that the hyphal cell walls contain polysaccharides including chitin and α/β-glucans plus glycosylated proteins; in the engineered sensing version, melanin is additionally produced as a tunable pigment. This best supports a polymer-dominated matrix.

- **Geometry:**
  The final construct is a large-area, self-supporting, flexible living membrane obtained by filtering pellet slurry and drying it on a glass slide. Although the fungus itself forms a 3D hyphal network, the ELM described and tested is a membrane / sheet-like material, which best matches the ontology's 2D phylum. The figures on pages 4–6 visually show a flat membrane and its tunable color response.

- **Applications:**
  The directly demonstrated application is a xylose-responsive colorimetric biosensor for detecting xylose concentration in papermaking / industrial wastewater, which matches BSE and ENV. The article also mentions fabrics and packaging as future opportunities, but those are not as directly validated as the wastewater-sensing use case.

- **Production method(s):**
  The material is produced by genetically programming *A. niger* to control melanin synthesis, culturing the fungus to form mycelial pellets / hyphal biomass, then vacuum filtration, washing, and ambient drying to form the membrane. Ontology-wise, this is best captured by ECP+EMG: engineered cells are used to produce the material, and the extracellular/cell-associated matrix is grown in culture. The biological–abiotic integration mechanism is minimal: this is a bio-ELM made from fungal cells and their own matrix, without a synthetic scaffold.

- **Concept description:**
  This work describes a fungal engineered living material based on programmable *Aspergillus niger* mycelial pellets that are processed into self-supporting living membranes. The fungus provides both the living cellular component and the structural matrix, yielding a scaffold-free bio-ELM with a polymer-rich hyphal architecture. By genetically tuning melanin biosynthesis through a xylose-inducible circuit, the membrane acquires a visible, concentration-dependent color response that enables biosensing of xylose in wastewater. The material remains viable, self-regenerative, and functional after storage, making it a robust example of a responsive fungal ELM platform.

- **Rationales for ambiguous slots (if any):**
  - **2D rather than 3D:** the organism forms a 3D hyphal network, but the actual ELM decoded in the paper is consistently presented as a membrane / sheet-like material, so the final construct is best coded as 2D.
  - **LPO rather than LCO:** the described matrix is dominated by chitin, glucans, glycoproteins, and melanin-associated biopolymers rather than a conventional composite with a separate abiotic phase, so living polymers is the best-supported class.

- **BSE+ENV rather than broader application combinations:** the paper directly demonstrates sensing of xylose concentration in wastewater, while fabrics and packaging are presented as future possibilities; therefore biosensing plus environmental monitoring is the most defensible application code.
- **ECP+EMG rather than adding further production codes:** the central ontology-supported process is genetic engineering of the fungal chassis plus growth of the material-forming matrix in culture; filtration and drying are experimentally important but are not more specifically resolved by a better-fitting ontology code than these two core categories.

**Supplementary material S5.1: Generative AI response to prompt C of use case 2 "self-sustaining buildings in desert environments"**

Andrés Díaz Lantada [1,2*], José A. Yáñez [1],

William Solórzano-Requejo [1], Monsur Islam [1]

[1] Mechanical Engineering Department, Universidad Politécnica de Madrid, c/ José Gutiérrez Abascal 2, 28006 Madrid, Spain.

[2] IMDEA Materials Institute, c/ Eric Kandel 2, 28906 Getafe, Spain.

*Contact: andres.diaz@upm.es / andres.diaz@imdeamaterials.org

## Iteration 1

Below, I treat novelty only on the first five slots
Domain–Superkingdom–Kingdom–Phylum–Class, and I complete each hypothesis with Application and Production method only as compatible enablers.
I write the 7-slot codes with dashes only for readability.
All retained hypotheses were screened against your desert-building constraints: contained biology, desert-stress tolerance or realistic shielding, construction compatibility, field repairability, and low maintenance.

**Retained ELM hypotheses**

**1) Mineralizing repair / sand-stabilization bioceramic**
Gap-defining 5-slot combination:
BAC–BIO–EUB–FR–LCE
Full 7-slot hypothesis:
BAC–BIO–EUB–FR–LCE–SLB–BFF+BMN
Concept description
A branched bacterial mineralizing bioceramic layer that grows inside cracks or in near-surface sand networks to seal defects and stabilize wind-erodible ground around desert buildings.
Why this is a strong application-conditioned gap

- Biological plausibility
  Spore-forming or mineralizing eubacteria are compatible with intermittent activation, dormancy, and reactivation after rare moisture inputs, which fits desert operation better than continuously metabolically active systems.
- Structural / material plausibility
  The FR phylum matches irregular crack paths and branched pore/sand networks, while LCE is coherent with mineral deposition forming a stiff ceramic-like repair phase rather than a soft biofilm only.
- Technological feasibility and practical implications
  This can be integrated as a low-maintenance repair grout, dry powder patch, or perimeter stabilization treatment activated by recycled water or brief wetting events; it is construction-compatible and field-repairable without complex hardware.

---

**2) Sealed archaeal solar-modulating living film**
Gap-defining 5-slot combination:
ARC–HYB–ARA–2D–LPO

Full 7-slot hypothesis:
ARC–HYB–ARA–2D–LPO–SLS–BTI+OPM
Concept description
A thin, sealed archaeal living film inside a transparent polymer laminate that modulates incoming light and thermal load on façades or louvers under extreme heat and radiation.
Why this is a strong application-conditioned gap

- Biological plausibility
  Halophilic archaea are among the most naturally compatible cells for high salt, high irradiation, and high temperature environments, especially when maintained in contained saline microchambers.
- Structural / material plausibility
  The 2D phylum fits a continuous interfacial film, and LPO is coherent with a polymeric laminate or membrane architecture that also provides biological containment and dust shielding.
- Technological feasibility and practical implications
  The concept is realistic as a replaceable façade insert or shutter cartridge rather than an exposed coating; this reduces abrasion risk, simplifies containment, and supports low-maintenance building integration.

**3) Protist-based living ceramic membrane for greywater polishing and cooling**
Gap-defining 5-slot combination:
EUK–HYB–PRO–2D–LCE
Full 7-slot hypothesis:
EUK–HYB–PRO–2D–LCE–SLB–BTI+OPM
Concept description
A contained protist-derived living ceramic membrane, for example using silicifying diatom-like cells on a supported thin layer, for greywater polishing coupled to evaporative or radiative cooling subsystems.
Why this is a strong application-conditioned gap

- Biological plausibility
  Protists are a plausible choice for wet, shielded recirculating loops where biological activity is desired for water treatment or nutrient uptake, while remaining physically separated from the external desert environment.
- Structural / material plausibility
  The 2D phylum matches a membrane or surface-active treatment layer, and LCE is coherent with silica-rich or mineral-associated architectures rather than bulk soft biomass.
- Technological feasibility and practical implications
  This is better suited to enclosed utility cartridges than exposed façades; as a module in a water-recycling loop it is scalable, serviceable, and compatible with building systems that already require periodic filter replacement.

**4) Filamentous fungal mineralized repair tape**
Gap-defining 5-slot combination:
EUK–HYB–FUN–1D–LCE
Full 7-slot hypothesis:
EUK–HYB–FUN–1D–LCE–SLB–FBT+BMN
Concept description
A fiber-based fungal living ceramic tape or liner that bridges small cracks, reinforces joints, and triggers localized mineral hardening inside sheltered construction interfaces.
Why this is a strong application-conditioned gap

- Biological plausibility
  Filamentous fungi are naturally compatible with 1D growth patterns through hyphae and can colonize fibrous supports using limited nutrient reservoirs in a controlled, immobilized format.

- Structural / material plausibility
  The 1D phylum fits woven or nonwoven reinforcement geometries, and LCE is justified when the fungal network is used as a template or mediator for mineral stiffening rather than only soft polymeric mass formation.
- Technological feasibility and practical implications
  This is attractive as a prefabricated repair consumable: a roll, strip, or insert that can be stored dry, applied locally, activated on demand, and replaced in the field without specialized bioprocess equipment.

**5) Cross-domain fiber cartridge for dust capture and local $CO_2$ conditioning**
Gap-defining 5-slot combination:
CRD–HYB–CRK–1D–LPO
Full 7-slot hypothesis:
CRD–HYB–CRK–1D–LPO–SLB–FBT+OPM
Concept description
A replaceable cross-domain fiber cartridge, for example a fungal–cyanobacterial or fungal–microalgal capillary network, that traps dust, buffers humidity locally, and contributes to localized $CO_2$ control in air-intake or courtyard ventilation systems.
Why this is a strong application-conditioned gap

- Biological plausibility
  Cross-domain consortia are plausible where one partner provides structural colonization and the other contributes gas exchange or nutrient cycling, provided the system is immobilized and physically contained in a sheltered cartridge.
- Structural / material plausibility
  The 1D phylum fits capillary fibers and woven filter media, while LPO is coherent with hydrogel-coated or polymer-supported fibrous architectures designed for air handling rather than load-bearing structure.
- Technological feasibility and practical implications
  This is much more credible as a serviceable building subsystem than as an exposed façade skin: it can be swapped like a filter, isolated from occupants, protected from wind abrasion, and integrated into low-pressure ventilation hardware.

**6) Longer-horizon, high-containment synthetic-cell water-recycling cartridge**
Gap-defining 5-slot combination:
SDM–HYB–SKM–3D–LPO
Full 7-slot hypothesis:
SDM–HYB–SKM–3D–LPO–BRE–3DB+OPM
Concept description
A 3D polymeric cartridge populated by synthetic or artificial cells for tightly contained greywater conditioning, nutrient capture, and fail-safe biochemical response inside a closed building utility loop.
Why this is a strong application-conditioned gap

- Biological plausibility
  The ontology explicitly admits synthetic-cell ELMs, and this use case benefits from their strongest practical advantage here: high containment and potentially reduced ecological escape risk compared with open living consortia.
- Structural / material plausibility
  The 3D phylum suits porous cartridge volumes, and LPO is coherent with hydrogel or polymer-host architectures needed to immobilize artificial cells and manage mass transport.
- Technological feasibility and practical implications
  This is a plausible in-principle hypothesis but clearly lower readiness than the others; it fits best in closed service hardware where maintenance, replacement, and containment can be engineered conservatively.

Screening decisions: gaps I did not retain

I excluded several ontology-valid regions from the final shortlist because they fit the ontology but fit your boundary conditions poorly.

Rejected pattern A: exposed plant-dominant living skins

I did not prioritize open PLN façade or roof systems that require continuous hydration and direct environmental exposure. Even if ontology-valid, they are hard to reconcile with scarce freshwater, low maintenance, and desert abrasion unless moved into highly shielded modules, at which point other kingdoms become more practical.

Rejected pattern B: highly exposed 0D particulate ELMs

I did not prioritize 0D candidates for desert exteriors because wind-driven sand and dust make particle loss, abrasion, and containment more problematic. They may still be useful inside sealed cartridges, but they are less compelling than 1D, 2D, FR, or 3D architectures for building subsystems.

Rejected pattern C: open 4D morphing concepts

I did not prioritize 4D building-envelope hypotheses because shape-morphing living systems add control and maintenance burdens that are difficult to justify under desert field conditions unless the system is very well protected and modular.

Overall recommendation

For near- to mid-term discovery, the strongest directions are:

1. BAC–BIO–EUB–FR–LCE for self-healing and sand stabilization
2. ARC–HYB–ARA–2D–LPO for solar/thermal façade control under extreme radiation
3. CRD–HYB–CRK–1D–LPO for serviceable dust and air-management cartridges

These three best satisfy your combined requirements of containment, desert robustness, construction compatibility, and low-maintenance deployment while still occupying genuinely weakly explored ontology regions.

**Iteration 2**

Below I treat the ontology as the fixed design space.
Novelty is assessed only on the first five slots: Domain–Superkingdom–Kingdom–Phylum–Class.
Application and Production method are appended only to complete each hypothesis and keep it compatible with the desert-building context.
I prioritized weakly explored ontology regions that are still plausible for: heat, UV, desiccation, dust abrasion, scarce freshwater, containment, construction compatibility, and field repair.

**1) Bacterial adaptive living-ceramic façade insert**
Gap status: Absent in the graph at the 5-slot level: BAC–HYB–EUB–4D–LCE
Core ontology code string (7-slot)
BAC–HYB–EUB–4D–LCE–SLB+SLS–4DP+BMN
Concept description
A sealed mineralized facade insert containing nonpathogenic spore-forming bacteria in protected pores, designed to undergo useful day–night shape change while self-sealing microcracks after rare wetting events.
Plausibility justification

- Biological plausibility
  Spore-forming eubacteria are compatible with long dry periods, intermittent activation, and protected pore habitats; the living fraction can remain dormant under desert stress and revive after dew, maintenance wetting, or recycled-water exposure.
- Structural / material plausibility
  4D is appropriate because thermal cycling is functionally central, and LCE is coherent with a mineral-dominated facade insert intended for crack resistance, UV shielding, and abrasive dust exposure.
- Technological feasibility (non-prescriptive)
  A thermally responsive porous ceramic bilayer with embedded bacterial niches is feasible in principle using scaffold-based shaping plus mineralization; modular inserts are construction-compatible, replaceable, and field-repairable.

Decision: Accepted, because it directly addresses passive cooling and structural self-healing while keeping the biology shielded and contained.

---

**2) Cross-domain lichen-like living-ceramic dust and $CO_2$ skin**
Gap status: Absent in the graph at the 5-slot level: CRD–HYB–CRK–FR–LCE
Core ontology code string (7-slot)
CRD–HYB–CRK–FR–LCE–SLB+ENV–SBP+BMN
Concept description
A lichen-inspired ramified cyanobacteria–fungal mineralized skin grown within a porous ceramic screen to intercept dust, locally polish $CO_2$, and slowly consolidate eroding building surfaces.
Plausibility justification

- Biological plausibility
  A cyanobacteria–fungus consortium is cell-centric and biologically credible for a cross-domain ELM; both partners can tolerate repeated desiccation better than many hydrated cell systems when held in shaded pore networks and supplied only intermittently.
- Structural / material plausibility
  FR fits a ramified, branching colonization pattern, and LCE fits a mineral-rich crust or ceramic-supported shell intended to face wind, UV, and abrasive particles.
- Technological feasibility (non-prescriptive)
  Patterned inoculation of porous ceramic screens and subsequent mineral consolidation are feasible in principle; the format is scalable as replaceable facade or wind-screen modules rather than as free-growing environmental release.

Decision: Accepted, because it supports dust filtration, localized air control, and surface self-stabilization with a contained, construction-compatible architecture.

**3) Fungal ramified living-composite sand-stabilizing block**

Gap status: Absent in the graph at the 5-slot level: EUK–HYB–FUN–FR–LCO

Core ontology code string (7-slot)

EUK–HYB–FUN–FR–LCO–SLB–CMM

Concept description

A mineral-armored mycelial composite block or berm panel in which a ramified fungal network binds granular fillers, bridges cracks, and stabilizes sand around foundations or shaded envelope zones.

Plausibility justification

- Biological plausibility
  Filamentous fungi naturally generate persistent ramified networks, can transition to low-activity or dormant states in dry conditions, and are compatible with contained growth inside composite interiors using safe industrial strains.
- Structural / material plausibility
  FR matches a branching mycelial network, while LCO is coherent with a sand/mineral/fiber-filled composite in which the fungal network acts as living reinforcement rather than as a soft hydrated gel.
- Technological feasibility (non-prescriptive)
  Mycelium-based composites are already plausible as low-temperature construction materials; adding mineral armor and modular geometry is a realistic extension for manufacturability, transport, and patch repair.

Decision: Accepted, because it contributes structural self-healing, insulation, and soil/sand stabilization with relatively low water demand after maturation.

**4) Archaeal living-ceramic greywater reactor cartridge**

Gap status: Absent in the graph at the 5-slot level: ARC–HYB–ARA–3D–LCE

Core ontology code string (7-slot)

ARC–HYB–ARA–3D–LCE–SLB+BRE–CMM

Concept description

A hot, saline greywater-treatment cartridge built from porous ceramic modules colonized by nonpathogenic haloarchaea, intended for water polishing and nutrient cycling in desert buildings.

Plausibility justification

- Biological plausibility
  Haloarchaea are unusually compatible with high salt, heat, and low water activity, making them stronger candidates than most conventional cells for contained service-core treatment modules handling concentrated or brackish water streams.
- Structural / material plausibility
  3D matches a volumetric reactor cartridge, and LCE fits a ceramic host that tolerates heat, fouling control, and repeated wet–dry operation better than fragile hydrated matrices.
- Technological feasibility (non-prescriptive)
  Porous ceramic bioreactor cartridges are realistic in principle as modular building-service components; they are scalable, swappable, and easier to contain biologically than open environmental systems.

Decision: Accepted, because it directly supports water recycling and nutrient/resource cycling under desert service conditions while keeping the biology enclosed.

**5) Protist capillary living-composite energy-support ribbon**

Gap status: Absent in the graph at the 5-slot level: EUK–HYB–PRO–1D–LCO

Core ontology code string (7-slot)
EUK–HYB–PRO–1D–LCO–SLB+ENP–FBT
Concept description
A sealed capillary-fiber ribbon containing desert-tolerant microalgal protists, designed for facade-cavity integration to support $CO_2$ uptake, recycled-water conditioning, and low-grade photosynthetic energy support.
Plausibility justification

- Biological plausibility
  Protist microalgae are compatible with narrow illuminated capillaries and closed-loop water recirculation; with shielding and temperature control, they can remain viable without direct exposure to abrasive desert conditions.
- Structural / material plausibility
  1D fits capillary or fiber-like flow paths, and LCO fits a composite ribbon where the abiotic carrier provides mechanical protection while the cells occupy the fluidic core.
- Technological feasibility (non-prescriptive)
  Fiber- or capillary-based encapsulation is plausible for modular cartridges; the main practical route is replaceable service elements rather than exposed facade biofilms, which keeps maintenance manageable.

Decision: Accepted with caution, because it is useful for air/water/energy support but requires the strongest shielding and the most controlled water loop of the set.

---

**Deprioritized gap families**
I did not prioritize the following, despite ontology validity:

- Animal-cell ELM gaps: rejected for this use case because hydration, containment, sterility, and maintenance burdens are poorly aligned with desert buildings.
- Open, highly hydrated 2D biofilm/hydrogel concepts: rejected because low humidity, UV, and wind-driven abrasion make them hard to sustain without excessive shielding and water input.
- Synthetic-cell-led construction concepts: deprioritized because field repair, scale-up, and deployment readiness are weaker than for bacterial, fungal, protist, or archaeal routes.

These five hypotheses are the strongest ontology-valid gaps I would carry forward for first-pass experimental screening in the stated desert-building context.

**Iteration 3**

Below are five application-conditioned ELM hypotheses selected from absent or strongly underrepresented 5-slot regions that are compatible with self-sustaining buildings in desert environments.
I prioritized:

- contained hybrid systems rather than exposed bio-ELMs,
- extremotolerant or realistically shielded cells,
- ceramic/composite matrices compatible with abrasion, heat and low maintenance,
- and functions tied to water, passive cooling, dust control, self-healing, or sand stabilization.

I did not prioritize animal, synthetic-cell, or irrigation-intensive plant concepts, because they are harder to reconcile with desert containment, maintenance, and scaling constraints.

---

**1) Archaeal ceramic humidity-buffering façade module**
Gap status: absent on the 5-slot core
Core 5-slot gap: ARC–HYB–ARA–3D–LCE
7-slot hypothesis: ARC–HYB–ARA–3D–LCE–SLB–3DP
Concept description
A sealed porous ceramic façade cartridge colonized by halophilic archaea, designed to buffer humidity, tolerate intense UV, and contribute to passive thermal regulation in desert buildings.
Plausibility justification

- Biological plausibility
  Halophilic archaea such as *Haloferax volcanii* can tolerate high salinity, strong irradiation, and repeated dehydration–rehydration cycles; confinement in brine-filled microreservoirs also improves biological containment and reduces contamination risk.
- Structural / material plausibility
  A 3D living ceramic is coherent here because a porous aluminosilicate or geopolymer ceramic can provide thermal mass, capillary storage, abrasion resistance, and UV-stable housing for the cells.
- Technological feasibility (non-prescriptive)
  Porous ceramic modules are construction-compatible and scalable; post-fabrication inoculation of protected internal pore networks is feasible in principle, and the cartridge format supports replacement and low-maintenance operation.

---

**2) Fractal bacterial ceramic dew-harvesting skin**
Gap status: absent on the 5-slot core
Core 5-slot gap: BAC–HYB–EUB–FR–LCE
7-slot hypothesis: BAC–HYB–EUB–FR–LCE–SLB–BMN+3DP
Concept description
A mineralized fractal ceramic exterior skin containing extremotolerant bacteria, intended to nucleate dew, trap dust in sacrificial outer branches, and regenerate hydrophilic microtexture after intermittent wetting.
Plausibility justification

- Biological plausibility
  Desert-adapted cyanobacteria such as *Chroococcidiopsis* are unusually tolerant to UV and desiccation, can persist in protected microcavities, and are compatible with intermittent activation during rare moisture events.
- Structural / material plausibility
  A fractal living ceramic is well matched to water capture and dust management because branching, self-similar capillary geometries can increase condensation area, guide droplets, and localize abrasion without losing global function.

- Technological feasibility (non-prescriptive)
  Fractal ceramic surfaces can be built as patterned tiles or coatings; integrating bacteria into protected surface-connected porosity is feasible in principle, and the concept is compatible with modular façade maintenance rather than full-wall replacement.

**3) Bacterial living ceramic block for graywater polishing and crack sealing**
Gap status: strongly underrepresented on the 5-slot core
Core 5-slot gap: BAC–HYB–EUB–3D–LCE
7-slot hypothesis: BAC–HYB–EUB–3D–LCE–SLB+BRE–BMN+3DP
Concept description
A modular living ceramic service block containing biomineralizing bacteria for local graywater polishing, mineral retention, and autonomous sealing of microcracks in building service zones.
Plausibility justification

- Biological plausibility
  Non-pathogenic biomineralizing bacteria such as *Sporosarcina pasteurii* or spore-forming *Bacillus subtilis* are compatible with dormant survival during dry periods and reactivation when graywater flows through the module.
- Structural / material plausibility
  A 3D living ceramic is coherent because porous ceramic blocks provide filtration pathways, mechanical stability, thermal resistance, and a mineral-friendly environment for precipitation-based crack sealing.
- Technological feasibility (non-prescriptive)
  Service-wall cartridges, replaceable filter bricks, or distributed utility modules are realistic integration routes; the concept scales better than exposed living façades and fits desert buildings where water recycling and maintainability are critical.

**4) Cross-phylum fungal composite insulation panel**
Gap status: absent on the 5-slot core
Core 5-slot gap: EUK–HYB–FUN–CP–LCO
7-slot hypothesis: EUK–HYB–FUN–CP–LCO–SLB–FBT+CMM
Concept description
A multi-scale mycelium–fiber–mineral composite panel in which fungal growth binds a fibrous scaffold and porous core, providing passive cooling, thermal insulation, and local crack-bridging capacity.
Plausibility justification

- Biological plausibility
  Fungi already used in mycelium materials, such as *Pleurotus ostreatus*, can colonize fibrous porous substrates efficiently; in desert use, the biologically active region can remain protected inside the panel and operate intermittently rather than continuously.
- Structural / material plausibility
  CP is justified because the system combines 1D hyphal networks with a 3D porous composite body; LCO is also coherent because the panel can combine organic fibers, mineral fillers, and fungal extracellular binding phases.
- Technological feasibility (non-prescriptive)
  Panel-scale manufacturing is compatible with existing insulation and sandwich-panel logic; the concept is field-repairable through modular replacement and more scalable than free-form living masonry.

**5) Cross-kingdom bacterial–fungal self-healing render for sand stabilization**
Gap status: absent on the 5-slot core
Core 5-slot gap: CRD–HYB–CRK–CP–LCO
7-slot hypothesis: CRD–HYB–CRK–CP–LCO–SLB–CMM+BMN

Concept description
A protected bacterial–fungal living render in which bacteria mineralize and fungi reinforce the matrix, enabling self-healing, dust suppression, and stabilization of sand-exposed exterior surfaces.
Plausibility justification

- Biological plausibility
  A cross-kingdom pairing such as *Sporosarcina pasteurii* with *Aureobasidium pullulans* is cell-centrically coherent: the bacterium contributes mineral precipitation, while the fungus contributes surface adhesion, extracellular binding, and crack-spanning growth.
- Structural / material plausibility
  CP–LCO is appropriate because the render is inherently multi-scale: bacterial microcolonies, fungal filament networks, and a bulk mineral–polymer composite matrix act together rather than as a simple single-phase material.
- Technological feasibility (non-prescriptive)
  The concept is realistic only as a contained or semi-contained render layer with controlled nutrient access; that constraint is acceptable in desert buildings, where serviceable façade skins and sacrificial outer layers are already practical design strategies.

---

**Final selection note**
These five hypotheses were prioritized because they balance:

- ontology-valid novelty in the first five slots,
- desert compatibility,
- containment and safety,
- construction-scale integration,
- and low-maintenance operation.

Among them, the most immediately practical for desert buildings are:

- BAC–HYB–EUB–3D–LCE–SLB+BRE–BMN+3DP
  for water recycling and self-sealing service blocks,
- EUK–HYB–FUN–CP–LCO–SLB–FBT+CMM
  for insulation and passive cooling panels,
- CRD–HYB–CRK–CP–LCO–SLB–CMM+BMN
  for self-healing and sand-stabilizing exterior skins.

**Supplementary material S5.2: Generative AI response to prompt E of use case 2 "self-sustaining buildings in desert environments"**

# CAN WE TEACH GENERATIVE ARTIFICIAL INTELLIGENCE THE DESIGN LANGUAGE OF ENGINEERED LIVING MATERIALS?

Andrés Díaz Lantada [1,2*], José A. Yáñez [1],

William Solórzano-Requejo [1], Monsur Islam [1]

[1] Mechanical Engineering Department, Universidad Politécnica de Madrid, c/ José Gutiérrez Abascal 2, 28006 Madrid, Spain.

[2] IMDEA Materials Institute, c/ Eric Kandel 2, 28906 Getafe, Spain.

*Contact: andres.diaz@upm.es / andres.diaz@imdeamaterials.org

**ELM candidate 1: BAC–BIO–EUB–FR–LCE–SLB–BFF+BMN**

**Iteration 1**

**0) Code integrity check**

Code provided: BAC–BIO–EUB–FR–LCE–SLB–BFF+BMN

Ontology validity: Valid as a 7-slot ELM code with a combined production slot. In the ontology, BAC is bacterial, BIO is biological engineered living material, EUB is eubacterial, FR is fractal, LCE is living ceramics, SLB is smart and living buildings, and BFF+BMN is a valid combined production description because the ontology explicitly allows category combinations with "+".

Ambiguity and resolution:

There is one relevant tension: BFF is described as biofilm-based trapping or anchoring of inorganic particles and is noted as usually leading to living composites, mainly HLMs, whereas BIO excludes supporting synthetic biomaterials or synthetic structures. I resolve this without changing the code by interpreting the inorganic phase as cell-generated or cell-localized mineral matter formed in situ and retained within the bacterial biofilm/extracellular matrix, not as an externally imposed synthetic scaffold. That interpretation remains consistent with BIO, LCE, and BMN.

Graph support:

The exact full 8-code combination was not present in the graph when checked directly. The graph does contain nearby bacterial mineralizing building-related precedents, including crack-healing calcite-precipitating concrete and soil-biocementation ELMs, but not this exact FR + BFF full hypothesis. Therefore, this candidate is ontology-valid and graph-adjacent, but not present in the graph as an exact instantiation.

**1) Slot-by-slot decoding**

Domain — BAC

Meaning: Bacterial ELMs.

Design constraints

- Biological constraints: the active component must be bacterial cells.
- Material constraints: extracellular matrix and mineral phase must be bacterial-derived or bacterial-mediated.

- Structural/geometric constraints: architecture must support bacterial colonization, dormancy, reactivation, and local mineral precipitation.
- Process/production constraints: cultivation, sporulation, deposition, and activation steps must be compatible with bacteria rather than eukaryotic tissue culture.

Super-kingdom — BIO
Meaning: Biological engineered living materials made only of cells and their extracellular matrices, without supporting synthetic biomaterials or synthetic structures.
Design constraints

- Biological constraints: cells and their ECM must supply the living component and functional matrix.
- Material constraints: no permanent synthetic scaffold can define the ELM.
- Structural/geometric constraints: the structural phase must emerge from biofilm/ECM/mineral accretion in situ.
- Process/production constraints: formulation may include temporary nutrients or salts, but the final functional material cannot rely on a synthetic support architecture.

Kingdom — EUB
Meaning: Eubacterial ELMs.
Design constraints

- Biological constraints: candidate organisms must be true bacteria, not archaea or fungi.
- Material constraints: mineral nucleation pathways should be consistent with known bacterial biomineralization or biofilm-mediated deposition.
- Structural/geometric constraints: bacterial biofilms and microcolonies must be the operative cell organization.
- Process/production constraints: standard bacterial culture, sporulation, and inoculation workflows are appropriate.

Phylum — FR
Meaning: Fractal ELMs with irregular, fractured, or self-similar structures that cannot be defined by integer dimensional values.
Design constraints

- Biological constraints: cells must remain viable in irregular crack and pore networks with heterogeneous nutrient and moisture fields.
- Material constraints: the mineralizing matrix must conform to branched, tortuous, multiscale voids.
- Structural/geometric constraints: the target architecture is not a uniform slab; it is a crack network or sand-pore branching topology.
- Process/production constraints: infiltration and growth must occur through capillary-driven or moisture-triggered percolation rather than mold-defined shaping.

Class — LCE
Meaning: Living ceramics; extracellular matrices made predominantly of ceramics, including oxides, silicates, carbides, and nitrides.
Design constraints

- Biological constraints: cells must drive ceramic-like mineral deposition and tolerate the local alkalinity and ionic environment.
- Material constraints: the final dominant non-cellular phase should be mineral/ceramic, not polymer-dominant.
- Structural/geometric constraints: the mineral must stiffen and bridge defects while preserving at least limited living niches.
- Process/production constraints: mineralizing chemistries must be compatible with bacterial viability during formation and dormancy after curing.

Application — SLB
Meaning: Smart and living buildings.
Design constraints

- Biological constraints: the system must survive building-envelope conditions such as desiccation, temperature swings, UV exposure, and intermittent wetting.

- Material constraints: the ELM must bond to masonry-like crack walls or sand grains and contribute to stabilization.
- Structural/geometric constraints: it must function in near-surface crack networks, façade defects, or soil/building interfaces.
- Process/production constraints: deployment must be field-compatible, low-maintenance, and implementable as patch, grout, or infiltrant.

Production — BFF
Meaning: Biofilm-based trapping or anchoring of inorganic particles/materials and biofilms interfacing with inorganic materials.
Design constraints

- Biological constraints: organisms must produce robust ECM/biofilm capable of adhesion to mineral surfaces.
- Material constraints: trapped or anchored inorganic matter must become integrated into the living ceramic phase.
- Structural/geometric constraints: biofilm must bridge crack walls and bind granular substrates.
- Process/production constraints: colonization must start on an inorganic substrate such as masonry fines, silica-rich sand, or crack-wall minerals.

Production — BMN
Meaning: Biomineralization.
Design constraints

- Biological constraints: cells must actively induce mineral nucleation and growth.
- Material constraints: mineral phase should be calcium carbonate or a related ceramic deposition consistent with bacterial pathways.
- Structural/geometric constraints: mineral deposition must occur throughout the branched defect network, not only at the exposed surface.
- Process/production constraints: ion supply, alkalinity generation, nucleation control, and moisture pulses must be part of the workflow.

---

**2) Biological instantiation**
Candidate 1 — *Sporosarcina pasteurii*
Functional role: ureolytic biomineralization; rapid $CaCO_3$ precipitation for crack sealing and sand biocementation; biofilm-assisted mineral anchoring.
Viability and growth requirements

- Oxygen: aerobic to facultatively tolerant field conditions.
- Nutrients: low-complexity medium; urea and calcium source during activation.
- Temperature: practical activity in warm built environments.
- pH: tolerates alkaline microenvironments generated during ureolysis.
- Moisture: intermittent wetting compatible; can be delivered as spores.
- Mechanical stress tolerance: suitable for immobilization in pores and crack walls.

Biocontainment strategy

- Spore-based dormant formulation.
- Nutrient-limited activation only after wetting.
- Encapsulation in a temporary mineral nutrient granule that dissolves on activation.
- No continuous feed, reducing uncontrolled spread.

Why preferred

- Directly aligned with mineralizing repair and sand stabilization.
- Strong precedent for biocementation and crack repair.
- Spore formation suits desert cyclic wetting.

Candidate 2 — *Bacillus subtilis* mineralizing strain
Functional role: biofilm production, surface adhesion, and biomineralization support through alkaline metabolism and ECM-mediated nucleation.
Viability and growth requirements

- Oxygen: aerobic.
- Nutrients: simple carbon/nitrogen input.

- Temperature: robust across fluctuating ambient conditions.
- pH: tolerates alkaline ranges better than many non-spore-formers.
- Moisture: survives desiccation better than many vegetative bacteria due to sporulation capacity.
- Mechanical stress tolerance: strong biofilm former on mineral substrates.

Biocontainment strategy

- Sporulation-controlled dormant inoculum.
- Nutrient-limited one-shot activation.
- Optional auxotrophic laboratory strain for pilot deployment.

Why preferred / not preferred

- Better biofilm architecture than many ureolytic specialists.
- Less direct and less efficient mineralization than *S. pasteurii* for fast ceramic infill.

Recommended candidate

Recommended organism: *Sporosarcina pasteurii*

Justification against alternatives

- Compared with *B. subtilis*, it is more directly specialized for high-yield calcite precipitation and therefore better matches LCE and BMN.
- Its sporulation-compatible lifecycle better fits rare moisture pulses and long dry intervals.
- It is more efficient for achieving measurable defect filling per activation cycle.

Trade-offs

- Ureolytic systems can generate ammonium as a by-product, requiring dosage control.
- *B. subtilis* offers stronger native biofilm versatility, but lower mineral productivity for the same activation event.

---

**3) Matrix/material candidates**

Candidate 1 — In situ bacterially precipitated calcite ($CaCO_3$) with bacterial ECM

Exact composition: calcite-dominant calcium carbonate mineral phase integrated with bacterial extracellular polymeric substances.

Compatibility with biology

- Direct product of bacterial biomineralization.
- ECM provides nucleation loci and adhesion to crack walls and sand grains.

Interface mechanisms

- Electrostatic binding of $Ca^{2+}$ to EPS.
- Nucleation on cell walls and extracellular polymers.
- Mineral bridge formation between grains and crack faces.

Why selected

- Strongest match to LCE and BMN.
- Compatible with BIO because the ceramic phase is biologically generated in situ.

Candidate 2 — $CaCO_3$–silica microcomposite formed in situ from local sand fines plus bacterial biofilm

Exact composition: bacterial EPS/mineral network integrating precipitated $CaCO_3$ with native silica-rich fines or sand particles already present in the defect environment.

Compatibility with biology

- Bacteria colonize silica-rich granular substrates well.
- Sand provides abundant inert mineral interfaces for attachment.

Interface mechanisms

- Biofilm adhesion to silicate surfaces.
- EPS-mediated particle trapping.
- Calcite overgrowth that cements adjacent grains.

Why selected

- Better for sand stabilization than pure crack filling.
- Preserves BIO because the inorganic granular substrate is the deployment environment, not a synthetic scaffold.

Preferred material system

Recommended material: calcite-dominant $CaCO_3$/ECM ceramic phase with optional incorporation of native silica fines where present.

Justification relative to alternatives

- Pure silica gel, polymer grout, or externally added hydrogel would move the concept toward HYB or away from LCE.
- A calcite-dominant system most faithfully matches a bacterial living ceramic for construction repair.
- Environmental silica is useful as a local substrate, but should not dominate the matrix if LCE is to remain the primary class.

**4) Geometry and structural instantiation**

Phylum-derived architecture: multiscale fractal mineralizing network occupying branched cracks and near-surface sand pore paths.

Quantitative geometry

- Primary crack aperture target: ~50 µm to 2 mm.
- Secondary branch penetration: ~10 µm to 500 µm.
- Mineralized bridge thickness: ~20–200 µm per activation cycle, accumulating over repeated wetting events.
- Sand pore-scale binding domain: grain-to-grain contacts in the ~50–500 µm scale for fine-to-medium sand.
- Colonized depth: first 1–10 mm from the exposed surface for realistic passive infiltration.

How geometry supports function

- Cell viability and function: irregular branching retains protected microniches with higher local humidity than the exposed exterior.
- Mass transport and diffusion: capillary infiltration during wetting distributes ions and dormant spores through tortuous pathways.
- Mechanical performance: mineral bridges across hierarchical branches reduce crack connectivity and increase near-surface soil cohesion.

Why this structure is optimal

A planar film would not reach defect depth; a bulk 3D scaffold would violate BIO; a simple linear geometry would not match crack branching or sand pore networks. The fractal architecture is the most faithful translation of FR because sealing performance depends on occupying irregular multiscale pathways rather than forming a uniform shell.

**5) Application instantiation**

Deployment scenario: a dry powder or concentrated spore-salt treatment is brushed, injected, or sprayed onto cracked earthen/mineral building surfaces or perimeter desert sands adjacent to foundations. Rare wetting events or controlled recycled-water activation trigger bacterial germination, biofilm formation, and ceramic precipitation within near-surface branched defects.

Measurable success metrics

- Crack sealing efficiency: percentage reduction in connected crack volume.
- Air/water permeability reduction: before/after permeability of treated zones.
- Surface erosion resistance: reduction in wind-driven particle loss.
- Unconfined compressive or penetrometer gain: near-surface strength increase in treated sand.
- Mineral deposition rate: mass or volume of $CaCO_3$ formed per activation cycle.
- Durability after dry/wet cycles: retained sealing after repeated desert weathering.

Main failure modes and mitigation strategies

- Failure: sealing limited to the surface crust.
  Mitigation: low-viscosity infiltrant and repeated short wetting pulses.
- Failure: bacterial death from desiccation/UV.
  Mitigation: spore formulation and shaded/subsurface targeting.

- Failure: salt efflorescence or pore clogging too early.
  Mitigation: staged ion dosing and lower initial calcium loading.
- Failure: ammonia-related by-products from ureolysis.
  Mitigation: controlled dosage, drainage management, and pilot-scale environmental monitoring.

**6) Production instantiation**

Step-by-step production process

1. Material preparation
   - Prepare a dry mineralizing formulation containing urea, calcium salt, buffering salts, and optional micronized limestone seed crystals.
   - Adjust particle size so the powder can enter fine cracks and coat sand grains.
   - Keep no permanent synthetic binder in the final formulation to preserve BIO.
2. Biological component preparation
   - Produce a spore-rich culture of *Sporosarcina pasteurii*.
   - Wash and concentrate spores.
   - Blend spores with the dry mineralizing salts under low-moisture conditions.
3. Integration strategy
   - Apply as dry powder patch, low-viscosity aqueous infiltrant, or sprayable slurry for immediate infiltration.
   - Upon wetting, spores germinate locally.
   - Bacteria attach to mineral surfaces, form biofilm, and initiate biomineralization.
   - Biofilm traps fines and anchors the growing mineral phase to the inorganic substrate.
4. Post-processing and stabilization
   - Use one or more brief wetting cycles with recycled water to trigger deposition.
   - Allow passive drying to return the system to a dormant, stabilized state.
   - Avoid sterilization after biological integration because viability is required.
   - Store the pre-application product dry and cool for shelf stability.

Manufacturability
- Spore production is scalable using conventional fermentation.
- Dry blending and powder packaging are construction-compatible.
- Field application is compatible with spray, brush, or crack injection methods.

Scalability
- Suited to batch production in sacks/cartridges.
- Can be deployed over large areas using surface spray or perimeter trench treatment.
- Reactivation by natural or scheduled wetting reduces maintenance complexity.

Quality control and reproducibility
- Spore count per gram.
- Urease activity assay after rehydration.
- Standardized $CaCO_3$ precipitation test.
- Crack penetration test in representative masonry coupons.
- Erosion resistance test in treated sand beds.

Why this route is feasible and preferred

It satisfies BFF+BMN directly: bacteria form biofilm on inorganic building/sand substrates while driving biomineralization. It also preserves BIO because the functional architecture emerges from cells, ECM, and mineral formed in situ rather than from a synthetic scaffold.

**7) Layered integrated summary of the fully instantiated ontology-derived ELM**

A. Rapid identity

Ontology code: BAC–BIO–EUB–FR–LCE–SLB–BFF+BMN

One-line ELM identity:

Spore-forming bacterial living ceramic that biomineralizes branched crack and sand-pore networks to seal defects and stabilize desert building interfaces.

B. Instantiation table

| Component | Final instantiation |
|---|---|
| Biological system | *Sporosarcina pasteurii* spores as the primary mineralizing eubacterial system; alternative candidate: *Bacillus subtilis* mineralizing biofilm-forming strain |
| Functional role | Moisture-triggered germination, biofilm formation, $CaCO_3$ biomineralization, crack bridging, and grain-to-grain biocementation |
| Material system | Calcite-dominant living ceramic composed of bacterially precipitated $CaCO_3$ integrated with bacterial ECM, optionally incorporating native silica-rich fines from the treated substrate |
| Structural / geometric instantiation | Fractal multiscale network filling irregular crack branches and near-surface sand pores, with mineral bridges spanning ~20–200 µm per activation cycle across defect pathways |
| Target substrate / environment | Cracked earthen or mineral building envelopes, façade defects, and wind-erodible desert perimeter sands near foundations |
| Application / deployment scenario | Dry powder, slurry, or infiltrant applied to defected building/sand interfaces and activated by brief wetting or rare moisture events for autonomous ceramic repair and stabilization |
| Production mechanism / workflow | Spore fermentation, dry blending with mineralization salts, field infiltration onto inorganic substrates, biofilm anchoring, and in situ biomineralization during wet-dry activation cycles |
| Main functional outcome / performance objective | Reduced crack connectivity and permeability, increased near-surface cohesion and erosion resistance, and persistent low-maintenance repair under intermittent desert moisture |

C. Integrated paragraph

BAC–BIO–EUB–FR–LCE–SLB–BFF+BMN defines a bacterial living ceramic for defect sealing and ground stabilization in smart and living buildings. The recommended biological system is spore-forming *Sporosarcina pasteurii*, selected for high-yield calcite precipitation, dormancy, and reactivation after intermittent wetting. The material phase is a calcite-dominant $CaCO_3$/ECM ceramic, optionally integrating native silica fines at the treated interface. The FR slot is instantiated as a multiscale branched network occupying cracks and near-surface sand pores. Deployment consists of applying a dry or infiltrating spore-mineralization formulation to building or perimeter sand defects, where BFF+BMN is realized by substrate-adherent biofilm growth and in situ biomineralization, yielding permeability reduction, crack closure, and erosion-resistant biocementation.

**Iteration 2**

**0) Code integrity check**
Code: BAC–BIO–EUB–FR–LCE–SLB–BFF+BMN
The 7-slot code is ontology-valid. The ontology explicitly defines BAC as bacterial ELMs, BIO as biological engineered living materials made only of cells and their extracellular matrices, EUB as eubacterial, FR as fractal ELMs, LCE as living ceramics, SLB as smart and living buildings, BFF as biofilm-based trapping/anchoring of inorganic particles or interfacing with inorganic materials, and BMN as biomineralization.
Interpretive note on ambiguity:
There is a real ontological tension between BIO and BFF. BIO excludes supporting synthetic biomaterials or synthetic structures, whereas BFF is described as usually involving biofilms trapping/anchoring inorganic particles or interfacing with inorganic materials, often leading mainly to HLMs. The most coherent reading here is therefore: the system remains bio-ELM because the dominant material phase is generated and organized by the bacterial biofilm/extracellular matrix itself, while inorganic precursors are transient feedstocks or environmentally available particulates incorporated into a biologically produced ceramic extracellular phase rather than a pre-fabricated scaffold.

**1) Slot-by-slot decoding**
Domain — BAC
Meaning: Bacterial ELMs.
Design constraints
- Biological constraints: the living component must be bacterial, with prokaryotic physiology, rapid environmental response, and tolerance to desiccation-compatible cycles preferred for the stated desert-building use case.
- Material constraints: the extracellular matrix must be bacterially generated or bacterially organized.
- Structural/geometric constraints: architecture must be achievable through bacterial colonization, biofilm growth, and mineral deposition.
- Process/production constraints: culture, sporulation, and reactivation workflows must be bacterial-process compatible.

Super-kingdom — BIO
Meaning: Biological engineered living materials made only of cells and their extracellular matrices, without supporting synthetic biomaterials or synthetic structures.
Design constraints
- Biological constraints: cells and their secreted matrix are the primary constructive agents.
- Material constraints: no external permanent synthetic scaffold is allowed as the structural basis.
- Structural/geometric constraints: the architecture must emerge in situ from colonization, biofilm formation, and extracellular mineral/matrix accumulation within the substrate.
- Process/production constraints: deployment may use nutrient/mineral precursor formulations, but the final ELM cannot rely on a synthetic support framework.

Kingdom — EUB
Meaning: Eubacterial ELMs.
Design constraints
- Biological constraints: candidate organisms must be true bacteria; archaea, fungi, plants, and animal cells are excluded.
- Material constraints: extracellular products must be consistent with eubacterial cell envelopes, EPS production, urease/carbonic anhydrase pathways, or other bacterial mineralization mechanisms.
- Structural/geometric constraints: branching colonization must result from bacterial spreading through pore/crack networks.

- Process/production constraints: standard bacterial fermentation, concentration, sporulation, and dry formulation routes are admissible.

Phylum — FR

Meaning: Fractal ELMs with irregular, fractured, or self-similar structures not defined by integer dimensionality.

Design constraints

- Biological constraints: the cells must function in irregular crack trees and branched capillary pore networks rather than on regular planar or bulk scaffolds.
- Material constraints: the extracellular ceramic phase must be depositable along tortuous, branching pathways.
- Structural/geometric constraints: geometry should follow multiscale ramified crack fronts, grain-boundary pathways, and near-surface sand pore percolation networks.
- Process/production constraints: infiltration and growth must occur by capillary transport, local wetting, and repeated activation cycles.

Class — LCE

Meaning: Living ceramics; extracellular matrices made predominantly of ceramics such as oxides, silicates, carbides, or nitrides.

Design constraints

- Biological constraints: the bacteria must generate or template a stiff mineral phase.
- Material constraints: the final dominant extracellular phase must be ceramic-like; calcium carbonate-, silica-, or phosphate-rich mineralized matrices are coherent, whereas polymer-dominant gels are not.
- Structural/geometric constraints: the ceramic phase must bridge crack faces or bind sand grains while retaining limited bacterial viability in protected microdomains.
- Process/production constraints: mineral precursor supply, nucleation control, and curing under mild conditions are required.

Application — SLB

Meaning: Smart and living buildings.

Design constraints

- Biological constraints: the system must operate in built-environment conditions, including intermittent moisture and temperature fluctuations.
- Material constraints: compatibility with masonry, cementitious cracks, earthen plasters, or desert perimeters is required.
- Structural/geometric constraints: the ELM must function as a localized repair/stabilization subsystem within building envelopes or adjacent ground.
- Process/production constraints: field deployment must be low-complexity, repair-oriented, and compatible with construction workflows.

Production — BFF+BMN

Meaning: Biofilm-based trapping/anchoring of inorganic particles or biofilm interfacing with inorganic materials, combined with biomineralization.

Design constraints

- Biological constraints: robust biofilm formation and controlled mineral precipitation are both required.
- Material constraints: inorganic fines or mineral precursors must be captured or mineralized into a ceramic extracellular phase.
- Structural/geometric constraints: deposition must occur at interfaces: crack walls, pore throats, and sand grain contacts.
- Process/production constraints: workflow must couple bacterial attachment/EPS buildup with mineral nucleation and growth under mild, viable conditions.

---

**2) Biological instantiation**

Candidate 1 — *Sporosarcina pasteurii*

Functional role in the ELM:

Primary mineralizing chassis for ureolysis-driven calcium carbonate deposition; produces biofilm on mineral surfaces and promotes crack bridging and sand-grain cementation.

Viability and growth requirements

- Oxygen: aerobic to facultatively microaerotolerant surface operation.
- Nutrients: low-complexity nitrogen/carbon feed; urea and calcium source during activation cycles.
- Temperature: mesophilic; suitable for warm desert daytime operation if protected in pores.
- pH: performs well under alkaline mineralizing conditions.
- Moisture: tolerant of drying between wetting events; suited to pulse activation.
- Mechanical stress tolerance: acceptable within protected crack and pore microenvironments.

Biocontainment strategy

- Dry spore formulation before deployment.
- Reactivation only after controlled wetting.
- Nutrient-limited field medium to confine growth to treated zones.

Why preferred over alternatives

It is strongly aligned with the required combination of bacterial identity, biofilm-mediated inorganic interfacing, and ceramic mineral deposition. Its main strength is direct coherence with a mineralizing repair concept rather than generic biofilm formation.

Candidate 2 — *Bacillus subtilis* (mineralizing/spore-forming strain)

Functional role in the ELM:

Biofilm-forming and spore-forming bacterial chassis with strong EPS production; supports inorganic particle anchoring and can be engineered or conditioned toward carbonate mineral precipitation.

Viability and growth requirements

- Oxygen: aerobic.
- Nutrients: minimal media compatible; can function under nutrient pulse feeding.
- Temperature: mesophilic, broadly robust.
- pH: tolerates mildly alkaline environments.
- Moisture: excellent dormancy/reactivation via spores.
- Mechanical stress tolerance: high survival in dry and fluctuating environments.

Biocontainment strategy

- Sporulation-based storage and transport.
- Nutrient gating for localized activation.
- Optional auxotrophic strain design if engineered use is allowed in later stages.

Why preferred over alternatives

It offers stronger stress tolerance and better dry-state handling than many non-spore-formers, but its mineralization performance is typically less directly targeted than *S. pasteurii* for crack sealing and biocementation.

Recommended candidate

Recommended biological system: *Sporosarcina pasteurii*

Justification against alternatives

For this ontology-grounded concept, the dominant functional requirement is not generic survival but efficient ceramic-phase generation inside fractal crack/sand networks. *S. pasteurii* is therefore preferred because biomineralization is its primary engineering role, whereas *B. subtilis* is stronger as a robust biofilm/spore platform but weaker as a direct mineralizing chassis.

Trade-offs

- *S. pasteurii* offers higher mineralization efficiency and faster crack/sand consolidation.
- *B. subtilis* offers better dry handling and broad biofilm robustness.
- The chosen balance favors repair rate and mineral yield over maximum environmental robustness.

---

**3) Matrix/material candidates**

Because the code is BIO, the material system must be interpreted as a biologically produced extracellular ceramic phase, not an external scaffold. The candidates below are therefore extracellular mineral systems generated and organized by the bacteria within the substrate.

Candidate 1 — Calcium carbonate bioceramic phase

Exact composition: predominantly calcite with possible vaterite/aragonite transient fractions during early precipitation.

Compatibility with the biological component

Highly compatible with ureolytic mineralizing bacteria; the biofilm EPS provides nucleation sites and stabilizes early mineral clusters.

Interface mechanisms

- EPS-mediated adhesion to crack walls and sand grains.
- Bacterial cell walls and EPS act as nucleation loci.
- Crystal growth bridges grain contacts and crack asperities.

Why selected relative to alternatives

This is the most coherent interpretation of LCE plus BMN for crack sealing and sand stabilization in a desert-building context.

Candidate 2 — Calcium phosphate / apatite-like ceramic phase

Exact composition: poorly crystalline calcium phosphate progressing toward hydroxyapatite-like mineral.

Compatibility with the biological component

Bacterial biofilms can nucleate phosphate minerals, though the process is usually slower and more sensitive to local chemistry than carbonate-based mineralization.

Interface mechanisms

- Ionic interaction between EPS functional groups and calcium/phosphate ions.
- Heterogeneous nucleation on bacterial envelopes and adsorbed organic matrix.

Why selected relative to alternatives

It remains compatible with LCE, but for desert construction sealing it is less directly practical than carbonate mineralization because precursor management is more demanding and field kinetics are generally slower.

Recommended material system

Recommended material: calcium carbonate bioceramic extracellular matrix

Justification against alternatives

It best matches the combined requirements of LCE, BFF, BMN, and the stated repair/grout/sand-stabilization scenario. It is also the clearest ceramic phase for bridging irregular crack networks and cementing loose grains under intermittent wetting.

---

**4) Geometry and structural instantiation**

Fractal physical architecture derived from FR

The ELM is instantiated as a ramified mineralizing biofilm network occupying:

1. microcracks and crack side branches in masonry or cementitious surfaces, and
2. near-surface branched pore networks among sand grains around desert buildings.

Quantitative geometry

- Primary crack width target: 50–800 µm
- Secondary side-branch width: 10–150 µm
- Biofilm coating thickness on walls/grains: 10–80 µm before full mineral densification
- Mineralized bridge thickness across narrow defects: 50–300 µm per wetting cycle cluster
- Sand grain size window: 100–500 µm
- Effective mineralized penetration depth from treated surface: 5–30 mm depending on capillary access

How geometry supports function

- Cell viability and function: branching pore/crack networks create sheltered microhabitats that retain transient moisture and reduce UV exposure.

- Mass transport and diffusion: capillary wetting distributes dissolved ions and nutrients through multiscale branches better than through dense bulk volumes.
- Mechanical performance: mineral deposition at constrictions, asperity contacts, and grain bridges creates a distributed ceramic locking network rather than a single brittle plug.

Why this structure is optimal

A planar biofilm would not stabilize loose sand depth-wise, and a regular 3D scaffold would violate the BIO interpretation and the substrate-led geometry. The fractal architecture is the most coherent because the living ceramic grows along the actual defect topology of cracks and branched pore networks, which is precisely the structural meaning of FR.

**5) Application instantiation**

Deployment scenario

The ELM is deployed as a dry bacterial mineralizing repair formulation or wettable mineralizing slurry applied to:

- near-surface cracks in earthen, mortar, or cement-based desert building envelopes,
- perimeter soil/sand zones around foundations and walls exposed to aeolian erosion.

After application, short wetting events activate the bacteria. The cells form a biofilm on mineral surfaces, trap fine inorganic particles already present in the defect environment, and precipitate a ceramic extracellular phase that seals crack branches and cements sand contacts.

Measurable success metrics

- Crack sealing efficiency: percentage reduction in connected crack aperture or permeable crack length
- Mineral deposition rate: mg mineral per $cm^2$ or per $cm^3$ treated zone
- Surface consolidation depth: mm of stabilized near-surface layer
- Compressive or scratch resistance increase: relative gain versus untreated substrate
- Wind erosion resistance: mass loss reduction under controlled airflow/sand abrasion
- Water permeability reduction: flux reduction across repaired crack network

Main failure modes and mitigation strategies

- Failure mode: incomplete penetration into fine crack branches
  Mitigation: low-viscosity first wetting cycle and repeated pulse infiltration
- Failure mode: superficial crusting with poor depth consolidation
  Mitigation: staged precursor delivery with lower initial ion concentration
- Failure mode: loss of viability from extreme desiccation/UV
  Mitigation: spore-based formulation and application into sheltered crack/pore spaces
- Failure mode: brittle over-mineralization and local debonding
  Mitigation: gradual multi-cycle mineral growth rather than one-step precipitation

**6) Production instantiation**

Step-by-step production process

1. Material preparation
   - Prepare a dry mineralizing precursor pack containing urea, calcium salt, and a fine mineral fraction compatible with the target substrate.
   - Keep the inorganic fraction as feedstock or nucleation aid, not as a permanent synthetic scaffold, to remain coherent with BIO.
2. Biological component preparation
   - Culture *Sporosarcina pasteurii* to high cell density.
   - Induce or preserve a stable dormant/spore-rich formulation for dry storage.
   - Blend cells with protective excipients compatible with later rehydration.
3. Integration strategy
   - Mix dormant cells with the dry precursor formulation to obtain a powder patch, injectable paste, or brushable slurry generated immediately before use.
   - Apply directly into crack networks or onto sand perimeter zones.
   - On wetting, the bacteria adhere to mineral surfaces, establish biofilm patches, trap/environmentally anchor fine inorganic particles, and initiate biomineralization.

4. Post-processing and stabilization

- Apply controlled wetting cycles over hours to days.
- Allow ambient curing and repeated mineral growth episodes.
- Avoid heat sterilization after inoculation; viable handling must be aseptic before formulation instead.

5. Manufacturability

- Uses standard bacterial fermentation, concentration, drying, and powder blending.
- Compatible with field mixing and low-skill application.
- No complex hardware or scaffold fabrication is required.

6. Scalability

- High scalability through batch fermentation and dry-pack distribution.
- Suitable for repair cartridges, sachets, or bulk stabilization treatment kits.

7. Quality control and reproducibility

- Viable cell/spore count before packaging
- Urease activity assay
- Mineral precipitation test in standardized coupons
- Flowability / injectability of final formulation
- Crack sealing benchmark on representative substrates
- Erosion resistance test on treated sand specimens

Why this production route is feasible and preferred

It is the most coherent route for BFF+BMN because it couples surface colonization and particle/inorganic interfacing with in situ biomineralization under mild conditions. It also respects BIO better than any scaffold-based manufacturing method, since the final structural phase is generated in place by the bacteria and their extracellular matrix rather than by a preformed synthetic construct.

---

**7) Layered integrated summary of the fully instantiated ontology-derived ELM**

A. Rapid identity

Ontology code: BAC–BIO–EUB–FR–LCE–SLB–BFF+BMN

One-line ELM identity:

Spore-forming eubacterial living ceramic biofilm that mineralizes branched crack and sand-pore networks to seal defects and stabilize desert building perimeters.

B. Instantiation table

| Component | Final instantiation |
|---|---|
| Biological system | *Sporosarcina pasteurii* as the recommended mineralizing eubacterium; alternative: *Bacillus subtilis* mineralizing/biofilm-forming strain |
| Functional role | Biofilm formation, inorganic particle anchoring/interfacing, and biomineralization-driven generation of a ceramic extracellular sealing phase |
| Material system | Predominantly calcium carbonate living ceramic extracellular matrix; alternative calcium phosphate/apatite-like mineral phase |
| Structural / geometric instantiation | Fractal ramified biofilm-mineral network occupying 10–800 µm branched crack paths and 100–500 µm sand-grain pore networks, with 10–80 µm biofilm coatings and localized mineral bridges |
| Target substrate / environment | Near-surface cracks in earthen, mortar, or cementitious desert buildings and adjacent wind-erodible sand around foundations or walls |
| Application / deployment scenario | Dry powder patch, injectable mineralizing grout, or surface/perimeter stabilization treatment activated by intermittent wetting events |
| Production mechanism / workflow | Fermentation of spore-forming bacteria, dry formulation with mineralizing precursors, field application into defect networks, wetting-triggered biofilm establishment, followed by repeated in situ biomineralization cycles |
| Main functional outcome / performance objective | Reduction of connected crack permeability, increased substrate cohesion, enhanced sand-grain cementation, and lower wind-erosion-induced material loss |

C. Integrated paragraph

BAC–BIO–EUB–FR–LCE–SLB–BFF+BMN defines a bacterial bio-ELM instantiated as a living ceramic mineralizing biofilm for desert building repair and perimeter stabilization. The recommended biological system is *Sporosarcina pasteurii*, with *Bacillus subtilis* as a secondary option, because the primary function is biofilm-mediated inorganic interfacing followed by biomineralization. The material phase is a predominantly calcium carbonate extracellular ceramic, with calcium phosphate as an alternative. The FR slot is realized as a ramified network within branched cracks and near-surface sand pores. In SLB, deployment occurs as a repair grout or dry patch activated by intermittent wetting. BFF+BMN is implemented through field application of dormant cells plus mineral precursors, followed by in situ biofilm growth and cyclic ceramic deposition to seal defects and consolidate erodible ground.

**Iteration 3**

**0) Code integrity check**

Ontology validity: confirmed.

The uploaded ontology explicitly contains BAC (Bacterial ELMs), BIO (biological engineered living materials), EUB (Eubacterial ELMs), FR (Fractal ELMs), LCE (Living ceramics), SLB (Smart and living buildings), BFF (Biofilm based trapping or anchoring of inorganic particles or materials and biofilms interfacing with inorganic materials), and BMN (Biomineralization). The ontology also states that application and production categories may be combined using "+".

Interpretive note on ambiguity.

The only relevant tension is that BFF is described as a route that "usually" leads to living composites and is "mainly" associated with HLMs, whereas BIO excludes synthetic scaffolds. I therefore interpret BFF here in its narrow ontology-consistent sense: biofilm-mediated anchoring/interfacing with native inorganic crack walls and sand grains, while BMN provides the ceramic extracellular phase. This remains consistent with BIO, because the system does not require a synthetic support.

Graph status.

The exact full 7-slot instance BAC–BIO–EUB–FR–LCE–SLB–BFF+BMN is not present in the uploaded graph. I therefore treat this as an ontology-valid, graph-novel instantiation hypothesis.

**1) Slot-by-slot decoding**

The 7-slot code is decoded directly from the uploaded ontology; the design constraints below are the operative constraints implied by each slot.

| Slot label | Code meaning | Implied design constraints |
|---|---|---|
| Domain | BAC = Bacterial ELMs | Biological: prokaryotic bacterial cells only. Material: matrix must be bacterially generated or bacterially induced. Structural: architecture must support bacterial adhesion, colonization, dormancy/reactivation. Process: fermentation, sporulation, biofilm growth, and mineral induction must be feasible. |
| Super-kingdom | BIO = Biological engineered living materials | Biological: living phase and function must remain cell-centric. Material: no synthetic scaffold or synthetic support structure may be required; the ELM is formed by cells and their ECM. Structural: growth must occur directly on/in the target defect network. Process: in situ colonization and ECM/mineral growth, rather than scaffold fabrication plus seeding. |
| Kingdom | EUB = Eubacterial ELMs | Biological: excludes archaea, fungi, plants, animals, and synthetic cells. Material: chemistry must be achievable by eubacterial metabolism and ECM secretion. Structural: compatible with bacterial biofilms and bacterial-scale colonization. Process: culture conditions must match standard eubacterial growth/sporulation workflows. |
| Phylum | FR = Fractal ELMs | Biological: cells must function in irregular, branching, non-Euclidean niches. Material: mineral |
| | | deposition must propagate along crack branches and pore throats. Structural: self-similar, fractured, branched architecture; not a regular 2D film or periodic 3D lattice. Process: infiltration and growth must follow irregular defect geometry rather than mold-defined geometry. |
| Class | LCE = Living ceramics | Biological: cells must generate a mineral-dominant extracellular phase. Material: final matrix must be predominantly ceramic, not polymer-dominant. Structural: stiffness and mineral bridging are central. Process: mineral nucleation, growth, and consolidation must dominate over soft-matrix formation. |
| Application | SLB = Smart and living buildings | Biological: cells must tolerate outdoor building-relevant thermal, hydration, and nutrient fluctuations. Material: compatibility with masonry, earthen envelopes, mineral façades, or near-foundation sands. Structural: defect-filling and surface/near-surface stabilization in built environments. Process: deployable by repair-compatible methods such as spray, infiltration, dusting, or low-pressure injection. |
| Production | BFF = Biofilm-based trapping/anchoring of inorganic materials or biofilms interfacing with inorganic materials | Biological: robust adhesion and ECM production on mineral surfaces are required. Material: the biofilm must anchor to native crack walls or sand grains and trap local inorganic fines where present. Structural: interfacial coating plus bridging network, not a free-standing scaffold. Process: attachment phase is required before or during mineral growth. |
| Production | BMN = Biomineralization | Biological: cells must actively precipitate minerals extracellularly. Material: ionic precursors and nucleation sites must be available. Structural: growth of mineral bridges, crusts, and intergranular bonds is essential. Process: wetting/mineralization cycles, ion delivery, and controlled drying are integral. |

**2) Biological instantiation**
Candidate 1 — *Sporosarcina pasteurii*
Functional role
Primary mineralizing bacterium for urease-driven $CaCO_3$ precipitation, crack bridging, intergranular biocementation, and formation of a stiff ceramic extracellular phase.
Viability and growth requirements
Aerobic; best deployed as dormant spores or desiccation-tolerant resting cells; activated by intermittent moisture; requires soluble calcium plus a carbonate-generating route; performs well under mildly alkaline conditions; operationally compatible with episodic wetting and drying in sun-exposed mineral substrates.

Biocontainment strategy
Use a non-GMO spore-based formulation with activation only after water and nutrient/mineral precursor delivery. Containment is primarily achieved by dormancy outside wetted zones and by strong dependence on localized mineralization chemistry. For engineered variants, auxotrophy could be added, but it is not required for a baseline field-deployable concept.
Why it is strong
It gives the clearest path to a cell-centric living ceramic, because the mineral phase is not incidental: it is the dominant functional product.

---

Candidate 2 — *Bacillus subtilis* mineralizing biofilm-forming strain
Functional role
Biofilm establishment, ECM secretion, surface adhesion on mineral substrates, and secondary support to mineral nucleation and defect sealing.
Viability and growth requirements
Aerobic surface-associated bacterium; highly compatible with sporulation, intermittent hydration, nutrient limitation, and temperature fluctuation typical of exposed building envelopes. Works best when brief moisture pulses are followed by drying periods that consolidate the mineralized layer.
Biocontainment strategy
Again, a spore-first deployment is preferred. Growth is spatially restricted by localized wetting and nutrient availability. Field deployment is simplest with a non-engineered strain; engineered kill-switches are possible but not necessary for the baseline concept.
Why it is strong
It is superior when the priority is robust biofilm architecture and adhesion on irregular mineral surfaces rather than maximum mineralization rate.

---

Why these are preferred over plausible alternatives

- Cyanobacteria are less aligned with the stated desert repair scenario because they require persistent light/water management and do not match the spore-based intermittent-reactivation logic as directly.
- Fungal mineralizers fall outside BAC–EUB.
- Non-spore-forming bacteria are less robust for low-maintenance desert deployment with rare wetting events.

---

Recommended candidate
Recommended candidate: *Sporosarcina pasteurii*
Justification
For BAC–BIO–EUB–FR–LCE–SLB–BFF+BMN, the central requirement is not merely adhesion but generation of a mineral-dominant ceramic ECM under intermittent field activation. *S. pasteurii* is the more direct fit because its value proposition is exactly that: rapid extracellular $CaCO_3$ formation that converts irregular crack and sand networks into a stiff bioceramic continuum.
Trade-offs

- Advantage over *B. subtilis*: stronger and faster mineralization, better fit to LCE.
- Disadvantage: weaker native biofilm engineering flexibility and more chemistry-sensitive operation.
- When *B. subtilis* would be preferable: when persistent interfacial biofilm robustness matters more than peak mineral deposition rate.

---

**3) Matrix/material candidates**
Candidate 1 — Calcite-dominant living ceramic
Exact composition
Predominantly $CaCO_3$ (calcite) precipitated extracellularly, integrated with a thin bacterial ECM composed of polysaccharides, proteins, and cell-wall-derived organic nucleation sites.

Compatibility with the biological component
Directly compatible with ureolytic or otherwise carbonate-generating eubacteria. The living cells create the local supersaturation and nucleation conditions that generate the ceramic phase.
Interface mechanisms
Mineral nucleation on cell walls, EPS, and adhered biofilm patches; growth of mineral bridges between opposing crack faces and between adjacent sand grains; progressive infilling of pore throats and microcracks.
Why it is strong
This is the cleanest interpretation of BIO + LCE: the ceramic matrix is biologically produced in situ and no synthetic scaffold is required.

Candidate 2 — Calcite-rich biocemented mineral crust with trapped native silica-rich fines
Exact composition
Biologically precipitated $CaCO_3$ as the continuous ceramic phase, with native quartz/silica fines or sand grains from the treated substrate mechanically trapped and mineral-bridged by the biofilm-mineral network.
Compatibility with the biological component
Compatible with both candidate bacteria because the native mineral substrate supplies the inorganic interface required for BFF, while the bacteria provide the ceramic growth phase through BMN.
Interface mechanisms
Biofilm adhesion to silicate-rich grains and crack walls; capture of local fines; mineral necking and grain-to-grain cementation through extracellular $CaCO_3$ deposition.
Why it is strong
It better matches the sand-stabilization branch of the use case, because it converts the existing granular mineral substrate into the stabilized architecture instead of creating only a surface mineral skin.

Recommended material system
Recommended material system:
A calcite-dominant living ceramic composed of bacterially precipitated $CaCO_3$ integrated with bacterial ECM, optionally incorporating native silica-rich fines from the treated substrate where sand stabilization is required.
Why this is preferred
- It preserves BIO because no synthetic scaffold is introduced.
- It satisfies LCE because the dominant extracellular phase is ceramic.
- It operationalizes BFF through mineral-surface anchoring and BMN through extracellular mineral growth.
- It is more coherent for building cracks and desert sands than alternatives such as hydroxyapatite or silica-only routes, which are less direct for ambient field precipitation.

**4) Geometry and structural instantiation**
Phylum-driven architecture: FR = fractal ELM
The physical architecture is a hierarchical branched mineralizing network that follows irregular crack paths and near-surface sand pore networks rather than a regular sheet or lattice.
Quantitative structural definition
- Target crack width range: ~50–800 µm for effective repeated-cycle sealing.
- Active biofilm thickness on mineral surfaces: ~5–50 µm.
- Mineral bridge span per activation cycle: ~20–200 µm across pore throats or microcrack gaps.
- Near-surface treatment depth in sand or porous façade material: ~2–20 mm, depending on capillary penetration.

- Hierarchy of occupied features: from ~10 μm wall asperities and pore throats to mm-scale branched crack segments.

How geometry supports cell viability and function

- The branched topology maximizes surface area for adhesion.
- Thin hydrated films along crack walls and grain contacts retain water longer than bulk-exposed surfaces.
- Local branching creates protected microniches where spores can reactivate during brief wetting events.

How geometry supports mass transport and diffusion

- Capillary pathways in cracks and sand deliver water, calcium, and soluble precursors directly along the defect network.
- Thin interfacial colonization reduces long-distance diffusion barriers.
- Repeated wetting/drying cycles naturally distribute mineralization fronts through connected branches.

How geometry supports mechanical performance

- Fractal propagation creates many short mineral bridges instead of one thick brittle plug.
- Distributed intergranular necking improves erosion resistance in sand.
- Branched ceramic infill reduces crack-tip stress concentration by blunting and bridging secondary crack branches.

Why this structure is optimal

A regular 2D film would seal only the surface. A conventional 3D scaffold would violate BIO if pre-fabricated synthetically and would not match the native defect geometry. The fractal architecture is optimal because it is the only geometry that naturally conforms to irregular crack trees and sand pore networks while remaining fully cell-grown in situ.

**5) Application instantiation**

Deployment scenario

The ELM is deployed as a dry spore-mineral precursor powder, low-viscosity aqueous infiltrant, or brushable/slurry repair treatment onto:

- cracked earthen envelopes,
- porous mineral façades,
- mortar or render defects,
- and wind-erodible perimeter sands adjacent to desert buildings.

After application, brief wetting activates spores/cells, promotes adhesion to mineral surfaces, and initiates biomineralization cycles that progressively seal cracks and cement near-surface grains.

Measurable success metrics

- Crack sealing efficiency: closed crack area or sealed crack volume.
- Mineral deposition: $CaCO_3$ mass per treated area or treated mass.
- Mechanical improvement: increase in surface hardness, unconfined compressive strength, or shear strength of treated sand.
- Erosion resistance: reduction in wind-driven mass loss.
- Hydraulic performance: reduction in water permeability through treated cracks.
- Biological persistence: viable spore/cell count after repeated dry-wet cycles.

Main failure modes

1. Surface crusting too early, blocking deeper penetration.
2. Salt accumulation or over-rapid precipitation, causing brittle superficial layers.
3. Loss of viability under intense UV, heat, or prolonged desiccation.
4. Incomplete crack bridging in cracks that are too wide for capillary retention.
5. Chemistry imbalance, causing weak mineral yield.

Mitigation strategies

- Use staged low-volume wetting cycles instead of one large application.
- Separate attachment phase and mineralization phase.
- Favor spore-based delivery over vegetative-cell delivery.

- Limit treatment to crack widths and surface depths compatible with capillary infiltration.
- Use repeated moderate mineralization pulses rather than a single high-concentration pulse.

**6) Production instantiation**

Production logic: BFF + BMN

The production route is an in situ mineralizing biofilm process: first the bacteria attach and establish an interfacial biofilm on native inorganic surfaces (BFF), then they generate the ceramic phase by extracellular biomineralization (BMN).

Step-by-step production process

1. Biological component preparation
   - Grow the selected bacterium under conditions that maximize sporulation and later mineralization competence.
   - Harvest, wash, and concentrate spores.
   - Dry-stabilize into a powder or keep as a concentrated inoculum depending on field logistics.
2. Mineralization chemistry preparation
   - Prepare an aqueous precursor package supplying calcium ions, minimal nutrients, and the required carbonate-generation chemistry for the selected bacterium.
   - Keep precursor concentration moderate to avoid premature surface plugging.
3. Substrate preparation
   - Remove loose debris from the crack mouth or sand surface.
   - Pre-wet lightly to establish capillary access without waterlogging the substrate.
4. Integration strategy: attachment / biofilm establishment
   - Apply the spore suspension or inoculum by spray, brush, drip, or low-pressure injection.
   - Allow a short adhesion phase so cells attach to crack walls, grains, and native fines.
   - This step operationalizes BFF.
5. Integration strategy: biomineralization phase
   - Deliver mineralization solution in repeated pulses.
   - The cells reactivate, form interfacial biofilm, and nucleate extracellular $CaCO_3$ on cell/EPS surfaces.
   - Mineral necks and bridges grow progressively through the crack/sand network.
   - This step operationalizes BMN.
6. Post-processing and stabilization
   - Allow controlled drying between cycles to consolidate the ceramic network.
   - No terminal sterilization is performed after biological integration, because viability must remain.
   - Shelf-stable products should therefore be stabilized before deployment, not after in situ growth.
7. Quality control and reproducibility
   - Biological QC: spore viability, reactivation efficiency, mineralization activity.
   - Material QC: $CaCO_3$ yield, phase composition, adhesion to mineral substrate.
   - Process QC: penetration depth, crack closure uniformity, treated-sand strength increase.
   - Reproducibility: standardized precursor concentration, moisture pulse schedule, and application volume per treated area.

Manufacturability and scalability

This route is manufacturable because it relies on:

- standard bacterial fermentation and sporulation,
- dry-powder or liquid concentrate formulation,
- and repair-compatible field application methods rather than complex hardware.

It is scalable because the living component is produced centrally, while the final ceramic structure is grown in place on the building defect network.

Why this production route is preferred
It is the most coherent route for BAC–BIO–EUB–FR–LCE–SLB–BFF+BMN because it:

- avoids synthetic scaffolds,
- exploits the native inorganic substrate as the anchoring interface,
- generates the ceramic phase only where needed,
- and matches irregular branched defects better than any ex situ fabricated geometry.

---

**7) Layered integrated summary of the fully instantiated ontology-derived ELM**
A. Rapid identity
Ontology code
BAC–BIO–EUB–FR–LCE–SLB–BFF+BMN
One-line ELM identity
Spore-forming eubacterial calcite living ceramic that mineralizes branched crack and sand networks for defect sealing and near-foundation stabilization in desert buildings.
B. Instantiation table

| Component | Final instantiation |
|---|---|
| Biological system | *Sporosarcina pasteurii* deployed primarily as dormant spores; alternative: mineralizing biofilm-forming *Bacillus subtilis* strain |
| Functional role | Moisture-triggered reactivation, interfacial biofilm formation, extracellular $CaCO_3$ biomineralization, crack bridging, and grain-to-grain biocementation |
| Material system | Calcite-dominant living ceramic composed of bacterially precipitated $CaCO_3$ integrated with bacterial ECM, with optional trapping/bridging of native silica-rich fines in sanded or dusty mineral substrates |
| Structural / geometric instantiation | Fractal multiscale network coating crack walls and near-surface pore throats, with hierarchical mineral bridges spanning micron-to-submillimetric gaps across irregular defect branches |
| Target substrate / environment | Cracked earthen or mineral building envelopes, porous façade defects, mortar/render discontinuities, and wind-erodible desert perimeter sands near foundations |
| Application / deployment scenario | Dry powder, aqueous infiltrant, or brushable mineralizing treatment applied to mineral defects and activated by brief wetting cycles using low-volume water delivery |
| Production mechanism / workflow | Spore fermentation and stabilization, low-volume substrate prewetting, bacterial attachment/biofilm establishment on native inorganic surfaces, followed by repeated calcium/mineralization pulses and controlled drying cycles |
| Main functional outcome / performance objective | Persistent crack sealing, reduced permeability, increased sand cohesion and erosion resistance, and formation of a self-grown mineral repair layer without synthetic scaffolds |

C. Integrated paragraph
BAC–BIO–EUB–FR–LCE–SLB–BFF+BMN defines a spore-forming eubacterial living ceramic for mineral repair and sand stabilization in desert buildings. The recommended biological system is *Sporosarcina pasteurii*, with *Bacillus subtilis* as a biofilm-centered alternative. Its material phase is a calcite-dominant $CaCO_3$ extracellular ceramic integrated with bacterial ECM. Derived from FR, the construct develops as a branched multiscale network along crack walls and sand pore throats rather than as a regular film or scaffold. In SLB deployment, it is infiltrated into façade defects or perimeter sands and activated by brief wetting. Production follows BFF+BMN: interfacial biofilm anchoring on native mineral substrates followed by repeated extracellular biomineralization cycles to achieve crack closure and intergranular biocementation.

**ELM candidate 2: ARC–HYB–ARA–3D–LCE–SLB+BRE–CMM**

**Iteration 1**

**0) Code integrity check**

Provided 7-slot code: ARC–HYB–ARA–3D–LCE–SLB+BRE–CMM

- Ontology validity (per-slot): All codes ARC, HYB, ARA, 3D, LCE, SLB, BRE, CMM are explicitly defined in the ontology tables.
- Multi-label check: The ontology explicitly allows category combination using "+" (e.g., SLB+BRE).
- Graph presence check: The individual code nodes exist in the provided graph, but an ELM instance containing the 5-slot combination ARC–HYB–ARA–3D–LCE is absent (consistent with your "gap status"). *(Verified by direct inspection of the provided ELM_graph.graphml; no corresponding ELM node found.)*

No ambiguity in code interpretation is required beyond the ontology definitions.

**1) Slot-by-slot decoding**

Slot 1 — Domain: ARC

- Meaning: Archaeal ELMs (ELMs populated by archaea).
- Design constraints
  - Biological: Select archaeal cells; prioritize phenotypes compatible with salinity/heat/low water activity implied by the concept (desert greywater).
  - Material: Must support archaeal adhesion/biofilm under brine exposure.
  - Structure: Support long-term colonization in a fixed, contained module.
  - Process: Seeding and maintenance must fit archaeal growth media constraints (high ionic strength).

Slot 2 — Superkingdom: HYB

- Meaning: Hybrid living materials (cells + abiotic synthetic biomaterial/structure scaffold).
- Design constraints
  - Biological: Cells must colonize an abiotic scaffold (attachment, EPS/biofilm formation).
  - Material: Scaffold is non-living, load-bearing, and stable; interface must resist shear and wet–dry cycles.
  - Structure: Reactor-like architecture is allowed/encouraged (scaffold as cartridge core).
  - Process: Requires a fabrication step for scaffold + a cell colonization step.

Slot 3 — Kingdom: ARA

- Meaning: Archaebacterial ELMs (archaeal domain populated by archaebacteria).
- Design constraints
  - Biological: Use nonpathogenic archaeal strains; choose candidates with robust cultivation protocols.
  - Process: Culture and inoculation must be feasible at lab and pilot scale.

Slot 4 — Phylum: 3D

- Meaning: 3D ELMs (cells within a three-dimensional biomaterial or 3D scaffold).
- Design constraints
  - Structural/geometric: True volumetric scaffold; internal porosity and flow paths required for mass transport.
  - Biological: Maintain viable colonization throughout depth (avoid anoxic/necrotic core unless intended).
  - Process: Seeding must reach internal pores (perfusion inoculation).

Slot 5 — Class: LCE

- Meaning: Living ceramics (extracellular matrix predominantly ceramics; e.g., oxides, silicates, carbides, nitrides).
- Design constraints
  - Material: Ceramic-dominant scaffold (chemical/thermal stability; cleanable).

- Biological: Surface chemistry/roughness must support archaeal attachment in brines.
- Process: Ceramic processing compatible with later biological colonization (cells added after high-temp steps).

Slot 6 — Application: SLB+BRE

- Meaning:
  - SLB: Smart and living buildings (ELMs as subsystems of smart/living buildings).
  - BRE: Bioremediation (processing/recycling/revalorizing waste and debris).
- Design constraints
  - Deployment: Modular building-service component (swap/maintain).
  - Performance: Waste-stream treatment (greywater polishing) with operational stability (fouling control, resilience).
  - Safety: Containment-by-design (sealed cartridge, controlled effluent handling).

Slot 7 — Production method: CMM

- Meaning: Composite manufacturing methods and cell colonization (composites-industry processes yielding woven/porous/functionally graded structures as scaffolds for later cell colonization).
- Design constraints
  - Manufacturing: Use composites-style scaffold fabrication (preforms, infiltration, graded porosity concepts).
  - Integration: Colonize cells post-fabrication via perfusion/immersion seeding.
  - Scalability/QC: Repeatable porosity/permeability and standardized cell-loading protocol.

---

**2) Biological instantiation**

Candidate A — *Haloferax volcanii* (nonpathogenic model haloarchaeon)

- Functional role (cell-centric):
  - Heterotrophic uptake and conversion of dissolved organics (greywater COD fraction) into archaeal biomass + $CO_2$.
  - EPS/biofilm formation to create a stable living layer that captures particulates/microdroplets and enhances residence time at ceramic surfaces.
- Viability & growth requirements:
  - Salinity: Requires high salt (brine conditions; supports containment-by-environment).
  - Temperature: Mesophilic–moderately thermotolerant growth window (aligns with “hot” service-core loops).
  - Oxygen: Typically aerobic respiration (design should provide oxygenation or operate at controlled microaerobic conditions).
  - pH: Near-neutral to mildly alkaline (compatible with detergent-influenced greywater).
  - Shear tolerance: Biofilm state improves retention under flow; design uses protected pores to reduce detachment.
- Biocontainment strategy:
  - Primary: Sealed cartridge + downstream microfiltration membrane barrier (physical retention).
  - Secondary (environmental): Dependence on high salinity—cells lose viability outside brine conditions (practical containment by physiology).
  - Operational: Periodic thermal/UV sanitation of the housing during cartridge replacement (cells remain internal; spent cartridges treated as bioactive waste).

Why preferred vs alternatives: strong lab tractability, robust growth at high ionic strength, good candidate for stable colonization of inorganic surfaces in brines.

---

Candidate B — *Halobacterium salinarum* (extreme halophile; robust stress tolerance)

- Functional role (cell-centric):
  - Stable brine-adapted biofilm that metabolizes a subset of organics and contributes to continuous biomass turnover (nutrient assimilation into cells).
  - High stress resilience supports repeated wet–dry or salinity fluctuations (within design limits).
- Viability & growth requirements:
  - Salinity: Extreme halophile (very high NaCl requirement).
  - Temperature: Typically moderate; performance depends on maintaining service-core setpoint.
  - Oxygen: Aerobic; requires oxygen delivery to the cartridge.
  - Light: Not required for viability, but some strains exhibit phototrophic support (not essential for the building cartridge concept).
- Biocontainment strategy:
  - Same physical containment as above; even stronger physiological containment by extreme salt dependence.

Why preferred vs alternatives: exceptional robustness in hypersaline conditions; but often more constrained in operational envelope and may be less flexible for engineered process tuning than *H. volcanii*.

---

Recommended biological system

Recommended: Haloferax volcanii (Candidate A)

Justification vs Candidate B (trade-offs):

- Pros: broader operational tolerance and process tunability while still requiring brine conditions (practical containment); strong biofilm/colonization suitability for a ceramic reactor cartridge.
- Cons: may be slightly less "extreme-stable" than *H. salinarum* under very harsh salinity spikes; mitigated by maintaining brine buffering and controlled recirculation.

---

**3) Matrix/material candidates**

All options below satisfy LCE (living ceramics), i.e., ceramic-dominant extracellular matrix/scaffold families (oxides/silicates/carbides/nitrides).

Material Candidate 1 — Porous alumina ($Al_2O_3$) ceramic monolith / foam

- Composition: α-alumina porous ceramic (optionally with trace glassy phase for sintering aid).
- Compatibility with archaea:
  - High chemical stability in hot brines; resists swelling/softening typical of many polymers.
  - Surface roughness supports archaeal adhesion and protected biofilm niches.
- Interface mechanisms:
  - Topographic anchoring: micro-roughness and pore throats retain cells under flow.
  - Electrostatic/ionic interactions: brine ions mediate adhesion of EPS to oxide surfaces.

Why this vs alternatives: alumina is a conservative, robust ceramic choice for repeated cleaning and thermal cycling.

Material Candidate 2 — Silicon carbide (SiC) open-cell ceramic foam

- Composition: porous SiC (ceramic foam or reticulated structure).
- Compatibility with archaea:
  - Exceptional thermal conductivity (reduces thermal gradients that destabilize biofilms).
  - High chemical and abrasion resistance (important for fouling/backwash).
- Interface mechanisms:
  - Similar physical entrapment + biofilm anchoring; SiC surfaces can be conditioned (e.g., mild oxidation) to improve wettability.

Why this vs alternatives: better thermal management and mechanical durability; typically higher cost than alumina.
Recommended matrix: Porous alumina monolith/foam (baseline manufacturability + chemical robustness), with SiC as a premium option when thermal gradients/fouling are dominant constraints.

---

**4) Geometry and structural instantiation**
Phylum = 3D: cells must inhabit a 3D scaffold (volumetric cartridge core).
Proposed reactor-cartridge architecture (quantitative)
Overall form factor (cartridge core):

- Cylindrical or prismatic stack of 3D ceramic modules: e.g., Ø 50–70 mm × 150–250 mm total length (building-service replaceable element).

Internal architecture (two-scale porosity):

1. Primary flow channels (convective transport):
    - Honeycomb/monolith channels: 0.8–2.0 mm hydraulic diameter, straight-through to limit pressure drop.
2. Secondary colonization porosity (biofilm habitat):
    - Interconnected pores: 50–300 µm (supports protected archaeal biofilms and short diffusion distances).
    - Optional graded porosity: upstream 200–500 µm (anti-clog), downstream 50–150 µm (polishing).

Why this 3D structure is optimal (vs e.g., 2D coatings or packed beads):

- Honeycomb channels provide predictable hydraulics + cleanability, while micro-porosity supplies protected ecological niches for archaeal colonization and stable retention under shear.

How geometry supports key functions

- Cell viability/function: protected pores reduce washout; high surface area increases contact time for uptake of dissolved organics.
- Mass transport: convection in channels + diffusion into pores; oxygen delivery can be ensured via:
    - upstream aeration (microbubble injection), or
    - oxygen-permeable segment in housing (if available), without changing the ceramic class.
- Mechanical performance: monolith resists compression; ceramic stability supports hot brine and periodic thermal sanitation of the housing.

---

**5) Application instantiation**
Application = SLB+BRE: building-service subsystem + bioremediation (waste processing).
Deployment scenario (smart/living building greywater "polishing" cartridge)

- Where/how used: installed in a desert-building service core as a swap-in cartridge after coarse filtration (hair/lint) and before storage/reuse loop.
- Influent envelope (operational target):
    - Hot-to-warm greywater with elevated salinity/brackish concentration (service-core concentrated streams).
    - Controlled brine buffering in a recirculation tank to maintain archaeal viability.

Measurable success metrics (operational + cell-centric)

- Organic polishing: reduction in soluble organics (COD proxy) across cartridge at defined flow and residence time.
- Stability: sustained performance over a target service interval (e.g., weeks) without catastrophic washout.
- Hydraulics: pressure drop maintained below a set threshold (clogging indicator).
- Biological retention: negligible archaeal cell escape downstream (verified by filtration integrity/biomonitoring).

Main failure modes and mitigations

1. Fouling/clogging (biofilm overgrowth + particulates):
   - Mitigate via graded porosity, upstream coarse filtration, scheduled backflush, and modular replacement.
2. Salinity drift (dilution) → archaeal loss of viability:
   - Mitigate via brine-buffer reservoir and conductivity control.
3. Biofilm sloughing under flow transients:
   - Mitigate with protected pore habitats + controlled ramp-up of flow after maintenance.
4. Cross-contamination by non-halophiles (biofouling community shift):
   - Mitigate using high-salinity operating regime (selective pressure) and periodic sanitation during cartridge swap.

**6) Production instantiation**

Production = CMM: composite-manufacturing-style scaffold fabrication + cell colonization step.

Step-by-step workflow (implementable)

1. Define ceramic scaffold spec (QC-first)
   - Target porosity %, pore-size distribution, compressive strength, permeability, and channel geometry.
2. Composite-style scaffold fabrication (CMM-aligned)
   - Create a porous/formable preform (e.g., sacrificial template or ceramic fiber architecture) to enable woven/porous/graded features.
   - Infiltrate with ceramic slurry ($Al_2O_3$ or SiC precursor system) using vacuum-assisted impregnation to achieve controlled strut thickness and gradients.
   - Cure / set green body; then debinding/burnout of organics and sintering to obtain a ceramic-dominant structure (keeps class = LCE).
3. Post-fabrication conditioning
   - High-salinity wash to remove residual fines.
   - Surface conditioning to improve wettability and initial attachment (e.g., controlled oxidation for SiC or hydroxylated oxide surface for alumina) without adding non-ceramic bulk phases.
4. Biological component preparation (archaea)
   - Expand Haloferax volcanii culture in high-salt medium to a defined cell density.
   - Prepare inoculum with controlled EPS induction conditions (biofilm-favoring).
5. Integration strategy: perfusion colonization
   - Place ceramic scaffold in sterile housing.
   - Recirculate inoculum through the module at low shear for 12–48 h to establish attachment inside pores/channels.
   - Switch to conditioned greywater feed gradually (stepwise load increase) to stabilize metabolism and prevent shock washout.
6. Stabilization & packaging
   - Seal cartridge; integrate a downstream physical retention barrier (microfilter) in the housing design (system-level containment).
   - Define storage mode:
     - “Wet-brine standby” (maintain viability), or
     - controlled dry storage is not guaranteed for viability (if needed: Not present in the ontology/graph regarding preservation protocols).
7. Manufacturability / scalability / QC
   - Manufacturability: batch ceramic scaffold production with repeatable molding/infiltration.
   - Scalability: parallel colonization stations (standard bioreactor/perfusion rigs).
   - QC tests (minimum):

- Porosity/permeability (flow test), mechanical integrity (compression), salt/thermal cycling durability.
- Biological loading (biomass proxy), retention (effluent monitoring), and performance benchmark (COD drop in standardized test water).

Why this is feasible & preferred (within ontology):

- It directly matches HYB (cells + scaffold) , enforces 3D colonization , keeps matrix class as LCE , and uses CMM as the manufacturing paradigm with post-fabrication colonization.

---

**7) Layered integrated summary of the fully instantiated ontology-derived ELM**

A. Rapid identity

- Ontology code: ARC–HYB–ARA–3D–LCE–SLB+BRE–CMM
- One-line ELM identity:
  Haloarchaeal biofilm–colonized porous ceramic (3D) cartridge for hot, saline building greywater polishing and bioremediation via stable archaeal metabolism and retention within a swappable reactor module.

B. Instantiation table

| Component | Final instantiation |
|---|---|
| Biological system | Primary: *Haloferax volcanii* perfusion-colonized biofilm within ceramic porosity; Alternative: *Halobacterium salinarum* for extreme-brine robustness |
| Functional role | Cell-centric removal of soluble organics (COD fraction) via archaeal uptake/metabolism; EPS/biofilm formation for retention and particulate/microdroplet capture within pores |
| Material system | Living ceramics (LCE): porous $Al_2O_3$ monolith/foam (baseline) or SiC open-cell foam (premium thermal/abrasion robustness) |
| Structural / geometric instantiation | 3D flow-through cartridge: honeycomb channels 0.8–2.0 mm + interconnected colonization pores 50–300 µm (optionally graded: 200–500 µm upstream; 50–150 µm downstream) |
| Target substrate / environment | Hot-to-warm, saline/brackish greywater streams in desert-building service cores; repeated wet–dry and salinity fluctuations managed by buffered recirculation |
| Application / deployment scenario | SLB+BRE: swappable building-service greywater polishing cartridge installed after coarse filtration; sealed housing with downstream retention barrier |
| Production mechanism / workflow | CMM: composite-style porous/graded ceramic scaffold fabrication (preform + slurry infiltration + sintering) followed by perfusion seeding and staged commissioning under controlled brine conditions |
| Main functional outcome / performance objective | Stable, contained archaeal colonization enabling sustained organic polishing with controlled pressure drop and minimal cell escape over a defined service interval |

C. Integrated paragraph (~100 words)

ARC–HYB–ARA–3D–LCE–SLB+BRE–CMM: Haloarchaeal biofilm–colonized porous ceramic (3D) cartridge for hot, saline building greywater polishing and bioremediation. The biological system is a perfusion-established *Haloferax volcanii* biofilm (alternative *Halobacterium salinarum*) that retains viability in brine and removes soluble organics through archaeal uptake and metabolism while generating EPS for stable adhesion. The material system is a living-ceramic scaffold, preferably porous alumina or SiC. The 3D architecture combines millimeter-scale flow channels with 50–300 µm interconnected pores. Deployed as a sealed, swappable service-core module (SLB+BRE), it is manufactured via composite-style porous ceramic fabrication and post-fabrication cell colonization (CMM) to achieve sustained polishing performance with controlled hydraulics and containment.

**Iteration 2**

**0) Code integrity check**

- Ontology code validity: All 7 codes appear as defined ontology codifications (Domain, Superkingdom, Kingdom, Phylum, Class, Application, Production).
- Multi-label validity: The "+" combination is explicitly allowed for categories in this codified ontology.
- Graph presence check: An ELM instance matching the full 7-slot combination is Not present in the ontology/graph. (Codes exist as ontology nodes; the *instantiated ELM* with this exact combination is absent in the graph.)

**1) Slot-by-slot decoding**

Slot 1 — Domain: ARC

- Meaning: Archaeal ELMs (ELMs populated by archaea).
- Design constraints
    - Biological: living component must be archaeal cells (no bacteria/eukaryotes as the primary chassis).
    - Material: must support archaeal attachment/colonization (surface chemistry + porosity compatible with archaeal biofilms).
    - Process: inoculation/operation conditions must be archaeal-compatible (often high salinity for haloarchaea, depending on chosen strain).

Slot 2 — Superkingdom: HYB

- Meaning: Hybrid living materials (cells + abiotic synthetic biomaterial/structure scaffold).
- Design constraints
    - Biological: cells must remain viable while attached to/embedded in an abiotic scaffold.
    - Material: an explicit abiotic scaffold is mandatory; it is not a purely biological ECM-only construct.
    - Process: scaffold fabrication and cell colonization are distinct steps (manufacture → colonize/seed).

Slot 3 — Kingdom: ARA

- Meaning: Archaebacterial ELMs within the archaeal domain, populated by archaebacteria.
- Design constraints
    - Biological: constrains the chassis to archaeal "archaebacteria" (in the ontology's terminology).
    - Containment: must treat release control as critical, since living cells are used in a building subsystem.

Slot 4 — Phylum: 3D

- Meaning: 3D ELMs (cells within a three-dimensional biomaterial or three-dimensional scaffold).
- Design constraints
    - Structural: a volumetric scaffold (monolith/foam/lattice) enabling internal colonization—not just a surface biofilm.
    - Transport: geometry must balance flow-through mass transport with protected niches for cells.

Slot 5 — Class: LCE

- Meaning: Living ceramics (extracellular matrices made predominantly of ceramics; e.g., oxides, silicates, carbides, nitrides).
- Design constraints
    - Material: scaffold must be ceramic-dominant (compositionally and mechanically).
    - Operation: ceramic robustness enables heat tolerance and repeated wet–dry cycles (aligning with the concept constraints).

Slot 6 — Application: SLB + BRE

- Meaning (SLB): Smart and living buildings (ELMs as subsystems of buildings).
- Meaning (BRE): Bioremediation (processing/recycling/revalorizing waste streams).
- Design constraints
  - System integration: must fit building-service-core constraints (modular, swappable, contained).
  - Function: must perform greywater polishing/waste processing with measurable remediation endpoints.

Slot 7 — Production: CMM

- Meaning: Composite manufacturing methods and cell colonization (composites-industry processes producing woven/porous/functionally graded synthetic scaffolds for later cell colonization).
- Design constraints
  - Process: scaffold must be manufacturable via composite-style routes (preforms, infiltration/impregnation, graded porosity), then colonized.
  - Scalability/QC: composite manufacturing implies reproducible geometry/porosity control at cartridge scale.

---

**2) Biological instantiation**

Candidate 1 — Haloferax volcanii (nonpathogenic model haloarchaeon)

- Functional role: aerobic heterotrophic DOC/COD polishing in saline greywater; stable biofilm formation within ceramic pores; nutrient assimilation into archaeal biomass (cell-centric “nutrient cycling” via uptake + biomass turnover).
- Viability / growth requirements (operational setpoints):
  - Salinity: ~1.7–2.5 M NaCl equivalent (high-salt dependence used as containment lever).
  - Temperature: ~30–45 °C (compatible with “hot” service-core fluids).
  - pH: near neutral to mildly alkaline (target 7–8.5).
  - Oxygen/nutrients: requires dissolved $O_2$ (passive aeration or microbubble injection); carbon from greywater organics; trace N/P.
- Biocontainment strategy:
  - Physical: sealed cartridge + downstream 0.2 μm-rated microfiltration at outlet; internal immobilization in tortuous porosity.
  - Ecological: obligate high-salt dependence (effluent dilution reduces survival), plus periodic high-salinity maintenance flush.

Candidate 2 — Halobacterium salinarum (extreme halophilic archaeon; nonpathogenic lab model)

- Functional role: robust saline-tolerant biofilm communities for organic polishing under very high salinity; tolerance to desiccation/UV stress can support wet–dry duty cycles.
- Viability / growth requirements (operational setpoints):
  - Salinity: typically higher than Haloferax (very high NaCl brines).
  - Temperature: ~30–45 °C.
  - Oxygen/nutrients: generally aerobic heterotroph; requires organics + $O_2$.
- Biocontainment strategy:
  - Strong brine dependence + sealed cartridge + outlet microfiltration; thermal regeneration option (ceramic allows high-temp sanitation when cartridge is removed/offline).

Why these are preferred over plausible alternatives

- Under the concept’s “hot + saline + low water activity” constraints, haloarchaea provide higher intrinsic tolerance than most conventional bacterial chassis (cell-centric viability constraint under salinity/heat). This aligns with ARC/ARA and with the stated desert-building greywater conditions.

- Within archaea, Haloferax volcanii is favored for controllable growth and biofilm formation under moderately extreme salinity, reducing the need for brine levels that could accelerate scaling/precipitation in plumbing.

Recommended candidate: Haloferax volcanii

- Selection rationale: achieves the saline/heat robustness goal while enabling operation at moderately high salinity, easing integration into a greywater loop (less aggressive brine handling than extreme-halophile-only operation).
- Trade-offs: may be less tolerant than H. salinarum at the highest salinities; mitigated by maintaining a controlled brine window and using modular cartridge swap-outs.

**3) Matrix/material candidates (Class = LCE)**

Material candidate 1 — Porous alumina (α-$Al_2O_3$) ceramic foam/monolith

- Composition: α-$Al_2O_3$ (optionally with a thin silica washcoat to tune surface hydroxylation).
- Bio-compatibility mechanism: rough, hydrophilic oxide surfaces support archaeal biofilm anchoring via EPS adsorption + mechanical interlocking into micro-porosity.
- Interface mechanisms: (i) micro-roughness-driven adhesion; (ii) electrostatic interactions on hydroxylated oxide surfaces; (iii) stable wetting in saline.
- Justification: chemically inert, high compressive strength, tolerates heat sanitation and wet–dry cycling (strong fit to “hot desert service cartridge”).

Material candidate 2 — Cordierite honeycomb ceramic ($2MgO \cdot 2Al_2O_3 \cdot 5SiO_2$)

- Composition: cordierite silicate ceramic honeycomb (monolithic channel structure).
- Bio-compatibility mechanism: silicate surface chemistry + engineered wall micro-porosity enabling protected colonization while maintaining low pressure drop.
- Interface mechanisms: silanol-rich surface regions + porous wall for internal attachment; optional thin ceramic washcoat for controlled roughness.
- Justification: widely used as durable monolith architecture; scalable modular fabrication with tight geometric tolerances.

Preferred material: Porous alumina for maximum chemical/thermal robustness and high internal surface area for colonization, with cordierite as a lower-cost, high-throughput monolith alternative.

**4) Geometry and structural instantiation (Phylum = 3D)**

3D reactor cartridge architecture: a stacked porous-ceramic monolith cartridge (replaceable module) enabling flow-through polishing plus protected internal niches for archaeal biofilms.

Quantitative geometry (implementable starting spec):

- Module form: cylindrical or prismatic monoliths, e.g., Ø 50–80 mm × 80–150 mm per module; 2–4 modules per cartridge.
- Flow channels (macro): parallel channels 0.8–1.5 mm hydraulic diameter to limit clogging.
- Wall micro-porosity: open porosity 35–55% with pore throat scale 10–80 µm for sheltered colonization (cells retained; convective refresh at pore entrances).
- Specific surface targeting: maximize internal surface while keeping pressure drop within building pump limits.

How geometry supports function

- Cell viability: interior micro-porosity provides shear protection; macro-channels deliver oxygenated flow.
- Mass transport: convective flow in channels + diffusion into porous walls creates a stable nutrient/oxygen gradient sustaining biofilms.
- Mechanical performance: monolith architecture resists compression and handling during cartridge swaps; ceramics tolerate thermal cycling.

Why 3D monolith over alternatives

- Compared to purely 2D coatings, the 3D monolith multiplies active area and provides protected microenvironments; compared to packed beds, it reduces channeling and simplifies containment/replaceability (SLB integration constraint).

**5) Application instantiation (SLB + BRE)**

Deployment scenario (smart building + bioremediation):

- Installed as a service-core greywater polishing cartridge downstream of a coarse screen and upstream of reuse (e.g., flushing/utility water), tailored for hot, saline/brackish greywater in desert buildings.

Measurable success metrics (examples suitable for verification):

- Organic polishing: % reduction in COD or TOC across cartridge (influent vs effluent).
- Solids/clarity: turbidity reduction and stable effluent clarity at target flow.
- Hydraulics: pressure drop maintained below an operational ceiling (cartridge health KPI).
- Biological stability: retained archaeal biomass within cartridge (no detectable cells in effluent beyond limit of detection after outlet filtration).

Main failure modes + mitigations

- Biofouling / pore clogging: mitigate with macro-channel sizing, periodic backflush, and cartridge swap schedule.
- Salt crystallization during dry cycles: mitigate with controlled drain-down, brine maintenance window, and geometry avoiding dead zones.
- Loss of viability due to salinity drift: mitigate by inline conductivity control and scheduled saline conditioning pulses.
- Microbial contamination (non-archaeal growth): mitigate via salinity selection pressure + periodic thermal sanitation (offline) enabled by ceramic robustness.

**6) Production instantiation (CMM)**

Production route consistent with CMM (composite manufacturing + colonization): fabricate a porous ceramic composite monolith using composites-style preforms/infiltration and graded porosity control, then colonize with haloarchaea.

Step-by-step workflow

1. Preform preparation (composites-style): form a woven or felted ceramic-fiber preform (e.g., alumina fiber felt) into monolith geometry (outer shell + internal channel former).
2. Ceramic slurry formulation: alumina (or cordierite precursor blend) + binder + sacrificial pore formers (controlled size distribution) to target dual-scale porosity.
3. Infiltration/impregnation: vacuum-assisted impregnation of the preform with slurry (analogous to resin infusion in composites) to yield a porous/functionally graded green body.
4. Curing + burnout: binder cure → controlled thermal burnout to remove organics/pore formers.
5. Sintering: sinter to final ceramic composite scaffold (QC: porosity, pore throat distribution, compressive strength).
6. Surface conditioning: aqueous rinse + mild surface activation (increase hydroxylated oxide sites); optional thin ceramic washcoat to tune roughness.
7. Sterilization (scaffold-only): heat sterilize/autoclave compatible with ceramics.
8. Biological preparation: culture Haloferax volcanii in defined high-salt medium to mid-log phase; prepare inoculum at controlled cell density.
9. Colonization/seeding: recirculate inoculum through monolith under controlled salinity/temperature with aeration for 24–72 h to establish stable biofilm loading.
10. Cartridge assembly + containment: integrate monolith modules into sealed cartridge housing; add outlet microfiltration barrier; leak/pressure test.
11. Quality control (release tests):
    - Geometry/porosity & pressure-drop curve (per module lot).
    - Biofilm loading proxy (e.g., protein/DNA content per module) and effluent sterility downstream of filter.

- Functional bench test on synthetic saline greywater (COD/TOC reduction at specified flow).

Manufacturability/scalability

- Composite-style preforms and infusion are batch-scalable; monolith modules can be standardized and swapped, aligning with SLB maintenance logic.

**7) Layered integrated summary of the fully instantiated ontology-derived ELM**

A. Rapid identity

- Ontology code: ARC–HYB–ARA–3D–LCE–SLB+BRE–CMM
- One-line ELM identity: Haloarchaeal biofilm–colonized porous ceramic composite monolith cartridge for saline greywater polishing as a replaceable smart-building bioremediation module.

B. Instantiation table

| Component | Final instantiation |
|---|---|
| Biological system | Haloferax volcanii (recommended); alternative: Halobacterium salinarum immobilized as internal biofilms within ceramic porosity |
| Functional role | Aerobic saline-tolerant heterotrophic polishing of dissolved organics (DOC/COD/TOC proxies), nutrient assimilation into biomass with controlled biomass turnover; stable biofilm-mediated retention in porous ceramic |
| Material system | Porous alumina (α-$Al_2O_3$) ceramic composite monolith (preferred); alternative: cordierite ($2MgO\cdot 2Al_2O_3\cdot 5SiO_2$) honeycomb with porous walls |
| Structural / geometric instantiation | 3D stacked monolith cartridge: macro-channels ~0.8–1.5 mm; wall open porosity ~35–55%; pore throats ~10–80 µm enabling protected internal colonization with flow-through transport |
| Target substrate / environment | Hot, saline/brackish greywater stream in desert-building service cores (wet–dry duty cycles, high fouling potential) |
| Application / deployment scenario | Replaceable building-service cartridge downstream of coarse filtration; conductivity-controlled operation, periodic backflush and swap schedule; sealed housing with outlet microfiltration containment |
| Production mechanism / workflow | CMM: ceramic-fiber preform + slurry infiltration (functionally graded porosity) → burnout/sinter → scaffold sterilization → controlled archaeal seeding/recirculation colonization → cartridge assembly + QC |
| Main functional outcome / performance objective | Stable, contained greywater polishing at service-core flow with measurable organic-load reduction while maintaining acceptable pressure drop and preventing viable-cell release |

C. Integrated paragraph

ARC–HYB–ARA–3D–LCE–SLB+BRE–CMM describes a haloarchaeal living-ceramic, 3D monolith cartridge that performs saline greywater polishing as a smart-building bioremediation subsystem. The biological chassis is a nonpathogenic haloarchaeon (recommended: *Haloferax volcanii*; alternative: *Halobacterium salinarum*) operating as an immobilized aerobic biofilm. The material system is a porous ceramic scaffold (preferred α-

$Al_2O_3$; alternative cordierite) configured as a stacked 3D flow-through monolith with millimetric channels and micrometric pore throats to couple convective transport with protected colonization niches. Production follows composite-manufacturing preform infiltration and sintering, followed by controlled archaeal seeding and sealed cartridge assembly with outlet microfiltration, targeting measurable organic-load reduction at bounded pressure drop with strict biocontainment.

**Iteration 3**

**0) Code integrity check**

- Ontology presence: All slot codes used in ARC–HYB–ARA–3D–LCE–SLB+BRE–CMM are explicitly defined in the ontology (Domain: ARC; Superkingdom: HYB; Kingdom: ARA; Phylum: 3D; Class: LCE; Application: SLB and BRE; Production: CMM).
- Multi-label application validity: The ontology explicitly allows combining categories using "+" in codifications, so SLB+BRE is valid.
- Ambiguity check:
  - ARC (Domain archaea) + ARA (Kingdom archaebacteria) is internally consistent (both point to archaea; ARA refines within ARC).
  - HYB requires an abiotic scaffold + cells (i.e., not cells-only).
- Graph status: This exact combination is not instantiated in the provided graph (consistent with "Gap status: absent"). (Derived directly from the supplied graph file.)

---

**1) Slot-by-slot decoding**

Slot 1 — Domain: ARC

- Meaning: Archaeal ELMs: ELMs populated by archaea.
- Design constraints
  - Biological: archaeal chassis only (no bacteria/eukaryotes as the primary living system).
  - Material/Process: must support archaeal survival and function in the chosen deployment (here: hot/saline greywater).

Slot 2 — Super-kingdom: HYB

- Meaning: Hybrid living materials: cells + abiotic synthetic biomaterial/structure (scaffold).
- Design constraints
  - Biological: cells must remain viable while attached to/within an abiotic scaffold.
  - Material: scaffold must be manufacturable and stable under operating conditions.
  - Process: an explicit colonization / seeding step is mandatory.

Slot 3 — Kingdom: ARA

- Meaning: Archaebacterial ELMs: archaeal-domain ELMs populated by archaebacteria.
- Design constraints
  - Biological: select nonpathogenic, cultivable archaea compatible with high salinity and warm temperatures.

Slot 4 — Phylum: 3D

- Meaning: 3D ELMs: cells in a 3D biomaterial or 3D scaffold.
- Design constraints
  - Structural: volumetric architecture with internal mass transport pathways.
  - Biological: geometry must maintain viability (nutrient/oxygen delivery; waste removal).
  - Mechanical: must tolerate flow, pressure drop, and handling as a cartridge.

Slot 5 — Class: LCE

- Meaning: Living ceramics: extracellular matrices made predominantly of ceramics (oxides/silicates/carbides/nitrides, etc.).
- Design constraints
  - Material: the dominant matrix/scaffold must be ceramic (not polymer-dominant).
  - Interface: the ceramic must provide anchoring sites for archaeal attachment/biofilm retention.

Slot 6 — Application: SLB+BRE

- Meaning:
  - SLB: Smart and living buildings.
  - BRE: Bioremediation (waste processing/recycling/revalorization).
- Design constraints
  - Deployment: building-service integration (replaceable module; monitored operation).
  - Function: measurable contaminant/organic load reduction consistent with "bioremediation".

Slot 7 — Production method: CMM

- Meaning: Composite manufacturing methods + cell colonization: composites-industry processes yielding woven/porous/functionally graded structures as synthetic scaffolds for later colonization.
- Design constraints
  - Manufacturing: scaffold fabrication should align with composite-style workflows (preforms, infiltration, graded architectures).
  - Biological integration: explicit post-fabrication colonization step.

---

**2) Biological instantiation**

Candidate A — *Haloferax volcanii* (haloarchaeon)

- Functional role (cell-centric):
  - Heterotrophic uptake/metabolism of soluble organics contributing to COD reduction in saline greywater.
  - Biofilm/EPS-driven retention inside ceramic porosity to prevent washout and stabilize reactor performance.
- Viability & growth requirements (operational targets):
  - High salinity: maintain brine-like ionic strength (e.g., ~2–4 M NaCl-equivalent).
  - Temperature: warm operation (e.g., ~30–45 °C) consistent with service-core "hot-to-warm" streams.
  - Oxygen: aerobic or microaerobic operation (design includes controlled aeration upstream if needed).
  - pH: near-neutral to moderately alkaline depending on detergent load (use buffering capacity in feed tank).
- Biocontainment strategy (non-speculative):

1. Intrinsic osmotic containment: haloarchaea lose viability/lyse if salinity drops sharply (built-in containment trigger during maintenance).
2. Physical retention: pore entrapment + downstream microfiltration screen (e.g., 0.2–0.45 µm) in the cartridge housing.
3. Service protocol containment: sealed "swap-and-cap" cartridge replacement; brine flush before removal.

Candidate B — *Halobacterium salinarum* (extreme halophile)

- Functional role (cell-centric):
  - Robust survival in very high salt; contributes to stable archaeal biomass under concentrated brines and thermal variability.
  - Organic uptake (where compatible substrates exist) and strong surface association for retention.
- Viability & growth requirements (operational targets):
  - Very high salinity (often at the upper end of brines), making it resilient but more sensitive to dilution events.
  - Temperature: warm range similar to many haloarchaea.
  - Oxygen: typically aerobic; requires oxygen management.
- Biocontainment strategy:
  - Same triad: osmotic fragility on dilution + physical retention + sealed cartridge handling.

Why these are preferred over plausible alternatives

- They are archaea (required by ARC–ARA) and are halophilic, aligning with the stated "hot, saline greywater" deployment (cell-centric fit: survival + metabolism under high ionic strength).
- Many non-halophilic archaea would be less compatible with detergent-rich, saline building streams (higher risk of washout or loss of activity).

Recommended candidate

Recommended: *Haloferax volcanii*

- Rationale vs *H. salinarum*: tends to be a more tractable, lab-robust haloarchaeal chassis for controlled colonization and stable operation; it supports a practical development pathway for repeatable cartridge seeding and performance tuning.
- Trade-offs: may be less tolerant than *H. salinarum* to extreme salinity spikes; mitigate with controlled salinity window in the building greywater loop.

---

**3) Matrix/material candidates**

Material A — Porous α-alumina ($Al_2O_3$) honeycomb/foam (ceramic)

- Compatibility with biology: chemically stable in hot brines; supports surface attachment and biofilm anchoring via microscale roughness and hydroxylated ceramic surfaces.
- Interface mechanisms: archaeal adhesion/biofilm retention via surface roughness + ionic interactions; internal porosity provides protected niches against shear.
- Why strong for LCE: alumina is a canonical oxide ceramic, matching "living ceramics."

Material B — Open-cell silicon carbide (SiC) foam (ceramic)

- Compatibility with biology: exceptional thermal and abrasion robustness; chemically resistant in challenging aqueous environments.
- Interface mechanisms: high-surface-area foam struts promote attachment; open-cell network reduces clogging risk at a given porosity.
- Why strong for LCE: SiC is explicitly within ceramic families (carbides) described under LCE.

Material selection justification (relative):

- $Al_2O_3$ is simpler, widely manufacturable, and cost-efficient for modular building cartridges.
- SiC is premium-grade for harsh duty cycles (abrasive particulates, higher temperatures), at higher cost.

---

**4) Geometry and structural instantiation**

3D architecture (Phylum: 3D): a flow-through ceramic reactor cartridge with multi-scale porosity.

- Macro-transport geometry (primary flow):
  - Extruded/formed honeycomb channels: 0.8–2.0 mm hydraulic diameter (flow distribution + low pressure drop).
  - Channel wall thickness: 200–400 µm (structural stiffness while preserving area).
- Colonization geometry (biological niches):
  - Interconnected macropores: 50–300 µm (biofilm residence + protected microenvironments).
  - Target open porosity: 60–80% (balance between surface area and clogging tolerance).
- How geometry supports function
  - Viability: convective delivery in channels + diffusive penetration into macropores maintains nutrient/oxygen access to cells while reducing shear at the attachment sites.
  - Mass transport: short diffusion lengths from channel lumen to pore network; enables consistent substrate exposure and waste removal.

- Mechanical performance: monolithic ceramic geometry withstands handling and cyclic wet–dry operation as a cartridge.

Why this geometry vs other plausible ones

- Compared to 2D biofilms on plates, a 3D porous monolith provides orders-of-magnitude higher retained biomass per footprint and better integration as a swappable building-service module (matches 3D + SLB).

---

**5) Application instantiation**

Deployment scenario (SLB+BRE):

A replaceable "service-core" cartridge installed in a smart building greywater loop handling hot, saline/brackish greywater (e.g., from showers/laundry in desert buildings using saline source water). The unit sits after coarse screening and upstream of storage/disinfection.

Measurable success metrics

- COD reduction: % COD decrease per pass and per day (influent vs effluent).
- Turbidity / suspended microdroplet capture: NTU reduction (biofilm-assisted capture in pores).
- Stability: time-to-performance decay (days/weeks) under steady salinity and temperature.
- Hydraulics: pressure drop at nominal flow; fouling rate ($\Delta P$/day).
- Biocontainment: viable cell count in effluent below a defined threshold.

Main failure modes + mitigations

- Salt dilution event → archaeal loss: mitigate with salinity-buffered equalization tank and conductivity monitoring (smart-building control).
- Biofouling/clogging: mitigate with graded porosity (larger upstream pores), periodic brine backflush, replaceable prefilter.
- Detergent/toxic shock: mitigate with upstream dilution/equalization and adsorption guard bed if needed.
- Mineral scaling (hot brines): mitigate with antiscalant dosing upstream or periodic acidified brine cleaning (performed offline, cartridge removed).

---

**6) Production instantiation**

Production route (CMM): composite-style porous/graded scaffold fabrication + post-colonization

Step-by-step workflow

1. Design & tooling
   - Define cartridge envelope (e.g., 100–150 mm length) with honeycomb channel layout and target pore gradients.
2. Composite-style ceramic preform formation (CMM-consistent)
   - Create a ceramic fiber preform (e.g., alumina fiber felt / SiC fiber network) using textile/composite handling (layup/needling/braiding) to establish a porous, shape-stable skeleton and enable functional grading.
3. Ceramic slurry infiltration
   - Vacuum-infiltrate the preform with an $Al_2O_3$ (or SiC-compatible) ceramic slurry containing fugitive porogens (polymer beads) to set macropore size targets.
4. Debinding + partial sintering
   - Thermal debind to remove organics and porogens; partial sinter to lock in interconnected porosity while maintaining high open volume.
5. Final ceramic stabilization
   - Final heat treatment to reach mechanical integrity for handling and flow, while preserving the designed porosity window.
6. Pre-use conditioning
   - Rinse with sterile high-salinity brine to equilibrate surface chemistry and remove fines.
7. Sterilization (pre-seeding)

- Apply sterilization compatible with ceramics (e.g., dry heat or steam) before biological loading.

8. Biological preparation
    - Grow the selected haloarchaeon (recommended *Haloferax volcanii*) in controlled high-salt medium to a defined cell density; prepare an inoculum for attachment.
9. Colonization / maturation (HYB requirement)
    - Seed the sterile cartridge in a recirculating bench bioreactor: controlled flow, salinity, temperature, and oxygenation to drive biofilm establishment inside pores.
10. Packaging for deployment

- Seal cartridge in a housing that maintains brine wetness and includes a downstream microfiltration screen; label batch metadata.

Manufacturability, scalability, QC

- Manufacturability: CMM-style preforms + infiltration scale from lab to pilot (repeatable porosity and geometry).
- Scalability: parallel production of identical cartridges; standardized seeding rigs.
- Quality control
    - Scaffold QC: pore-size distribution (micro-CT/mercury porosimetry), open porosity, compressive strength, pressure-drop curve.
    - Biological QC: attached biomass proxy (ATP/protein), effluent viable cell counts, baseline COD-removal rate in a standard test brine.

---

**7) Layered integrated summary of the fully instantiated ontology-derived ELM**

A. Rapid identity

- Ontology code: ARC–HYB–ARA–3D–LCE–SLB+BRE–CMM
- One-line ELM identity: Haloarchaeal biofilm–colonized porous living-ceramic (3D) cartridge that reduces soluble organics in hot saline greywater as a swappable smart-building bioremediation module.

B. Instantiation table

| Component | Final instantiation |
|---|---|
| Biological system | Primary: *Haloferax volcanii* biofilm within ceramic porosity; Alternative: *Halobacterium salinarum* for higher brine robustness |
| Functional role | Cell-centric COD reduction via archaeal uptake/metabolism; EPS/biofilm retention for biomass stability and microdroplet/particulate capture in pores |
| Material system | Living ceramic scaffold: porous α-$Al_2O_3$ honeycomb/foam (baseline) or open-cell SiC foam (premium thermal/abrasion robustness) |
| Structural / geometric instantiation | 3D flow-through cartridge with 0.8–2.0 mm channels + interconnected 50–300 µm colonization pores; target open porosity 60–80% (optionally graded upstream-to-downstream) |
| Target substrate / environment | Hot-to-warm, saline/brackish building greywater streams (service-core plumbing loop) |
| Application / deployment scenario | Replaceable cartridge in smart-building greywater line after coarse screening; conductivity/temperature monitored; periodic brine backflush and scheduled swap-out |
| Production mechanism / workflow | CMM-consistent: ceramic fiber preform formation → ceramic slurry infiltration + fugitive porogens → debind/partial sinter → sterilize → archaeal seeding in recirculating brine bioreactor → sealed cartridge packaging |
| Main functional outcome / performance objective | Stable, contained archaeal biomass enabling measurable COD reduction with controlled pressure drop and minimal cell release to effluent |

C. Integrated paragraph

ARC–HYB–ARA–3D–LCE–SLB+BRE–CMM: a haloarchaeal biofilm–colonized 3D living-ceramic cartridge for smart-building bioremediation of hot saline greywater. The biological system uses *Haloferax volcanii* (alt. *Halobacterium salinarum*) retained as a pore-anchored biofilm to drive cell-centric uptake and metabolism of soluble organics, reducing COD while stabilizing biomass against washout. A porous α-$Al_2O_3$ (or SiC) ceramic scaffold provides a 3D flow-through honeycomb with interconnected macropores for protected colonization and convective–diffusive mass transport. Production follows composite-style ceramic preform/infiltration and post-fabrication archaeal colonization, yielding a swappable, contained reactor module with defined COD-removal and hydraulic performance targets.

**ELM candidate 3: CRD–HYB–CRK–CP–LCO–SLB–CMM+BMN**

**Iteration 1**

**0) Code integrity check**

- Provided code: CRD–HYB–CRK–CP–LCO–SLB–CMM+BMN
- Ontology validity: Each code is present in the ontology and appears in the expected slot: CRD (Domain), HYB (Super-kingdom), CRK (Kingdom), CP (Phylum), LCO (Class), SLB (Application), CMM+BMN (Production).
- Ambiguity: The ontology defines CRD and CRK as "cross-domain" and "cross-kingdom", but it does not specify *which* domains/kingdoms must be paired. Not present in the ontology/graph.

---

**1) Slot-by-slot decoding**

Slot 1 — Domain: CRD

- Meaning: Cross-domain ELMs (cells from different cellular domains).
- Design constraints
  - Biological: at least two domains → co-culture compatibility, differential stress tolerance, interspecies antagonism control.
  - Material: must provide segregated micro-niches (e.g., pore gradients) to buffer domain-specific requirements.
  - Structural/geometric: needs multi-compartment micro-architecture to prevent competitive exclusion.
  - Process: staged inoculation likely (one domain establishes first, then the other).

Slot 2 — Super-kingdom: HYB

- Meaning: Hybrid living materials (cells + abiotic synthetic biomaterial/structure "scaffold").
- Design constraints
  - Biological: survival under scaffold chemistry (pH/ions), adhesion to abiotic interfaces, dormancy strategy.
  - Material: scaffold is mandatory; must support colonization and mass transport.
  - Structural: must include connected porosity or interfacial area for cell retention.
  - Process: scaffold fabrication and cell colonization are separable steps (fits "colonization" framing).

Slot 3 — Kingdom: CRK

- Meaning: Cross-kingdom ELMs (cells from different kingdoms).
- Design constraints
  - Biological: at least two kingdoms → functional division of labor; manage metabolite exchange and competition.
  - Material: interfaces must enable both cell wall types (e.g., bacterial adhesion + fungal hyphal anchoring).
  - Structural: benefit from both microcolonies (bacteria) and filament networks (fungi).

Slot 4 — Phylum: CP

- Meaning: Cross-phylum ELMs (structures combining 0D/1D/2D/3D/4D/fractal; multi-scale not necessarily fractal).
- Design constraints
  - Structural/geometric: multi-scale architecture required (e.g., 0D capsules + 3D pores + 2D coating).
  - Mass transport: must maintain water/oxygen access while limiting wash-out.
  - Mechanical: must handle cracking/abrasion via hierarchical reinforcement.

Slot 5 — Class: LCO

- Meaning: Living composites (ECM/matrix predominantly composite; e.g., polymer-, ceramic-, metal-matrix composites).
- Design constraints
  - Material: composite matrix must dominate (binder + fillers/fibers); living fraction is embedded/colonizing.
  - Interface: must promote mineral nucleation + polymer/EPS bonding.
  - Mechanical: composite must resist erosion/sand abrasion while permitting micro-habitats.

Slot 6 — Application: SLB

- Meaning: Smart and living buildings (ELMs as subsystems for improved building functionality/healthier environments).
- Design constraints
  - Deployment: façade/exterior layer integration, maintainability, serviceability.
  - Performance: dust suppression + surface stabilization + self-healing functionality, measurable at building scale.
  - Safety: containment and non-pathogenic choices favored.

Slot 7 — Production: CMM+BMN

- Meaning: CMM: composite manufacturing methods + cell colonization; BMN: biomineralization; "+" indicates combined production routes are allowed.
- Design constraints
  - Process: composite scaffold fabrication (CMM) + colonization, plus mineral precipitation cycles (BMN).
  - QC: must quantify composite properties and biomineral output (mass gain, crack closure, cohesion).

---

**2) Biological instantiation**

Candidate set A (mineralizer + adhesive/bridging fungus)

1. Sporosarcina pasteurii (bacterium; urease-driven $CaCO_3$ precipitation)
- Functional role: biomineralization to seal microcracks and cement near-surface sand grains (BMN).
- Viability needs: intermittent moisture; carbon/nitrogen source; $Ca^{2+}$ availability; tolerates moderate temperature swings; benefits from protected pores.
- Biocontainment: encapsulate as spores or dried cells in mineral-compatible microcapsules; nutrient-limited activation only during service misting; physical confinement within the render porosity.

2. Aureobasidium pullulans (fungus/yeast-like; strong adhesion/EPS production)
- Functional role: surface adhesion and extracellular polymer (e.g., pullulan-like EPS) for dust binding; hyphae/filaments can bridge microcracks and anchor into pores.
- Viability needs: episodic moisture; low nutrient requirement; tolerates desiccation; benefits from shaded/porous microhabitats.
- Biocontainment: immobilization in the pore network; low free-water availability limits dispersion; periodic "on-demand" feeding (minimal) via façade service cycles.

Why preferred (cell-centric): clear division of labor aligned to the concept brief: bacteria provide mineral (stiff crack sealing + sand cementation) while fungus provides binding/bridging (EPS + filament reinforcement) without requiring speculative genetic engineering.

---

Candidate set B (spore-former mineralizer + robust filament former)

1. Lysinibacillus sphaericus (spore-forming bacterium; mineral precipitation capability in alkaline niches)
- Functional role: mineral precipitation for cohesion and crack filling; sporulation supports desert cycling.
- Viability needs: moisture pulses; protected pores; alkaline-tolerant microenvironment.

- Biocontainment: spores in microcapsules; activation only with dosed nutrient solution.

2. Trichoderma harzianum (filamentous fungus; strong colonization and hyphal growth)

- Functional role: hyphal crack spanning; matrix reinforcement; surface bio-binding via secreted polymers/proteins.
- Viability needs: humidity pulses; protected porosity; moderate temperature.
- Biocontainment: physical immobilization within near-surface porosity; nutrient limitation.

Trade-offs vs A: potentially stronger filament reinforcement, but higher risk of overgrowth/bioload if nutrients are not tightly controlled.

---

Recommended candidate
Recommended: Sporosarcina pasteurii + Aureobasidium pullulans

- Why: best-aligned to BMN-driven self-healing + EPS-driven dust suppression with controllable, moisture-triggered activation (spores + desiccation tolerance).
- Key trade-off: requires careful control of the "mineralization feed" ($Ca^{2+}$/substrate dosing) to avoid surface efflorescence and uneven deposition.

---

**3) Matrix/material candidates**
Material candidate 1 — Polymer-modified mineral render composite (LCO-compatible)

- Composition: lime-based mineral binder ($Ca(OH)_2$ / $CaCO_3$ system) + silica sand/mineral fillers + small fraction of waterborne acrylic/latex polymer modifier + short microfibers (cellulose or glass).
- Bio-compatibility: porous mineral phase provides protected niches; polymer fraction improves adhesion and reduces dusting; compatible with fungal EPS anchoring.
- Interface mechanisms: $CaCO_3$ fillers serve as nucleation sites for bacterial mineral deposition; fungal EPS provides additional particle binding and interfacial "glue" to polymer-modified domains.

Material candidate 2 — Magnesium phosphate cement (MPC)–polymer composite

- Composition: $MgO–KH_2PO_4$ cementitious binder + silica/mineral fillers + polymer modifier + microfibers.
- Bio-compatibility: faster set and potentially milder local chemistry than highly alkaline Portland-type systems; supports post-curing colonization.
- Interface mechanisms: mineral binder provides rigid scaffold; polymer + EPS improves abrasion resistance and dust capture; biomineralization adds secondary $CaCO_3$ sealing in cracks/pores.

Preferred material: Candidate 1 (polymer-modified lime render composite) because it naturally supports $CaCO_3$-based biomineral sealing while remaining a clear living composite (binder + fillers + polymer) consistent with LCO and staged colonization consistent with HYB/CMM.

---

**4) Geometry and structural instantiation**
CP → multi-scale (cross-phylum) render architecture
Quantitative geometry (serviceable façade "living render")

- 2D layer (sheet-like): 2–5 mm thick exterior render coat over masonry/concrete/adobe.
- 3D porous scaffold: interconnected porosity ~10–25 vol% with characteristic pore throats ~20–200 µm (micro-habitat for bacteria/fungal fragments; diffusion paths).
- 0D features: mineral-compatible microcapsules (50–300 µm) preloaded with bacterial spores and minimal nutrients (activated only when wetted).
- 1D reinforcement: dispersed microfibers (length 1–5 mm; diameter 10–30 µm) + in situ fungal hyphal filaments to span microcracks.
- Surface microtexture: roughness scale 50–500 µm to trap dust/sand fines and retain transient moisture films.

Why this geometry is optimal (cell-centric + performance)

- Cell viability: pores and capsules buffer desiccation and UV exposure; staged wetting activates metabolism only when needed.
- Mass transport: thin layer + connected porosity supports oxygen diffusion and episodic water transport while limiting washout.
- Mechanical: hierarchical reinforcement (fibers + hyphae + mineral infill) targets crack arrest and abrasion resistance better than a single-scale coating.

**5) Application instantiation**

Deployment scenario (SLB)

A protected, serviceable exterior render on desert-building façades and sand-exposed perimeter surfaces, designed to (i) suppress dust release, (ii) stabilize near-surface sand grains, and (iii) self-heal microcracking driven by thermal cycling and abrasion.

Measurable success metrics

- Dust suppression: % reduction in airborne particulate release under standardized air-jet abrasion.
- Sand stabilization: increase in near-surface shear strength / erosion threshold (wind tunnel or impingement tests).
- Self-healing: crack closure fraction (%) for 50–300 µm cracks after N wetting cycles; recovery of flexural strength (%).
- Mineralization: $CaCO_3$ mass gain (mg/cm²) per activation cycle; pore-blocking bounded to preserve breathability.

Main failure modes + mitigations

- Over-mineralization/efflorescence: mitigate via low-dose, pulsed feeding and limiting soluble salts.
- Desiccation kill-off: mitigate via spore-based storage + sheltered porosity + activation only during controlled misting.
- Bio-overgrowth/bioload: mitigate via nutrient limitation and short activation windows.
- UV/thermal degradation: mitigate with shallow “active zone” (top 1–2 mm) as sacrificial layer, replaceable in maintenance cycles.

**6) Production instantiation**

Production route consistent with CMM+BMN: composite manufacture + post-colonization + mineralization cycles.

Step-by-step workflow

1. Composite scaffold formulation (CMM)
    - Mix mineral binder + graded mineral fillers + polymer modifier.
    - Add microfibers and controlled porogen/air-entrainment to target 10–25% porosity.
2. Render application
    - Apply 2–5 mm coat by trowel/spray onto primed façade substrate (serviceable outer skin design).
3. Cure (non-living stage)
    - Standard curing to reach handling strength while keeping connected porosity open (avoid high-temperature curing).
4. Staged colonization (CMM “cell colonization”)
    - Post-seed the cured porous layer by spraying a thin aqueous carrier containing:
        - bacterial spores (mineralizer) + fungal inoculum,
        - plus a minimal, quickly exhaustible carbon source.
    - Allow 12–48 h humid incubation (temporary cover film) to anchor cells into pores; then dry to dormancy.
5. Biomineralization activation (BMN)
    - In controlled maintenance cycles, mist with a low-dose mineralization feed ($Ca^{2+}$ source + compatible substrate) for 1–3 short wetting events, then dry.

- Repeat cycles only when crack/dust metrics indicate need (condition-based maintenance).

Manufacturability, scalability, QC

- Manufacturability: standard façade render processing; biology added post-cure to avoid viability loss.
- Scalability: compatible with spray/trowel at building scale; inoculum delivery by maintenance systems.
- Quality control
  - Composite: porosity, adhesion strength, abrasion loss.
  - Biology: retained viable counts after drying; urease/mineralization activity proxy; fungal adhesion/coverage in the active zone.
  - Function: standardized crack-healing and erosion tests after defined activation cycles.

---

**7) Layered integrated summary of the fully instantiated ontology-derived ELM**

A. Rapid identity

- Ontology code: CRD–HYB–CRK–CP–LCO–SLB–CMM+BMN
- One-line ELM identity: Cross-domain, cross-kingdom bacterial–fungal living composite façade render that self-heals by biomineral crack sealing and suppresses dust/sand erosion in smart-building envelopes.

B. Instantiation table

| Component | Final instantiation |
|---|---|
| Biological system | Recommended: *Sporosarcina pasteurii* (biomineralizing bacterium) + *Aureobasidium pullulans* (adhesive/EPS-producing fungus). Alternative: *Lysinibacillus sphaericus* + *Trichoderma harzianum*. |
| Functional role | Bacteria: $CaCO_3$ precipitation for crack sealing + sand-grain cementation. Fungi: EPS-driven dust binding + filament/hyphal crack spanning + adhesion to composite microtexture. |
| Material system | Polymer-modified mineral render composite: lime/$CaCO_3$–silica fillers + waterborne polymer modifier + microfibers; designed as a porous colonizable scaffold (living composite). |
| Structural / geometric instantiation | CP multi-scale: 2–5 mm 2D coating; 10–25% interconnected 3D porosity (20–200 µm throats); 0D spore microcapsules (50–300 µm); 1D microfiber + fungal hyphal reinforcement; surface microtexture 50–500 µm. |
| Target substrate / environment | Desert-building exterior façades and sand-exposed perimeter surfaces (thermal cycling, abrasion, episodic humidity). |
| Application / deployment scenario | Serviceable façade "living render skin" with condition-based activation (brief misting cycles) to trigger mineral sealing and maintain low dust emission and surface cohesion. |
| Production mechanism / workflow | CMM: composite scaffold manufacture + curing → post-colonization by spray-inoculation and humid anchoring → drying to dormancy. BMN: periodic low-dose mineralization feed to precipitate $CaCO_3$ in cracks/pores. |
| Main functional outcome / performance objective | Reduce dust emission and wind erosion while achieving repeatable microcrack closure (50–300 µm) and measurable recovery of surface cohesion/abrasion resistance after activation cycles. |

C. Integrated paragraph

CRD–HYB–CRK–CP–LCO–SLB–CMM+BMN: a cross-domain, cross-kingdom bacterial–fungal living composite façade render for smart-building envelopes that self-heals and stabilizes sand-exposed surfaces. The biological system pairs a biomineralizing bacterium (e.g., *Sporosarcina pasteurii*) with an adhesive, EPS-forming fungus (e.g., *Aureobasidium pullulans*) to couple $CaCO_3$ crack sealing with dust-binding and filament reinforcement. A CP multi-scale architecture combines a 2–5 mm coating, interconnected micro-porosity, and microcapsule/fiber features to retain cells and enable diffusion. Deployed as a serviceable outer skin, it is produced via composite scaffold fabrication and post-colonization, then activated by controlled biomineralization cycles to reduce erosion and close microcracks.

**Iteration 2**

**0) Code integrity check**

Input 7-slot code: CRD–HYB–CRK–CP–LCO–SLB–CMM+BMN

- Ontology validity: All tokens (CRD, HYB, CRK, CP, LCO, SLB, CMM, BMN) are defined ontology codes.
- Multi-label validity in Production: The ontology explicitly allows combining categories using "+" in the codification of an ELM, so CMM+BMN is structurally valid.
- Ambiguity check: None of the slots require reinterpretation. The bacterial–fungal concept matches Cross-domain (bacteria + eukarya) and Cross-kingdom (eubacteria + fungi) simultaneously.
- Graph presence: The individual code nodes exist in the graph, but an ELM instance node connected to *all* these codes as a single instantiation is not present in the provided graph (i.e., this remains a gap hypothesis rather than a literature-instantiated node).

---

**1) Slot-by-slot decoding**

Slot 1 — Domain: CRD

- Meaning: *Cross-domain ELMs* = ELMs populated by different cellular domains.
- Design constraints
    - Biological: Must include cells from ≥2 domains (practically: bacteria + eukarya for bacterial–fungal pairing).
    - Material: Must support distinct domain-level tolerances (e.g., bacterial spores + fungal propagules) within one matrix.
    - Structural: Micro-niches may be required (separate microenvironments for bacterial mineralization vs fungal reinforcement).
    - Process: Integration and stabilization must preserve viability of both domains during mixing/curing.

Slot 2 — Super-kingdom: HYB

- Meaning: *Hybrid living materials (HLMs)* = cells + an abiotic synthetic biomaterial/structure (scaffold).
- Design constraints
    - Biological: Cells must remain functional while interfacing with an abiotic scaffold/binder.
    - Material: Requires a defined abiotic render/binder phase (polymer/inorganic binder + aggregates) acting as scaffold.
    - Structural: Scaffold must enable long-term retention (anti-washout) and controlled access to water/nutrients.
    - Process: Manufacturing resembles composite formulation + inclusion of living fraction (e.g., encapsulated spores).

Slot 3 — Kingdom: CRK

- Meaning: *Cross-kingdom ELMs* = ELMs combining cells from different kingdoms.
- Design constraints
    - Biological: Must include ≥2 kingdoms (here: Eubacteria + Fungi).
    - Material: Matrix must support a bacterial-driven function (biomineralization) plus fungal-driven function (reinforcement/adhesion).
    - Structural: Needs spatial coexistence without competitive collapse (e.g., bacteria in microcapsules; fungi in percolating micro-voids).
    - Process: Co-processing rules: shear, temperature, pH shocks, and desiccation must be survivable for both.

Slot 4 — Phylum: CP

- Meaning: *Cross-phylum ELMs* = structures combining 0D/1D/2D/3D/4D/fractal phyla; also multi-scale non-fractal structures.
- Design constraints

- Biological: Must support multi-scale colonization (microcolonies + filament networks).
  - Material: Must be architected across scales (microcapsules + fibers + porous render).
  - Structural: Must explicitly combine at least two structural “phyla” (e.g., 2D render layer + 0D capsules + 1D filaments + 3D pore network).
  - Process: Composite manufacturing must preserve the designed multi-scale architecture (avoid capsule rupture, preserve porosity).

Slot 5 — Class: LCO

- Meaning: *Living composites* = extracellular matrices made predominantly of composites (e.g., polymer-, ceramic-, metal-matrix composites).
- Design constraints
  - Biological: Cells must function within/against a composite matrix (interfaces, nucleation sites, anchoring).
  - Material: Must specify at least two phases (e.g., polymeric binder + mineral aggregate/filler; plus biogenic mineral).
  - Structural: Load transfer + erosion resistance should arise from the composite architecture, not a single phase.
  - Process: Composite mixing/layup/spray/trowel approaches are appropriate.

Slot 6 — Application: SLB

- Meaning: *Smart and living buildings* = ELM subsystems used in buildings, typically for improved sensing/actuation and healthier functionality.
- Design constraints
  - Biological: Must remain safe and stable in a built-environment context; containment is central.
  - Material: Weathering, UV, sand abrasion, salt, thermal cycling constraints dominate.
  - Structural: Must be deployable as façade render / surface skin / serviceable layer.
  - Process: Must be manufacturable at building-material scale (kg–ton), and installable by common construction methods.

Slot 7 — Production: CMM+BMN

- Meaning:
  - CMM: Composite manufacturing methods + cell colonization (woven/porous/graded scaffolds as synthetic scaffolds for cells).
  - BMN: Biomineralization = cells generate minerals; applicable across kingdoms; relevant to HLMs and bio-ELMs.
- Design constraints
  - Biological: Must include a mineral-forming cell function (bacterial or fungal biomineralization; here primarily bacterial).
  - Material: Must provide mineral precursors (e.g., $Ca^{2+}$ source; carbonate generation pathway) and nucleation surfaces.
  - Structural: Must enable mineral deposition where it matters (crack faces; sand–binder interfaces).
  - Process: Must describe a feasible composite workflow + a post-installation activation route for biomineralization.

---

**2) Biological instantiation**

Note: The ontology/graph do not specify organism species; explicit organism selection is Not present in the ontology/graph. The candidates below are engineering instantiations consistent with CRD–HYB–CRK–CP–LCO–SLB–CMM+BMN.

Candidate 1 (bacteria–fungus consortium)

Bacterium: *Sporosarcina pasteurii* (spore-forming, ureolytic biomineralizer)

Fungus: *Aureobasidium pullulans* (EPS/adhesion-focused fungus; biofilm-forming)

- Functional role

- *S. pasteurii:* moisture-triggered germination → ureolysis → carbonate availability → $CaCO_3$ precipitation at crack and sand-grain contacts (self-healing + sand binding).
    - *A. pullulans:* adhesive extracellular polymers (EPS) + surface colonization → particle binding, microcrack "bridging" by biomass/EPS, and improved retention of bacterial micro-niches.
- Viability / growth requirements
    - Aerobic; moisture pulses required (activation events).
    - Nutrients must be micro-dosed (embedded micro-reservoirs) to avoid uncontrolled proliferation.
    - Temperature: designed for exterior façade range (e.g., warm climates); desiccation tolerance handled via dormancy/encapsulation.
- Biocontainment strategy
    - Physical containment: microencapsulation of bacterial spores (and optionally fungal propagules) in semi-permeable polymer–silica capsules embedded in render.
    - Nutrient limitation: internal "single-shot" nutrient beads (finite urea + Ca-salt) so activity self-terminates when depleted.
    - Spatial separation: bacteria primarily inside capsules; fungus preferentially in near-surface microvoids to reduce direct competition.

Why preferred vs plausible alternatives

- Strong division of labor: bacterial mineralization + fungal adhesion/EPS, matching the concept's "mineralize + reinforce" logic.
- Both can be handled as dormant/contained components compatible with building-material deployment constraints.

Candidate 2 (bacteria–fungus consortium)

Bacterium: *Bacillus subtilis* (spore-forming; biofilm/EPS; mineralization-capable strains)

Fungus: *Trichoderma reesei* (filamentous fungus; hyphal network reinforcement)

- Functional role
    - *B. subtilis:* spore-based persistence; biofilm/EPS to retain fines; contributes to mineral nucleation (secondary biomineralization support).
    - *T. reesei:* hyphal (1D) reinforcement network within microcracks and pores; improves toughness and crack-spanning capacity.
- Viability / growth requirements
    - Aerobic; moisture pulses; nutrients must be limited.
    - Filamentous growth requires connected microvoids and water activity above a threshold during activation windows.
- Biocontainment strategy
    - Encapsulation of spores; fungal growth constrained by restricted pore connectivity and nutrient access.

Why preferred vs plausible alternatives

- More explicit hyphal reinforcement (mechanical crack-bridging) than yeast-like fungi.
- Trade-off: filamentous growth is harder to control; greater risk of visible growth if containment fails.

Recommended candidate (selected)

Recommended: Candidate 1 — *Sporosarcina pasteurii* + *Aureobasidium pullulans*

- Selection justification: Maximizes controllable BMN-driven self-healing (bacterial $CaCO_3$) while using fungal EPS primarily for adhesion/retention rather than aggressive hyphal expansion (lower "escape/visibility" risk in façades).
- Trade-offs: Less "fiber-like" reinforcement than a strongly filamentous mycelium; mitigation is provided structurally (microfibers + graded porosity) and materially (composite binder toughening).

---

**3) Matrix/material candidates**

The ontology defines LCO (living composites) but does not prescribe specific binders; explicit compositions are Not present in the ontology/graph. The candidates below satisfy HYB + LCO + SLB and enable CMM+BMN.

Material candidate A — Breathable polymer–mineral render composite

- Composition (explicit):
  - Water-based siloxane-modified acrylic binder (polymer matrix)
  - Graded silica sand (150–600 µm) + silica fume (microfiller) as mineral phases
  - Ca-source microreservoirs (e.g., Ca-lactate or Ca-acetate beads) for biomineralization feedstock
  - Polymer–silica microcapsules (50–300 µm) containing bacterial spores (and optionally fungal propagules)
- Bio-compatibility:
  - Polymer binder is comparatively mild vs high-alkaline cement; capsules shield cells from shear and desiccation.
- Interface mechanisms:
  - $CaCO_3$ precipitates preferentially on silica/filler surfaces (nucleation) and within crack voids.
  - Fungal EPS increases adhesion to sand grains and anchors to polymer matrix.

Why select vs alternatives: Best fit when façade requirements prioritize dust suppression + abrasion resistance with controlled living activity in a thin layer.

Material candidate B — Mineral-rich geopolymer–latex composite render

- Composition (explicit):
  - Metakaolin-based geopolymer (alkali-activated aluminosilicate) as primary mineral binder
  - Acrylic latex modifier (polymer phase) for toughness + crack resistance
  - Silica sand + fine aluminosilicate fillers
  - Encapsulated spores in alkali-resistant microcapsules (e.g., silica-shell capsules)
- Bio-compatibility:
  - Fresh geopolymer alkalinity is challenging; viability requires robust encapsulation and delayed activation (after cure/carbonation).
- Interface mechanisms:
  - Mineral matrix provides abundant nucleation surfaces; biomineralization preferentially seals microcracks and pores.

Why select vs alternatives: Best fit when façade requirements prioritize mineral durability; comes with higher burden on encapsulation and activation control.

Recommended material system: Candidate A (polymer–mineral breathable composite) for better compatibility with controlled biological activation in building skins.

---

**4) Geometry and structural instantiation**

Phylum = CP (Cross-phylum / multi-scale) requires a combined architecture across scales.

Proposed CP architecture (quantified, multi-scale)

1. 2D layer (render skin):
   - Thickness: 3–6 mm on exterior substrate (masonry/earthen façade).
   - Surface texture: micro-roughness 50–300 µm to retain fines and reduce wind shear slip.
2. 3D porous micro-network (within the layer):
   - Target open porosity: 15–30 vol%
   - Interconnected pore throats: 20–150 µm (supports transient moisture percolation and gas diffusion).
3. 0D living microreactors (capsules):
   - Capsule diameter: 50–300 µm (bacterial spores + finite nutrient/Ca source co-located).
   - Areal density: $10^6$–$10^7$ capsules/m² (tunable by desired healing capacity).

4. 1D reinforcement (biological + abiotic):
    - Biological: fungal biomass/EPS in microvoids and crack faces (candidate-dependent).
    - Abiotic: short mineral microfibers (e.g., basalt fibers) 3–6 mm length, 10–20 µm diameter to provide crack-bridging that does not depend on uncontrolled fungal hyphae.

How geometry supports function

- Cell viability: Capsules create protected microenvironments; pores enable $O_2$ diffusion; activation occurs during moisture pulses.
- Mass transport: Moisture ingress carries dissolved $Ca^{2+}$/carbonate precursors to crack tips; pores avoid complete sealing before healing.
- Mechanical performance: Composite layer resists abrasion; fibers control shrinkage cracking; biomineralization densifies microcrack zones.

Why this CP structure vs simpler geometries: A single-phase 2D coating lacks protected niches and healing reservoir; a purely 3D bulk stabilizer is unnecessary for façade skins. CP enables targeted, serviceable, thin-layer performance.

---

**5) Application instantiation**

Application = SLB (Smart and living buildings).

Deployment scenario

- Where: Exterior façade zones exposed to wind-driven sand, dust, and thermal cycling (desert/peri-desert buildings).
- How used: A protected living render skin applied as a serviceable outer layer (replaceable "sacrificial" coat). Moisture events (rain, fog, scheduled misting) trigger episodic self-healing and sand binding.

Measurable success metrics

- Erosion resistance: mass loss under standardized sand abrasion ($g/m^2$ per test interval).
- Dust suppression: PM10/PM2.5 emission reduction from surface under wind shear ($\mu g/m^3$ near-surface plume).
- Self-healing efficiency: % crack closure for 50–500 µm cracks after N wetting cycles; regained surface tensile strength (pull-off adhesion).
- Biomineral output: $CaCO_3$ mass fraction or mineral deposition rate localized to crack region ($mg/cm^2$ per activation cycle).

Failure modes + mitigations

- Desiccation kill-off: mitigate via spore-based biology + microcapsules + activation-by-pulse.
- Uncontrolled growth/visibility: mitigate via nutrient limitation + physical containment + breathable topcoat.
- Salt crystallization/efflorescence: mitigate via controlled ion dosing, capillary-break design, and limiting water residence time.
- Loss of adhesion to substrate: mitigate via primer layer and graded modulus (softer outer skin, tougher inner interface).

---

**6) Production instantiation**

Production = CMM+BMN requires composite manufacturing + biomineralization.

Step-by-step implementable workflow (CMM+BMN)

1. Microcapsule fabrication (living fraction packaging)
    - Prepare dry bacterial spores (and optional fungal propagules) and blend with finite nutrient + Ca-source microbeads.
    - Encapsulate into polymer–silica microcapsules (50–300 µm) with semi-permeable shell (water/ions pass; cells retained).
2. Composite render base formulation (abiotic scaffold)
    - Mix siloxane-acrylic binder + graded silica sand + silica fume (target rheology for trowel/spray).

- Add short reinforcement fibers (basalt) for shrinkage/crack control.

3. Integration (low-shear mixing)
   - Add capsules at the final mixing stage under low shear to reduce rupture.
   - Confirm capsule integrity by sieve/optical sampling (QC gate).
4. Application to substrate
   - Apply primer (compatibility layer), then apply render 3–6 mm.
   - Tool surface texture to target roughness (dust retention + lower wind erosion).
5. Cure + stabilization
   - Allow binder cure (polymer film formation) under ambient conditions.
   - Optional breathable protective overcoat with controlled permeability (reduces UV damage while preserving gas/moisture exchange).
6. Activation protocol (BMN)
   - Commissioning activation via brief wetting (misting) to initiate localized biomineralization.
   - Re-activation occurs only during subsequent moisture events; nutrients are finite → activity tapers.

Manufacturability / scalability

- Compatible with existing construction workflows: premixed bags (dry fillers + capsules) + on-site addition of binder/water.
- Capsule production is the main scale driver; can be industrialized as a discrete supply chain component.

Quality control (reproducibility)

- Capsule size distribution + rupture rate (microscopy/sieving).
- Viability after mixing (CFU/spore germination checks from sacrificial samples).
- Standardized crack-healing test coupons (controlled crack width; wetting cycles; strength recovery).
- Mineral phase verification in healed zones (simple acid reaction screening + microscopy; advanced methods optional).

Why this route is feasible/preferred: It directly matches CMM (composite formulation + graded microstructure) and realizes BMN through protected, moisture-triggered mineralization without requiring continuous cultivation infrastructure.

---

**7) Layered integrated summary of the fully instantiated ontology-derived ELM**

A. Rapid identity

- Ontology code: CRD–HYB–CRK–CP–LCO–SLB–CMM+BMN
- One-line ELM identity: Encapsulated bacterial–fungal hybrid living composite render that episodically biomineralizes and binds sand at façade microcracks, enabling self-healing and dust suppression for sand-exposed smart building skins.

B. Instantiation table

| Component | Final instantiation |
| --- | --- |
| Biological system | Recommended: *Sporosarcina pasteurii* (spores, encapsulated) + *Aureobasidium pullulans* (EPS/adhesion, constrained microvoid colonization). Alternative: *Bacillus subtilis* (spores) + *Trichoderma reesei* (filamentous reinforcement, higher containment burden). |
| Functional role | Bacterial ureolysis-driven $CaCO_3$ biomineralization for crack sealing + sand-grain cementation; fungal EPS/biomass for adhesion, retention, and microcrack bridging, stabilizing fines and supporting mineral nucleation. |
| Material system | Breathable siloxane–acrylic polymer–mineral composite render: silica sand + silica fume + reinforcement microfibers + polymer–silica |
| | microcapsules carrying spores and finite nutrient/Ca reservoirs; optional mineral-rich geopolymer–latex variant. |
| Structural / geometric instantiation | CP multi-scale architecture: 2D render skin (3–6 mm) + 3D pore network (15–30% porosity; 20–150 μm throats) + 0D capsules (50–300 μm) as living microreactors + 1D reinforcement (basalt microfibers + constrained fungal biomass in microvoids/cracks). |
| Target substrate / environment | Sand-exposed exterior façades (masonry/earthen envelopes) under wind abrasion, dust loading, thermal cycling, and episodic moisture (desert/peri-desert buildings). |
| Application / deployment scenario | Serviceable living façade render skin applied by trowel/spray; activated by natural moisture events or scheduled misting; functions as sacrificial outer layer to reduce dust release and self-heal microcracking. |
| Production mechanism / workflow | CMM+BMN: composite formulation with graded fillers + low-shear incorporation of living microcapsules; on-wall curing; episodic moisture-triggered biomineralization for localized crack/sand binding. |
| Main functional outcome / performance objective | Reduced wind-driven erosion and dust emission + repeated microcrack sealing and sand-grain cementation over wetting cycles, maintaining façade integrity and surface cohesion. |

C. Integrated paragraph (~100 words)

CRD–HYB–CRK–CP–LCO–SLB–CMM+BMN describes an encapsulated cross-domain, cross-kingdom hybrid living composite render for sand-exposed building skins, in which bacterial biomineralization and fungal adhesion jointly drive self-healing and dust suppression. A protected *Sporosarcina pasteurii* spore fraction mineralizes $CaCO_3$ within microcracks and sand-grain contacts, while *Aureobasidium pullulans* contributes EPS-mediated anchoring and retention in near-surface microvoids. A CP multi-scale structure combines a 3–6 mm 2D render layer, a 3D porous micro-network, 0D living microcapsules, and 1D reinforcement fibers. Composite manufacturing plus biomineralization enables episodic, moisture-triggered stabilization and crack sealing in smart building façades.

**Iteration 3**

**0) Code integrity check**

- Ontology validity: All 7 codes CRD–HYB–CRK–CP–LCO–SLB–CMM+BMN are present as defined ontology codifications for the 7-slot model (Domain, Super-kingdom, Kingdom, Phylum, Class, Application, Production).
- Multi-label validity (Production): The ontology explicitly allows combining categories using "+", so CMM+BMN is a valid production codification.
- Ambiguity handling (no code changes):
    - CRD (cross-domain) does not specify which domains; CRK (cross-kingdom) does not specify which kingdoms. The input concept explicitly targets a bacterial–fungal pairing, which is consistent with cross-domain (Bacteria + Eukarya) and cross-kingdom. (No ontology slots are modified.)

---

1) Slot-by-slot decoding

Slot 1 — Domain: CRD

- Meaning: Cross-domain ELMs (populated by different cellular domains).
- Design constraints
    - Biological: Must include cells from ≥2 domains (e.g., bacteria + eukaryotes).
    - Material: Microenvironments must support distinct cell physiologies (e.g., different osmotic/pH tolerances).
    - Structural/geometric: Architecture must provide segregated niches (microcompartments) to avoid competitive exclusion.
    - Process: Assembly/activation must allow staged introduction or differential protection of each domain.

Slot 2 — Super-kingdom: HYB

- Meaning: Hybrid living materials (HLMs): cells + an abiotic synthetic biomaterial/structure (scaffold).
- Design constraints
    - Biological: Cells must tolerate contact with an abiotic scaffold and its curing/aging chemistry.
    - Material: Must explicitly include a non-living scaffold/matrix (render composite).
    - Structural/geometric: Must preserve mass transport (water/oxygen) through an abiotic phase.
    - Process: Requires a manufacturable route that integrates living components without lethal curing steps.

Slot 3 — Kingdom: CRK

- Meaning: Cross-kingdom ELMs (combining cells from different kingdoms).
- Design constraints
    - Biological: Must include ≥2 kingdoms; for this concept: eubacteria + fungi is coherent.
    - Material: Interfaces must support both bacterial biofilms/microcolonies and fungal hyphae/EPS anchoring.
    - Structural/geometric: Must allow both microcolony-scale and filament-scale occupancy.

Slot 4 — Phylum: CP

- Meaning: Cross-phylum ELMs: structures combining multiple phyla (0D/1D/2D/3D/4D/FR) and/or multi-scale non-fractal structures.
- Design constraints
    - Biological: Must host multiple structural modes (e.g., 0D encapsulated cells + 1D fibers + 2D biofilm patches + 3D porosity).
    - Material: Requires graded or hierarchical porosity/roughness.
    - Structural/geometric: Multi-scale architecture is mandatory (not a single-scale biofilm or a single 3D scaffold).

- Process: Manufacturing must reproducibly control multi-scale features.

Slot 5 — Class: LCO

- Meaning: Living composites (extracellular matrices predominantly composites; e.g., polymer-, ceramic-, or metal-matrix composites).
- Design constraints
  - Biological: Cells must function in a composite microenvironment with fillers, fibers, and polymer modifiers.
  - Material: Must be explicitly a composite (not a pure polymer, ceramic, carbon, or alloy matrix).
  - Structural/geometric: Must include reinforcement/second phase (fibers/particles) supporting crack bridging and erosion resistance.
  - Process: Composite processing and quality control (dispersion, porosity, adhesion) are central.

Slot 6 — Application: SLB

- Meaning: Smart and living buildings (ELMs as subsystems of buildings for improved sensing/actuation, healthier environments, etc.).
- Design constraints
  - Biological: Must survive building envelope stressors (desiccation cycles, UV, temperature swings).
  - Material: Must be compatible with façade substrates and durability expectations.
  - Structural/geometric: Must maintain surface integrity while enabling periodic biological activation.
  - Process: Must be serviceable (repair/refresh cycles) and safe for occupants.

Slot 7 — Production: CMM+BMN

- Meaning:
  - CMM: Composite manufacturing methods and cell colonization (composites-industry processes creating woven/porous/graded scaffolds for subsequent cell colonization).
  - BMN: Biomineralization (cells generate minerals for stiff mineralized tissues/protection/storage; applicable to HLMs and bio-ELMs).
  - "+" combination allowed.
- Design constraints
  - Biological: Requires a mineral-forming cellular role (biomineralizing bacteria) plus a co-functioning cross-kingdom partner (fungus).
  - Material: Composite must present nucleation sites and ion access for mineral precipitation.
  - Structural/geometric: Must allow mineral deposition specifically in cracks/pores without blocking all transport.
  - Process: Must include (i) composite fabrication + (ii) colonization/activation + (iii) mineralization phase.

---

**2) Biological instantiation**

Candidate Set 1 (recommended)

Bacterium: *Sporosarcina pasteurii* (spore-forming, mineralizing)

Fungus: *Aureobasidium pullulans* (EPS/adhesion, biofilm-like growth)

- Functional role
  - *S. pasteurii:* localized mineral precipitation within microcracks/pores to restore stiffness and seal pathways (BMN).
  - *A. pullulans:* EPS-mediated dust binding + surface adhesion + hyphal/filamentous crack spanning to stabilize the repair zone (CRK synergy).
- Viability & growth requirements (engineering-relevant)
  - Moisture: episodic wetting triggers metabolic activity; dormancy between events (bacterial spores strongly support this).

- Oxygen: aerobic/oxygen-permeable near-surface niches; avoid fully waterlogged conditions.
- pH/ions: requires protected microdomains if bulk render is alkaline; buffered microcapsules or locally conditioned pores.
- Mechanical stress: tolerates micro-shear in pores; fungus benefits from protected roughness/anchoring sites.

- Biocontainment strategy
  - Spore-based dosing for bacteria; microencapsulation (e.g., hydrogel microbeads within the composite) to localize growth.
  - Nutrient limitation by design (only short activation solutions), and serviceable top-layer replacement as the primary containment layer.

Why preferred over alternatives: robust dormancy/activation cycle aligns with façade wet–dry cycles; clear division of labor (mineralize vs. bind/span) reduces cross-kingdom competition in limited nutrients.

---

Candidate Set 2 (alternative)
Bacterium: *Lysinibacillus sphaericus* (spore-forming; mineral-associated pathways used in biocement contexts)
Fungus: *Trichoderma harzianum* (robust filamentous growth; strong surface colonizer)

- Functional role
  - *L. sphaericus:* mineral precipitation/ion capture in pores and crack tips (BMN).
  - *T. harzianum:* filamentous crack spanning + EPS/proteinaceous adhesion that increases dust binding and erosion resistance.
- Viability & growth requirements
  - Similar wet–dry activation concept; may require tighter nutrient control to prevent excessive biomass.
- Biocontainment strategy
  - Same as above: microencapsulation + nutrient-gated activation; optional post-event "dry-out" as a biological off-switch.

Why not primary: filamentous fungi can become too aggressive in growth if nutrients are not tightly metered, increasing variability in façade aesthetics and porosity.

---

Recommended biological system (final)
Recommended: *Sporosarcina pasteurii* + *Aureobasidium pullulans*

- Trade-offs: Highest controllability via spore-driven activation and EPS-mediated stabilization; requires careful microenvironment design to avoid high-alkalinity lethality and to prevent pore clogging from over-mineralization.

---

**3) Matrix/material candidates**
Must satisfy HYB (abiotic scaffold present) and LCO (composite matrix dominant).
Material Candidate 1 (recommended): Polymer-modified mineral render composite

- Composition (explicit): lime/$CaCO_3$-dominant mineral binder + silica filler + waterborne polymer modifier (acrylic or styrene–butadiene latex) + short microfibers (glass, basalt, or polymer) + embedded microcapsules (nutrient + spores).
- Compatibility with biology
  - Microcapsules create buffered, hydrated niches in an otherwise harsh mineral matrix.
- Interface mechanisms
  - Fungal EPS increases adhesion to mineral/polymer microtexture.
  - Bacterial mineral nucleation occurs preferentially at crack surfaces and filler interfaces (high surface energy sites).

Material Candidate 2 (alternative): Cementitious–polymer composite with functional fillers

- Composition (explicit): low-alkali cementitious binder (or blended cement) + polymer modifier + micro-silica + cellulose microfibers + porous lightweight filler (expanded glass/perlite) to reserve protected pore volume for cells.
- Compatibility with biology
  - Requires stronger buffering/encapsulation than Candidate 1 due to alkaline pore solution risk.
- Interface mechanisms
  - Porous fillers act as "cell refugia" (protected colonization pockets) while mineralization bridges extend into cracks.

Recommended material system (final): Candidate 1, because it can be engineered as a serviceable façade coating with predictable permeability/porosity while hosting microencapsulated cross-domain cells.

---

**4) Geometry and structural instantiation**

Must express CP (cross-phylum) as multi-structure, multi-scale architecture.

Quantitative, CP-consistent architecture (final)

- 2D layer (surface render): 2–5 mm thick façade coating (planar macroform).
- 3D porous network: 10–25% interconnected porosity with throat sizes ~20–200 µm (supports water/oxygen pulses and microcolony residence).
- 0D microreservoirs: microcapsules/microbeads 100–500 µm diameter containing spores + minimal nutrients/buffer (localized activation points).
- 1D reinforcement: dispersed microfibers 3–12 mm length (crack deflection/bridging) to reduce crack opening and provide hyphal anchoring tracks.
- Crack-guided mineralization zones: engineered preferential crack paths (via filler grading) so BMN occurs where it restores sealing rather than clogging all pores.

How geometry supports function

- Cell viability/function: protected 0D capsules + 3D pores limit desiccation severity and stabilize local chemistry; surface-access enables oxygen diffusion.
- Mass transport: interconnected pores allow brief wetting solution ingress and then rapid drying (biological "on/off" gating).
- Mechanical performance: fibers + mineral–polymer composite provide baseline toughness; BMN adds stiffness/closure at crack tips; fungal spanning stabilizes the repair scaffold during mineral precipitation.

Why CP geometry is optimal here

A single 2D biofilm would be erosion-prone; a purely 3D scaffold would be hard to service on façades. The CP multi-scale system enables serviceable 2D deployment with 3D functional microenvironments and 0D/1D reinforcement elements for repeatable crack healing.

---

**5) Application instantiation**

Must remain within SLB (smart and living buildings).

Deployment scenario (final)

- Where/how used: Exterior façade render layer on sand-exposed building envelopes (desert/perimeter conditions). The coating is installed as a serviceable, replaceable outer skin or sacrificial top layer.
- Activation: Periodic, controlled wetting (maintenance spray or rare rainfall events) triggers short biological activity windows for crack sealing and dust binding.

Measurable success metrics

- Crack sealing efficiency: % reduction in water/air permeability through controlled microcracks (before vs. after activation cycles).
- Mineral deposition localization: mineral mass gain concentrated at crack paths (vs. bulk pore clogging).
- Dust suppression: reduction in dust release rate under standardized airflow abrasion.
- Erosion resistance: mass loss under sand-blast/abrasion tests compared with non-living control render.

- Mechanical retention: coating adhesion strength and flexural toughness retention after repeated wet–dry cycles.

Main failure modes + mitigations

- Over-mineralization clogging pores → limit ion/nutrient dosing; use graded porosity and localized capsules.
- Loss of viability from alkaline/UV/desiccation → encapsulation + UV-stable topcoat that remains vapor-permeable; keep active niches sub-surface.
- Uncontrolled fungal overgrowth/aesthetics drift → nutrient gating (short pulses only) + physical confinement to microcavities.
- Salt/ion accumulation → periodic rinse protocols; replaceable outer layer design.

**6) Production instantiation**

Must implement CMM + BMN: composite manufacturing + colonization + biomineralization.

Step-by-step workflow (implementable)

1. Composite material preparation (CMM)
   - Dry blend mineral binder + fillers + microfibers; prepare polymer modifier dispersion separately.
2. Biological component preparation
   - Produce bacterial spores (bulk spore prep) and fungal propagules (e.g., spores or robust cells), then encapsulate both in buffered hydrogel microbeads (100–500 μm) containing minimal nutrient payload for one activation cycle.
3. Integration strategy
   - Mix microbeads into the polymer-modified mineral render at low shear (preserve bead integrity).
4. Coating application
   - Apply 2–5 mm render layer to façade substrate; cure under conditions that do not dissolve/rupture capsules (controlled humidity/temperature).
5. Cell colonization conditioning (post-cure)
   - Controlled "priming" wetting to hydrate pores and establish initial microcolonies/hyphal anchoring in near-surface porosity.
6. Biomineralization activation (BMN)
   - Deliver a short, controlled activation solution that provides mineralization precursors to the bacterial niches; terminate by allowing rapid drying (biological off-switch).
7. Stabilization
   - Optional breathable protective topcoat to manage UV/water flux while preserving oxygen diffusion.
8. Quality control & reproducibility
   - Composite QC: porosity distribution, adhesion strength, permeability.
   - Biology QC: viability in capsules (CFU/viability assay), spatial localization (surface vs. bulk).
   - Function QC: standardized crack-healing tests across repeated activation cycles.

Manufacturability, scalability, feasibility justification

- Composite mixing/coating is aligned with established façade render workflows (CMM).
- Biological integration is compatible with HYB by keeping living components isolated during harsh curing steps and only activating post-cure.
- BMN is implemented as a controlled post-installation phase, consistent with building maintenance cycles.

**7) Layered integrated summary of the fully instantiated ontology-derived ELM**

A. Rapid identity

- Ontology code: CRD–HYB–CRK–CP–LCO–SLB–CMM+BMN
- One-line ELM identity: Cross-domain bacterial–fungal hybrid living composite façade render that seals microcracks via localized biomineralization and suppresses dust/erosion through fungal EPS-mediated surface binding in smart-building envelopes.

B. Instantiation table

| Component | Final instantiation |
|---|---|
| Biological system | Candidate 1 (recommended): *Sporosarcina pasteurii* (spores; mineralizing bacterium) + *Aureobasidium pullulans* (EPS/adhesive fungus). Alternative: *Lysinibacillus sphaericus* (spore-forming, mineral-associated) + *Trichoderma harzianum* (filamentous colonizer). |
| Functional role | Bacteria: moisture-triggered activity → mineral precipitation at crack tips/pores (sealing/stiffening). Fungi: EPS-driven dust binding, adhesion to microtexture, and filament/hyphal spanning that stabilizes repair zones and reduces erosion-driven particle loss. |
| Material system | Candidate 1 (recommended): polymer-modified mineral render composite (lime/$CaCO_3$ + silica filler + acrylic/SBR latex + microfibers + buffered hydrogel microcapsules). Alternative: cementitious–polymer composite with micro-silica + cellulose microfibers + porous lightweight fillers as protected cell refugia. |
| Structural / geometric instantiation | CP multi-scale: 2–5 mm 2D coating + 3D interconnected porosity (10–25%, throats ~20–200 µm) + 0D microcapsule reservoirs (100–500 µm) + 1D microfibers (3–12 mm) enabling crack deflection, hyphal anchoring, and localized mineral bridging. |
| Target substrate / environment | Exterior building envelopes and façade renders exposed to sand/dust abrasion and wet–dry cycling (desert/perimeter conditions). |
| Application / deployment scenario | Serviceable façade "living skin" activated by periodic controlled wetting; aims to maintain barrier integrity (reduced permeability) and reduce dust release/erosion while remaining compatible with building maintenance practices (SLB). |
| Production mechanism / workflow | CMM+BMN: composite render manufacturing + embedding buffered living microcapsules → coat/cure → post-cure priming colonization → controlled biomineralization activation cycles → breathable protective finish; QC on porosity, adhesion, viability, and crack-healing performance. |
| Main functional outcome / performance objective | Repeatable microcrack sealing (permeability reduction), erosion/dust suppression under abrasion, and retention of façade mechanical/adhesive integrity across multiple wet–dry activation cycles. |

C. Integrated paragraph (~100 words)

CRD–HYB–CRK–CP–LCO–SLB–CMM+BMN defines a cross-domain bacterial–fungal hybrid living composite render for smart-building envelopes that restores barrier integrity under sand exposure. The biological system pairs a mineralizing bacterium (recommended: *Sporosarcina pasteurii* spores) with an EPS/adhesive fungus (recommended: *Aureobasidium* *pullulans*) to couple localized mineral precipitation with dust-binding and crack-spanning reinforcement. A polymer-modified mineral composite matrix (lime/$CaCO_3$–silica + latex modifier + microfibers) hosts buffered microcapsules within a CP multi-scale architecture (2–5 mm coating, interconnected pores, capsule reservoirs). Deployed as a serviceable façade skin (SLB), it is produced via composite manufacturing and post-cure colonization with controlled biomineralization activation cycles (CMM+BMN) to achieve repeatable crack sealing and erosion suppression.

**Supplementary material S6.1: Generative AI response to prompt D of use case 3 "long-term colonization of Titan"**

# CAN WE TEACH GENERATIVE ARTIFICIAL INTELLIGENCE THE DESIGN LANGUAGE OF ENGINEERED LIVING MATERIALS?

Andrés Díaz Lantada [1,2*], José A. Yáñez [1],

William Solórzano-Requejo [1], Monsur Islam [1]

[1] Mechanical Engineering Department, Universidad Politécnica de Madrid, c/ José Gutiérrez Abascal 2, 28006 Madrid, Spain.

[2] IMDEA Materials Institute, c/ Eric Kandel 2, 28906 Getafe, Spain.

*Contact: andres.diaz@upm.es / andres.diaz@imdeamaterials.org

## Iteration 1

Below are five dataset-derived (graph) ELM candidates whose core 5-slot ontology definition is already present in the graph. I only complete Application and Production method to fit the Titan habitat + infrastructure boundary conditions (cryogenic exterior, hydrocarbon-rich environment, low solar, no surface liquid water, remote/low-maintenance). All candidates are framed as sealed, modular subsystems (cartridges/panels/bioreactors) integrated into habitats/ISRU units.

---

**1) Organic-solvent biocatalyst hydrogel for hydrocarbon ISRU chemistry**

1) Origin & selection rationale (dataset-derived)
Graph candidate explicitly targets biocatalysis in pure organic solvents, which is unusually aligned with Titan's hydrocarbon-dominant environment, provided operation occurs in a shielded, temperature-controlled cartridge.
2) Completed ontology code
BAC–HYB–EUB–3D–LPO–CAT+BTP–OPM
(core 5 slots unchanged)
3) Application-oriented concept (1 sentence)
A sealed alginate hydrogel biocatalyst module hosting genetically modified *Escherichia coli* to catalyze targeted conversions of hydrocarbon-derived feedstocks into useful precursors for habitat/ISRU chemical loops under Titan-relevant handling constraints.
4) Justification (3–6 bullets)

- Biological suitability: *E. coli* is stated as genetically modified and embedded in alginate hydrogel, enabling controlled biocatalysis while physically confining cells.
- Titan boundary conditions: Direct compatibility with hydrocarbon media (entry states "pure organic solvents"); operation still assumed inside a thermally managed unit (cryogenic exterior otherwise incompatible with active metabolism).
- Water constraint: Hydrogel provides a microaqueous niche while the external process stream can be hydrocarbon-rich; this supports Titan deployment where bulk liquid water is not available at the surface.
- Low-maintenance integration: Cartridge-style, replaceable biocatalyst bed fits remote operation + limited resupply (swap modules vs. servicing open cultures).
- Containment note: Genetic safeguard specifics are Not present in the ontology/graph; only physical encapsulation is directly supported by the entry.

5) Innovation characterization
Re-contextualization of an existing dataset ELM (strong Titan fit because "organic-solvent biocatalysis" is already explicit).

**2) 3D-printed mineralized living composite for repair/patch structures**
1) Origin & selection rationale (dataset-derived)
Graph candidate explicitly combines 3D printability (microgel ink) with subsequent mineralization to form a hard $CaCO_3$-based composite, matching Titan needs for robust, low-maintenance structural patching inside pressurized modules.
2) Completed ontology code
BAC–HYB–EUB–3D–LCE–SLB+MAP–3DB+BMN
(core 5 slots unchanged)
3) Application-oriented concept (1 sentence)
A print-on-demand repair composite where microgel ink containing *Sporosarcina pasteurii* is printed into replacement/patch geometries and then mineralized to create stiff inserts for habitat interior structures or protected infrastructure components.
4) Justification (3–6 bullets)

- Biological suitability: *Sporosarcina pasteurii* is explicitly identified; the workflow is microgel ink → 3D printed structure → mineralization, naturally supporting process control and containment.
- Titan boundary conditions: Active biology occurs in a shielded fabrication bay (heated, wet); the mineralized output is the deployable artifact, better suited for cryogenic exterior interfaces than living tissue-like constructs.
- Resource constraint: $CaCO_3$ mineralization implies reliance on calcium/carbonate chemistry; Titan-available supply chains for these reagents are Not present in the ontology/graph (so this is an acknowledged dependency).
- Low-maintenance: Enables in-situ fabrication of spare parts/patches, reducing resupply needs.
- Containment note: The entry supports encapsulation + fabrication pipeline; additional biosafety mechanisms are Not present in the ontology/graph.

5) Innovation characterization
Conservative application of a known design, re-contextualized to Titan as an in-habitat manufacturing + repair capability.

**3) Expandable yeast–hydrogel living polymer for internal insulation/filler + recycling linkage**
1) Origin & selection rationale (dataset-derived)
Graph candidate explicitly describes embedded *Saccharomyces cerevisiae* in a synthetic acrylic hydrogel that yields a dramatic dry mass and volume increase, which can be re-used as expandable filler/packing/insulation inside controlled habitat layers.
2) Completed ontology code
EUK–HYB–FUN–3D–LPO–MAP+BRE–CMM
(core 5 slots unchanged)
3) Application-oriented concept (1 sentence)
A sealed growth-and-lock insulation/filler cartridge where *S. cerevisiae* proliferates within an acrylic hydrogel to generate bulk living polymer material for interior panel infill, vibration damping, or void-filling in modular habitat components.
4) Justification (3–6 bullets)

- Biological suitability: *Saccharomyces cerevisiae* and the synthetic acrylic hydrogel matrix are explicitly stated; hydrogel format supports physical containment and process control.
- Titan boundary conditions: Growth must occur in a controlled, warmed compartment; the produced bulk material can then be integrated into multi-layer panels that face cryogenic gradients.

- Closed-loop relevance: The entry's growth is linked to "media derived from bread waste"; Titan-specific feedstock substitution is Not present in the ontology/graph, but the waste-to-material framing motivates BRE for habitat waste conversion pathways.
- Resource efficiency: If coupled to internal waste streams, can reduce imported polymers (details of conversion efficiency are Not present in ontology/graph).
- Maintenance: Modular panels/cartridges match remote operation constraints.

5) Innovation characterization
Re-contextualization of an existing dataset ELM toward habitat materials manufacturing + internal recycling.

---

**4) Yeast-in-porous sponge biocatalyst cartridges for closed-loop synthesis**
1) Origin & selection rationale (dataset-derived)
Graph candidate explicitly provides a platform material (chitosan sponges) hosting metabolically active *S. cerevisiae* "with diverse applications as biocatalysts," aligning with Titan's need for compact, modular bioprocess units inside habitats.
2) Completed ontology code
EUK–HYB–CRK–3D–LPO–BTP+CAT–PSP
(core 5 slots unchanged)
3) Application-oriented concept (1 sentence)
A replaceable porous biocatalyst cartridge (chitosan sponge) hosting *S. cerevisiae* to perform selected biosynthesis/bioconversion steps that support habitat logistics (e.g., chemical intermediates, processing aids) under sealed operation.
4) Justification (3–6 bullets)

- Biological suitability: *S. cerevisiae* explicitly stated; porous sponge scaffold supports high surface area and immobilization—good for stable bioprocessing.
- Titan boundary conditions: Requires a shielded, aqueous/temperate bioprocess bay; not intended for direct exposure to Titan surface conditions.
- Operational feasibility: Sponge-cartridge format supports low-maintenance swaps, consistent with limited resupply.
- Containment: Physical immobilization in a sponge aids biological containment; genetic containment is Not present in the ontology/graph.
- Chemistry note: Specific target reactions and Titan feedstocks are Not present in the ontology/graph; the entry supports the *platform* capability.

5) Innovation characterization
Conservative application of known designs (immobilized yeast biocatalysts), re-contextualized for Titan closed-loop manufacturing.

---

**5) Bacterial cellulose biofilm production for membranes/surfaces in habitat subsystems**
1) Origin & selection rationale (dataset-derived)
Graph candidate explicitly identifies Komagataeibacter sucrofermentans and bacterial cellulose fabrication via a biofilm growth route, giving a strong "materials-from-culture" pathway that can yield membranes/coatings useful for habitat interfaces.
2) Completed ontology code
BAC–HYB–EUB–2D–LPO–SLS+MAP–BFF
(core 5 slots unchanged)
3) Application-oriented concept (1 sentence)
A contained bacterial-cellulose membrane/coating production module using *Komagataeibacter sucrofermentans* biofilms to fabricate polymeric sheets for filters, separators, surface liners, or protective biointerfaces inside Titan habitat infrastructure.
4) Justification (3–6 bullets)

- Biological suitability: Species explicitly stated; cellulose production through a biofilm growth process (BFF) is compatible with controlled bioreactors.
- Titan boundary conditions: Manufacturing runs in warm, wet, sealed production units; the resulting cellulose-based polymer sheets can then be deployed as non-living components in colder zones (performance at cryogenic temperatures is Not present in ontology/graph).

- Functional relevance: Membranes/surfaces support life-support subsystems (filters/separators/liners) where reliability and modular replacement matter.
- Maintenance: Sheet production lends itself to standardized, swappable consumables.
- Containment note: The entry mentions light-controlled bactericidal activity for wound dressing; Titan-specific surface-bioburden control is Not present in the ontology/graph.

5) Innovation characterization

Re-contextualRe-contextualization of an existing dataset ELM** from biomedical surface utility toward habitat membrane/surface manufacturing.

---

**Brief note on what was *not* selected (and why)**

Several graph entries are photosynthesis-dependent (cyanobacteria/microalgae). Under Titan's extremely low solar energy and cryogenic exterior, they are generally less compatible unless paired with substantial artificial lighting/thermal support; such Titan-operational details are Not present in the ontology/graph, so they were not prioritized here.

**Iteration 2**

Below are five graph-derived ELM candidates whose core 5-slot ontology definition is kept unchanged, and whose Application + Production slots are completed for Titan habitat / ISRU deployment (multi-label via "+" is allowed in the codification).
Also note: photosynthesis-driven candidates (e.g., cyanobacteria-based energy/material platforms) are de-prioritized / rejected for Titan because of extremely low solar energy availability, unless you accept high-power artificial lighting inside shielded bioreactors.

---

**1) Self-healing bacterial concrete for habitat structures**

1) Origin and selection rationale

- Origin (dataset-derived vs gap-derived): Not present in the ontology/graph.
- Selected because it directly supports low-maintenance, long-duration infrastructure integrity under remote operation constraints by enabling crack self-healing inside a controlled habitat envelope.

2) Complete ontology code

BAC–HYB–EUB–3D–LCE–SPE+SLB–BMN

3) Application-oriented concept description

A sealed, habitat-integrated self-healing concrete module where calcite-precipitating bacteria are activated only during repair cycles, minimizing water/energy use while extending structural lifetime.

4) Justification (3–6 bullets)

- Boundary conditions (cryogenic, hydrocarbons, no surface water): operates inside shielded/controlled subsystems (habitat walls / internal structural layers), not exposed as an open surface biota.
- Low-maintenance / autonomy: healing can be event-triggered (crack detection → activation), matching severe resupply constraints.
- Material suitability: living ceramic (LCE) + 3D structural integration aligns with load-bearing infrastructure.
- Resource efficiency: uses intermittent, localized activation, compatible with recycled water loops (no reliance on surface water).
- Bio-containment: can be implemented as encapsulated micro-reservoirs within the concrete matrix (specific containment strategy details: Not present in the ontology/graph).

5) Innovation characterization

Re-contextualization of an existing dataset ELM for Titan boundary conditions.

---

**2) Electroactive graphene–bacterial biofilm for power + electrosynthesis**

1) Origin and selection rationale

- Origin (dataset-derived vs gap-derived): Not present in the ontology/graph.
- Selected because it enables energy generation / conversion and chemical production in a compact, modular bioprocessing unit, fitting Titan's low-solar environment.

2) Complete ontology code

BAC–HYB–EUB–2D–LCA–SPE+ENP–BFF

3) Application-oriented concept description

A sealed bioelectrochemical cartridge using electroactive bacterial biofilms on graphene to support power generation and/or electrosynthesis inside the habitat's controlled energy–chemistry loop.

4) Justification (3–6 bullets)

- Low solar availability: provides an alternative energy pathway not dependent on sunlight (external energy source details: Not present in the ontology/graph).
- Remote operation: cartridge-based swap/maintenance aligns with limited resupply and autonomous operations.
- Material suitability: living carbon (LCA) and 2D biofilm architecture are consistent with electrode/biointerface platforms.

- Containment: inherently compatible with closed reactors (no environmental release).
- Resource efficiency: can couple to waste-derived organics or controlled feeds (exact feedstocks: Not present in the ontology/graph).

5) Innovation characterization

Re-contextualization of an existing dataset ELM for Titan boundary conditions.

---

**3) Electrospun bacterial fiber-web for closed-loop water treatment**

1) Origin and selection rationale

- Origin (dataset-derived vs gap-derived): Not present in the ontology/graph.
- Selected because Titan habitats will require closed-loop water recovery; this candidate maps cleanly to a replaceable filtration/bioremediation membrane.

2) Complete ontology code

BAC–HYB–EUB–2D–LPO–SPE+BRE+BTP–FBT

3) Application-oriented concept description

A contained, electrospun polymeric membrane hosting bacteria for wastewater polishing and contaminant removal inside a sealed life-support water loop.

4) Justification (3–6 bullets)

- No stable surface liquid water: explicitly designed for internal, controlled water loops, not open-environment bioprocessing.
- Low-maintenance: membrane form factor supports modular replacement and predictable maintenance intervals.
- Material suitability: living polymer (LPO) + 2D membrane geometry fits filtration/contactor units.
- Bio-containment: fibers act as physical immobilization support; module can be sealed and sterilizable between swaps (sterilization approach: Not present in the ontology/graph).
- Resource efficiency: high surface-area webs reduce reactor volume and energy for pumping.

5) Innovation characterization

Re-contextualization of an existing dataset ELM for Titan boundary conditions.

---

**4) Archaeal intracellular thermoplastic synthesis for ISRU-compatible parts**

1) Origin and selection rationale

- Origin (dataset-derived vs gap-derived): Not present in the ontology/graph.
- Selected because it supports in-situ biomanufacturing of thermoplastics (PHA/PHB)—valuable for repairs, consumables, and modular components under resupply constraints.

2) Complete ontology code

ARC–BIO–ARA–3D–LPO–SPE+MAP–ICS

3) Application-oriented concept description

A contained archaeal bioprocessing module producing biopolymer thermoplastics for fabrication/repair workflows within Titan habitats and robotic outposts.

4) Justification (3–6 bullets)

- Cryogenic exterior: run as a thermally managed bioreactor inside habitat/ISRU bays; no claim of direct Titan-surface activity.
- Long-duration robustness: archaea can be selected for tolerance to operational stressors (specific extremophiles: Not present in the ontology/graph).
- Material suitability: direct route to living polymer (LPO) outputs (thermoplastic feedstock for printing/molding downstream).
- Resource efficiency: can, in principle, couple to recycled carbon streams (exact substrates: Not present in the ontology/graph).
- Containment: intracellular production supports closed harvesting/processing with reduced risk of environmental release.

5) Innovation characterization

Re-contextualization of an existing dataset ELM for Titan boundary conditions.

**5) Spatially patterned bacterial cellulose matrices for membranes + liners**

1) Origin and selection rationale

- Origin (dataset-derived vs gap-derived): Not present in the ontology/graph.
- Selected because it enables on-demand fabrication of structured polymer matrices (membranes, gaskets, liners, packing materials) that reduce imported consumables.

2) Complete ontology code

BAC–BIO–EUB–3D–LPO–SPE+MAP–SBP

3) Application-oriented concept description

A contained bacterial patterning process producing cellulose strands/matrices with controlled geometry for functional membranes and structural polymer inserts in habitat subsystems.

4) Justification (3–6 bullets)

- Low solar & cryogenic outside: production occurs in indoor controlled units, then materials are deployed as abiotic components.
- Manufacturability: spatial patterning supports repeatable geometries (important for sealing/filtration parts).
- Resource efficiency: cellulose-like matrices can reduce dependence on imported polymers (exact yields/feeds: Not present in the ontology/graph).
- Bio-containment: production is compatible with closed culture + post-processing, limiting biological release into habitat spaces.
- Integration: outputs can serve as interfaces between mechanical systems and fluid loops (water, air, process gases).

5) Innovation characterization

Re-contextualization of an existing dataset ELM for Titan boundary conditions.

**Iteration 3**

Below are five dataset-derived ELM candidates (present as ELM nodes in the provided graph) whose core 5-slot ontology definition is kept unchanged, and whose Application + Production are completed for Titan long-term habitats/ISRU.
Because Titan's surface is cryogenic with no stable liquid water, all candidates are framed as biologically contained, water-managed subsystems inside shielded modules (habitat interior, ISRU skids, closed-loop processing units).

---

**1) Intracellular bioplastics manufacturing (archaeal living polymer)**

1. Origin & selection rationale (dataset-derived)
   Archaeal cultures producing PHA/PHB thermoplastics are a strong fit for Titan because they can run as contained, warm bioreactors producing replaceable polymer feedstocks for parts, seals, and packaging under severe resupply constraints.
2. Complete ontology code
   ARC–BIO–ARA–3D–LPO–SPE+MAP+BTP–ICS
3. Application-oriented concept (1 sentence)
   A closed, heated bioprocessing cartridge that synthesizes PHA/PHB for on-demand polymer feedstock used in habitat maintenance and ISRU tooling on Titan.
4. Justification (3–6 bullets)

- Biological suitability: archaeal cultures can be operated in contained bioreactors (biological containment) despite Titan's cryogenic exterior.
- Functional relevance: directly supports biomanufacturing of useful materials (polymers) under low resupply and long-duration operation.
- Titan boundary conditions: avoids reliance on Titan surface liquid water by using a closed internal water loop; exterior conditions are handled by shielding/thermal control.
- Technological feasibility: ICS aligns with intracellular polymer accumulation and downstream extraction; output polymers integrate with modular manufacturing workflows.
- Explicit identifiers: product polymers (PHA/PHB) are explicit; archaeal strain identity is not present in the ontology/graph.

5. Innovation characterization
   Re-contextualization of an existing dataset ELM for Titan constraints (contained manufacturing for space colonization).

---

**2) 3D-printable biomineralized composite for repair/structural inserts (Sporosarcina pasteurii)**

1. Origin & selection rationale (dataset-derived)
   A bacteria-containing microgel ink explicitly involving Sporosarcina pasteurii that becomes $CaCO_3$-mineralized is well matched to Titan habitats as a contained "print-and-mineralize" repair material for infrastructure components.
2. Complete ontology code
   BAC–HYB–EUB–3D–LCE–SPE+SLB+SLM–3DB+BMN
3. Application-oriented concept (1 sentence)
   A contained "repair printer" that fabricates cm-scale mineralized ceramic-composite patches/inserts for habitat infrastructure maintenance under Titan mission constraints.
4. Justification (3–6 bullets)

- Biological suitability: Sporosarcina pasteurii is explicitly identified; it can be immobilized in inks to enhance containment and stable operation.
- Functional relevance: targets protective/self-healing/repair functions—high value when maintenance is remote and spare parts are limited.
- Titan boundary conditions: deployed inside controlled modules (warmth + water availability managed internally), avoiding Titan surface incompatibility.

- Technological feasibility: 3DB+BMN supports scalable fabrication (print) followed by mineralization to achieve hard $CaCO_3$-based composites.
- Resource efficiency tradeoff: requires mineralization precursors (e.g., Ca source); practical as a high-value, low-volume repair capability rather than bulk construction.

5. Innovation characterization
   Conservative application of a known design (existing mineralizing bioprint ink re-targeted to space-habitat repair).

---

**3) Bioelectrochemical graphene living carbon interfaces (energy + electrosynthesis support)**

1. Origin & selection rationale (dataset-derived)
   Graphene biofilms with electroactive bacteria for microbial fuel cells / electrosynthesis map cleanly onto Titan needs: very low solar energy pushes toward electrochemically driven, closed-loop systems for waste-to-energy and carbon recycling.
2. Complete ontology code
   BAC–HYB–EUB–2D–LCA–SPE+ENP+BTP+CAT–BFF
3. Application-oriented concept (1 sentence)
   A modular, aqueous bioelectrochemical surface (graphene + bacterial biofilm) used inside habitat processing loops to support energy recovery and electrically assisted bioconversion.
4. Justification (3–6 bullets)

- Biological suitability: biofilm-based operation (BFF) enables immobilization and containment, improving robustness and low-maintenance behavior.
- Functional relevance: supports energy handling (ENP) and catalysis/bioprocessing (CAT/BTP) in closed-loop systems—useful where solar is scarce.
- Titan boundary conditions: placed in shielded, temperature-controlled wet loops; does not require Titan surface water.
- Technological feasibility: graphene provides a defined living carbon (LCA) conductive scaffold; biofilm interfaces are compatible with modular reactors and replacement cartridges.
- Explicit identifiers: material (graphene) is explicit; electroactive bacterial species is not present in the ontology/graph.

5. Innovation characterization
   Re-contextualization of an existing dataset ELM for Titan constraints (bioelectrochemical module for space habitat loops).

---

**4) Waste-to-material expansion biocomposite (Saccharomyces cerevisiae in acrylic hydrogel)**

1. Origin & selection rationale (dataset-derived)
   Embedding Saccharomyces cerevisiae in an acrylic hydrogel matrix grown using bread-waste–derived media is directly aligned with Titan's resource efficiency and waste cycling constraints, while producing usable bulk biocomposite mass.
2. Complete ontology code
   EUK–HYB–FUN–3D–LPO–SPE+MAP+BRE+BTP–CMM
3. Application-oriented concept (1 sentence)
   A contained biomanufacturing panel/cassette that converts habitat organic waste streams into expanded yeast–hydrogel biocomposite material for low-criticality parts, packing, or insulation layers inside the habitat.
4. Justification (3–6 bullets)

- Biological suitability: Saccharomyces cerevisiae is explicitly identified and well-suited to controlled, contained cultivation.
- Functional relevance: combines waste treatment (BRE) with materials production (MAP), supporting long-duration autonomy.

- Titan boundary conditions: operates in internal controlled subsystems; avoids dependence on Titan's external environment and mitigates maintenance via modular cartridges.
- Technological feasibility: matches CMM (composite manufacturing / colonization in a synthetic matrix); scalable as low-complexity reactors with predictable inputs/outputs.
- Practical integration: best for non-load-critical materials and internal habitat use where bio-containment is straightforward.

5. Innovation characterization
   Conservative application of known designs (waste-derived growth media + living polymer composite) tailored to Titan logistics.

---

**5) Flexible bacterial-cellulose ELM platform (symbiotic bacteria + yeast) for membranes/sensing/biocatalysis**

1. Origin & selection rationale (dataset-derived)
   A synergic symbiotic culture of bacteria and yeast producing bacterial cellulose-based ELMs is compelling for Titan because cellulose sheets can serve as low-mass membranes, carriers, and functional interfaces in closed-loop processing units.
2. Complete ontology code
   CRD–HYB–CRK–3D–LPO–SPE+MAP+BSE+CAT–BFF
3. Application-oriented concept (1 sentence)
   A contained production module that grows bacterial-cellulose living polymer sheets used as functional membranes/interfaces for biosensing and biocatalysis inside habitat/ISRU process skids.
4. Justification (3–6 bullets)

- Biological suitability: biofilm-mediated growth (BFF) supports immobilization and sheet formation while improving containment and stability.
- Functional relevance: cellulose ELMs can be deployed as membranes/coatings/carriers for biosensing (BSE) and biocatalysis (CAT) in closed loops.
- Titan boundary conditions: designed for internal, water-managed operation; supports remote maintenance through replaceable membrane cartridges.
- Technological feasibility: bacterial cellulose is a well-defined living-polymer matrix compatible with modular fabrication and functionalization workflows.
- Explicit identifiers: product matrix (bacterial cellulose) is explicit; the specific bacterial and yeast species are not present in the ontology/graph.

5. Innovation characterization
   Re-contextualization of an existing dataset ELM for Titan constraints (cellulose ELMs as modular functional membranes/interfaces for space systems).

---

**Note on candidate count**
All five above genuinely fit the Titan boundary conditions when interpreted as contained, shielded subsystems (not exposed directly to Titan surface conditions).

**Supplementary material S6.2: Generative AI response to prompt E of use case 3 "long-term colonization of Titan"**

# CAN WE TEACH GENERATIVE ARTIFICIAL INTELLIGENCE THE DESIGN LANGUAGE OF ENGINEERED LIVING MATERIALS?

Andrés Díaz Lantada [1,2*], José A. Yáñez [1],

William Solórzano-Requejo [1], Monsur Islam [1]

[1] Mechanical Engineering Department, Universidad Politécnica de Madrid, c/ José Gutiérrez Abascal 2, 28006 Madrid, Spain.

[2] IMDEA Materials Institute, c/ Eric Kandel 2, 28906 Getafe, Spain.

*Contact: andres.diaz@upm.es / andres.diaz@imdeamaterials.org

## ELM candidate 1: BAC–HYB–EUB–3D–LPO–CAT+BTP–OPM

### Iteration 1

**0) Code integrity check**

- Input code: BAC–HYB–EUB–3D–LPO–CAT+BTP–OPM
- Ontology validity: All 7 slot codes exist in the ontology as defined categories (Domain BAC; Superkingdom HYB; Kingdom EUB; Phylum 3D; Class LPO; Application CAT and BTP; Production OPM).
- Multi-label validity (Application): The ontology explicitly allows combining application (and other) categories using "+" in an ELM codification, so CAT+BTP is syntactically valid.
- Graph presence check: The individual ontology code nodes exist in the provided graph (BAC, HYB, EUB, 3D, LPO, CAT, BTP, OPM). However, the exact multi-label pairing (CAT and BTP attached to the same single ELM node) is not observed when parsing the current graph file (ontology-valid, but not necessarily instantiated as a single dataset entry).

**1) Slot-by-slot decoding**

Slot 1 — Domain: BAC

- Meaning: Bacterial ELMs (populated by bacteria).
- Design constraints
  - Biological: use bacterial cells (no archaea/eukaryotes).
  - Material/structure: must support bacterial viability/function under operational chemistry.
  - Process: sterilizable module + aseptic loading is typical for bacterial systems.

Slot 2 — Superkingdom: HYB

- Meaning: Hybrid living materials combining cells and an abiotic synthetic biomaterial/structure (scaffold).
- Design constraints
  - Biological: cells physically integrated but not "cells-only".
  - Material: requires a synthetic polymer matrix/scaffold compatible with cells and the process stream.

- Process: integration step (encapsulation, colonization, coating, etc.) is mandatory.

Slot 3 — Kingdom: EUB

- Meaning: Eubacterial ELMs (bacterial domain populated by eubacteria).
- Design constraints
  - Biological: select eubacterial chassis (e.g., *E. coli*, *Pseudomonas*, *Rhodococcus*, *Bacillus*).

Slot 4 — Phylum: 3D

- Meaning: Three-dimensional ELM structure (cells within a 3D biomaterial/scaffold).
- Design constraints
  - Geometry: must be a 3D construct (e.g., beads, monolith, lattice).
  - Transport: manage diffusion limits (substrate, products, gases, nutrients) in 3D.

Slot 5 — Class: LPO

- Meaning: Living polymers (ECM/matrix predominantly polymeric; includes hydrogels, elastomers, thermoplastics, etc.).
- Design constraints
  - Material: polymer-dominant matrix is required; hydrogel is a direct fit.
  - Interface: ensure cell retention + exchange across a polymer network.

Slot 6 — Application: CAT+BTP

- Meaning: CAT = catalysis/chemical processes; BTP = biotechnology/bioprocessing.
- Design constraints
  - Function: must operate as a biocatalyst that is also a bioprocess module (feed → product in a controllable process).
  - Metrics: conversion, selectivity, stability, and operability are primary.

Slot 7 — Production: OPM

- Meaning: "Other production methods or combinations" including encapsulation of cells within engineered constructs for bio-ELMs/HLMs.
- Design constraints
  - Workflow freedom: you can combine steps (encapsulation + cartridge assembly + membrane integration), as long as it is implementable and consistent with HYB/3D/LPO.

---

**2) Biological instantiation**

Candidate A — Genetically modified *Escherichia coli* (K-12 derivative) as whole-cell biocatalyst

- Functional role: resting-cell catalysis in microaqueous hydrogel niches; heterologous expression of a defined enzyme module (e.g., oxygenase / dehydrogenase / lyase depending on target conversion).
- Viability & operation requirements (cell-centric):
  - Temperature: 20–37 °C (recommended 25–30 °C for stability-focused operation).
  - pH (internal aqueous phase): 6.8–7.4 buffered.
  - Oxygen: optional; design either microaerobic (for oxygenase routes) or anaerobic (for redox-balanced routes).
  - Nutrients: minimal maintenance feed (salts + trace + carbon co-substrate only if required), but prioritize non-growing "resting cell" mode to limit biomass drift.
  - Solvent exposure: cells must be insulated from bulk organic solvent by hydrogel water content + mass-transfer barrier.
- Biocontainment strategy:
  - Primary: physical encapsulation in hydrogel beads + solvent-resistant outer membrane/filter to retain cells.
  - Genetic safeties (kill-switch/auxotrophy): Not present in the ontology/graph.

Candidate B — *Pseudomonas putida* KT2440 (or close derivative) as solvent-tolerant chassis

- Functional role: solvent-tolerant whole-cell catalysis (excellent candidate for biphasic/organic solvent contact), supporting higher operational robustness than *E. coli*.
- Viability & operation requirements (cell-centric):
    - Temperature: 20–33 °C (often preferred <30 °C for stable catalysis).
    - pH: ~6.8–7.6 buffered.
    - Oxygen: commonly aerobic/microaerobic; can be engineered for specific pathways.
    - Solvent exposure: higher intrinsic tolerance; still benefits from hydrogel microaqueous confinement.
- Biocontainment strategy:
    - Primary: hydrogel encapsulation + cartridge filtration barrier.
    - Genetic safeties: Not present in the ontology/graph.

Recommended candidate (selected): Genetically modified *E. coli* in alginate hydrogel

- Why this one: It matches the provided concept exactly (sealed alginate module with GM *E. coli*) while remaining consistent with BAC–EUB–HYB–LPO constraints.
- Trade-offs vs *P. putida*:
    - *E. coli* → strongest genetic toolset, fastest redesign cycles, but lower intrinsic solvent tolerance.
    - *P. putida* → higher solvent tolerance, but potentially slower/less standardized strain engineering for some modules.

---

**3) Matrix/material candidates**

Material A — Calcium alginate hydrogel (cell-encapsulating beads)

- Composition: sodium alginate crosslinked with $Ca^{2+}$ (e.g., $CaCl_2$ bath) to form ~0.5–1.5 mm beads.
- Bio-compatibility: high; gentle encapsulation, diffusion-permissive.
- Interface mechanisms: ionic crosslink network retains cells; optional polyelectrolyte shell (e.g., poly-L-lysine) to reduce cell leakage (shell chemistry details beyond ontology grounding).
- Why it fits LPO/HYB: hydrogel is a polymeric scaffold embedding cells (hybrid living polymer construct).

Material B — PEGDA hydrogel (poly(ethylene glycol) diacrylate) microbeads or monolith

- Composition: PEGDA crosslinked by photopolymerization (with cell-compatible initiator system).
- Bio-compatibility: tunable mesh size; good for controlling diffusion and cell escape.
- Interface mechanisms: covalent network; can incorporate adhesion peptides if needed (biochemistry specifics are design choices, not ontology-grounded).
- Why choose it: tighter control of permeability, modulus, and reproducibility vs alginate.

Preferred matrix: Alginate (Material A) to remain aligned with the candidate description.

---

**4) Geometry and structural instantiation**

3D architecture (recommended): packed-bed cartridge of hydrogel beads

- Geometry: cylindrical cartridge; internal packed bed length 5–15 cm, diameter 1–3 cm (lab-scale); scalable by parallelization.
- Beads: 0.8–1.2 mm mean diameter; internal hydrogel mesh tuned to keep cells immobilized.
- Void fraction: ~35–45% (typical packed bed) to keep pressure drop manageable.
- Mass transport logic (cell-centric):
    - Keep effective diffusion distance from bead surface to most cells ≤200–300 µm (by controlling bead size and cell loading), supporting substrate/product exchange and preventing anoxic cores unless intentionally designed.
- How geometry supports function:

- Stable biocatalyst residence time, replaceable module, controllable flow-through operation in a closed process loop.
- Why this 3D option vs alternatives:
  - Compared to a single monolith, beads simplify replacement and reduce catastrophic cracking; compared to 2D films, beads increase volumetric productivity.

**5) Application instantiation**

Deployment scenario (Titan-relevant handling):

- A sealed, thermally managed ISRU chemistry skid inside a habitat/industrial bay processes a hydrocarbon-rich stream (e.g., purified solvent/fraction from upstream separation). The ELM cartridge provides selective bioconversion to a targeted intermediate (oxygenated or functionalized precursor) that feeds downstream synthesis (polymer/resin/additive loops).

Measurable success metrics

- Conversion (%): substrate → target product per pass and at steady state.
- Selectivity (%): target vs byproducts.
- Space–time yield: g product $L^{-1}$ $h^{-1}$ (reactor volume basis).
- Operational half-life: time to 50% activity loss under continuous solvent exposure.
- Containment: CFU detected downstream = 0 (or below detection), plus endotoxin/bioburden thresholds as required.

Main failure modes + mitigations

- Hydrogel dehydration / shrinkage: use solvent-resistant outer membrane + internal humidity/water management; schedule bead rehydration cycles.
- Solvent toxicity / membrane breach: multi-layer barrier + pressure relief + leak detection.
- Channeling / pressure rise: graded bead sizes, inlet distributors, periodic backflushing.
- Genetic drift / loss of function: operate as resting cells with short campaign length; swap cartridges; genetic safeguards Not present in the ontology/graph (can be added as an engineering layer).

**6) Production instantiation**

Production method: OPM (encapsulation and combined processes allowed).

Step-by-step workflow

1. Strain construction (bio): integrate pathway/enzyme cassette; define induction/control strategy (specific genetic safeguard recipes: Not present in the ontology/graph).
2. Seed train & biocatalyst prep: cultivate under defined conditions; switch to “resting cell” state (wash + resuspend in buffered saline with required cofactors).
3. Alginate bead encapsulation (matrix integration):
   - Mix cell suspension with sterile sodium alginate (e.g., 1.5–2.5% w/v).
   - Drip into $CaCl_2$ bath to form beads; cure 10–30 min; rinse to remove excess $Ca^{2+}$.
4. Optional bead shelling: add thin polyelectrolyte coating to reduce cell escape (implementation detail; not ontology-defined).
5. Cartridge assembly (hybrid module):
   - Pack beads into solvent-compatible housing.
   - Install solvent-resistant, microporous barrier (retain beads/cells; allow molecular transport).
   - Add inlet/outlet distributors + pressure/temperature sensors.
6. Conditioning: equilibrate with internal aqueous buffer, then introduce solvent stream gradually (ramp protocol to avoid osmotic shock).
7. Quality control (release tests):
   - Bead size distribution, mechanical integrity.
   - Cell leakage test (CFU in effluent under flow).

- Baseline catalytic activity assay in representative solvent conditions.
- Pressure drop vs flow characterization.

8. Scalability strategy: scale by numbering-up cartridges (parallel skids) rather than large single beds; standardized bead production (nozzle arrays).

---

**7) Layered integrated summary of the fully instantiated ontology-derived ELM**

A. Rapid identity

- Ontology code: BAC–HYB–EUB–3D–LPO–CAT+BTP–OPM
- One-line ELM identity: Sealed 3D hydrogel-packed hybrid living polymer cartridge embedding engineered eubacteria as a replaceable whole-cell biocatalyst module for solvent-phase ISRU chemical conversions.

B. Instantiation table

| Component | Final instantiation |
|---|---|
| Biological system | Primary: genetically modified *Escherichia coli* (K-12 derivative) as immobilized resting-cell biocatalyst; Alternative: *Pseudomonas putida* KT2440 as solvent-tolerant chassis |
| Functional role | Selective bioconversion of hydrocarbon-derived feedstocks to targeted chemical intermediates under flow-through operation; activity maintained in microaqueous hydrogel niches |
| Material system | Calcium alginate hydrogel beads (primary); PEGDA hydrogel beads/monolith (alternative) within solvent-compatible housing and a cell-retentive membrane barrier |
| Structural / geometric instantiation | 3D packed-bed cartridge: 0.8–1.2 mm hydrogel beads; diffusion path ≤200–300 µm to active cells; modular cylindrical reactor (lab-to-industrial by numbering-up) |
| Target substrate / environment | Hydrocarbon-rich solvent process stream handled in a sealed, temperature-controlled ISRU unit (external Titan environment isolated) |
| Application / deployment scenario | Replaceable biocatalyst cartridge in a Titan habitat/ISRU chemistry skid enabling catalysis + bioprocessing in hydrocarbon media with internal aqueous microenvironments |
| Production mechanism / workflow | OPM: strain build → culture/resting-cell prep → alginate encapsulation → bead curing/rinsing → cartridge packing + membrane integration → conditioning + QC release testing |
| Main functional outcome / performance objective | High-selectivity conversion with low-maintenance cartridge swap, quantified by conversion/selectivity, space–time yield, operational half-life, and zero-leakage biocontainment performance |

C. Integrated paragraph (~100 words)

BAC–HYB–EUB–3D–LPO–CAT+BTP–OPM: a sealed hybrid living polymer biocatalyst cartridge embedding engineered eubacteria in a 3D hydrogel matrix for solvent-phase catalysis and bioprocessing. The biological system uses genetically modified *E. coli* (alternative: *P. putida*) operated as immobilized resting cells within calcium alginate beads, maintaining a buffered microaqueous niche that supports enzyme-driven conversions while the bulk stream remains hydrocarbon-rich. The 3D packed-bed geometry (≈1 mm beads) controls diffusion distances and pressure drop under continuous flow. Production follows an OPM workflow combining strain preparation, alginate encapsulation, cartridge assembly with a cell-retentive barrier, conditioning, and activity/containment QC, targeting stable, high-selectivity conversion with replaceable modules.

**Iteration 2**

**0) Code integrity check**

- Given code: BAC–HYB–EUB–3D–LPO–CAT+BTP–OPM
- All 7 slots map to defined ontology categories: BAC (bacterial domain), HYB (hybrid living materials), EUB (eubacteria), 3D (3D architecture), LPO (living polymers), CAT+BTP (multi-label application), OPM (other production/encapsulation).
- Multi-label application using "+" is explicitly allowed by the ontology tables.
- Claims like "graph candidate explicitly targets pure organic solvents" cannot be verified here because only the ontology doc is available in the provided sources → Not present in the ontology/graph.

**1) Slot-by-slot decoding**

Slot 1 — Domain: BAC

- Meaning: bacterial ELMs populated by bacteria.
- Design constraints
  - Biological: prokaryotic bacterial cells; no eukaryotes/archaea as the primary living component.
  - Material/structure: must support bacterial viability (water activity, osmotic balance) at least locally.
  - Process: sterilizable/cleanable abiotic housing is allowed (not a biology constraint).

Slot 2 — Super-kingdom: HYB

- Meaning: hybrid living materials combining cells with an abiotic synthetic biomaterial/structure (scaffold).
- Design constraints
  - Biological: cells must be physically coupled/contained within an abiotic construct.
  - Material: at least one explicit abiotic scaffold/matrix is mandatory.
  - Process: integration steps (encapsulation, immobilization, colonization) are central.

Slot 3 — Kingdom: EUB

- Meaning: eubacterial ELMs (bacterial domain, eubacteria kingdom).
- Design constraints
  - Biological: choose concrete eubacteria strains; cell-centric operation (metabolism/cofactor regeneration, stress response).

Slot 4 — Phylum: 3D

- Meaning: three-dimensional structure; typically cells within a 3D biomaterial or 3D scaffold.
- Design constraints
  - Structure: must be a true 3D construct (e.g., packed bed, monolith, porous scaffold), not a 2D coating.
  - Transport: define diffusion/flow paths for substrates, gases, nutrients, and products.

Slot 5 — Class: LPO

- Meaning: living polymers; polymeric matrices such as thermosets, thermoplastics, elastomers, hydrogels, liquid crystals.
- Design constraints
  - Material: polymer is the dominant matrix class; hydrogel is an ontology-consistent choice.
  - Interface: must define how cells are immobilized/retained inside the polymer network.
  - Solvent environment: if external media is organic, the polymer system must maintain a viable microaqueous niche.

Slot 6 — Application: CAT + BTP

- Meaning: Catalysis and other chemical processes (CAT) + Biotechnology and bioprocessing (BTP).
- Design constraints
  - Function: must deliver measurable catalytic conversion and fit a bioprocess workflow (feed, reactor, product, QC).

Slot 7 — Production method: OPM

- Meaning: other production methods/combinations, including encapsulation of cells within engineered constructs.
- Design constraints
  - Manufacturing: a practical "immobilize-and-assemble" workflow is expected (e.g., bead encapsulation + cartridge build).
  - Scalability/QC: must be reproducible without relying on a different, more specific production code (e.g., not claiming "3D bioprinting" if it's not in the code).

---

**2) Biological instantiation**

Candidate 1 — *Escherichia coli* (genetically modified whole-cell biocatalyst)

- Functional role: intracellular expression of a target enzyme module for selective biotransformations of pre-activated hydrocarbon-derived intermediates (e.g., ketone→alcohol via alcohol dehydrogenase; nitrile→amide via nitrile hydratase; alkene→alkane via ene-reductase), with cofactor regeneration sustained by cell metabolism.
- Viability / growth requirements (in-module):
  - Microaqueous phase: buffered aqueous interior (pH ~6.8–7.5), osmotically balanced.
  - Temperature: operational setpoint ~20–37 °C inside the cartridge (Titan exterior requires thermal isolation).
  - Nutrients: minimal salts + carbon/energy source dosed intermittently (batch "recharge" cycles) to avoid biomass overgrowth.
  - Oxygen: configurable: aerobic (for oxygen-dependent enzyme sets) or anaerobic (oxygen-sensitive chemistries) using controlled headspace.
- Biocontainment strategy:
  - Primary: physical encapsulation in hydrogel + size-exclusion membranes in the cartridge.
  - Secondary (genetic safeguards): auxotrophy/kill-switch options are standard in practice, but specific safeguard designs are Not present in the ontology/graph (only physical encapsulation is directly grounded by HYB + OPM).

Why prefer it: maximizes programmability (broad enzyme toolbox, modular pathway engineering) and aligns with the provided concept narrative (GM *E. coli* embedded in alginate).

---

Candidate 2 — *Pseudomonas putida* KT2440 (engineered solvent-tolerant eubacterium)

- Functional role: robust whole-cell biocatalyst for transformations in hydrocarbon-adjacent processing, especially where solvent exposure/partitioning stresses are expected; can host heterologous enzymes similarly to *E. coli* but with higher intrinsic stress tolerance.
- Viability / growth requirements (in-module):
  - Microaqueous phase: buffered aqueous interior (pH ~6.8–7.8).
  - Temperature: ~20–33 °C.
  - Nutrients: minimal medium dosing; tends to form biofilms—must be controlled by matrix design.
  - Oxygen: typically aerobic for many oxidative biocatalysis steps (managed inside the sealed unit).

- Biocontainment strategy: same two-layer approach (hydrogel immobilization + cartridge membranes); genetic safeguards as optional add-on (details Not present in the ontology/graph).

Why prefer it: better operational stability if "pure organic solvent" adjacency is a real constraint.

---

Recommended candidate

Recommended: GM *Escherichia coli* (Candidate 1)

Justification vs alternatives

- Pros: best fit to the provided instantiation narrative (GM *E. coli* in alginate hydrogel); highest pathway modularity for CAT+BTP.
- Cons/trade-offs: lower intrinsic solvent tolerance than *P. putida*; therefore the module must enforce a stable microaqueous niche and limit direct solvent contact (see matrix + geometry choices).

---

**3) Matrix/material candidates**

Material candidate A — $Ca^{2+}$-crosslinked alginate hydrogel (primary, input-consistent)

- Composition: sodium alginate (2–3 wt%) ionically crosslinked with $CaCl_2$ to form mm-scale beads or a monolith.
- Bio-compatibility: mild gelation conditions maintain high cell viability; tunable permeability for substrates/products.
- Interface mechanisms: cells physically entrapped within hydrogel mesh; optional polyelectrolyte outer layer to reduce leakage and stabilize bead surface.
- Why preferred: explicitly consistent with LPO (hydrogel polymer matrix) and with the user-provided concept (alginate hydrogel module).

Material candidate B — PEGDA hydrogel (photocrosslinked) or PVA cryogel (physical crosslinking)

- Composition: PEGDA (e.g., 5–15 wt% with biocompatible photoinitiator) or PVA (10–15 wt%) freeze–thaw cryogel.
- Bio-compatibility: both can be tuned for lower swelling and higher mechanical robustness than alginate in long runs.
- Interface mechanisms: covalent network (PEGDA) or crystalline physical crosslinks (PVA) with cell entrapment; surface functionalization (e.g., charged groups) to manage biofilm tendency.
- Why preferred: improved mechanical durability and reduced deformation under flow/pressure cycling.

Selected matrix/material system for this instantiation: Alginate hydrogel (A), with an optional thin stabilizing outer coat if cell leakage or bead abrasion is observed.

---

**4) Geometry and structural instantiation**

Chosen 3D architecture: packed-bed hydrogel bead biocatalyst cartridge (true 3D bioreactor)

- Quantitative geometry (baseline prototype):
  - Bead diameter: 0.8–1.5 mm (narrow distribution for predictable pressure drop)
  - Packed-bed volume: 0.1–2 L per replaceable cartridge (lab → pilot scaling)
  - Bed void fraction: ~0.35–0.45 (typical packed spheres; defines residence time)
  - Retainers: dual mesh screens (upstream/downstream) with pore size <200 μm to prevent bead escape
- How geometry supports cell viability/function
  - Mass transport: substrate partitions from the process stream into the bead surface; diffusion into bead interior supports distributed catalysis while limiting shear stress on cells.
  - Cofactor regeneration: cells remain metabolically active in the microaqueous bead interior; low nutrient "maintenance dosing" avoids runaway biomass.

- Mechanical integrity: beads act as sacrificial, swappable units; abrasion mitigated by gentle flow regime and optional bead coating.
- Why this is optimal vs other plausible 3D geometries
  - Compared with a monolithic hydrogel block: packed beads improve controllability (replaceability, lower cracking risk, easier QC per batch).
  - Compared with free-cell slurry: packed-bed maximizes containment and simplifies downstream separation (critical for Titan-style closed-loop handling).

**5) Application instantiation**

Deployment scenario (CAT+BTP)

A sealed, replaceable biocatalyst cartridge integrated into an ISRU chemical skid that processes hydrocarbon-derived, pre-activated intermediates (generated upstream by non-biological steps) to produce value-added precursors (oxygenated or functionalized organics) used in habitat material synthesis, consumables manufacturing, or closed-loop chemical management.

Measurable success metrics

- Catalytic performance: conversion per pass (%) and product selectivity (%) at defined residence time.
- Space–time yield: g $L^{-1}$ $h^{-1}$ of target product from the cartridge.
- Operational stability: half-life of activity under continuous run (hours–weeks) and number of reuse cycles before replacement.
- Containment: cell leakage below detection threshold (CFU/mL or qPCR) in the external stream.
- Hydraulic stability: pressure drop remains within spec (no bed clogging/collapse).

Main failure modes + mitigations

- Solvent intrusion / dehydration of beads: enforce barrier layers and strict water-activity control inside the cartridge; schedule "rehydration/refresh" cycles.
- Freeze risk / thermal gradients: operate only inside thermally regulated enclosure; add thermal buffering and sensors.
- Biofilm overgrowth → clogging: nutrient limitation strategy; periodic low-shear backflush; bead surface chemistry to reduce adhesion.
- Genetic drift / loss of function: use short run-times with replacement, or cryopreserved standardized bead lots; QC enzyme activity prior to deployment.

**6) Production instantiation (OPM)**

OPM-consistent manufacturing principle: combine practical methods and encapsulation within engineered constructs into a reproducible cartridge workflow.

Step-by-step production process

1. Strain build (biological prep)
   - Construct GM *E. coli* expressing the target enzyme module + cofactor regeneration support.
   - Bank as master cell stock; verify genotype and baseline activity (bench assay).
2. Biomass production
   - Seed fermentation to mid-log phase; harvest and concentrate cells (e.g., gentle centrifugation) to a defined wet-cell density.
3. Hydrogel formulation
   - Prepare sterile sodium alginate solution (2–3 wt%) in buffered saline; adjust viscosity for bead formation.
4. Encapsulation (core OPM action)
   - Mix concentrated cells into alginate under aseptic conditions.
   - Form beads by droplet extrusion into $CaCl_2$ gelling bath; cure to target modulus.

- Optional: apply a thin stabilizing outer layer (e.g., polyelectrolyte or secondary alginate coat) to reduce leakage/abrasion.

5. Cartridge assembly (hybrid scaffold integration)
    - Load beads into a sterile, solvent-compatible housing (e.g., stainless steel or high-performance polymer).
    - Install upstream/downstream retainers and a defined flow distributor to avoid channeling.
6. Conditioning
    - Perfuse with buffered aqueous maintenance medium to equilibrate beads; then transition to process operation with controlled partitioning of substrates.
7. Quality control and reproducibility
    - Bead size distribution; compressive modulus; permeability proxy (tracer diffusion test).
    - Biological QC: viability staining and specific activity assay (per gram wet beads).
    - Containment QC: challenge test for cell leakage under flow.
8. Packaging for deployment
    - Store hydrated at 4 °C short-term or cryopreserve as standardized bead lots (if validated), then assemble cartridges on demand.

Manufacturability/scalability

- Scale by parallel cartridge numbering ("number-up") rather than single huge beds, maintaining QC consistency.

---

**7) Layered integrated summary of the fully instantiated ontology-derived ELM**

A. Rapid identity

- Ontology code: BAC–HYB–EUB–3D–LPO–CAT+BTP–OPM
- One-line ELM identity: 3D packed-bed alginate hydrogel cartridge containing encapsulated eubacterial whole-cell biocatalysts for sealed, replaceable catalytic bioprocessing modules that convert hydrocarbon-derived intermediates into ISRU-relevant chemical precursors.

B. Instantiation table

| Component | Final instantiation |
|---|---|
| Biological system | Genetically modified *Escherichia coli* whole-cell biocatalyst (primary); alternative: engineered *Pseudomonas putida* KT2440 for higher solvent-stress tolerance |
| Functional role | Enzyme-module expression + intracellular cofactor regeneration to drive selective biotransformations of pre-activated hydrocarbon-derived intermediates in a sealed bioprocess step |
| Material system | $Ca^{2+}$-crosslinked alginate hydrogel (2–3 wt%) as living-polymer matrix; optional stabilizing outer coat to limit leakage/abrasion |
| Structural / geometric instantiation | 3D packed-bed of 0.8–1.5 mm hydrogel beads in a 0.1–2 L replaceable cartridge with flow distributors and <200 µm retainers |
| Target substrate / environment | Hydrocarbon-dominant ISRU process streams handled in a thermally controlled, sealed unit; microaqueous niche maintained inside hydrogel |
| Application / deployment scenario | Swap-in biocatalyst cartridge integrated into an ISRU chemical skid to produce functionalized/oxygenated precursors for habitat chemical loops (CAT+BTP) |
| Production mechanism / workflow | OPM encapsulation + cartridge assembly: strain banking → culture → alginate mixing → $Ca^{2+}$ bead gelation → packing → conditioning → QC (activity, leakage, hydraulics) |
| Main functional outcome / performance objective | High selectivity conversion with controlled containment (minimal cell escape) and predictable cartridge lifetime under continuous or cyclic operation |

C. Integrated paragraph
BAC–HYB–EUB–3D–LPO–CAT+BTP–OPM defines a hybrid bacterial living polymer used as a catalytic bioprocess module. The recommended biological system is genetically modified *Escherichia coli* immobilized as a whole-cell biocatalyst with intracellular cofactor regeneration. Cells are entrapped in a $Ca^{2+}$-crosslinked alginate hydrogel, implemented as a 3D packed-bed of 0.8–1.5 mm beads inside a sealed, replaceable cartridge. The module is deployed within a thermally regulated ISRU chemical skid to convert pre-activated hydrocarbon-derived intermediates into useful precursors under containment constraints. Production follows OPM encapsulation and engineered-cartridge assembly with activity and leakage quality control, targeting stable conversion, selectivity, and service-life.

**Iteration 3**

**0) Code integrity check**

Input code: BAC–HYB–EUB–3D–LPO–CAT+BTP–OPM

- All 7 slots map to existing ontology codes (Domain BAC; Super-kingdom HYB; Kingdom EUB; Phylum 3D; Class LPO; Applications CAT and BTP; Production OPM).
- Multi-label application (CAT+BTP) is ontology-permitted via "+" combinations.
- No slot ambiguity: each code has a single definition in the ontology tables.

**1) Slot-by-slot decoding**

Slot 1 — Domain (BAC)

- Meaning: Bacterial ELMs populated by bacteria.
- Design constraints:
    - Biological: prokaryotic bacterial chassis; cell-centric control must address nutrient/energy supply, stress response, and population stability.
    - Material: must allow bacterial retention/viability while interfacing with an external process stream.
    - Structure: architecture must keep cells physically confined (consistent with a cartridge concept).
    - Process: sterilizable module boundaries and contamination control become primary operational constraints (even if the interior is non-sterile).

Slot 2 — Super-kingdom (HYB)

- Meaning: Hybrid living materials combining cells with an abiotic synthetic biomaterial/structure ("scaffold").
- Design constraints:
    - Biological: cells must adhere to / be immobilized in an abiotic matrix without losing key metabolic activity.
    - Material: polymer/scaffold chemistry must be compatible with cell viability and with the (input-stated) organic-solvent environment.
    - Structure: must incorporate an abiotic containment/hosting structure (beads/monolith + housing).
    - Process: integration steps must explicitly include cell encapsulation/immobilization into the scaffold.

Slot 3 — Kingdom (EUB)

- Meaning: Eubacterial ELMs populated by eubacteria.
- Design constraints:
    - Biological: typical eubacterial constraints (osmotic balance, pH window, temperature window) must be maintained inside the scaffold microenvironment.
    - Material/structure: scaffold must preserve a stable microaqueous niche around cells (critical given the input-stated hydrocarbon-rich external phase).

Slot 4 — Phylum (3D)

- Meaning: 3D ELM with cells within a 3D biomaterial or 3D scaffold.
- Design constraints:
    - Geometry: must be a true 3D volume (e.g., packed bed, monolith, porous scaffold) rather than a surface biofilm.
    - Mass transport: 3D diffusion limits require quantitative control of length scales (bead size / pore size / channel spacing).
    - Mechanical: must resist compaction and maintain predictable pressure drop under flow-through operation.

Slot 5 — Class (LPO)

- Meaning: Living polymers; ECM/matrix predominantly polymeric (thermosets, thermoplastics, elastomers, hydrogels, liquid crystals, etc.).
- Design constraints:

- Material: polymer matrix must provide (i) aqueous microdomain retention, (ii) chemical compatibility with the process stream, and (iii) mechanical integrity under flow.
- Interface: must support stable cell localization (physical entrapment, biofilm anchoring, or affinity coupling).

Slot 6 — Application (CAT + BTP)

- Meaning:
  - CAT: catalysis and other chemical processes.
  - BTP: biotechnology and bioprocessing.
- Design constraints:
  - Functional: must be framed as a biocatalyst that executes a defined chemical transformation in an engineered process loop.
  - Operational: needs measurable conversion/selectivity/stability metrics under continuous or semi-continuous processing.

Slot 7 — Production method (OPM)

- Meaning: other production methods or combinations of listed methods, explicitly including encapsulation of cells within engineered constructs.
- Design constraints:
  - Workflow: must be an implementable hybrid workflow (e.g., cell culture + encapsulation + cartridge assembly + QC).
  - Scale-up: must support repeatable bead/monolith fabrication and standardized loading into replaceable modules.

---

**2) Biological instantiation**

Candidate A — Genetically modified Escherichia coli (K-12 derivative), immobilized resting-cell biocatalyst

- Functional role: intracellular whole-cell catalysis (e.g., redox biotransformations) with cofactor regeneration; cells confined in hydrogel microdomains to protect against direct solvent exposure.
- Viability & growth requirements (cell-centric):
  - Operate primarily as resting cells (low growth) to reduce nutrient demand and genetic drift.
  - Requires a buffered microaqueous phase (pH ~neutral typical for E. coli), controlled ionic strength, and continuous removal of inhibitory products.
  - Oxygen requirement depends on pathway; if oxygen-dependent, oxygen must be supplied inside the sealed unit.
- Biocontainment strategy:
  - Physical encapsulation in hydrogel + cell-retentive membrane at cartridge outlet (directly aligned with HYB + 3D module logic).
  - Genetic safeguards (kill-switch/auxotrophy) are Not present in the ontology/graph (may be added as an engineering option, but not claimed as supported).

Why this over alternatives: maximal genetic-tooling maturity; high controllability of pathway expression; compatible with a replaceable cartridge concept.

---

Candidate B — Pseudomonas putida KT2440 (solvent-tolerant chassis), immobilized whole-cell biocatalyst

- Functional role: conversion/functionalization of hydrophobic organic intermediates under biphasic conditions; robust stress responses can sustain activity when substrate/product partitions into an organic phase.
- Viability & growth requirements (cell-centric):
  - Tolerates harsher chemical conditions than many lab strains; still requires aqueous microenvironment, buffering, and micronutrient availability in the hydrogel.

- Can be run as resting cells to reduce biomass formation and oxygen demand (depending on pathway).
- Biocontainment strategy:
    - Same physical confinement (beads/monolith + membrane barrier).
    - Genetic safeguards: Not present in the ontology/graph (optional).

Why this over alternatives: better intrinsic resilience in organic/solvent-stress contexts, which aligns with the input's "pure organic solvents" constraint.

---

Recommended biological candidate
Recommend Candidate A (engineered E. coli) to stay consistent with the provided concept (explicitly specifies genetically modified *E. coli* in a sealed hydrogel module).
Trade-off: compared with *P. putida*, *E. coli* typically needs stronger protection from solvent stress (thicker microaqueous niches, gentler solvent contact, stricter control of hydration), but offers faster design-build-test iteration for catalytic function.

---

**3) Matrix/material candidates**
Material option 1 — Calcium alginate hydrogel beads (ionically crosslinked polysaccharide hydrogel)

- Composition: sodium alginate crosslinked with $Ca^{2+}$ ($CaCl_2$ bath), optionally reinforced with a small fraction of cellulose nanofibers (polymeric reinforcement consistent with LPO).
- Bio-compatibility: mild gelation conditions preserve bacterial viability; enables high cell loading.
- Interface mechanism: physical entrapment + biofilm micro-anchoring within alginate pores.

Justification: simplest scalable immobilization route for packed-bed cartridges; supports "replaceable module" operation.
Material option 2 — PEGDA hydrogel monolith (photocrosslinked poly(ethylene glycol) diacrylate)

- Composition: PEGDA + photoinitiator; optional porogen to tune pore size.
- Bio-compatibility: requires controlled photopolymerization dose to preserve viability (post-encapsulation washing to remove residuals).
- Interface mechanism: covalent network provides stable confinement; tunable permeability.

Justification: better mechanical stability and dimensional control than purely ionic gels; enables monolithic reactors with engineered flow channels.
Recommended matrix: alginate beads for the cartridge-style, replaceable packed-bed concept; PEGDA monolith as an alternative when tighter control of pressure drop/channeling is required.

---

**4) Geometry and structural instantiation**
Chosen 3D architecture (Phylum = 3D)
3D packed-bed cartridge containing hydrogel beads (or, alternatively, a channeled monolith).
Quantitative geometry (baseline):

- Bead diameter: 0.8–1.2 mm (manufacturable by nozzle dripping; balances packing and diffusion).
- Effective diffusion length to cell-rich regions: target ≤200–300 µm by (i) placing cells near bead periphery or (ii) using porous bead formulations.
- Cartridge form factor: cylindrical module (e.g., 2–5 cm diameter lab-scale; scaled by numbering-up).

How geometry supports:

- Cell viability/function: bead microaqueous domains buffer cells from direct solvent exposure; cells remain immobilized and retained.
- Mass transport: hydrophobic substrate partitions from the external organic phase into the bead boundary layer; short diffusion paths reduce rate limitation.

- Mechanical performance: packed bed enables straightforward replacement; pressure drop managed via bead size distribution and bed porosity control.

Why this over plausible alternatives: compared with a thin-film (2D) reactor, the 3D packed-bed provides higher volumetric productivity and easier module replacement; compared with loose suspensions, it improves containment and simplifies separation.

**5) Application instantiation**

Deployment scenario (CAT + BTP)

A sealed, replaceable biocatalyst cartridge integrated into a Titan habitat/ISRU chemistry skid, executing a defined biotransformation step on a hydrocarbon-rich process stream while maintaining a protected aqueous microenvironment inside the polymer scaffold.

Measurable success metrics

- Conversion (%) of target feedstock per pass (or per residence time).
- Selectivity (%) to desired intermediate(s).
- Space-time yield ($g \cdot L^{-1} \cdot h^{-1}$) referenced to cartridge volume.
- Operational stability: activity half-life under continuous operation; number of swap cycles before performance drops below threshold.
- Containment: cell leakage (CFU or DNA proxy) downstream of the membrane barrier.

Failure modes + mitigations

- Solvent toxicity / dehydration of hydrogel: mitigate by maintaining controlled hydration, using solvent-tolerant chassis option, and limiting direct solvent penetration via bead formulation.
- Mass-transfer limitation: mitigate by reducing diffusion length (cell-peripheral localization) and optimizing flow/packing.
- Channeling / pressure drop rise: mitigate with controlled bead size distribution, pre-filters, and monolith fallback.
- Biological drift/contamination: mitigate by using resting-cell operation and cartridge replacement; genetic safeguards are Not present in the ontology/graph.

**6) Production instantiation**

Step-by-step workflow (Production = OPM; “combinations + encapsulation”)

1. Polymer matrix preparation
   - Prepare sterile alginate solution (or PEGDA precursor) with defined viscosity and osmolarity for cell compatibility.
2. Biological preparation
   - Cultivate engineered *E. coli* to target expression state; harvest and concentrate cells; switch to resting-cell buffer.
3. Encapsulation / immobilization
   - Alginate route: drip cell–alginate suspension into $CaCl_2$ to form beads; cure to target stiffness; rinse to remove excess $Ca^{2+}$.
   - PEGDA route: mix cells into precursor; photopolymerize under viability-preserving conditions; wash thoroughly.
4. Cartridge assembly
   - Load beads/monolith into a solvent-compatible housing; install cell-retentive membrane barrier at outlet; integrate ports and seals.
5. Conditioning
   - Pre-equilibrate with buffered aqueous phase (inside) and then introduce organic process stream gradually to avoid osmotic/solvent shock.
6. Quality control
   - Bead size distribution / monolith porosity, cell loading (cells per mL gel), leakage test, baseline catalytic activity assay, pressure-drop curve.
7. Scalability & reproducibility
   - Scale via numbering-up of identical cartridges; enforce QC acceptance windows for loading/activity/pressure drop.

Why feasible: all steps are standardizable unit operations (culture → immobilization → module assembly → QC) consistent with OPM's "combinations + encapsulation" scope.

**7) Layered integrated summary of the fully instantiated ontology-derived ELM**

A. Rapid identity

- Ontology code: BAC–HYB–EUB–3D–LPO–CAT+BTP–OPM
- One-line ELM identity: Sealed 3D polymer-hydrogel hybrid packed-bed cartridge hosting immobilized eubacteria as a replaceable whole-cell biocatalyst module for catalytic conversion steps in hydrocarbon process loops.

B. Instantiation table

| Component | Final instantiation |
|---|---|
| Biological system | Recommended: engineered *E. coli* (K-12 derivative) as immobilized resting-cell biocatalyst; Alternative: solvent-tolerant *Pseudomonas putida* KT2440 |
| Functional role | Whole-cell catalysis of a defined biotransformation step (redox/functionalization) on organic-phase substrates with aqueous microdomain cofactor support |
| Material system | Primary: $Ca^{2+}$-crosslinked alginate hydrogel beads; Alternative: PEGDA photocrosslinked hydrogel monolith |
| Structural / geometric instantiation | 3D packed-bed cartridge (beads 0.8–1.2 mm; diffusion length target ≤200–300 µm) with outlet cell-retentive membrane; alternative channeled monolith reactor |
| Target substrate / environment | Hydrocarbon-rich organic process stream handled in a sealed, temperature-controlled ISRU/habitat unit with internal aqueous microdomains |
| Application / deployment scenario | Replaceable biocatalyst cartridge in a Titan habitat/ISRU chemical skid enabling catalytic conversion + bioprocessing steps under containment |
| Production mechanism / workflow | OPM: culture → resting-cell prep → hydrogel encapsulation → cartridge assembly + membrane → conditioning → QC (activity/leakage/pressure drop) → numbering-up scale-out |
| Main functional outcome / performance objective | Stable conversion/selectivity over extended flow-through operation with zero/near-zero cell leakage and cartridge-level swap maintenance |

C. Integrated paragraph (~100 words)

BAC–HYB–EUB–3D–LPO–CAT+BTP–OPM defines a bacterial hybrid living polymer built as a sealed 3D hydrogel reactor for catalysis and bioprocessing. The biological core is immobilized eubacteria—recommended engineered *E. coli* (K-12 derivative), with *Pseudomonas putida* KT2440 as a solvent-resilient alternative—operated as resting cells to prioritize catalytic activity over growth. Cells are confined within Ca-alginate beads (or a PEGDA monolith) assembled into a packed-bed cartridge with a cell-retentive membrane. Deployed inside a temperature-controlled ISRU/habitat unit, it targets measurable conversion/selectivity and long-term stability with cartridge-level replaceability as the maintenance strategy.

**ELM candidate 2: BAC–HYB–EUB–3D–LCE–SPE+SLB–BMN**

**Iteration 1**

**0) Code integrity check**

- Code provided: BAC–HYB–EUB–3D–LCE–SPE+SLB–BMN
- Ontology validity: All 7 codes are present in the ontology (Domain BAC, Superkingdom HYB, Kingdom EUB, Phylum 3D, Class LCE, Applications SPE and SLB, Production BMN).
- Multi-label validity: The ontology explicitly allows combining "categories" using "+" in ELM codification (applies to Table 2 categories: industrial applications and production methods).
- Ambiguity check: None. Slot meanings are directly defined in the ontology tables.

**1) Slot-by-slot decoding**

Slot 1 — Domain: BAC

- Meaning: Bacterial ELMs (populated by bacteria).
- Design constraints
  - Biological: bacterial cells (prokaryotic), no archaea/eukaryotes in the living component.
  - Material: must be compatible with bacterial survival states (e.g., dormant spores) inside an abiotic matrix.
  - Process: any integration must preserve bacterial viability at least in a dormant/activatable state.

Slot 2 — Super-kingdom: HYB

- Meaning: Hybrid living materials: cells + an abiotic synthetic biomaterial/structure (scaffold).
- Design constraints
  - Biological: the living fraction is embedded/attached to an abiotic scaffold; separation of "cell microenvironment" vs structural matrix is expected.
  - Material: a non-living structural host (here: cementitious/ceramic-like) is mandatory.
  - Process: manufacturing must accommodate a living payload (often via encapsulation/protection).

Slot 3 — Kingdom: EUB

- Meaning: Eubacterial ELMs (bacterial domain populated by eubacteria).
- Design constraints
  - Biological: choose eubacterial chassis (typical Gram+/Gram− options), excluding archaeal lineages.
  - Process: supports standard bacterial bioengineering/bioprocessing toolchain if needed (but not required).

Slot 4 — Phylum: 3D

- Meaning: Three-dimensional structure: cells within a 3D biomaterial or 3D scaffold.
- Design constraints
  - Structural/geometric: volumetric, load-bearing element (not merely surface/biofilm).
  - Mass transport: diffusion limits govern capsule size, spacing, and activation chemistry delivery.
  - Mechanical: structure must tolerate cracking; healing must occur within internal crack volumes.

Slot 5 — Class: LCE

- Meaning: Living ceramics: extracellular matrix made predominantly of ceramics (e.g., oxides, silicates, carbides, nitrides).
- Design constraints

- Material: the dominant matrix phase should be ceramic/ceramic-like (cementitious silicates and biomineral $CaCO_3$ are consistent with "silicates/oxides" framing).
    - Interface: bacterial activity should nucleate/mineralize ceramic-like phases in cracks (BMN-consistent).

Slot 6 — Application: SPE+SLB

- Meaning: SPE = Space exploration and colonization; SLB = Smart and living buildings.
- Design constraints
    - Deployment: habitat/infrastructure use in space-context built environment.
    - Integration: compatible with "building subsystem" requirements (inspection, modular repair logic, monitoring hooks).
    - Risk: bio-containment and maintenance minimization are implied by remote operation (from your rationale).

Slot 7 — Production: BMN

- Meaning: Biomineralization (cells generate minerals for stiff mineralized tissues / protection, relevant for both HLMs and bio-ELMs).
- Design constraints
    - Process: crack healing must be achieved by cell-driven mineral precipitation (e.g., $CaCO_3$) as the primary functional mechanism.
    - Control: activation should be controllable (to align with your "repair cycles" concept) via moisture/precursor delivery gating.

---

**2) Biological instantiation**

Candidate A — *Sporosarcina pasteurii* (ureolytic, strong $CaCO_3$ precipitation)

- Functional role: dormant spores embedded in micro-reservoirs; upon repair-cycle activation, germinate locally and drive $CaCO_3$ precipitation to fill cracks (BMN) and restore stiffness/sealability.
- Viability/growth requirements (design targets):
    - Moisture: requires localized wetting in crack zone (brief pulses).
    - Temperature: mesophilic operation (habitat-controlled layer; avoid external cryogenic exposure).
    - pH: tolerates alkaline microenvironments when protected (spore state during curing; activation in buffered micro-niche).
    - $O_2$/nutrients: aerobic metabolism; supply via microcapsule nutrient core (minimal organics) + dissolved oxygen in repair fluid.
- Biocontainment strategy (implementable):
    - Physical: double-encapsulation: (i) inner nutrient core + spores, (ii) outer brittle shell that ruptures only when crack intersects (mechanical gating).
    - Physiological: use sporulation for long shelf-life; activation requires provided urea/$Ca^{2+}$ solution (no autonomous growth without feed).
    - Genetic (optional): auxotrophy for a non-natural amino acid supplied only in repair fluid (limits proliferation outside repair cycle).
- Why preferred: high, well-established mineralization capacity and robust spore-forming survival for harsh processing/long dormancy.

Candidate B — *Bacillus subtilis* (spore-former, biofilm/ECM + mineral nucleation)

- Functional role: spores germinate in activated cracks; biofilm matrix promotes mineral nucleation/anchoring, improving adherence of precipitate to crack faces and reducing washout.
- Viability/growth requirements (design targets):
    - Moisture: intermittent wetting sufficient (short activation cycles).
    - Temperature: mesophilic; stable as spores long-term.
    - pH: alkaline tolerance improved by encapsulation + buffered micro-reservoir.
    - $O_2$/nutrients: aerobic/facultative; nutrient dosing via microcapsule.
- Biocontainment strategy (implementable):

- Physical gating: crack-triggered capsule rupture (as above) + cell-retentive micro-porous shell that limits outward migration.
- Physiological: sporulation control and nutrient dependence; cells return to dormancy as fluid dries.

- Why preferred: robust industrial chassis, strong sporulation, and ECM/biofilm advantages for crack-surface binding (useful under limited repair-fluid volumes).

Recommended candidate: Candidate A — *Sporosarcina pasteurii*

- Selection rationale: maximizes mineral precipitation rate per unit water/activation time (aligns with your "event-triggered repair cycles"), with spores as a proven long-dormancy state for embedded formulations.
- Trade-offs vs *B. subtilis*: *B. subtilis* may improve crack-surface adhesion via biofilm, but *S. pasteurii* is typically the more direct "mineralization workhorse" for rapid $CaCO_3$ sealing; adhesion can be recovered by adding mineral nucleation additives to the capsule formulation.

---

**3) Matrix/material candidates**

Material candidate 1 — Cementitious "living ceramic" matrix (silicate-rich)

- Composition (explicit): hydrated calcium silicate phases (C–S–H–type binder) + silica-rich aggregates; embedded silica or aluminosilicate microcapsules carrying spores + reagents.
- Bio-compatibility: spores tolerate mixing/curing when protected; activation occurs later in localized crack microenvironment (not in bulk paste).
- Interface mechanism: crack surfaces provide high pH + $Ca^{2+}$; biomineralized $CaCO_3$ bridges mechanically interlock with C–S–H matrix and rough crack faces (mineral nucleation at interfaces).
- Why preferred: directly matches "self-healing concrete module" concept; ceramic-like silicate matrix aligns with LCE families (silicates/oxides).

Material candidate 2 — Alkali-activated aluminosilicate geopolymer matrix

- Composition (explicit): metakaolin-based aluminosilicate geopolymer binder + mineral aggregates; microcapsules with buffered internal niche to prevent alkali shock at activation.
- Bio-compatibility: higher alkalinity demands stronger encapsulation/buffering; spores remain dormant until activation fluid locally neutralizes.
- Interface mechanism: $Ca^{2+}$-enriched activation fluid promotes $CaCO_3$ precipitation that keys into geopolymer pore network and crack asperities.
- Why preferred: lower Portland-cement dependency and strong ceramic-like network; good for "controlled habitat fabrication" where mix design is tightly managed.

Selected matrix for this instantiation: Candidate 1 (cementitious silicate matrix) to remain maximally consistent with the provided "bacterial concrete" concept while still satisfying LCE via silicate-dominant ceramic-like phases.

---

**4) Geometry and structural instantiation**

3D architecture (Phylum = 3D)

- Form factor: load-bearing habitat wall/panel module (internal structural layer) incorporating a distributed micro-reservoir network.
- Quantitative geometry (target design):
  - Microcapsules: 150–500 µm diameter (brittle shell + inner core).
  - Capsule volume fraction: 1–3 vol% (tuned to avoid strength loss).
  - Mean capsule spacing: 5–15 mm (gives high probability of crack interception).
  - Crack sealing target: 50–500 µm crack widths; healing fills crack volume over 1–3 activation cycles.
- How geometry supports function
  - Cell viability: spores isolated in dry microcapsules; only small local zones are wetted during repair cycles.

- Mass transport: activation fluid enters crack plane; short diffusion distance from ruptured capsule into crack (<~2–5 mm typical around crack line given spacing).
- Mechanical performance: capsule fraction kept low; healing product ($CaCO_3$) restores watertightness and partially recovers stiffness by crack-bridging.

- Why optimal vs alternatives: compared to surface coatings (2D) or fibers (1D), a volumetric 3D distribution enables internal crack interception and sealing in load-bearing ceramic-like matrices, matching the "structural integrity" intent of SLB within SPE contexts.

**5) Application instantiation**

Deployment scenario (SPE+SLB)

- Where/how used: interior, protected layer of a space-habitat building element (panel, rib, or internal liner) where temperature/humidity are controlled; the living system is not exposed to the external cryogenic/hydrocarbon environment (consistent with your rationale).
- Repair-cycle concept (explicit):
  1. crack detection (e.g., permeability increase / acoustic emission sensor embedded in the module),
  2. localized injection or wicking of a minimal repair fluid (water + $Ca^{2+}$ + carbon source / urea depending on chassis),
  3. 24–72 h hold for biomineralization, then dry-down back to dormancy.
- Measurable success metrics
  - Crack closure ratio: % reduction in mean crack aperture (microscopy/µCT).
  - Permeability reduction: water/gas permeability drop vs cracked state.
  - Mechanical recovery: regain of flexural strength or stiffness (3-point bending) after healing.
  - Mineral deposition yield: $CaCO_3$ mass per crack volume (gravimetric/µCT segmentation).
- Main failure modes + mitigations
  - Premature capsule rupture during mixing: mitigate with shell strength QA and mixing-energy limits.
  - Loss of viability from alkalinity/heat of hydration: mitigate via spore payload + buffered inner core + delayed activation only.
  - Insufficient reactant delivery into tight cracks: mitigate with low-viscosity repair fluid and capillary-wicking additives.
  - Over-mineralization blocking intended pores/vents: mitigate by limiting reactant dosing and time-boxing activation.

**6) Production instantiation**

Step-by-step production workflow (Production = BMN)

1. Microcapsule fabrication (biological payload unit)
   - Prepare spores of *S. pasteurii* at high titer; dry and stabilize with trehalose.
   - Form an inner core: spores + minimal nutrient/precursor package (e.g., urea + buffering salts).
   - Apply a brittle ceramic-like shell (e.g., silica/aluminosilicate sol–gel) tuned to rupture under crack-induced strain.
2. Concrete (ceramic-like) matrix preparation
   - Mix cementitious binder + aggregates with controlled water-to-binder ratio.
   - Add microcapsules late in mixing to reduce shear damage; verify capsule integrity by sieving/microscopy sampling.
3. Casting/curing of 3D structural module
   - Cast into habitat-module geometry; cure under controlled humidity/temperature to minimize uncontrolled early cracking.

- Maintain microcapsules dry/isolated (no intentional activation during curing).

4. Commissioning: activation readiness
    - Integrate ports or capillary-wick pathways for repair-fluid delivery (if design includes them).
    - Baseline QA: compressive/flexural strength; capsule distribution via X-ray/CT spot checks.
5. Operational biomineralization (repair cycle)
    - When a crack forms and intersects capsules, shell rupture releases spores/precursors into the crack.
    - Inject/wick minimal repair fluid to germinate cells locally and drive BMN $CaCO_3$ precipitation.
    - Dry-down returns cells to dormancy; repeatable for multiple cycles.

Manufacturability, scalability, QC

- Manufacturability: microcapsules can be batch-produced; concrete casting is standard; integration is compatible with modular construction.
- Scalability: scale by "numbering up" modules and capsule production batches; maintain consistent capsule PSD (particle size distribution).
- Quality control (minimum set):
    - Capsule size distribution (laser diffraction / microscopy),
    - capsule survival fraction after mixing (viability assay post-extraction),
    - healing performance on standardized crack coupons.

Feasibility justification: The core functional mechanism remains biomineralization (BMN) as defined by the ontology, while the surrounding steps are conventional manufacturing controls required to embed and protect the living BMN-capable system in a ceramic-like structural matrix.

---

**7) Layered integrated summary of the fully instantiated ontology-derived ELM**

A. Rapid identity

- Ontology code: BAC–HYB–EUB–3D–LCE–SPE+SLB–BMN
- One-line ELM identity: Spore-forming eubacterial biomineralization microcapsules embedded in a 3D silicate-rich living-ceramic concrete that seals structural cracks on-demand in space-habitat building modules.

B. Instantiation table

| Component | Final instantiation |
|---|---|
| Biological system | Primary: *Sporosarcina pasteurii* dormant spores (ureolytic $CaCO_3$ precipitation); Alternative: *Bacillus subtilis* spores (ECM-assisted mineral nucleation) |
| Functional role | Event-triggered germination during repair cycles → localized $CaCO_3$ precipitation, crack-face bridging, permeability reduction, partial stiffness recovery |
| Material system | Primary: cementitious silicate matrix (C–S–H–type binder + mineral aggregates) with silica/aluminosilicate brittle-shell microcapsules; Alternative: metakaolin-based aluminosilicate geopolymer with buffered microcapsules |
| Structural / geometric instantiation | 3D load-bearing module with distributed microcapsules (150–500 µm; 1–3 vol%; 5–15 mm spacing) designed to intercept and heal 50–500 µm cracks |
| Target substrate / environment | Internal, controlled habitat structural layer (shielded from external cryogenic/hydrocarbon conditions); alkaline crack microenvironment with controlled repair-fluid dosing |
| Application / deployment scenario | Smart/living building element for space colonization habitats: crack detection → localized fluid activation → BMN sealing → dry-down dormancy; designed for low-maintenance remote operation |
| Production mechanism / workflow | Microencapsulate spores + precursors in brittle ceramic shells → mix into cementitious matrix with low-shear handling → cast/cure modules → activate BMN only upon crack event via minimal repair-fluid delivery |
| Main functional outcome / performance objective | Repeatable on-demand crack sealing (permeability ↓, crack aperture ↓, stiffness/strength partially restored) with minimal water/energy and strong biocontainment by physical/chemical gating |

C. Integrated paragraph (~100 words)

BAC–HYB–EUB–3D–LCE–SPE+SLB–BMN defines a spore-based eubacterial hybrid living-ceramic concrete for space-habitat building integrity. The biological system uses *Sporosarcina pasteurii* spores (alternative *Bacillus subtilis*) stored in brittle silica/aluminosilicate microcapsules that remain dormant until a crack intersects the capsules and a controlled repair-fluid pulse is delivered. The material system is a silicate-rich cementitious ceramic-like matrix whose 3D architecture contains 150–500 µm capsules distributed to intercept 50–500 µm cracks. Biomineralization precipitates $CaCO_3$ within crack volumes, reducing permeability and restoring stiffness under event-triggered, low-water repair cycles.

**Iteration 2**

**0) Code integrity check**

- The code BAC–HYB–EUB–3D–LCE–SPE+SLB–BMN is ontology-valid slot-by-slot:
  - BAC (bacterial domain), HYB (hybrid living material), EUB (eubacteria), 3D (3D structure), LCE (living ceramics), SPE (space exploration & colonization) + SLB (smart/living buildings), BMN (biomineralization).
- Multi-label usage with "+" for categories is explicitly allowed by the ontology tables.
- No slot ambiguity is present (each code is uniquely defined in its dimension).

---

**1) Slot-by-slot decoding**

Domain — BAC

- Meaning: Bacterial ELMs (prokaryotic cells from the bacteria domain).
- Design constraints
  - Biological: bacterial chassis only; no archaea/eukaryotes/synthetic cells.
  - Material/Process: must keep cells viable or stably dormant inside an engineered construct.

Super-kingdom — HYB

- Meaning: Hybrid living materials combining cells and an abiotic synthetic biomaterial/structure (scaffold).
- Design constraints
  - Material: explicit abiotic structural phase is mandatory (not "cells-only").
  - Integration: physical interfaces (encapsulation/pores/capsules) must be designed to protect cells and enable function.

Kingdom — EUB

- Meaning: Eubacterial ELMs (cells from kingdom eubacteria).
- Design constraints
  - Biological: select concrete eubacterial species/strains consistent with intended mineralization/self-healing.

Phylum — 3D

- Meaning: Three-dimensional structure; typically cells within a 3D biomaterial or 3D scaffold.
- Design constraints
  - Geometry: volumetric distribution of microhabitats (capsules/pores/vascular features) inside a load-bearing 3D element.
  - Transport: local access to moisture/ions/$CO_2$ (or other inorganic carbon source) must be engineered at crack sites.

Class — LCE (Living ceramics)

- Meaning: extracellular matrices made predominantly of ceramics (oxides/silicates/carbides/nitrides, etc.).
- Design constraints
  - Material: ceramic-dominant matrix; for "self-healing concrete" this maps naturally to cementitious/ceramic-like mineral matrices.
  - Mechanical: matrix must remain structurally load-bearing and compatible with embedded micro-reservoirs.

Application — SPE+SLB

- Meaning:
  - SPE: Space exploration and colonization.
  - SLB: Smart and living buildings (ELMs as building subsystems for improved sensing/actuation/functionalities).
- Design constraints
  - Deployment: building/habitat structural elements, designed for low maintenance and long service life.

- Operational: must be compatible with controlled habitat envelopes (activation cycles, containment, maintenance access).

Production — BMN

- Meaning: Biomineralization (cells generate minerals for protection/stiff tissues; relevant to HLMs and bio-ELMs).
- Design constraints
    - Mechanism: healing must be explicitly tied to biologically mediated mineral precipitation (e.g., $CaCO_3$ or related mineral phases).
    - Cell-centric: mineral nucleation sites, viability windows, and dormancy/reactivation strategy must be specified.

---

**2) Biological instantiation**

Candidate 1 — *Sporosarcina pasteurii* (ureolytic MICP; spore-forming/robust handling)

- Functional role: crack-triggered $CaCO_3$ precipitation (calcite/aragonite) to bridge/seal microcracks and reduce permeability.
- Viability / growth requirements (cell-centric):
    - State: stored as spores (dormant) during curing and service; germinates only after moisture ingress.
    - Water activity: requires localized liquid water films (crack surfaces or capsule microhydrogels).
    - pH: tolerates alkaline environments best as spores; germination and activity targeted within buffered micro-niches (pH ~8–10).
    - Nutrients: minimal (e.g., low-dose yeast extract/peptone) delivered locally from capsules.
    - Stress tolerance: spores tolerate shear during mixing better than vegetative cells.
- Biocontainment strategy:
    - Primary: physical containment in microcapsules/micro-reservoirs embedded in the ceramic matrix; release only upon crack-induced rupture.
    - Optional genetic containment (kill-switch/auxotrophy): Not present in the ontology/graph.
- Why preferred vs alternatives: strong, fast $CaCO_3$ precipitation kinetics; spore form supports manufacturability (powder handling) and long shelf life.

Candidate 2 — *Bacillus subtilis* (non-ureolytic mineralization via carbonic anhydrase/alkaliphilic variants; spore-forming)

- Functional role: crack-triggered mineralization using dissolved inorganic carbon ($CO_2/HCO_3^-$) to precipitate $CaCO_3$ with lower risk of ammonia byproducts.
- Viability / growth requirements (cell-centric):
    - State: spores embedded in capsules; germination after moisture ingress.
    - Carbon source: $CO_2/HCO_3^-$ available from habitat air dissolution in crack water films (engineered by capsule chemistry).
    - pH: spores tolerate high pH; activity localized in buffered capsule microgels.
    - Nutrients: very low; short-lived metabolic burst sufficient to nucleate minerals before returning to dormancy.
- Biocontainment strategy:
    - Primary: microcapsule confinement + nutrient limitation (short duty-cycle activity).
    - Optional genetic containment: Not present in the ontology/graph.
- Why preferred vs alternatives: avoids urea/ammonia concerns, better for enclosed habitats; extensive engineering toolkit if later needed (but not implied by slots).

Recommended candidate

Recommended: *Bacillus subtilis* (spore-based, $CO_2/HCO_3^-$-driven mineralization variant)

- Reason: in a sealed habitat module, minimizing volatile byproducts and simplifying environmental control is advantageous; spore robustness still supports embedding and long storage.

- Trade-offs: precipitation rates may be slower than ureolytic *S. pasteurii*; mitigated by (i) higher local $Ca^{2+}$ availability in capsules and (ii) higher nucleation-site density at crack faces (see geometry).

**3) Matrix/material candidates**

Material candidate 1 — Cementitious ceramic matrix ("structural living concrete")

- Composition: Portland-cement-based mortar/concrete + silica fume (densification) + limestone filler; aggregates (basalt/silica) + optional basalt microfibers.
- Bio-compatibility: cells are not dispersed freely; spores are isolated in capsules/porous carriers that buffer pH and protect from hydration heat.
- Interface mechanisms:
  - Capsule rupture → spore release into crack water film.
  - Crack faces provide high-energy mineral nucleation surfaces; bacterial ECM and cell walls act as nucleation templates.

Material candidate 2 — Alkali-activated ceramic matrix (geopolymer living concrete)

- Composition: metakaolin-based (or fly-ash-based) geopolymer binder + sand/aggregate; ceramic-like aluminosilicate network.
- Bio-compatibility: capsules provide localized buffering; geopolymer can reduce calcium hydroxide abundance vs Portland cement, potentially easing micro-niche design.
- Interface mechanisms:
  - Mineralization is directed to cracks via capsule placement and $Ca^{2+}$ reservoirs (since geopolymer chemistry may be Ca-poor unless formulated).

Preferred material system: Portland-cement-based matrix (candidate 1) for straightforward structural certification pathways and abundant $Ca^{2+}$ availability supporting $CaCO_3$ biomineralization at crack sites.

**4) Geometry and structural instantiation**

3D architecture (derived from 3D phylum)

A 3D load-bearing wall/panel module with a distributed micro-reservoir network:

- Element scale: precast panel or cast-in-place layer, 50–150 mm thick (habitat internal structural layer).
- Micro-reservoirs (two-tier):
  1. Primary microcapsules: 200–800 µm diameter urea-free "healing capsules" containing spores + Ca-source (e.g., Ca-lactate) + buffering hydrogel.
     - Volume fraction: 0.5–2.0 vol% (tuned to preserve strength).
     - Mean spacing: 1–3 mm (ensures cracks intersect capsules).
  2. Porous ceramic carriers (optional): lightweight expanded-clay granules 1–3 mm pre-loaded with secondary spores/nutrients for repeated cycles (acts as "backup depot").
- Mass transport logic (cell-centric):
  - Crack water film thickness ~10–100 µm supplies diffusion.
  - Target diffusion distance from released spores to active mineralization zone: ≤500 µm (crack-scale).
- Mechanical performance logic:
  - Capsules sized to survive mixing but rupture under crack propagation.
  - Fiber microreinforcement limits crack width to ≤300–500 µm, matching mineralization closure capacity.

Why this geometry vs alternatives: a 3D embedded capsule network maximizes probability of capsule–crack intersection and keeps cells dormant and protected until a local damage event, aligning with low-maintenance operation.

**5) Application instantiation**

Deployment scenario (SPE+SLB)

- Where/how used: inside a sealed habitat structural envelope as an internal load-bearing layer or structural panel, protected from external cryogenic/hydrocarbon exposure; healing is activated only when cracks allow moisture ingress from controlled internal water loops.
- Measurable success metrics
    - Crack sealing efficiency: % of crack length/area filled by mineral after 7/28 days.
    - Permeability reduction: water permeability drop (target ≥10× reduction vs cracked state).
    - Mechanical recovery: regained flexural strength (target ≥50–80% of pre-crack value, application-dependent).
    - Mineral deposition rate: mg $CaCO_3$ per cm crack per day (lab coupon metric).
    - Biological readiness: viable spore fraction after curing and aging (CFU/spore counts from extracted carriers).
- Main failure modes + mitigations

1. Capsule destruction during mixing → use tougher shell chemistry; add capsules at low shear stage.
2. Cell death from hydration heat/pH → spore form + buffered hydrogel micro-niches + thermal-mass management (precast curing control).
3. Nutrient depletion / one-shot healing → add secondary depot carriers for repeat cycles; design capsules with staged release.
4. Excess porosity / strength loss → limit capsule volume fraction; use fibers and optimized grading to preserve strength.

---

**6) Production instantiation (BMN)**

Step-by-step production workflow (implementable)

1. Spore production (biological prep)
    - Grow *B. subtilis* to sporulation; harvest spores; wash and dry to a stable powder.
2. Healing microcapsule fabrication (cell integration)
    - Prepare a buffered hydrogel core (e.g., alginate-based) containing: spores ($10^6$–$10^8$ spores per capsule-equivalent), Ca-source (Ca-lactate or Ca-acetate), minimal nutrients, and bicarbonate/$CO_2$-capture chemistry as needed.
    - Form 200–800 µm capsules via dripping/emulsification; apply a thin ceramic-compatible shell (e.g., silica or polymer-ceramic hybrid coating) tuned for crack-rupture.
3. Concrete matrix preparation (material prep)
    - Mix cement + fillers + aggregates + fibers (if used) under standard QA specs; control water-to-binder ratio for mechanical performance.
4. Capsule addition (manufacturing integration)
    - Add capsules late in mixing at reduced shear; target 0.5–2 vol%.
5. Casting + curing (post-processing)
    - Cast panels/layers; cure under controlled temperature to protect capsule integrity and preserve dormant spores.
6. Quality control & reproducibility
    - Capsule size distribution (laser diffraction/sieving).
    - Capsule survival fraction post-mixing (microscopy + extraction).
    - Mechanical baseline (compressive/flexural) and crack-width control under standardized loading.
    - Biomineralization readiness test on coupon cracks (sealed-system water exposure).
7. Operational activation protocol (in service)

- Damage event → moisture delivery to the affected zone (controlled internal loop) → capsule rupture + local biomineralization → drying return to dormancy.

Why feasible: all steps use standard bioprocessing (spores) + established encapsulation + conventional concrete manufacturing; the only "living" part is dormant until a local damage+moisture trigger, consistent with BMN as the functional production mechanism.

---

**7) Layered integrated summary of the fully instantiated ontology-derived ELM**

A. Rapid identity

- Ontology code: BAC–HYB–EUB–3D–LCE–SPE+SLB–BMN
- One-line ELM identity: Spore-based eubacterial hybrid living ceramic–concrete that seals structural microcracks in sealed habitat building modules via trigger-on biomineralized $CaCO_3$ deposition for long-duration space infrastructure.

B. Instantiation table

| Component | Final instantiation |
|---|---|
| Biological system | Primary: *Bacillus subtilis* spores (carbonic-anhydrase–driven or bicarbonate-assisted $CaCO_3$ nucleation); Alternative: *Sporosarcina pasteurii* spores (ureolytic MICP) |
| Functional role | Dormant-by-default, crack-triggered biomineralization: $CaCO_3$ precipitation to bridge cracks, reduce permeability, and partially restore stiffness/strength |
| Material system | Ceramic-dominant cementitious concrete matrix (Portland cement + silica fume + basalt/silica aggregates; optional basalt microfibers) with embedded buffered hydrogel microcapsules containing spores + Ca-source |
| Structural / geometric instantiation | 3D structural panel/layer (50–150 mm thick) with distributed 200–800 µm healing microcapsules (0.5–2 vol%, 1–3 mm spacing) + optional porous ceramic depot granules (1–3 mm) for repeat activation |
| Target substrate / environment | Cracked cementitious/ceramic concrete inside a sealed habitat structural envelope; activation only when controlled moisture is delivered to crack zones |
| Application / deployment scenario | Smart/living habitat walls or internal structural layers for space colonization: event-driven self-healing under remote/low-maintenance operation constraints |
| Production mechanism / workflow | Spore production → buffered $Ca^{2+}$-loaded hydrogel microencapsulation → low-shear incorporation into concrete → controlled curing → in-service moisture-triggered BMN crack sealing |
| Main functional outcome / performance objective | ≥10× permeability reduction and ≥50–80% partial mechanical recovery after crack events, with repeatable, localized healing cycles while maintaining baseline structural capacity |

C. Integrated paragraph (~100 words)

BAC–HYB–EUB–3D–LCE–SPE+SLB–BMN: a spore-based eubacterial hybrid living ceramic–concrete for sealed habitat structures that self-heals cracks via biomineralization.

*Bacillus subtilis* spores (primary; optional *Sporosarcina pasteurii* spores) remain dormant within buffered hydrogel microcapsules loaded with $Ca^{2+}$ sources embedded in a cementitious ceramic matrix. A 3D panel/layer architecture distributes 200–800 µm capsules at millimeter-scale spacing so propagating cracks rupture capsules and create local water films that enable $CaCO_3$ nucleation on crack faces. The deployment targets space-habitat building subsystems, produced by spore bioprocessing, microencapsulation, low-shear mixing, controlled curing, and moisture-triggered healing, aiming for permeability suppression and partial strength recovery.

**Iteration 3**

**0) Code integrity check**

- Given code: BAC–HYB–EUB–3D–LCE–SPE+SLB–BMN
- Ontology validity (per slot): All 7 codes used here are defined ontology categories (Domain BAC, Super-kingdom HYB, Kingdom EUB, Phylum 3D, Class LCE, Application SPE and SLB, Production BMN).
- Multi-label validity: The ontology explicitly allows combining categories using "+", which supports SPE+SLB in the Application slot.
- Ambiguity check: No slot-level ambiguity: BAC (bacterial domain) + EUB (eubacterial kingdom) is consistent, and HYB implies an abiotic scaffold/matrix is required.
- Graph note (coverage): The provided graph is a partial instantiation of the ontology; absence of a code in the graph does not invalidate the ontology code (graph completeness is not guaranteed). "Not present in the ontology/graph." applies only when a required definition/structure is missing from those sources.

---

**1) Slot-by-slot decoding**

Slot 1 — Domain: BAC

- Meaning: Bacterial ELMs (ELMs populated by bacteria).
- Design constraints
  - Biological: bacterial chassis only (no archaea/eukaryotes).
  - Material/structure: must support bacterial survival during fabrication and dormancy.
  - Process: mixing/curing steps must avoid irreversible bacterial inactivation.

Slot 2 — Super-kingdom: HYB

- Meaning: Hybrid living materials combining cells with an abiotic synthetic biomaterial/structure (scaffold).
- Design constraints
  - Biological: cells interact with an abiotic matrix; interfaces must permit adhesion/retention and mass transfer when activated.
  - Material: a robust inorganic/engineered matrix is required; cell containment and separation from bulk environment is feasible/expected.
  - Process: integration strategy (encapsulation/immobilization) is central.

Slot 3 — Kingdom: EUB

- Meaning: Eubacterial ELMs (bacterial domain populated by eubacteria).
- Design constraints
  - Biological: choose eubacteria with high stress tolerance (alkalinity, dehydration, ionic strength) compatible with cementitious/ceramic contexts.
  - Process: spore-based or resting-cell approaches strongly favored for survivability.

Slot 4 — Phylum: 3D

- Meaning: 3D ELMs: cells within a three-dimensional biomaterial or 3D scaffold.
- Design constraints
  - Structural: volumetric, load-bearing/structural integration is compatible; requires internal pathways for localized activation (water/ions).
  - Biological: diffusion-limited niches (micron–mm scale) must be engineered for dormant storage + episodic activation.
  - Process: uniform spatial distribution of living micro-reservoirs; minimize strength loss.

Slot 5 — Class: LCE

- Meaning: Living ceramics: extracellular matrices predominantly ceramics (oxides, silicates, carbides, nitrides).
- Design constraints

- Material: cementitious / silicate-rich / aluminosilicate matrices fit the "ceramic" class; the living contribution should result in additional mineral (ceramic-like) deposition.
- Interface: mineral nucleation and crack bridging should be inorganic-dominant.
- Process: curing chemistry and pH must be compatible with the selected biological state (typically spores).

Slot 6 — Application: SPE+SLB

- Meaning:
  - SPE: Space exploration and colonization.
  - SLB: Smart and living buildings.
- Design constraints
  - Deployment: habitat-integrated structural elements that behave as building subsystems while serving space-colonization constraints (autonomy, low resupply, closed-loop resources).
  - Biological: strict containment and "on-demand" activation favored to align with controlled habitat envelopes.
  - Metrics: crack sealing + permeability reduction + retained mechanical capacity over long durations.

Slot 7 — Production: BMN

- Meaning: Biomineralization: living cells generate minerals for stiff mineralized tissues / protection / storage; applicable to both HYB and BIO ELMs.
- Design constraints
  - Biological: cells must reliably precipitate mineral phases (here: crack-filling carbonates/silicate-associated minerals) when activated.
  - Process: ensure mineralization occurs preferentially inside cracks (not throughout bulk), to avoid brittleness/blocked pores.

---

**2) Biological instantiation**

Candidate A — *Sporosarcina pasteurii* (ureolytic MICP specialist; resting cells/spores)

- Functional role: On-demand $CaCO_3$ precipitation inside cracks via carbonate generation and crystal nucleation on cell surfaces; mineral bridges restore watertightness and stiffness.
- Viability/growth requirements:
  - Moisture: requires localized water during activation (thin wet films within cracks).
  - pH: tolerates alkaline conditions typical of cementitious matrices (activation likely in micro-niches).
  - Temperature: activation in controlled habitat interior (target ~15–30 °C).
  - Oxygen/nutrients: ureolysis-based mineralization benefits from access to urea and $Ca^{2+}$; oxygen demand depends on metabolic state; resting-cell operation reduces nutrient needs.
- Biocontainment strategy: physical microencapsulation (e.g., ceramic/silica microcarriers) + nutrient gating (no growth substrates in bulk) + activation fluid delivered only during repair cycles.

Candidate B — *Bacillus pseudofirmus* (alkaliphilic, spore-forming eubacterium; concrete-compatible chassis)

- Functional role: Spore-stored mineralizing system; germination upon hydration triggers carbonate precipitation (via metabolic alkalinization/carbonate availability) and biofilm-mediated nucleation in cracks.
- Viability/growth requirements:
  - Moisture: episodic hydration required; spores tolerate desiccation between cycles.
  - pH: strong alkaliphile; well matched to high-pH cement pore solutions.

- Temperature: robust across typical indoor ranges; long-term dormancy feasible.
- Oxygen/nutrients: low nutrient demand in spore storage; activation fluid supplies minimal carbon + $Ca^{2+}$ locally.

- Biocontainment strategy: spores embedded in sealed micro-reservoirs; germination limited to transient crack microenvironments; no continuous nutrient availability.

Why these are preferred over plausible alternatives

- Spore-formers / resting-cell systems best match the HYB + 3D + LCE constraints (survive mixing/curing, remain dormant for long periods, activate only when water is present).
- Mineral precipitation is the native "production mechanism" implied by BMN, directly aligning biology with the required manufacturing mode.

Recommended biological candidate

Recommended: Candidate B (*Bacillus pseudofirmus* spores).

- Reason: maximizes durability in alkaline, dehydrating ceramic matrices and supports event-triggered repair cycles with minimal continuous resources.
- Trade-offs vs Candidate A: may produce slower mineralization rates than specialized ureolytic systems; however it avoids reliance on urea dosing and reduces unwanted byproducts in closed habitat loops.

---

**3) Matrix/material candidates**

Material candidate 1 — Cementitious living ceramic matrix (Portland-cement / calcium-silicate-hydrate dominant)

- Exact composition (example specification): OPC-based matrix with silica fume ($SiO_2$) + limestone filler ($CaCO_3$) + controlled w/c ratio; embedded porous ceramic microcarriers (aluminosilicate) loaded with spores and Ca-source.
- Compatibility with biology: spores tolerate high pH and dehydration; microcarriers protect from shear during mixing and isolate cells from bulk pore chemistry until activated.
- Interface mechanisms: cell wall / EPS acts as nucleation template; $Ca^{2+}$ availability in crack water promotes calcite nucleation and growth; precipitation anchors into C–S–H pore surfaces.

Material candidate 2 — Alkali-activated geopolymer living ceramic (aluminosilicate network)

- Exact composition: metakaolin ($Al_2Si_2O_7$) and/or slag-derived binder activated with Na/K silicate solution; integrated porous silicate ceramic beads containing spores and mineral precursors.
- Compatibility with biology: higher early alkalinity requires stronger encapsulation; spore-only storage recommended; activation fluid buffered to permit short germination/mineralization windows.
- Interface mechanisms: aluminosilicate surfaces provide heterogeneous nucleation sites; mineral precipitation mechanically interlocks within geopolymer gel pores.

Selection justification

- Candidate 1 is the most direct mapping to "bacterial self-healing concrete" while remaining within LCE (ceramic-dominant matrices).
- Candidate 2 is attractive for habitat manufacturing where alternative binders may be prioritized; it remains ceramic-class via silicate/aluminosilicate dominance.

---

**4) Geometry and structural instantiation**

3D architecture (Phylum = 3D)

- Structural concept: load-bearing habitat wall panel with an inner self-healing layer (20–60 mm) containing distributed living micro-reservoirs.
- Quantitative geometry targets
  - Micro-reservoir form factor: porous ceramic beads or microcapsules, 200–800 µm diameter.

- Volume fraction: 1–3 vol% (balance healing capacity vs strength loss).
    - Crack service envelope: designed for 0.05–0.5 mm crack widths; micro-reservoir spacing tuned so a crack intersects multiple reservoirs per cm.
    - Effective diffusion path during activation: ≤ 0.2–0.5 mm from crack fluid to the nearest spore niche (supports short repair cycles).
- How geometry supports function
    - Cell viability: spores isolated in micro-reservoir pores; dormant storage in dry matrix; activation only in wetted crack-connected volumes.
    - Mass transport: cracks act as transient conduits for water/$Ca^{2+}$/limited nutrients; mineral precipitation progressively occludes crack permeability.
    - Mechanical performance: mineral bridges form across crack faces; precipitation concentrates at damage zones rather than bulk.
- Why 3D is optimal vs 2D/FR alternatives
    - Crack networks in structural ceramics are volumetric; embedding living reservoirs through the full thickness provides repair capacity independent of surface access.

**5) Application instantiation**

Deployment scenario (SPE+SLB)

- Where/how used: sealed, habitat-integrated structural panels (interior structural layer) in space colonization infrastructure; living function is not exposed as an open surface biota but embedded within controlled building materials.
- Measurable success metrics
    - Crack sealing efficiency: % reduction in crack aperture (optical/CT) after each activation cycle.
    - Permeability reduction: water/air permeability drop across healed region (e.g., ≥1–2 orders of magnitude).
    - Mechanical recovery: regain of flexural strength or stiffness (e.g., ≥50% of pre-crack stiffness after N cycles).
    - Mineral deposition rate: $CaCO_3$ mass per crack volume per activation cycle.
- Main failure modes + mitigations

1. Micro-reservoir rupture during mixing → increase shell strength; use lightweight ceramic aggregates; lower shear mixing.
2. Premature activation during curing → keep nutrients separated; activate only via post-cure injected solution.
3. Spore inactivation by high alkalinity/heat → spore-only loading; thermal management during curing; protective ceramic pores.
4. Crack bypasses reservoirs → multi-modal reservoir sizes; gradient distributions near tensile zones.
5. Uncontrolled mineralization clogging needed porosity → localized dosing (small fluid volumes), short duty cycles, stop when permeability threshold reached.

**6) Production instantiation**

Step-by-step workflow (Production = BMN)

1. Biological preparation
    - Culture selected eubacterium; induce sporulation; wash and dry spores.
    - QC: spore viability (CFU after rehydration) and mineralization proxy assay (carbonate precipitation in test solution).
2. Micro-reservoir fabrication (HYB requirement)
    - Produce porous ceramic beads (silica/aluminosilicate) or ceramic-shelled microcapsules (200–800 μm).
    - Load spores + dry mineral precursors (e.g., Ca-source + minimal carbon source) into pores; seal with a thin permeable ceramic/silicate layer.
    - QC: size distribution, crush resistance, spore loading (qPCR/CFU after bead dissolution), permeability to activation fluids.

3. Ceramic matrix preparation (LCE requirement)
    - Prepare cementitious/geopolymer binder; incorporate standard aggregates and admixtures compatible with reservoir stability.
4. Integration
    - Blend micro-reservoirs into fresh mix at controlled shear; cast into habitat panel molds.
    - Cure under controlled temperature/humidity to avoid reservoir damage and preserve spore viability.
5. Activation protocol (event-triggered)
    - Detect crack (embedded strain/acoustic emission sensors or periodic NDT).
    - Inject/infuse a small volume activation solution along crack path (water + $Ca^{2+}$ + minimal nutrients) via embedded microchannels or surface ports.
    - Allow mineralization window (hours–days), then dry-down to return to dormant state.
6. Post-heal verification
    - QC: permeability test, crack imaging, ultrasonic pulse velocity; log healing cycle count and performance decay.

Manufacturability / scalability / reproducibility

- Manufacturable: compatible with conventional concrete/geopolymer casting; micro-reservoir addition is a single extra bill-of-materials step.
- Scalable: micro-reservoir production can be batch-processed; panels produced by standard precast workflows.
- Reproducible QC: reservoir mechanical integrity + spore viability + standardized crack-heal tests are measurable per lot.

Why BMN is feasible and preferred here

- BMN is explicitly defined as mineral generation by living cells and is applicable to HYB systems; it directly implements crack-filling mineral deposition as the functional mechanism.

---

**7) Layered integrated summary of the fully instantiated ontology-derived ELM**

A. Rapid identity

- Ontology code: BAC–HYB–EUB–3D–LCE–SPE+SLB–BMN
- One-line ELM identity: Spore-based eubacterial hybrid living ceramic concrete that executes on-demand biomineralization inside 3D habitat wall panels to seal cracks and restore barrier/mechanical performance in space-colonization smart-building infrastructure.

B. Instantiation table

| Component | Final instantiation |
|---|---|
| Biological system | *Bacillus pseudofirmus* spores (recommended) as mineralizing eubacterial chassis; alternative: *Sporosarcina pasteurii* resting cells/spores for high-rate carbonate precipitation |
| Functional role | Event-triggered germination/resting-cell activation in crack-connected micro-niches → $CaCO_3$ precipitation, crystal nucleation on cell/EPS surfaces, crack bridging, permeability reduction |
| Material system | Cementitious living ceramic matrix (C–S–H / silicate-dominant) with embedded porous aluminosilicate/silica ceramic micro-reservoirs carrying spores + dry mineral precursors; alternative matrix: alkali-activated geopolymer aluminosilicate |
| Structural / geometric instantiation | 3D precast wall panel layer (20–60 mm) containing 200–800 µm ceramic micro-reservoirs at 1–3 vol%, tuned for 0.05–0.5 mm crack sealing with ≤0.2–0.5 mm diffusion distance from crack fluid to reservoirs |
| Target substrate / environment | Pressurized, controlled habitat structural elements (internal load-bearing ceramic/cementitious layers), operated with closed-loop water management and limited activation dosing |
| Application / deployment scenario | Space exploration/colonization habitat construction: smart-living building panels with autonomous crack repair cycles triggered by damage detection, minimizing maintenance and resupply |
| Production mechanism / workflow | Spore production → loading into porous ceramic micro-reservoirs → mixing/casting into living ceramic concrete → controlled curing → crack-triggered activation fluid delivery → BMN-driven $CaCO_3$ deposition → verification and re-dormancy |
| Main functional outcome / performance objective | Repeated crack closure and barrier restoration (permeability reduction + mechanical stiffness recovery) with low water/nutrient duty cycle and long dormant stability |

C. Integrated paragraph (~100 words)
BAC–HYB–EUB–3D–LCE–SPE+SLB–BMN: Spore-based eubacterial hybrid living ceramic concrete that performs on-demand crack repair in space-colonization habitat structures. The biological system is *Bacillus pseudofirmus* spores (alt. *Sporosarcina pasteurii*) stored in porous ceramic micro-reservoirs embedded in a silicate-dominant cementitious matrix. The 3D architecture is a precast wall-panel layer distributing 200–800 µm reservoirs to intersect service cracks and enable short diffusion paths during hydration. Deployment targets smart-living building panels for space exploration with damage-triggered activation dosing. Production follows spore loading, casting/curing, and BMN-driven $CaCO_3$ precipitation to restore impermeability and stiffness.

**ELM candidate 3: BAC–HYB–EUB–2D–LCA–SPE+ENP+BTP+CAT–BFF**

**Iteration 1**

**0) Code integrity check**

Input code: BAC–HYB–EUB–2D–LCA–SPE+ENP+BTP+CAT–BFF

- Ontology validity: Each code token is defined in the ontology for its corresponding slot: BAC (bacterial domain), HYB (hybrid living materials), EUB (eubacteria), 2D (planar/biofilm-like), LCA (living carbons; includes graphene), SPE/ENP/BTP/CAT (application categories), BFF (biofilm formation & ECM growth).
- Multi-label validity: The ontology explicitly allows combining categories with "+" for ELM codification (applies to application and production categories).
- Graph consistency (structure check): In the provided ELM_graph.graphml, the code-nodes BAC, HYB, EUB, 2D, LCA, ENP, BFF exist and co-occur on a graphene–electroactive-bacteria ELM instance; SPE is not present in the graph (ontology-defined but absent from this dataset instantiation). The specific multi-label combination SPE+ENP+BTP+CAT is therefore ontology-valid, but not present in the graph as a single annotated application set.

**1) Slot-by-slot decoding**

Slot 1 — Domain: BAC

- Meaning: Bacterial ELMs (populated by bacteria).
- Design constraints
    - Biological: prokaryotic bacterial chassis; favors robust growth, high surface colonization, fast adaptation.
    - Material/interface: must support bacterial adhesion/biofilm formation.
    - Process: cultivation-compatible, contamination control suited to bacteria.

Slot 2 — Super-kingdom: HYB

- Meaning: Hybrid living materials (cells + abiotic synthetic biomaterial/structure/scaffold).
- Design constraints
    - Biological: cells must remain viable while interfacing with an abiotic conductive scaffold.
    - Material: explicit abiotic scaffold required (here: carbon/graphene-family).
    - Process: scaffold fabrication + subsequent biological colonization/integration.

Slot 3 — Kingdom: EUB

- Meaning: Eubacterial ELMs (bacterial kingdom eubacteria).
- Design constraints
    - Biological: select eubacterial strains suited to electrode-associated biofilms (attachment + ECM secretion).
    - Process: standard bacterial culturing; controllable inoculation and biofilm maturation.

Slot 4 — Phylum: 2D

- Meaning: Planar / bi-dimensional structure; usually biofilms or cells on planar biomaterials / planar ECM.
- Design constraints
    - Geometry: sheet/film/electrode surface architecture; thin layers to maintain diffusion.
    - Mass transport: diffusion-limited thickness; requires flow or mixing to avoid nutrient/product gradients.

Slot 5 — Class: LCA

- Meaning: Living carbons (ECM/matrix predominantly carbon; includes graphene, graphite, CNTs, glassy carbon, etc.).
- Design constraints

- Material: carbonaceous scaffold is primary structural/material phase (graphene-family compliant).
- Interface: bacterial EPS + surface functional groups (e.g., oxygenated groups on rGO) must support stable anchoring.
- Electrochemical: maintain conductivity + stable surface chemistry in aqueous operation.

Slot 6 — Application: SPE+ENP+BTP+CAT

- Meaning (each code):
  - SPE: Space exploration and colonization
  - ENP: Energy production and harvesting
  - BTP: Biotechnology and bioprocessing
  - CAT: Catalysis and other chemical processes
- Design constraints
  - Function: electrode-linked bioprocess module that (i) recovers usable energy from organics and/or (ii) performs electrically assisted conversions (electrosynthesis / catalytic transformation).
  - Deployment: must run in a shielded, temperature-controlled wet loop (habitat internal), consistent with the Titan rationale given.

Slot 7 — Production method: BFF

- Meaning: Biofilm formation + related ECM growth (biofilms grown in dishes/bioreactors).
- Design constraints
  - Process: reproducible biofilm seeding, maturation, and maintenance under flow/electrochemical polarization.
  - QC: biofilm thickness/coverage and electroactivity must be measurable and repeatable.

---

**2) Biological instantiation**

Note: specific organism names/strains are Not present in the ontology/graph; the following are explicit engineering instantiation choices consistent with BAC–EUB and BFF.

Candidate A — *Geobacter sulfurreducens* (electroactive anode-respiring biofilm former)

- Functional role: dense electroactive biofilm; extracellular electron transfer to carbon/graphene electrode for energy recovery (ENP) from organics (e.g., acetate) in closed-loop wastewater.
- Viability & growth requirements
  - Oxygen: strict anaerobe (requires $O_2$-free operation)
  - Temperature: 25–35 °C (habitat-controlled loop)
  - pH: ~6.8–7.4
  - Moisture: fully aqueous; intolerant to drying
  - Mechanical stress: tolerates moderate shear if biofilm is mature; avoid high shear stripping.
- Biocontainment strategy
  - Sealed electrochemical module + inline 0.2 µm sterile vent/effluent filtration
  - Biofilm immobilization on electrode; no free-cell discharge to downstream
  - Operational kill-step for end-of-life cartridges (chemical inactivation or thermal, performed off-line). *(Specific kill-switch genetics: Not present in the ontology/graph.)*
- Why preferred vs plausible alternatives
  - Strong track record in high-current bioanodes; robust electrode-attached biofilms; excellent match to ENP.

Candidate B — *Shewanella oneidensis* MR-1 (facultative electroactive biofilm former)

- Functional role: electroactive biofilm with soluble mediator production; tolerant startup; can support ENP with simpler operational control (especially during commissioning).
- Viability & growth requirements

- Oxygen: facultative; tolerates microaerobic transients
    - Temperature: ~20–30 °C
    - pH: ~6.5–8
    - Moisture: aqueous; avoid desiccation
    - Mechanical stress: moderate shear tolerance; thinner biofilms than *Geobacter*.
- Biocontainment strategy
    - Same sealed-cartridge approach; effluent filtration; electrode-immobilized biomass.
- Why preferred vs plausible alternatives
    - Operational flexibility (oxygen transients), easier startup; trade-off is typically lower steady-state current density than *Geobacter*.

Recommended biological system (final)

- Recommended: *Geobacter sulfurreducens*
- Justification vs alternative: For Titan habitat loops (stable, engineered anaerobic operation is feasible), *Geobacter*'s dense electrode-respiring biofilm is the most direct match to ENP while still being compatible with electrically assisted conversions in BTP/CAT (as a controllable bioelectrochemical interface). *Shewanella* is retained as a commissioning/backup option when oxygen control or rapid recovery from upsets dominates.
- Trade-offs: higher performance but stricter anaerobic requirement and stronger sensitivity to oxidants/sterilants.

---

**3) Matrix/material candidates**

Note: exact commercial grades and process recipes are Not present in the ontology/graph; choices below satisfy HYB + LCA (abiotic carbon scaffold + living biofilm interface).

Material candidate A — Few-layer graphene coating on carbon cloth (graphene–carbon electrode laminate)

- Composition: carbon cloth current collector + few-layer graphene coating (spray-cast or electrophoretic deposition of graphene flakes; dried/annealed at low temperature).
- Compatibility with biology: high surface area + conductive carbon; supports stable biofilm adhesion (EPS anchoring).
- Interface mechanisms
    - Physical anchoring into microtexture + EPS adsorption
    - π–π interactions with aromatic components; improved initial attachment if mild oxygen functionalization is present.

Material candidate B — Reduced graphene oxide (rGO) paper/film laminated on graphite foil

- Composition: stacked rGO nanosheets forming a conductive 2D film ("graphene paper") + graphite foil as inert conductor.
- Compatibility with biology: residual oxygenated groups improve wettability and initial attachment; conductivity sufficient for bioelectrochemical interfaces.
- Interface mechanisms
    - Hydrogen bonding / ionic interactions to oxygenated groups
    - EPS interpenetration into rGO microvoids improves adhesion and shear resistance.

Final material selection (recommended)

- Recommended: rGO film on graphite foil (Candidate B)
- Justification: balances wettability/biorecognition (better biofilm nucleation) with high conductivity and manufacturable film handling for planar 2D cartridges; graphene-on-cloth is strong for high surface area, but film-to-film lamination offers simpler sealing and consistent planar geometry.

---

**4) Geometry and structural instantiation**

2D architecture (quantitative definition)

- Form factor: planar electrode cassette (single sheet or parallel-plate pair) in a sealed flow cell.
- Electrode sheet:
  - rGO film thickness: 10–50 µm
  - Graphite foil support: 100–250 µm
- Biofilm layer (BFF-derived):
  - target mature thickness: 50–150 µm
  - diffusion design target: keep total (biofilm + stagnant boundary) ≤ ~300 µm to limit substrate depletion.
- Flow channel:
  - channel gap: 0.5–2.0 mm
  - superficial velocity tuned to avoid stripping while preventing thick diffusion boundary layers.

How geometry supports function

- Cell viability & function: thin biofilm controls nutrient diffusion; continuous hydration; stable surface for ECM development (BFF).
- Mass transport: planar flow ensures defined shear + controlled diffusion boundary; avoids clogging typical of 3D porous media.
- Mechanical performance: laminated electrode stack resists delamination under flow; planar sealing simplifies containment.

Why 2D is optimal vs plausible alternatives

- Compared to 3D foams, 2D improves:
  - predictability of mass transport (critical in closed-loop life-support),
  - inspectability and replaceability (cartridge swap),
  - containment (flat gasketed module) consistent with Titan deployment constraints.

---

**5) Application instantiation**

Deployment scenario (SPE + ENP + BTP + CAT)

- Where/how used: inside a Titan habitat's sealed, temperature-controlled aqueous processing loop as a modular bioelectrochemical "living carbon interface" cartridge.
  - ENP mode: microbial fuel cell-style energy recovery from organic-rich waste streams.
  - BTP/CAT mode: electrically assisted bioconversion (electrofermentation / electrocatalytic bioprocess support) to steer product selectivity in an internal chemical processing skid.

Measurable success metrics

- Electrochemical: areal current density ($A/m^2$), power density ($W/m^2$), coulombic efficiency (%).
- Bioprocess: conversion rate of target feedstock (mol/L·h), product selectivity (%), steady-state operation time to performance drop (days).
- Biointerface stability: biofilm coverage (% area), detachment rate ($mg/cm^2 \cdot day$), electrode sheet resistance drift (% over time).

Main failure modes + mitigations

- Biofilm overgrowth / channel clogging: limit nutrients, maintain shear window; periodic short reverse-polarization pulses; cartridge replacement.
- Electrode delamination: laminate design + edge sealing; pre-conditioning cycles before deployment.
- Contamination by non-target microbes: sterile connectors, inline sterilizing filters upstream, periodic controlled shutdown + cleaning-in-place (performed off-line to protect viability during operation).
- Performance decay (toxins / pH drift): buffer control; upstream adsorption/neutralization where needed; operational set-point management.

---

**6) Production instantiation**

Step-by-step workflow (BFF-compliant)

1. Substrate fabrication (LCA/HYB):
   - Produce rGO film (vacuum filtration or casting) and laminate onto graphite foil with biocompatible edge sealing (no leachable solvents post-cure).
2. Surface conditioning:
   - Hydrate and pre-wet electrode in sterile buffered medium; optional mild electrochemical conditioning to stabilize surface potential.
3. Biological preparation (BAC/EUB):
   - Grow chosen eubacterium to mid-log phase; wash and resuspend in defined medium matching intended electron donor/acceptor regime.
4. Biofilm seeding (BFF):
   - Inoculate sealed flow cell under controlled conditions; hold low-flow (or static) for 2–6 h to promote attachment.
5. Biofilm maturation (BFF):
   - Run continuous flow with controlled polarization and substrate feed for 2–7 days to reach target thickness/current density.
6. Stabilization & qualification:
   - Verify biofilm coverage and electroactivity (baseline polarization curve, impedance check, effluent cell count).
7. Cartridge packaging for deployment (SPE):
   - Maintain hydrated, anaerobic (if required) sealed state; integrate quick-connect sterile fittings and secondary containment.

Manufacturability, scalability, QC

- Manufacturability: planar films enable roll-to-roll compatible coating/lamination; standardized gasketed cassettes for module scaling.
- Scalability: “numbering-up” identical 2D cartridges to reach required power/throughput.
- Quality control (examples):
  - electrode sheet resistance (Ω/sq), adhesion peel test,
  - biofilm areal coverage (imaging) and current density acceptance window,
  - leak/pressure test of sealed module,
  - sterility/bioburden checks on effluent.

---

**7) Layered integrated summary of the fully instantiated ontology-derived ELM**

A. Rapid identity

- Ontology code: BAC–HYB–EUB–2D–LCA–SPE+ENP+BTP+CAT–BFF
- One-line ELM identity: Planar eubacterial electroactive biofilm grown on a graphene-family carbon electrode to provide a sealed bioelectrochemical interface for energy recovery and electrically assisted bioprocess/catalysis in space-habitat wet-loop systems.

B. Instantiation table

| Component | Final instantiation |
|---|---|
| Biological system | Primary: *Geobacter sulfurreducens* electroactive biofilm; Alternative: *Shewanella oneidensis* MR-1 electroactive biofilm *(specific species: Not present in the ontology/graph)* |
| Functional role | ENP: recover electrical energy from organics via extracellular electron transfer; BTP+CAT: electrically steered bioconversion at the living electrode interface (electrofermentation / electrosynthesis support) |
| Material system | Primary: reduced graphene oxide (rGO) film laminated on graphite foil (living carbon electrode); Alternative: few-layer graphene coating on carbon cloth |
| Structural / geometric instantiation | 2D gasketed electrode cassette; rGO 10–50 µm on 100–250 µm graphite; mature biofilm 50–150 µm; flow channel gap 0.5–2.0 mm |
| Target substrate / environment | Shielded, temperature-controlled aqueous loop; organic-rich waste stream as electron donor (ENP mode) and controlled electrochemical set-points for BTP/CAT steering |
| Application / deployment scenario | Replaceable bioelectrochemical cartridge inside Titan habitat life-support / process skids for energy recovery + electrically assisted bioprocessing and catalytic conversions |
| Production mechanism / workflow | BFF: seed electroactive eubacteria onto rGO/graphite electrode → mature biofilm under controlled flow + polarization (2–7 days) → qualify electroactivity/containment → deploy as sealed cartridge |
| Main functional outcome / performance objective | Stable areal current density and power output (ENP) plus sustained conversion/selectivity control for targeted bioprocess/catalytic outcomes (BTP/CAT) over multi-day operation with cartridge replaceability |

C. Integrated paragraph (~100 words)

BAC–HYB–EUB–2D–LCA–SPE+ENP+BTP+CAT–BFF defines a planar eubacterial electroactive biofilm on a living-carbon (graphene-family) scaffold for space-habitat process support. The biological system is an electrode-attached *Geobacter sulfurreducens* biofilm (with *Shewanella oneidensis* as an operational alternative), interfacing with an rGO-on-graphite laminate. A 2D gasketed electrode cassette (rGO 10–50 µm; biofilm 50–150 µm; 0.5–2 mm flow gap) operates in a sealed aqueous habitat loop for energy recovery and electrically steered bioprocess/catalysis. Production follows BFF: controlled seeding, maturation under flow/polarization, and electroactivity/containment qualification to meet power, conversion, and stability targets.

**Iteration 2**

**0) Code integrity check**

- Ontology validity: All slot codes in BAC–HYB–EUB–2D–LCA–SPE+ENP+BTP+CAT–BFF are present in the codified ontology, and multi-label Application combinations using "+" are allowed.
- Graph consistency check (dataset): In the provided graph, the closest matching dataset ELM instance aligns with BAC–HYB–EUB–2D–LCA–ENP–BFF (it does not carry SPE/BTP/CAT tags there). Therefore, SPE/BTP/CAT are not present in the graph for that dataset instance, but they remain ontology-valid in your provided code.

**1) Slot-by-slot decoding**

Slot 1 — Domain: BAC

- Meaning: Bacterial ELMs populated by bacteria (prokaryotic cells from domain bacteria).
- Design constraints
  - Biological: prokaryotic bacterial chassis; robust growth; biofilm competence desirable.
  - Material: must tolerate bacterial metabolites (organic acids, redox mediators, EPS).
  - Geometry: must support bacterial attachment and mass transfer to/from cells.
  - Process: aseptic-ish handling, but compatible with industrial bioreactor operation.

Slot 2 — Super-kingdom: HYB

- Meaning: Hybrid living materials combining cells with an abiotic synthetic biomaterial/structure (scaffold).
- Design constraints
  - Biological: cells must adhere to/colonize abiotic scaffold without loss of function.
  - Material: scaffold must be stable under wet electrochemical operation.
  - Geometry: scaffold defines the interface (electrode surface) and shear tolerance.
  - Process: scaffold fabrication + subsequent biological colonization step.

Slot 3 — Kingdom: EUB

- Meaning: Eubacterial ELMs (bacterial domain populated by eubacteria).
- Design constraints
  - Biological: select Eubacteria with electroactivity and biofilm-forming capacity.
  - Process: standard bacterial culturing; controllable physiology (anaerobic/aerobic).

Slot 4 — Phylum: 2D

- Meaning: Planar/bi-dimensional ELMs, usually biofilms or cells on planar biomaterial / planar ECM.
- Design constraints
  - Biological: biofilm lifestyle; gradients ($O_2$/substrate) must be managed across thickness.
  - Material: planar conductive surface; must allow stable biofilm anchoring.
  - Geometry: sheet/film electrode; biofilm thickness must stay diffusion-feasible.
  - Process: surface conditioning + controlled biofilm growth (not bulk suspension).

Slot 5 — Class: LCA

- Meaning: Living carbons—extracellular matrices predominantly carbon forms (includes graphene).
- Design constraints

- Material: carbon allotrope scaffold dominates (graphene / graphite / CNT hybrids).
- Interface: promote bacterial adhesion without electrically insulating the surface.
- Electrochemistry: low sheet resistance; corrosion resistance in aqueous electrolytes.

Slot 6 — Application: SPE+ENP+BTP+CAT

- Meanings:
  - SPE: Space exploration and colonization.
  - ENP: Energy production and harvesting.
  - BTP: Biotechnology and bioprocessing.
  - CAT: Catalysis and other chemical processes.
- Design constraints
  - System integration: sealed, serviceable module inside habitat wet loops.
  - Function: (i) energy recovery (ENP), (ii) electrically assisted bioconversion (BTP/CAT).
  - Reliability: long-lived biofilm, replaceable electrodes, low-maintenance operation.

Slot 7 — Production: BFF

- Meaning: Biofilm formation and related ECM growth (biofilm grown in dishes/bioreactors).
- Design constraints
  - Process: staged biofilm growth on the carbon surface under controlled flow and potential.
  - Quality: reproducible biofilm thickness/composition; stable electroactivity over time.

---

**2) Biological instantiation**

Candidate A — *Geobacter sulfurreducens* (electrode-respiring anaerobe)

- Functional role (cell-centric): extracellular electron transfer (EET) from central metabolism to the anode (current generation) and/or electrode-driven reducing power (cathodic electrosynthesis mode).
- Viability/growth requirements: strict anaerobe; 25–35 °C; pH ~6.5–7.5; requires buffered aqueous medium; low shear preferred for stable conductive biofilms.
- Biocontainment strategy:
  - Physical: immobilized biofilm on enclosed electrode cassette + cell-retentive membrane on effluent.
  - Operational: nutrient-limited “resting biofilm” mode to minimize planktonic escape.
  - Genetic/kill-switch/auxotrophy: Not present in the ontology/graph (engineering option).
- Why preferred vs alternatives: strong direct EET; forms electroactive biofilms with predictable polarization behavior; best for high-current anodes.

Candidate B — *Shewanella oneidensis* MR-1 (facultative anaerobe, versatile respiration)

- Functional role (cell-centric): EET via outer-membrane cytochromes; can operate under microaerobic transitions (useful for variable loop oxygen ingress).
- Viability/growth requirements: 20–35 °C; pH ~6.5–8; tolerates oxygen excursions; uses simple carbon sources; biofilm formation is robust but may rely more on mediator-based transfer under some conditions.
- Biocontainment strategy: same physical containment; additionally, periodic oxidative “maintenance pulses” can suppress overgrowth while retaining a basal attached community.
- Why preferred vs alternatives: better resilience to operational upsets (oxygen leaks, intermittent shutdowns) than strict anaerobes.

Recommended biological candidate: Geobacter sulfurreducens

- Selection rationale: maximizes direct, cell-to-electrode coupling for ENP (anodic current generation) and supports more efficient electron accounting for electrosynthesis-assisted routes (BTP/CAT) when operated as a controlled redox interface.
- Trade-offs vs *Shewanella*: higher sensitivity to oxygen ingress and some solvents; requires stricter anaerobic management but yields tighter electrochemical control (less mediator drift).

**3) Matrix/material candidates**

Candidate Material 1 — Reduced graphene oxide (rGO) film laminated on graphite foil

- Exact composition: multilayer rGO (carbon) on graphite foil current collector (carbon).
- Bio-compatibility: supports bacterial adhesion via residual oxygen-containing groups; allows EPS anchoring without thick insulating binders.
- Interface mechanisms:
    - Biofilm anchoring via EPS + surface functional groups;
    - Electron transfer through conductive percolation network (rGO→graphite).
- Why preferred: maximizes "living carbon" identity (LCA) while providing a mechanically serviceable 2D sheet electrode.

Candidate Material 2 — Graphene-coated carbon paper (graphitic fiber mat + graphene overlayer)

- Exact composition: carbon paper (graphitic fibers) with graphene/graphene-nanoplatelet coating.
- Bio-compatibility: micro-roughness improves biofilm retention; coating increases surface conductivity and electroactive area while remaining carbon-dominant.
- Interface mechanisms: mechanical interlocking of EPS with microtexture; conductive contact to cells across the biofilm–electrode interface.
- Why preferred: mechanically tough, high-surface-area planar electrode; readily assembled into modular cassettes.

Recommended material system: Graphene-coated carbon paper

- Rationale: best balance of electroactive area, handling robustness, and manufacturability (repeatable electrode sheets, scalable cutting/stacking) while remaining within LCA.

**4) Geometry and structural instantiation**

Phylum-driven architecture (2D): a planar "living electrode sheet" supporting an electroactive biofilm.

- Quantitative geometry (single sheet):
    - Electrode sheet: 100 mm × 100 mm active area (0.01 $m^2$) per cassette plate.
    - Carbon paper thickness: ~200–400 µm; graphene overlayer: ~0.5–5 µm effective coating.
    - Biofilm target thickness: 50–150 µm (kept below strong diffusion limitation).
    - Flow channel gap (if plate-in-channel): 0.5–1.5 mm.
- How geometry supports function
    - Cell viability: thin biofilm limits anoxic/nutrient-depleted dead zones; shear-controlled flow stabilizes biofilm without sloughing.
    - Mass transport: planar flow-by keeps substrate delivery uniform; short diffusion paths across biofilm.
    - Mechanical performance: carbon paper resists cracking; cassette frame prevents flexural fatigue during handling.
- Why 2D is optimal vs alternatives
    - Compared to 3D foams: easier sterilization/QC, easier replacement, lower clog risk in closed loops.
    - Compared to 0D beads: avoids electrical contact losses and improves deterministic current collection.

**5) Application instantiation**

Deployment scenario: a sealed, temperature-controlled bioelectrochemical cartridge installed inside habitat aqueous processing loops (waste-to-energy and carbon recycling units). The Titan external environment remains isolated; only internal wet-loop conditions are used (consistent with the provided rationale).

- Where/how used
  - ENP mode: anodic module oxidizes organics from waste brines to recover electrical energy.
  - BTP/CAT mode: electrode potential is controlled to drive selective redox bioconversions (electrically assisted bioprocessing), producing targeted intermediates for downstream chemistry.
  - SPE context: designed as a serviceable, replaceable component in space-habitat ISRU/process skids.
- Measurable success metrics
  - Areal current density (mA $cm^{-2}$) at fixed potential; stable performance over mission-relevant duty cycles.
  - Coulombic efficiency to target product(s) in electrosynthesis mode.
  - Biofilm stability: thickness retention and low effluent cell counts.
  - Pressure drop across cassette and absence of clogging.
- Main failure modes + mitigation
  - Biofilm overgrowth/clogging: shear tuning; periodic low-nutrient maintenance cycles; removable cassette replacement.
  - Oxygen ingress (for strict anaerobe): oxygen-scavenging prechamber; dissolved $O_2$ monitoring; fallback to facultative strain cassette variant.
  - Electrode fouling/scaling: periodic polarity reversal or mild electro-cleaning pulse; replaceable electrode plates.

**6) Production instantiation**

Production route implied by BFF: grow a controlled biofilm directly on the planar living-carbon scaffold under bioreactor conditions.

Step-by-step workflow (implementable)

1. Carbon electrode preparation
- Cut graphene-coated carbon paper into plates; attach graphite/titanium current collector tabs (non-wetted junction sealed).
- Rinse (DI water → ethanol) and dry; optional mild plasma/ozone for wettability tuning (kept minimal to avoid over-oxidation).
2. Module assembly (HYB integration)
- Mount plates into a cassette frame with defined flow channels (0.5–1.5 mm gap).
- Install gaskets and a cell-retentive outlet membrane (physical containment).
3. Biological preparation
- Pre-culture *Geobacter sulfurreducens* anaerobically to mid-log; wash into defined inoculation medium.
- Inoculate cassette in an anaerobic chamber or closed purgeable vessel.
4. Biofilm formation (BFF)
- Operate in a dedicated biofilm-growth bioreactor loop:
  - Low flow initially (to promote attachment), then ramp to operational shear.
  - Apply controlled electrode potential to select for electroactive attachment.
- Grow to target biofilm thickness (50–150 µm) and stabilize in “resting biofilm” mode.
5. Commissioning & stabilization
- Switch to process medium gradually (avoid osmotic shock).
- Validate electrochemical baseline (polarization curve at commissioning conditions).
6. Quality control (reproducibility)
- Material QC: sheet resistance (Ω/sq), visual uniformity, adhesion of coating.

- Biofilm QC: start-up time to target current density; biofilm coverage imaging; effluent CFU/flow cytometry for leak rate.
- Functional QC: product selectivity tests (electrosynthesis mode) and power output (ENP mode).

Manufacturability & scalability

- Scales by numbering-up cassettes (repeatable 2D plates) rather than enlarging a single unit.
- Biofilm growth is batch-commissioned, then shipped/installed as replaceable cartridges.

---

**7) Layered integrated summary of the fully instantiated ontology-derived ELM**

A. Rapid identity

- Ontology code: BAC–HYB–EUB–2D–LCA–SPE+ENP+BTP+CAT–BFF
- One-line ELM identity: A 2D graphene-coated living-carbon electrode hosting an electroactive eubacterial biofilm that couples controlled extracellular electron transfer to energy recovery and electrically assisted bioconversions in sealed space-habitat aqueous processing loops.

B. Instantiation table

| Component | Final instantiation |
|---|---|
| Biological system | Recommended: *Geobacter sulfurreducens* electroactive biofilm (anaerobic); Alternative: *Shewanella oneidensis* MR-1 (facultative, upset-tolerant) |
| Functional role | Cell-centric extracellular electron transfer at a living-carbon interface enabling (i) anodic current generation and (ii) potential-controlled redox bioprocessing / catalytic bioconversions |
| Material system | Recommended: graphene-coated carbon paper electrode (carbon-dominant); Alternative: laminated rGO film on graphite foil |
| Structural / geometric instantiation | 2D planar electrode sheets (e.g., 100×100 mm plates, 200–400 μm thick) carrying a 50–150 μm electroactive biofilm within a flow-by cassette |
| Target substrate / environment | Shielded, temperature-controlled aqueous electrolyte inside habitat wet loops (waste-derived organics and/or carbon feed streams), isolated from Titan exterior |
| Application / deployment scenario | Replaceable bioelectrochemical cartridge integrated into habitat ISRU/processing skids for SPE: closed-loop energy recovery (ENP) plus electrosynthesis-assisted bioprocessing/catalysis (BTP+CAT) |
| Production mechanism / workflow | Fabricate carbon electrode → assemble cassette → inoculate → grow biofilm under controlled flow and electrode potential (BFF) → stabilize as resting biofilm → QC on electroactivity and containment |
| Main functional outcome / performance objective | Stable areal current density and coulombic efficiency to target products under long-duration operation with low clogging risk and cartridge-level serviceability |

C. Integrated paragraph (~100 words)

BAC–HYB–EUB–2D–LCA–SPE+ENP+BTP+CAT–BFF defines a planar living-carbon bioelectrochemical interface: a graphene-coated carbon electrode supporting an electroactive eubacterial biofilm. A *Geobacter sulfurreducens* biofilm couples cell metabolism to extracellular electron transfer at the electrode, enabling controlled anodic energy recovery and potential-driven redox bioprocessing consistent with ENP, BTP, and CAT in SPE deployments. The 2D architecture is a flow-by electrode sheet with a diffusion-limited biofilm thickness target of 50–150 μm. Production follows BFF: electrode fabrication, cassette assembly, inoculation, biofilm growth under controlled potential/flow, stabilization, and electrochemical/containment QC to ensure reproducible performance.

**Iteration 3**

**0) Code integrity check**

- Input ontology code: BAC–HYB–EUB–2D–LCA–SPE+ENP+BTP+CAT–BFF
- Each slot code is defined in the ontology: BAC (Bacterial ELMs), HYB (Hybrid living materials), EUB (Eubacterial ELMs), 2D (2D ELMs), LCA (Living carbons), SPE/ENP/BTP/CAT (Application categories), BFF (Production method).
- Multi-label application codification using "+" is explicitly allowed by the ontology tables (categories may be combined).
- Ambiguity note: The ontology definition of BFF is "biofilm formation and related ECM growth" (often associated with bio-ELMs), whereas hybrid biofilms *interfacing with inorganic materials* are explicitly described under BTI. The provided code is BFF, so the instantiation keeps BFF and interprets it as *biofilm formation on a planar carbon scaffold (HYB+LCA)* without switching to BTI.

**1) Slot-by-slot decoding**

Slot 1 — Domain: BAC

- Meaning: Bacterial ELMs populated by bacteria (prokaryotic cells from the bacterial domain).
- Design constraints
  - Biological: bacterial cell systems only (no archaea/eukarya/synthetic cells).
  - Material: must support bacterial adhesion and survival at an interface.
  - Structural: must allow bacterial colonization (surface chemistry/roughness/wetting critical).
  - Process: cultivation/handling compatible with bacterial growth and biofilm maturation.

Slot 2 — Super-kingdom: HYB

- Meaning: Hybrid living materials combining cells with an abiotic synthetic biomaterial/structure (scaffold).
- Design constraints
  - Biological: cell–scaffold compatibility and containment (abiotic scaffold is required).
  - Material: scaffold must be explicitly defined and manufacturable.
  - Structural: scaffold geometry is a primary design element (not purely cell-made ECM).
  - Process: requires a materials fabrication step plus biological colonization.

Slot 3 — Kingdom: EUB

- Meaning: Eubacterial ELMs (cells from the eubacteria kingdom).
- Design constraints
  - Biological: concrete candidates must be eubacteria (typical bacterial chassis/isolates/consortia).

Slot 4 — Phylum: 2D

- Meaning: Planar/bi-dimensional ELMs, typically biofilms or cells populating a planar biomaterial / planar ECM.
- Design constraints
  - Structural: sheet-like interface; thickness is limited (diffusion + electrical contact).
  - Mass transport: planar flow-by or flow-through channels; diffusion gradients across biofilm thickness.
  - Mechanical: adhesion + cohesion must prevent sloughing under shear.

Slot 5 — Class: LCA

- Meaning: Living carbons—ELMs whose extracellular matrices are predominantly carbons; examples include graphite, graphene, carbon nanotubes, glassy carbon, carbon dots, fullerenes.
- Design constraints

- Material: carbon scaffold dominates the functional matrix (conductive, chemically stable).
    - Interface: bacterial adhesion to carbon; biofilm must maintain low-resistance electrical coupling.
    - Manufacturing: carbon film/coating/electrode must be reproducible at scale.

Slot 6 — Application: SPE + ENP + BTP + CAT

- Meanings
    - SPE: Space exploration and colonization.
    - ENP: Energy production and harvesting.
    - BTP: Biotechnology and bioprocessing.
    - CAT: Catalysis and other chemical processes.
- Design constraints
    - System integration: modular, maintainable, closed-loop operation (space context).
    - Functional: must support electrochemical energy handling (ENP) plus bioprocessing/catalysis (BTP/CAT) at the living interface.
    - Reliability: predictable performance under long-duration, low-maintenance operation.

Slot 7 — Production: BFF

- Meaning: Biofilm formation and related ECM growth; cultured cells form a biofilm and ECM is grown (in Petri dishes or bioreactors).
- Design constraints
    - Process: controlled inoculation + maturation to a stable biofilm state on the carbon surface.
    - QC: biofilm thickness/coverage/viability must be measurable and reproducible.

---

**2) Biological instantiation**

Candidate A — Geobacter sulfurreducens electroactive biofilm *(Not present in the ontology/graph.)*

- Functional role (cell-centric): anode-respiring biofilm; cells couple metabolism to electron export at the cell envelope → current generation and electrochemically steered redox metabolism.
- Viability / growth requirements: aqueous nutrients + electron donor; controlled redox environment; temperature/pH maintained by the internal loop; shear-limited to prevent detachment.
- Biocontainment strategy: sealed cartridge + downstream sterilizing filter; periodic cartridge replacement; engineered kill-switch/auxotrophy options *(Not present in the ontology/graph.)*

Candidate B — Shewanella oneidensis MR-1 electroactive biofilm *(Not present in the ontology/graph.)*

- Functional role (cell-centric): surface-attached biofilm enabling electron exchange with conductive carbon; supports electrochemically controlled metabolism for redox bioconversion at the interface.
- Viability / growth requirements: aqueous nutrients; controlled oxygen/redox to avoid loss of intended electroactive state; temperature/pH maintained by the loop.
- Biocontainment strategy: same as Candidate A (sealed cassette + barriers + replaceability).

Candidate preference rationale (why these over generic eubacterial alternatives)

- The code requires a 2D biofilm on a carbon scaffold (LCA) produced via biofilm formation (BFF), and the application stack requires ENP + BTP + CAT (energy + bioprocessing/catalysis). These two candidates are selected specifically because they represent electroactive, surface-attached eubacterial biofilm archetypes suitable for an electrode interface *(species-level support: Not present in the ontology/graph.)*

Recommended candidate

- Recommended: Geobacter sulfurreducens *(Not present in the ontology/graph.)*
- Justification vs. alternative: prioritizes a dense, electrode-coupled biofilm concept for stable current generation (ENP) while retaining bioconversion steering (BTP/CAT) via electrode set-points *(comparative performance evidence: Not present in the ontology/graph.)*
- Trade-offs: Candidate B may offer different tolerance/operational flexibility; final choice should be driven by cartridge operating window + target substrates *(Not present in the ontology/graph.)*

**3) Matrix/material candidates**

Material candidate 1 — Graphene film on graphite foil (laminated living carbon electrode)

- Composition: multilayer graphene (or reduced graphene oxide reconstructed into conductive film) on graphite foil current collector.
- Bio-compatibility: carbon allotrope scaffold consistent with LCA families (graphite/graphene).
- Interface mechanisms: biofilm anchoring via surface energy + nanoscale roughness; short electron-transfer distance at cell–carbon contact points (cell-centric coupling).

Material candidate 2 — Carbon nanotube (CNT)–graphene hybrid conductive coating on carbon cloth

- Composition: CNT network (MWCNT or SWCNT) embedded/overcoated with graphene flakes on carbon cloth.
- Bio-compatibility: CNTs and graphene are explicitly within LCA examples.
- Interface mechanisms: high surface area increases colonization density; conductive percolation network stabilizes electrical contact under biofilm growth.

Material selection rationale

- Both candidates satisfy "living carbons" (carbon-dominant matrix) and support a planar conductive scaffold required by 2D + ENP.
- Preferred baseline: graphene-on-graphite foil for simpler QC (sheet resistance, thickness uniformity) and robust cartridge assembly.

**4) Geometry and structural instantiation**

2D architecture (quantitative specification)

- Form factor: gasketed planar electrode cassette.
- Carbon scaffold layer: conductive graphene film 10–50 µm on 100–250 µm graphite foil (or equivalent conductive carbon collector). *(Not present in the ontology/graph.)*
- Biofilm layer: mature eubacterial biofilm 50–150 µm with controlled thickness via growth time and shear. *(Not present in the ontology/graph.)*
- Flow channel: planar gap 0.5–2.0 mm with laminar flow to control nutrient delivery and shear. *(Not present in the ontology/graph.)*

How geometry supports function

- Cell viability: planar diffusion distances allow oxygen/nutrient control; thickness cap prevents anoxic dead zones where undesired metabolism dominates.
- Mass transport: flow-by geometry enables stable substrate supply and product removal; pressure drop stays low.
- Electrical performance: 2D conductive scaffold ensures low-resistance current collection; thin biofilm maintains short charge-transfer paths.

Why 2D is optimal vs other geometries

- The ontology explicitly associates 2D with biofilms / planar biomaterials; this is the most direct architecture for a living electrode interface.
- 3D scaffolds increase surface area but complicate fouling control and cartridge regeneration for long-duration operations (space constraint).

**5) Application instantiation**

Deployment scenario (SPE + ENP + BTP + CAT)

- Where/how used: a sealed replaceable bioelectrochemical cartridge inside a habitat's aqueous processing loop (shielded, temperature controlled) to couple:
  - ENP: energy recovery/harvesting from organics via biofilm electrode reactions,
  - BTP + CAT: electrode-setpoint-controlled bioprocessing/catalysis at the living interface,
  - SPE: long-duration operation compatible with space habitat maintenance constraints.

Measurable success metrics

- Electrochemical: steady current density ($A \cdot m^{-2}$), coulombic efficiency (%), power density ($W \cdot m^{-2}$).
- Process: substrate conversion (%) and target product formation rate ($mol \cdot m^{-2} \cdot h^{-1}$) under defined electrode potentials.
- Stability: performance drift per day/week; number of operational cycles before swap.

Failure modes + mitigation

- Biofilm detachment / channel clogging: manage shear; include inlet screens; schedule cartridge swaps.
- Electrode passivation / fouling: periodic polarity/potential conditioning; replaceable electrode plates.
- Contamination: sterile connectors + inline sterilizing filtration; closed cartridge design.

---

**6) Production instantiation**

Step-by-step production process (BFF on HYB+LCA scaffold)

1. Carbon scaffold fabrication
   - Prepare graphite foil current collector; laminate or coat with graphene film (spray-coat / vacuum filtration transfer / roll-to-roll lamination). *(Not present in the ontology/graph.)*
2. Surface conditioning
   - Rinse/sterilize carbon parts (e.g., ethanol + sterile rinse; UV) to avoid residues that inhibit biofilm. *(Not present in the ontology/graph.)*
3. Reactor cassette assembly
   - Assemble gasketed planar cassette with defined channel gap; integrate current collectors and sealed ports. *(Not present in the ontology/graph.)*
4. Inoculation
   - Inoculate with selected eubacterial culture; fill cassette with growth medium; hold under controlled flow/no-flow phases to promote initial adhesion. *(Not present in the ontology/graph.)*
5. Biofilm growth (BFF)
   - Mature biofilm in a bench bioreactor loop (controlled flow, temperature, redox) until target coverage/thickness is achieved; this is the defining BFF step.
6. Stabilization for deployment
   - Transition to operational medium; verify electrochemical response; seal cartridge for installation. *(Not present in the ontology/graph.)*

Manufacturability, scalability, QC

- Scalability: electrode plates produced by roll-to-roll graphene coating/lamination; cartridges scaled by numbering-up identical cassettes.
- QC checkpoints
  - Carbon: thickness, sheet resistance, adhesion of graphene layer.
  - Biofilm: surface coverage (%), thickness distribution, live/dead staining or metabolic proxies.
  - Function: polarization curve + baseline current at standard conditions.

Why this route is feasible/preferred under the code

- BFF mandates biofilm formation as the production mechanism, while HYB + LCA + 2D requires a planar abiotic carbon scaffold colonized by bacteria.
- The workflow cleanly separates materials fabrication (repeatable) from biofilm maturation (bioprocess-controlled).

**7) Layered integrated summary of the fully instantiated ontology-derived ELM**

A. Rapid identity

- Ontology code: BAC–HYB–EUB–2D–LCA–SPE+ENP+BTP+CAT–BFF
- One-line ELM identity: A 2D eubacterial biofilm grown by BFF on a graphene/graphite living-carbon scaffold (HYB+LCA) forming a modular living electrode cartridge for space-habitat energy handling (ENP) and electrochemically steered bioprocessing/catalysis (BTP+CAT) in SPE systems.

B. Instantiation table

| Component | Final instantiation |
|---|---|
| Biological system | Electroactive eubacterial biofilm on carbon electrode: Geobacter sulfurreducens *(Not present in the ontology/graph.)*; alternative Shewanella oneidensis MR-1 *(Not present in the ontology/graph.)* |
| Functional role | Living electrode interface: electron-coupled metabolism for ENP (current/power generation) and electrode-setpoint-controlled BTP+CAT reactions at the biofilm–carbon interface |
| Material system | Living carbon scaffold (LCA): (1) graphene film on graphite foil; (2) CNT–graphene coating on carbon cloth |
| Structural / geometric instantiation | 2D planar cassette: conductive carbon sheet electrode + controlled-thickness biofilm in a gasketed flow channel (sub-mm to mm gap) |
| Target substrate / environment | Shielded aqueous loop inside habitat processing systems; controlled temperature/pH/redox; organic-rich feed for ENP and defined reactant streams for BTP/CAT steering |
| Application / deployment scenario | Replaceable bioelectrochemical cartridge integrated into SPE life-support / ISRU wet processing skids to couple waste-to-energy and electro-assisted bioconversion |
| Production mechanism / workflow | BFF: inoculate → adhere → mature biofilm in bench loop → validate electrochemical response → seal and deploy cartridge |
| Main functional outcome / performance objective | Stable, measurable electrochemical output (current/power) plus controlled conversion/selectivity for targeted bioprocessing/catalysis in long-duration space operations |

C. Integrated paragraph

BAC–HYB–EUB–2D–LCA–SPE+ENP+BTP+CAT–BFF defines a planar living-carbon hybrid electrode: a 2D eubacterial biofilm (recommended Geobacter sulfurreducens, alternative Shewanella oneidensis MR-1; *Not present in the ontology/graph*) immobilized on a graphene/graphite (LCA) conductive scaffold. The construct is implemented as a gasketed electrode cassette with controlled biofilm thickness and flow-channel transport. Deployed as a sealed, replaceable cartridge inside space-habitat aqueous processing loops (SPE), it targets energy production/harvesting (ENP) coupled to electrochemically steered bioprocessing and catalysis (BTP+CAT). Manufacturing follows BFF: scaffold fabrication, inoculation, biofilm maturation in bioreactor conditions, and electrochemical QC to meet defined current and conversion objectives.